\documentclass[11pt]{article}
\usepackage[utf8]{inputenc}
\usepackage[letterpaper, portrait, margin=1in]{geometry}
\usepackage{amsmath,amsfonts,amssymb,amsthm,mathtools,booktabs,enumitem,float,tabularx,longtable}
\usepackage[dvipsnames]{xcolor}
\usepackage{pgfplots}
\pgfplotsset{compat=1.18}
\newcommand{\compact}{2} %
\usepackage[T1]{fontenc}
\ifcase\compact\relax
  \usepackage{mathpazo}                 %
\or
  \usepackage{newtxtext,newtxmath}      %
\or
  \usepackage[tt=false]{libertine}      %
  \usepackage[libertine]{newtxmath}
  \usepackage[varqu]{zi4}
\else
  \usepackage{mathpazo}                 %
\fi
\usepackage{etoolbox}
\usepackage{hyperref}
\usepackage{authblk}
\makeatletter
\AddToHookNext{cmd/@maketitle/before}{%
  \def\@makefnmark{\hbox{\@textsuperscript{\normalfont\@thefnmark}}}}
\makeatother
\hypersetup{colorlinks=true,linkcolor=blue,citecolor=teal}
\newtheorem{theorem}{Theorem}[section]
\newtheorem{lemma}[theorem]{Lemma}
\newtheorem{proposition}[theorem]{Proposition}
\newtheorem{corollary}[theorem]{Corollary}
\newtheorem{conjecture}[theorem]{Conjecture}
\newtheorem{result}{Result}
\newtheorem{question}{Question}
\newtheorem{maintheorem}{Main Theorem}
\theoremstyle{definition}
\newtheorem{definition}[theorem]{Definition}
\newtheorem{hypothesis}[theorem]{Hypothesis}
\theoremstyle{remark}
\newtheorem{remark}[theorem]{Remark}

\def\ket#1{\mathinner{|{#1}\rangle}}
\def\bra#1{\mathinner{\langle{#1}|}}
\DeclareMathOperator{\tr}{tr}
\DeclareMathOperator{\poly}{poly}
\DeclareMathOperator{\rank}{rank}
\newcommand{\Z}{\mathbb Z}
\newcommand{\E}{\mathbb E}
\newcommand{\unif}{\mathsf u}

\newcount\Comments
\newcommand{\kibitz}[2]{\ifnum\Comments=1\textcolor{#1}{#2}\fi}

\title{No Exponential Quantum Advantage for Decoded Quantum Interferometry with Local Quantum Decoders}
\author{Brandon Augustino}
\author{Shouvanik Chakrabarti\thanks{\texttt{shouvanik.chakrabarti@jpmchase.com}}}
\author{Enrico Fontana}
\author{Dylan Herman}
\author{Guneykan Ozgul}
\author{Ruslan Shaydulin\thanks{\texttt{ruslan.shaydulin@jpmchase.com}}}
\author{Shreehari Sureshbabu}
\author{Jacob Watkins}
\affil{Global Technology Applied Research, JPMorganChase, New York, NY, 10001}
\date{}

\makeatletter
\patchcmd{\l@section}{\addvspace{1.0em \@plus\p@}}{\addvspace{0.3em \@plus\p@}}{}{}
\makeatother

\begin{document}
\maketitle
{\renewcommand{\thefootnote}{}%
 \footnotetext{\copyright~2026 JPMorgan Chase \& Co. All rights reserved.}}

\begin{abstract}
Decoded quantum interferometry (DQI) reduces optimization of sparse linear constraint problems, and sampling from their Gibbs measures, to decoding a linear code from a quantum state. We study DQI on random $D$-regular $k$-sparse instances with $k\ge3$, for which local decoders such as belief propagation with quantum messages (BPQM) are commonly used. Such decoders estimate a typical codeword symbol using only the qudits within logarithmic distance of it. Our main insight is that the thresholds of the decoding problem induced by DQI correspond to phase transitions of the Gibbs measure that are well studied in statistical physics: tree reconstruction for local decoders and, for Max-$k$-XORSAT, condensation for the optimal decoder. This insight allows us to prove two main results. First, we prove an information-theoretic obstruction: no local decoder can recover the codeword with high probability above the tree reconstruction threshold. Conversely, we show that a decoder based on BPQM reaches this threshold. Since the cavity method predicts that local Markov chains equilibrate rapidly up to the same threshold, we expect DQI with local decoders to give no exponential speedup for sampling and no advantage in approximate optimization with amplitude functions that depend only on the number of violated constraints. Second, we consider the optimal decoder (the pretty-good measurement, for which no efficient implementation is known) and show that it succeeds exactly below the condensation threshold. The condensation threshold thus provides a tight upper bound on the performance of DQI with these amplitude functions, providing a strict refinement of the limit that Jordan et al.\ derived from the Holevo bound. Our results point to efficient nonlocal decoders that succeed between the two thresholds as the only viable path to a quantum advantage with DQI on these instances.
\end{abstract}

\clearpage
{\small\tableofcontents}
\clearpage

\section{Introduction}\label{sec:intro}

Decoded quantum interferometry (DQI) was introduced by Jordan et al.~\cite{jordan2024optimization} as a quantum algorithm for approximate optimization of problems whose objective is a sum of functions of linear forms of the candidate solution. We consider problems of the form
\begin{equation}\label{eq:problem}
  \min_{z\in\Z_q^n}H(z),\qquad H(z)=\sum_{j=1}^m f\big((Bz)_j+x_{0,j}\big),
\end{equation}
where $B\in\{0,1\}^{m\times n}$ is viewed as a matrix over $\Z_q$, $x_0\in\Z_q^m$ is an offset, and $f:\Z_q\to\mathbb R$ is a cost function. We study random $(k,D)$-regular instances, in which $B$ has $k$ ones in each row and $D$ ones in each column for fixed $2\le k<D$, so that each constraint contains $k$ variables and each variable lies in $D$ constraints. Problem~\eqref{eq:problem} is then an instance of Max-$k$-LINSAT, and for $q=2$ and $f(z)=z$ it is Max-$k$-XORSAT, in which $H(z)$ is the number of violated equations $(Bz)_j=x_{0,j}$. We draw $B$ from one of two ensembles (Section~\ref{sec:typical}): the configuration model, which is standard in coding theory and in which $B$ is uniformly random among such matrices, and Gallager's ensemble, from which the benchmark families of Shutty et al.~\cite{shutty2026locallyquantumdecoders} are drawn. We call~\eqref{eq:problem} the primal problem, to distinguish it from the decoding problem that DQI induces, and we write $\mu_\beta(z)\propto e^{-\beta H(z)}$ for its Gibbs measure at inverse temperature $\beta\ge0$.

DQI finds approximate solutions of~\eqref{eq:problem} by preparing and measuring the ``target state'', a superposition of candidate solutions in which the amplitude of $z$ is proportional to $\Phi(Bz+x_0)$ for a function $\Phi$ on $\Z_q^m$, which we call the ``amplitude function''. To obtain the target state, DQI prepares the state of $m$ qudits with amplitude $\Phi(x)$ at each $x\in\Z_q^m$, together with a uniform superposition of the codewords $c$ of the linear code $C$ with parity-check matrix $B^\top$, and multiplies the amplitude of each pair $(c,x)$ by $e^{2\pi i\,c\cdot(x-x_0)/q}$. Once the register that holds $c$ has been uncomputed, these phases interfere destructively at every $x$ except the vectors $Bz+x_0$, which leaves the target state. Uncomputing $c$ is a decoding problem for $C$: a decoder must recover $c$ from the qudits that hold $x$, whose state in the Fourier basis is $c$ subjected to a coherent superposition of errors, with a distribution that depends on the amplitude function. We say that the decoder succeeds if it recovers a uniformly random codeword with probability tending to one as $m\to\infty$. The performance of DQI therefore depends on two choices: the amplitude function and the decoder.

The amplitude function determines the target state, and therefore the output of DQI when the decoder succeeds. Many amplitude functions that have been used with DQI are symmetric, that is, $\Phi(x)$ depends on $x$ only through the total cost $\sum_jf(x_j)$, which equals $H(z)$ for $x=Bz+x_0$. Our main results concern such symmetric amplitude functions. Examples include the polynomial amplitude functions of Jordan et al.\, which are polynomials of degree $\rho m$ in the number of satisfied constraints, where $\rho$ is the error rate. For Max-XORSAT, with the best such polynomial and a decoder that corrects every error of weight at most $\rho m$, where $\rho\le\tfrac12$, DQI reaches the expected satisfied fraction $\tfrac12+\sqrt{\rho(1-\rho)}$ as $m\to\infty$, a relation that Jordan et al.\ call the semicircle law. Zhou et al.~\cite{zhou2026gibbs} and Shutty et al.\ instead use the product amplitude function $\Phi_\beta(x)=\prod_j\sqrt{p_\beta(x_j)}$, where $p_\beta(z)\propto e^{-\beta f(z)}$ is the Gibbs weight of one constraint. With this amplitude function and a decoder that succeeds, DQI samples from a law close to $\mu_\beta$, and for Max-XORSAT it reaches the same satisfied fraction as the best polynomial at the same error rate. 

At a given error rate, optimizing the symmetric amplitude function for the satisfied fraction, as done by Jordan et al., therefore does not improve on the product amplitude function. This suggests that DQI can be analyzed as a Gibbs sampler also for approximate optimization: since the error rate increases with $\beta$, the threshold of the decoder would then determine both the largest inverse temperature at which DQI samples and the largest satisfied fraction that it reaches. This does not simply follow from the above observation however, because amplitude functions with the same error rate induce different decoding problems. In this paper, we make precise the link between Gibbs sampling and approximate optimization with DQI.

The decoding problem induced by DQI is a problem of quantum state discrimination, so the decoder may be quantum. A classical decoder measures each symbol in a fixed basis and uses only the resulting classical channel~\cite{jordan2024optimization,zhou2026gibbs}, so its threshold is at most the Shannon limit of this channel. A quantum decoder may measure several symbols jointly, and its threshold can exceed this limit, so quantum decoders offer a way to improve the performance of DQI. A commonly studied example is belief propagation with quantum messages (BPQM)~\cite{renes2017bpqm}, which passes quantum states along the factor graph of the code (Section~\ref{sec:bpqm}). BPQM performs well in numerical studies of DQI~\cite{mandal2026blockerror}.

A central question is whether DQI gives a quantum advantage, that is, whether it outperforms classical algorithms such as Glauber dynamics, simulated annealing, and Prange's algorithm. Previous results on random sparse instances show a curious pattern: for each efficient quantum decoder that has been analyzed, a classical algorithm that processes the instance in a similar way reaches about the same values as DQI with this decoder. For example, Mandal et al.~\cite{mandal2026blockerror} observe numerically that DQI with BPQM, which estimates each symbol from its neighborhood in the factor graph, reaches nearly the same satisfied fraction as simulated annealing on the benchmark families of Shutty et al. Decoders that apply Gaussian elimination to the whole code can exceed the satisfied fraction of simulated annealing, but they follow the same pattern~\cite{shutty2026locallyquantumdecoders,zhou2026gibbs}. Since local Markov chains are predicted to sample efficiently from the primal Gibbs measure exactly up to one of its phase transitions~\cite{mezard2006reconstruction}, these results point to a correspondence between the thresholds of the decoding problem induced by DQI and thresholds of the primal Gibbs measure.

Our main insight is that the thresholds of the decoding problem induced by DQI correspond to phase transitions of the Gibbs measure that are well studied in statistical physics. We consider these thresholds for two classes of decoders. We call a decoder \emph{local} if its estimate of a codeword symbol is the outcome of a quantum measurement on the qudits within distance $2c_0\log m$ in the factor graph, for $c_0>0$ a small enough constant that this neighborhood typically is a tree. The measurement may be collective and may have any complexity. Our precise definition (Section~\ref{sec:ov_setting}) requires this only at more than logarithmically many symbols with disjoint neighborhoods, so a local decoder may use all the qudits to estimate the other symbols.
In contrast, a decoder whose estimate at every symbol depends on the whole code, such as unambiguous state discrimination followed by Gaussian elimination~\cite{chailloux2024quantumdecoding}, is \emph{nonlocal}. 
Similarly, we call Glauber dynamics and simulated annealing local classical algorithms, because each of their steps updates one variable according to its constraints. Prange's algorithm and its variants, which solve linear systems over the whole instance, are nonlocal classical algorithms.

This insight motivates the main question of this paper, which concerns the potential for an advantage with DQI using local decoders.
\begin{question}\label{q:main}
Consider random $(k,D)$-regular instances in the limit where the number $m$ of constraints tends to infinity, and local decoders that recover the codeword with probability tending to one. What is the largest inverse temperature at which DQI with such a decoder samples from the Gibbs measure? What is the largest expected satisfied fraction that DQI reaches for Max-XORSAT with such a decoder? Do local classical algorithms, such as Glauber dynamics and simulated annealing, reach the same values?
\end{question}

Question~\ref{q:main} has a counterpart for arbitrary decoders, local or nonlocal, for which the only known limit was the Holevo bound~\cite{jordan2024optimization}: no decoder succeeds above an inverse temperature $\beta_{\mathrm H}$ determined by the rate of the code, and DQI reaches at most the corresponding satisfied fraction $s_H$. For uniformly random linear codes, whose parity-check matrices are dense, this limit is attained by the optimal decoder, the pretty-good measurement (PGM)~\cite{chailloux2024quantumdecoding,blanvillain2026tight}, which is nonlocal and for which no efficient implementation is known in general. For sparse instances, we ask the following question.

\begin{question}\label{q:any}
Consider the same instances and limit, and decoders, local or nonlocal, that recover the codeword with probability tending to one. What is the largest inverse temperature at which DQI with such a decoder samples from the Gibbs measure? What is the largest expected satisfied fraction that DQI reaches for Max-XORSAT with such a decoder?
\end{question}

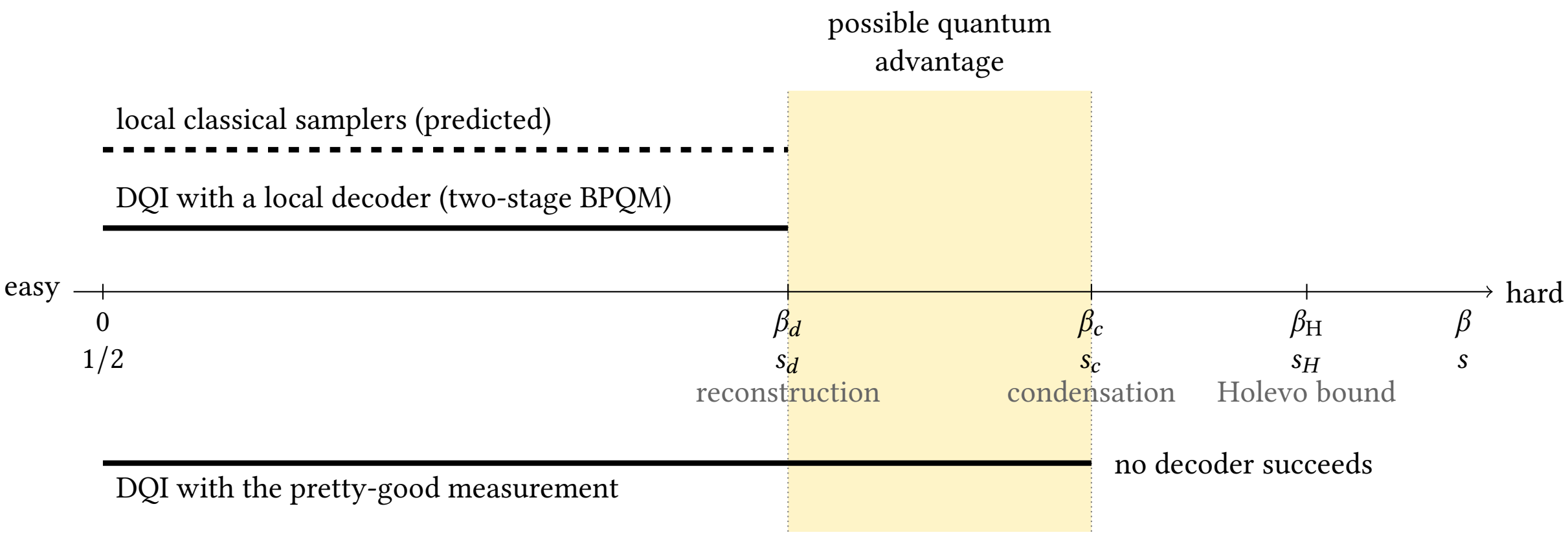
\begin{figure}[t]
\centering
\begin{tikzpicture}[x=1cm,y=1cm,font=\footnotesize]
\fill[Goldenrod!30] (7.3,-2.45) rectangle (10.4,2.05);
\node[align=center,anchor=south] at (8.85,2.05) {possible quantum\\advantage};
\draw[black!50,densely dotted] (7.3,-2.45) -- (7.3,2.05) (10.4,-2.45) -- (10.4,2.05);
\draw[->] (0,0) -- (14.5,0);
\node[left] at (0,0) {easy};
\node[right] at (14.5,0) {hard};
\foreach \x in {0.3,7.3,10.4,12.6}
  \draw (\x,-0.08) -- (\x,0.08);
\foreach \x/\b/\s/\n in {0.3/{$0$}/{$1/2$}/{}, 7.3/{$\beta_d$}/{$s_d$}/{reconstruction}, 10.4/{$\beta_c$}/{$s_c$}/{condensation}, 12.6/{$\beta_{\mathrm H}$}/{$s_H$}/{Holevo bound}, 14.2/{$\beta$}/{$s$}/{}} {
  \node[anchor=base] at (\x,-0.4) {\b};
  \node[anchor=base] at (\x,-0.78) {\s};
  \node[anchor=base,black!60] at (\x,-1.12) {\n};
}
\draw[line width=1.6pt,dashed] (0.3,1.45) -- (7.3,1.45);
\node[above right] at (0.3,1.45) {local classical samplers (predicted)};
\draw[line width=1.6pt] (0.3,0.65) -- (7.3,0.65);
\node[above right] at (0.3,0.65) {DQI with a local decoder (two-stage BPQM)};
\draw[line width=1.6pt] (0.3,-1.75) -- (10.4,-1.75);
\node[below right] at (0.3,-1.75) {DQI with the pretty-good measurement};
\node[right] at (10.5,-1.75) {no decoder succeeds};
\end{tikzpicture}
\caption{Thresholds for Max-XORSAT on random instances from the configuration model with $3\le k<D$ (not to scale). The inverse temperature $\beta$ increases from left to right, and so does the satisfied fraction $s=\tfrac12+\tfrac12\tanh(\beta/2)$ that DQI reaches with the product amplitude function at $\beta$. Solid lines show what is proved (Main Theorems~\ref{thm:main_local} and~\ref{thm:main_pgm}), and the dashed line shows the prediction of the cavity method (Conjecture~\ref{conj:classical} below $\beta_d$). If this prediction holds, DQI can give a quantum advantage in Gibbs sampling, or in approximate optimization with symmetric amplitude functions, only with an efficient nonlocal decoder that succeeds in the shaded range. For approximate optimization, DQI must then also exceed the satisfied fractions of classical algorithms such as simulated annealing and variants of Prange's algorithm, which can lie above $s_d$ (Figure~\ref{fig:fractions}).}
\label{fig:thresholds}
\end{figure}

\paragraph{Our results.} We answer Questions~\ref{q:main} and~\ref{q:any} by proving our main insight for both classes of decoders. In both cases the correspondence takes the same form: some decoder of the class succeeds exactly when an inference problem for the primal Gibbs measure is information-theoretically impossible. For local decoders, this problem is \emph{tree reconstruction}~\cite{mezard2006reconstruction}, which asks whether distant variables carry information about the root variable on the infinite tree with the same degrees, so the threshold of local decoders is the reconstruction threshold $\beta_d$ (Main Theorem~\ref{thm:main_local}). The reason is that, on a tree, the overlap between the two possible states of a BPQM message and the bias of a message of belief propagation for the primal Gibbs measure evolve by recursions with the same law, once the roles of the two kinds of nodes are exchanged. The states that a local decoder must distinguish therefore become orthogonal exactly when distant variables carry no information about the root (Section~\ref{sec:ov_duality}).

For arbitrary decoders and Max-XORSAT, the inference problem is to distinguish right-hand sides generated from a planted assignment, by violating each equation independently with probability $p_\beta(1)$, from uniformly random ones. Decoding and this inference problem are related because the block success of the PGM, an optimal decoder, equals the squared Bhattacharyya coefficient between the two laws of the right-hand side; the PGM therefore succeeds exactly when the two laws are close in total variation (Section~\ref{sec:ov_condensation}). On the configuration model with $3\le k<D$, the two laws become distinguishable exactly above the \emph{condensation} threshold $\beta_c$~\cite{cojaoghlan2018information,cojaoghlan2021inference}, so the PGM succeeds below $\beta_c$ and no decoder succeeds above it (Main Theorem~\ref{thm:main_pgm}). Unlike for uniformly random linear codes, this threshold is strictly smaller than the limit $\beta_{\mathrm H}$ given by the Holevo bound (Proposition~\ref{prop:condensation_holevo}). Figure~\ref{fig:thresholds} shows these thresholds and the ranges in which the decoders succeed.

A consequence of the correspondence we prove is that DQI with local decoders samples from the Gibbs measure only up to $\beta_d$. For $k\ge3$, a two-stage decoder based on BPQM~\cite{mandal2026blockerror} reaches this threshold. The reconstruction threshold is also the transition up to which the cavity method predicts that local Markov chains sample efficiently~\cite{mezard2006reconstruction}, and we state its algorithmic consequence as Conjecture~\ref{conj:classical}. Under this conjecture, DQI with local decoders gives no exponential speedup for Gibbs sampling, and our numerical results provide strong evidence for the conjecture.

For approximate optimization, DQI may use amplitude functions whose target laws are not Gibbs measures. For $k\ge3$, however, the expected cost per constraint that DQI reaches with local decoders and any symmetric amplitude function is asymptotically at least the Gibbs average of the cost at the reconstruction threshold, for every cost that is integer-valued up to an affine rescaling. This class includes the costs of unweighted Max-LINSAT, which take two values. With product amplitude functions, DQI approaches this bound, so optimizing a symmetric amplitude function does not improve the asymptotic expected cost. Under Conjecture~\ref{conj:classical}, classical algorithms approach the same bound. In our simulations, simulated annealing even does better than this bound on every degree pair and alphabet that we tested, which supports the absence of an advantage in approximate optimization with local decoders on these random instances.

Exact optimization depends on the probability of measuring an optimal assignment, rather than on the expected cost. Quantum minimum finding~\cite{boyer1998tight,durr1996minimum} could yield a superquadratic speedup if DQI substantially increased this probability relative to efficiently sampleable classical distributions. We provide partial evidence against this possibility. Under Conjecture~\ref{conj:classical}, the speedup is at most quadratic for Gibbs states below $\beta_d$ (Corollary~\ref{cor:quadratic}). For binary Max-XORSAT, we extend this comparison to target states of polynomial amplitude functions~\cite{jordan2024optimization} whose degree corresponds to a fixed inverse temperature below $\beta_d$.

\paragraph{Limitations.} The general approximation bounds assume an offset-independent symmetric amplitude function and reliable block decoding. For the amplitude function of Shutty et al.~\cite{shutty2026locallyquantumdecoders} that is a product over disjoint blocks of constraints, which is not symmetric, the bound $s_d$ is proved on four degree pairs for every decoder (Proposition~\ref{prop:block_four}) and conjectured for local decoders on the other pairs (Conjecture~\ref{conj:block}). Decoders with constant rather than vanishing failure probability can still be used with postselection; this is a separate setting from reliable decoding (Section~\ref{sec:scope}).

The exact-search comparison concerns ideal target states. Exact decoding prepares these states, but success tending to one alone does not control the probability of an exponentially rare optimum in the actual output. Extending the comparison to imperfectly prepared states requires additional control at that probability scale (Section~\ref{sec:prepared}).

\paragraph{Related work.} The binary case of our correspondence for local decoders is closely related to Renes' duality between channels and codes~\cite{renes2018duality}. Piveteau, Quek, and Franco Garrido~\cite{piveteau2026gibbs} estimated the threshold of BPQM for Gibbs-state preparation by density evolution, and the values that they report agree with our reconstruction thresholds (Table~\ref{tab:pqf}). Anschuetz, Gamarnik, and Lu~\cite{anschuetz2025decoded} bound the satisfied fraction that DQI reaches through the overlap gap property, for algorithms that are stable under small changes of the instance. Their results complement ours, which rest on locality and reconstruction. Kramer et al.~\cite{kramer2026provable} prove a separation in query complexity between DQI with a nonlocal decoder and classical algorithms with oracle access, and Gil-Fuster et al.~\cite{gilfuster2026approximate} compare DQI numerically with a classical Markov chain. A detailed comparison is given in Section~\ref{sec:related}.

\paragraph{Concurrent work.} During the preparation of this manuscript we became aware that the central part of our first main result, in particular the correspondence between DQI with BPQM and the reconstruction threshold, was independently and simultaneously obtained by Jordan, Nguyen, and Zhou~\cite{JNZ26}. We point readers to that upcoming manuscript for a more complete perspective on their results.

\section{Overview of results}\label{sec:overview}

This section states our results and outlines their proofs. In Section~\ref{sec:ov_main}, we state the main results informally and discuss their consequences. In Sections~\ref{sec:ov_setting}--\ref{sec:ov_condensation}, we state them precisely and outline their proofs. We present the arguments for Max-XORSAT, except that Section~\ref{sec:ov_approx} states the approximation bounds for all the costs that they cover, and in Section~\ref{sec:ov_alphabets} we explain what else extends to larger alphabets and other costs. Table~\ref{tab:results} lists the results with the problems and ensembles for which they are proved.

\subsection{Main results}\label{sec:ov_main}

We characterize the inverse temperatures at which DQI can sample from the primal Gibbs measure, first with local decoders and then without a locality restriction. The local-decoding result holds for both the configuration model and Gallager's ensemble. For Max-XORSAT from the configuration model, we also determine the threshold of the optimal decoder, which need not be efficient.

\paragraph{Local decoders and tree reconstruction.} Our first result relates local decoding to reconstruction for the primal Gibbs measure on the infinite regular tree~\cite{mezard2006reconstruction}. Reconstruction means that distant variables retain information about the value at the root, even as their distance from it tends to infinity. We write $\beta_d$ for the supremum of the inverse temperatures without reconstruction.

\begin{maintheorem}[informal version of Corollary~\ref{cor:typical_nogo}, Theorem~\ref{thm:only_nonrecon}, Corollary~\ref{cor:first_local_nonrecon}, and Proposition~\ref{prop:sharp}]\label{thm:main_local}
Fix an alphabet $\Z_q$, a cost $f$, and degrees $2\le k<D$. Let $B$ be a random $(k,D)$-regular instance from the configuration model or Gallager's ensemble, and consider the decoding problem induced by the product amplitude function at a fixed inverse temperature $\beta$.

For a suitable constant $c_0>0$, the largest probability of recovering a typical codeword symbol from the qudits within distance $2c_0\log m$ tends to one if and only if the primal Gibbs measure on the infinite regular tree has no reconstruction at $\beta$. BPQM attains this optimal local success probability.

No local decoder recovers the whole codeword with probability tending to one at an inverse temperature with reconstruction. For $k\ge3$ and every finite inverse temperature without reconstruction, a two-stage decoder based on BPQM~\cite{mandal2026blockerror} does recover the whole codeword with probability tending to one.
\end{maintheorem}

The obstruction in Main Theorem~\ref{thm:main_local} is information-theoretic: it concerns the information available in the qudits within distance $2c_0\log m$ of a typical symbol in the factor graph of the code, rather than the particular choice of quantum operations used to process them. Above $\beta_d$, the error probability at this symbol remains bounded away from zero even for an optimal collective measurement, so changing the local BPQM operations cannot remove the obstruction.

The obstruction applies as soon as the decoder retains one local estimate in its output, since a wrong retained estimate makes the recovered codeword wrong. It therefore also applies to implementations of BPQM that estimate the symbols one at a time and retain their first estimate, which is computed from the original input, even when later steps use the whole code (Section~\ref{sec:ov_setting}). We explain the correspondence with reconstruction in Section~\ref{sec:ov_duality} and extend it from trees to random instances in Section~\ref{sec:ov_locality}.

\paragraph{Approximate optimization.} For $k\ge3$, the reconstruction threshold also determines the smallest expected cost per constraint attainable with local decoding and a symmetric amplitude function. This holds for unweighted Max-LINSAT and, more generally, for any fixed cost that is integer-valued up to an affine rescaling. With any symmetric amplitude function chosen independently of the offset, the expected cost per constraint, averaged over a uniformly random offset, is at least its single-constraint Gibbs average at $\beta_d$, up to $o(1)$. For Max-XORSAT, this is equivalent to an upper bound on the expected satisfied fraction of
$$
  s_d+o(1),\qquad s_d=\frac12+\frac12\tanh\!\left(\frac{\beta_d}{2}\right).
$$
Product amplitude functions and the two-stage BPQM decoder approach these bounds. Thus the freedom to choose a symmetric amplitude function does not improve the asymptotic expected cost. To establish this, we show that the reduced state on a neighborhood of a codeword symbol is close to a mixture of product states, to which the local-decoding bound can be applied separately (Section~\ref{sec:ov_approx}).

The comparison with classical algorithms depends on Conjecture~\ref{conj:classical}, which predicts efficient classical Gibbs sampling at every fixed inverse temperature without reconstruction. Under this conjecture, DQI with local decoders gives no exponential speedup for Gibbs sampling at a fixed inverse temperature. Classical sampling also approaches the same expected-cost bound, including $s_d$ for Max-XORSAT. Together with our numerical results, this provides evidence against an exponential speedup for approximate optimization with symmetric amplitude functions and reliable local decoders, including BPQM with at most $c_0\log m$ iterations, on these random instances.

Quantum decoding does improve DQI: with BPQM, DQI reaches a larger satisfied fraction than with classical belief propagation on every benchmark family, and on some families a larger one than any classical decoder permits (Table~\ref{tab:benchmark}). This improvement ends at $s_d$, however, which local classical algorithms are predicted to reach without solving a decoding problem. We discuss the evidence for classical sampling in Section~\ref{sec:ov_sampling} and compare optimization performance in Section~\ref{sec:ov_approx}.

Exact optimization requires a different comparison, because distributions with similar expected costs can assign very different probabilities to an optimal assignment. Under Conjecture~\ref{conj:classical}, quantum minimum finding on ideal Gibbs states below $\beta_d$ provides at most a quadratic speedup over classical sampling. For binary Max-XORSAT, the comparison holds for every complex polynomial target state whose degree corresponds to a fixed inverse temperature below $\beta_d$ (Corollary~\ref{cor:poly_quadratic}). The comparisons, including the distinction between ideal target states and states prepared with imperfect decoding, are discussed in Section~\ref{sec:ov_exact}.

\paragraph{Optimal decoding and condensation.} Our second result determines the limit when the decoder is unrestricted. For Max-XORSAT from the configuration model, this limit is the condensation threshold $\beta_c$ of the primal Gibbs measure. It is also the threshold for distinguishing planted instances from instances with uniformly random right-hand sides~\cite{cojaoghlan2018information,cojaoghlan2021inference}. We relate this inference problem to optimal decoding in Section~\ref{sec:ov_condensation}.

\begin{maintheorem}[informal version of Corollary~\ref{cor:two_thresholds}]\label{thm:main_pgm}
Consider Max-XORSAT from the configuration model with $3\le k<D$, and the decoding problem induced by the product amplitude function at a fixed inverse temperature $\beta$.

If $\beta<\beta_c$, the pretty-good measurement, which is an optimal decoder, recovers the whole codeword with probability tending to one. If $\beta>\beta_c$, every decoder has exponentially small block success with high probability over the instance.
\end{maintheorem}

Main Theorem~\ref{thm:main_pgm} gives a tight limit for optimal decoding and refines the upper bound derived from the Holevo bound~\cite{jordan2024optimization}. The three thresholds satisfy
$$
  \beta_d\le\beta_c<\beta_{\mathrm H},
$$
and we certify $\beta_d<\beta_c$ on each of the fifteen benchmark families. For symmetric amplitude functions and any decoder whose block success tends to one, the corresponding upper bound on the expected satisfied fraction is
$$
  s_c=\frac12+\frac12\tanh\!\left(\frac{\beta_c}{2}\right).
$$
Product amplitude functions and the pretty-good measurement approach this bound, so optimizing over symmetric amplitude functions does not improve it.

Reaching $s_c$ efficiently would require more than the existence of an optimal measurement. A coherent implementation of the pretty-good measurement and its inverse is equivalent, up to polynomial overhead, to coherent preparation of the primal Gibbs states for all right-hand sides below $\beta_c$ (Corollary~\ref{cor:pgm_equiv}). Under a prediction of the cavity method, efficient implementation near $\beta_c$ would also yield an efficient quantum algorithm for weak recovery in regular noisy XOR systems, meaning recovery of an assignment with nonvanishing correlation with the planted assignment. The analogous problem with uniformly random equations has no known polynomial-time algorithm and is conjectured hard; the distinction between these ensembles is discussed in Section~\ref{sec:window}.

\paragraph{Possible quantum advantage.} Under Conjecture~\ref{conj:classical}, obtaining an advantage in approximate optimization with symmetric amplitude functions therefore requires an efficient nonlocal decoder that succeeds between $\beta_d$ and $\beta_c$. Such a decoder would have to improve on nonlocal classical algorithms as well, including Prange's algorithm and Turbo Prange. The two thresholds identify the range in which an advantage remains possible, while the classical comparisons determine the performance that DQI would have to exceed.

\subsection{The decoding problem and local decoders}\label{sec:ov_setting}

The decoding problem induced by DQI is determined by the amplitude function and the linear code with parity-check matrix $B^\top$. We describe the connection here for Max-XORSAT; the changes for larger alphabets and other costs are collected in Section~\ref{sec:ov_alphabets}.

\paragraph{The decoding task.} For an assignment $z\in\Z_2^n$, the corresponding violation pattern is $x=Bz+x_0$: its $j$-th entry is one precisely when constraint $j$ is violated. The possible patterns form the coset $x_0+K$, where $K=\operatorname{im}B$. DQI aims to prepare the normalized projection of the state $\ket\Phi=\sum_x\Phi(x)\ket x$ onto this coset. Measuring the target state produces a violation pattern $x$, from which an assignment is obtained by solving $Bz=x+x_0$.

DQI prepares $\ket\Phi$ together with a uniform superposition of the codewords of $C=K^\perp$, and applies the phases $(-1)^{c\cdot(x-x_0)}$. For each codeword $c$, the state in the received register is, up to a global phase,
$$
  \ket{\psi^\Phi_c}=\sum_{x\in\Z_2^m}\Phi(x)(-1)^{c\cdot x}\ket x.
$$
The codeword register is then uncomputed by recovering its label from this state. Thus the decoder must distinguish the states $\ket{\psi^\Phi_c}$ when $c$ is uniformly distributed on $C$. We write $1-\varepsilon$ for its block success, the probability of recovering the whole codeword.

For the product amplitude function $\Phi_\beta$, the received state is a product over the symbols. Each bit $y$ is transmitted through the pure-state channel
$$
  y\longmapsto\sqrt{1-\alpha}\,\ket0+(-1)^y\sqrt\alpha\,\ket1,\qquad\alpha=p_\beta(1)=\frac1{1+e^\beta}.
$$
Measuring the qubits individually in the Fourier basis instead produces independent bit flips with probability $\alpha^\perp=\tfrac12-\sqrt{\alpha(1-\alpha)}$, which is also the error rate of the decoding problem. Quantum decoding retains the pure channel outputs rather than first converting them into these classical outcomes.

\paragraph{From decoding to the objective.} Decoding success controls the distribution of the samples produced by DQI and their expected objective value. If $\Phi$ has positive mass on every coset of $K$, the output distribution is within $\sqrt\varepsilon$ in expected total variation of the target distribution, averaged over a uniformly random offset. For every amplitude function and every objective with values in $[0,1]$, its expectation under the output distribution differs by at most $2\sqrt\varepsilon$ from its expectation under $|\Phi(x)|^2$ on all of $\Z_2^m$. Shutty et al.\ proved these guarantees over finite fields~\cite{shutty2026locallyquantumdecoders}, and Theorem~\ref{thm:guarantee} gives the form we use over $\Z_q$. The latter guarantee allows us to study the expected cost in $\ket\Phi$ without analyzing its separately normalized projection onto each coset.

\paragraph{Local estimates.} Let $N_t(i)$ be the symbols within distance $2t$ of symbol $i$ in the factor graph of the code. An estimate of $c_i$ is $t$-local if it is the outcome of a quantum measurement on the received qubits of $N_t(i)$ (Definition~\ref{def:local}). The measurement may be collective and may use ancillas. Locality refers to the qubits used to form the estimate, rather than to how many qubits each individual operation acts on or to a tree representation of the computation.

Fix a positive constant $c_0$ with
$$
  c_0<\frac{1}{2\log((k-1)(D-1))}.
$$
For $t\le c_0\log m$, with probability tending to one, the ball around a typical symbol of either random ensemble is a tree, and the checks outside it impose no additional constraint on the codeword values inside. These are the local-tree condition $\mathrm{LT}_t(i)$ and the tree-projection condition $\mathrm{TP}_t(i)$, proved in Section~\ref{sec:ov_locality}. The choice of $c_0$, which depends only on the degrees, ensures that both conditions hold with this probability throughout the radius range.

A single local estimate already limits the whole decoder. If a decoder outputs a $t$-local estimate at a symbol whose neighborhood satisfies both conditions, then its block success is at most the probability that this estimate is correct, whatever the decoder does at the other symbols, and Sections~\ref{sec:ov_duality} and~\ref{sec:ov_locality} bound this probability. The same holds when the symbol is chosen with randomness that does not depend on the input, and when the estimate is computed from the original input and retained in the output while later steps use all the received qubits (Section~\ref{sec:strategy}). Our bounds for local decoders use only such an estimate.

\paragraph{Local decoders.} Our definition of a local decoder (Definition~\ref{def:local_decoder}) requires $t$-local estimates, with $t\le c_0\log m$, jointly at a set of at least $s_m$ symbols, where $s_m/\log m\to\infty$, whose neighborhoods are pairwise disjoint and satisfy both local conditions. Given the codeword, the estimates must have the joint law of separate POVMs on these neighborhoods; the received qubits may themselves be entangled. The decoder may first make a random choice that does not depend on its input, and it must make these estimates except for choices of total probability at most $2^{-s_m}$. At every other symbol, it may use all the received qubits. We call decoders that do not satisfy this definition nonlocal.

Requiring more than logarithmically many estimates keeps the definition meaningful for decoders that are used with postselection (Section~\ref{sec:scope}). Replacing the outputs of any decoder at $s$ symbols by independent uniformly random values makes these estimates trivially local. For a single symbol, this reduces the block success by at most a factor $q$, a loss that repeated attempts with postselection recover. For $s/\log m\to\infty$, it reduces the block success below every inverse polynomial (Section~\ref{sec:strategy}).

\paragraph{BPQM variants and locality.} BPQM was introduced by Renes~\cite{renes2017bpqm}. Subsequent work developed sequential receivers~\cite{rengaswamy2021bpqm}, efficient coherent message-passing and finite-precision implementations~\cite{piveteau2022messagepassing}, paired-measurement BPQM~\cite{brandsen2022bpqm}, and approximate cloning for factor graphs with cycles~\cite{piveteau2022messagepassing}. Extensions to prime alphabets and finite abelian groups were introduced in~\cite{mandal2026qary,mandal2026abelian} and are discussed in Section~\ref{sec:ov_alphabets}. For calls with $t\le c_0\log m$ on our random instances, the definition of locality above and the bounds on block success cover most BPQM variants studied in the literature under the operational conditions described below.

A BPQM call with $t$ iterations, which we call BPQM of depth $t$, estimates a symbol from channel outputs within distance $2t$ of that symbol. The bound applies to every quantum measurement on these inputs, so changing the operations within the neighborhood cannot improve on it. For example, approximate-cloning variants create approximate copies of a received qubit for use in several branches of the decoding computation~\cite{piveteau2022messagepassing}. These copies are created from qubits already inside the neighborhood and supply no information from outside it. The BPQM operations and their implementations are described in Section~\ref{sec:bpqm}.

Sequential BPQM estimates the codeword symbols one at a time~\cite{rengaswamy2021bpqm,piveteau2022messagepassing}. For each symbol, it runs the BPQM circuit on that symbol's neighborhood, records the estimate, and reverses the node unitaries before moving to the next symbol. This reversal does not undo the disturbance caused by recording the estimate. Later calls may therefore act on qubits already affected by earlier calls, so their estimates need not be measurements of the original input restricted to their own neighborhoods.

For a uniformly random processing order, the joint-locality requirement can be established directly. Suppose a set of symbols of linear size is determined by the instance before the order is drawn. Each call uses a fixed local instrument and fresh independent ancillas, does not depend on estimates at other symbols, and retains its recorded estimate on every branch. Among the symbols near the beginning of the order, more than logarithmically many have suitable neighborhoods untouched by any earlier call. Their neighborhoods are disjoint, so their estimates have exactly the required joint measurement law. With high probability over the instance, and with the exceptional probability permitted by Definition~\ref{def:local_decoder}, this schedule is local. No reordering of the decoder is needed.

For a coordinate order fixed independently of the instance, and for an information set selected greedily in such an order, the first processed symbol has a suitable neighborhood with probability tending to one. Its estimate is computed from the original input, so if the output retains it, the block success is at most the probability that it is correct, even when later calls overlap or depend on earlier outcomes and classical completion uses the whole code. A fixed default output on failure contributes only an exponentially small correction. These implementations therefore satisfy the decoding and approximation bounds even when their membership in Definition~\ref{def:local_decoder} has not been established.

The random-instance guarantees used above concern balls of graph radius at most $2c_0\log m$ around the symbol being estimated, so they do not cover BPQM of unrestricted depth (Section~\ref{sec:ov_locality}). Likewise, a wrong first estimate implies block failure only if the decoder does not correct it later. Finally, quantum operations on few qudits followed by classical computation over the whole code need not give local estimates. Shutty et al.~\cite{shutty2026locallyquantumdecoders} apply fine-grained unambiguous measurements (FGUM)~\cite{buzet2025fine} to small blocks of qudits and then Gaussian elimination to the whole code. They call such decoders locally-quantum, but each resulting estimate depends on the whole code, so these decoders are nonlocal in our sense.

\subsection{Optimal local decoding and tree reconstruction}\label{sec:ov_duality}

We first solve the decoding problem for the root of a tree code. Optimizing over all quantum measurements, rather than a prescribed BPQM circuit, is important: the resulting success probability will bound every local estimate whose input has the same tree law. Section~\ref{sec:ov_locality} establishes when this is true for a neighborhood of a finite random instance.

In the $(k,D)$ tree of depth $t$, the root symbol lies in $k$ checks, each check contains $D-1$ further symbols, and each non-root symbol lies in $k-1$ further checks. A uniformly random codeword of the tree code is sent through the channel. We write $P_t$ for the largest probability, over all quantum measurements, of recovering the root value. To make the definition of reconstruction in Section~\ref{sec:ov_main} precise, we exchange the roles of the two node types: checks become primal variables, each in $D$ clauses, and symbols become primal clauses, each with $k$ variables. We root this tree at a variable and draw the variables from the primal Gibbs measure at inverse temperature $\beta$. The primal model has non-reconstruction if, as $t\to\infty$, the posterior law of the root variable given the values of the variables at distance $t$ from the root tends to the uniform law in probability, and reconstruction otherwise (Section~\ref{sec:primal_recon}).

\begin{result}[Optimal tree decoding and tree reconstruction]\label{res:duality}
For every $\beta\ge0$, depth-$t$ BPQM attains $P_t$ (Proposition~\ref{prop:tree_optimal}). Moreover, $P_t\to1$ as $t\to\infty$ if and only if the primal model has non-reconstruction (Theorem~\ref{thm:duality}).
\end{result}

Result~\ref{res:duality} holds at every tree depth. For Max-XORSAT, non-reconstruction holds on an interval starting at zero (Corollary~\ref{cor:binary}). Above $\beta_d$, the optimal success is at most $P_\infty<1$ even as $t\to\infty$, so increasing the tree depth cannot make root decoding reliable.

The optimal measurement can be constructed by combining messages from the leaves towards the root. For binary channels, a message consists of one of two pure states, according to the symbol value, and their overlap determines how well the states can be distinguished. The received qubit has overlap $\sigma_\beta=\tanh(\beta/2)$. At a symbol node, the BPQM operation combines two messages about the same bit into one message with overlap $ab$, without losing information. At a check node, the outgoing message concerns the parity of the incoming bits. BPQM records a label whose law is independent of the codeword: with probability $(1+ab)/2$ the outgoing overlap is $(a+b)/(1+ab)$, and otherwise it is $(b-a)/(1-ab)$.

Conditional on these labels, the root qubit is in one of two pure states, according to the root value, and the states have overlap $U_t$. Their Helstrom measurement has success $\tfrac12(1+\sqrt{1-U_t^2})$. Since each preceding operation preserves the information about the root, its average over the labels is $P_t$, the optimum for the original tree input. The recursion for the random overlaps is the density evolution of BPQM.

The primal recursion has the same law when posterior messages are represented by their biases, the differences between their probabilities of zero and one. At a primal clause, the message to a variable has bias $\sigma_\beta$ times the product of the biases of the other variables. This is the multiplication of overlaps at a code symbol node.

At a primal variable, two messages with bias magnitudes $a$ and $b$ combine to $(a+b)/(1+ab)$ when they favor the same value, and to $|b-a|/(1-ab)$ otherwise. By the Nishimori property, conditional on its bias magnitude $a$, a message favors the true value with probability $(1+a)/2$. Hence independent messages favor the same value with probability $(1+ab)/2$, exactly the probability of the BPQM label at a code check node. The two recursions agree up to signs. Consequently, the quantum overlap tends to zero precisely when the primal posterior bias tends to zero, which is non-reconstruction. This is the tree form of the channel/code correspondence discussed in Section~\ref{sec:related}. To use it on finite instances, we must next establish that the neighborhood supplies the same tree decoding input.

\subsection{From trees to random instances}\label{sec:ov_locality}

The tree computation alone does not determine the input seen by a local decoder on a finite graph. Even when the neighborhood has no cycle, checks outside it can restrict the allowed codeword values inside. We therefore require both $\mathrm{LT}_t(i)$, which says the radius-$2t$ ball is a tree, and $\mathrm{TP}_t(i)$, which says the code projects onto all assignments satisfying the internal checks. Restriction then sends a uniformly random codeword to a uniformly random codeword of the tree code, so the decoder receives exactly the input of Result~\ref{res:duality}.

\begin{result}[Locality bound]\label{res:locality}
If $\mathrm{LT}_t(i)$ and $\mathrm{TP}_t(i)$ hold, every decoder that makes a $t$-local estimate at $i$ has symbol error at least $1-P_t$ and block success at most $P_t$ (Theorem~\ref{thm:locality}); depth-$t$ BPQM attains the optimal symbol success $P_t$ (Corollary~\ref{cor:typical_nogo}). For random $(k,D)$-regular instances with $k<D$ and $D\ge3$, from the configuration model or Gallager's ensemble, fix $0<c_0<1/(2\log((k-1)(D-1)))$. For $t\le c_0\log m$, both conditions hold at a uniformly random symbol with probability $1-o(1)$ (Theorem~\ref{thm:tp_typical}).
\end{result}

The projection proof starts from the vectors of $K$ supported inside the ball. An additional such vector would impose an additional parity constraint there. Lemma~\ref{lem:tp} characterizes the absence of these constraints, and Lemma~\ref{lem:tp_kernel} reduces it to the kernel of the matrix outside the ball. Some kernel vectors are restrictions of global kernel vectors, such as the all-ones vector for even $k$ in the binary configuration model. They impose no new constraint inside the ball and must not be excluded by the counting argument.

We show that additional kernel vectors are unlikely by counting them in two ranges. A small support would require an unusually poorly expanding set: every constrained symbol meeting it must meet at least two nonzero entries. For a large support, assign values to the external checks. At each symbol their sum must be zero, so one value is determined by the others. This reduces the entropy by at least one coordinate per symbol. Because there are more symbols than checks when $k<D$, the reduction outweighs the number of possible check assignments. Boundary effects are sublinear in the prescribed radius range, and a first-moment argument then proves tree projection in both ensembles.

The conditional bound in Result~\ref{res:locality} applies at any radius satisfying LT and TP. The logarithmic restriction is used to guarantee those conditions on typical finite neighborhoods. It does not restrict the all-depth tree correspondence, and going beyond it does not establish successful decoding above reconstruction. Whether sufficiently deep BPQM can exploit the additional constraints of a finite cyclic graph remains a separate question.

\subsection{Sampling}\label{sec:ov_sampling}

We can now use the tree correspondence and its transfer to random instances to study the limits of DQI for sampling and optimization. Gibbs sampling is the natural starting point, since with the product amplitude function $\Phi_\beta$ the distribution of the target state is exactly $\mu_\beta$.

The sampling result is sharp. For $k\ge3$ and both ensembles, a reliable local decoder permits DQI to sample from a law approaching $\mu_\beta$, averaged over the offset, at exactly the finite inverse temperatures without reconstruction. Necessity follows from the bound on one local estimate, including its input-independent randomized form (Theorem~\ref{thm:only_nonrecon} and Corollary~\ref{cor:first_local_nonrecon}). The two-stage BPQM decoder attains the converse (Proposition~\ref{prop:sharp}).

Block recovery requires more than a correct estimate at a typical symbol. The decoder must also recover the few symbols whose neighborhoods fail LT or TP. It first applies depth-$t_m$ BPQM to the symbols satisfying both conditions, with $t_m=O(\log\log m)$, records each estimate, and reverses the node operations between calls. It then treats the other symbols as erasures and completes them by linear equations.

Without reconstruction, the tree error decreases doubly exponentially with depth. At depth $t_m$, the original-input errors are therefore small enough that the sum of their square roots tends to zero. This sum bounds the probability that any record is incorrect, despite overlap between successive neighborhoods; undoing the node operations does not undo the measurement disturbance. The resulting reliability bound is uniform over the processing order. Disjoint-first and uniformly random schedules satisfy the joint-locality conditions of Section~\ref{sec:strategy}.

Finally, the code has linear minimum distance, while the erased set has size $o(m)$. Two codewords agreeing on the recorded coordinates therefore cannot differ on the erased set. The completion is unique and can be found in polynomial time. These arguments, including finite-precision implementation, are given in Appendix~\ref{app:sharp}.

The comparison with classical algorithms depends on a separate prediction. The cavity method identifies reconstruction with the dynamical transition~\cite{mezard2006reconstruction}, and predicts efficient Gibbs sampling at fixed inverse temperatures without reconstruction. Rigorous results establish efficient dynamics in some high-temperature regimes and relate relaxation to point-to-set correlations~\cite{montanari2006rigorous}; they do not prove efficient sampling throughout the phase for $k\ge3$. We state the required algorithmic conclusion as Conjecture~\ref{conj:classical}. Across $62$ binary degree pairs, the fixed-temperature experiments show weak size dependence of equilibrium relaxation below $\beta_d$ and rapid growth near it (Table~\ref{tab:dynamics} and Appendix~\ref{app:sim_glauber}). They support the prediction without establishing polynomial-time mixing. Under the conjecture, local quantum decoding does not give an exponential speedup for Gibbs sampling at these fixed inverse temperatures.

\subsection{Approximate optimization}\label{sec:ov_approx}

For the product amplitude functions $\Phi_\beta$, the sampling threshold also bounds the expected cost. Approximate optimization, however, allows a general symmetric amplitude function to be chosen to favor low-cost assignments. We must therefore compare the local reduced states of these more general states with mixtures of Gibbs factors.

Optimizing the symmetric amplitude function does not improve the asymptotic expected cost attainable with reliable local decoding. For fixed $3\le k<D$, zero-one matrices from either regular ensemble, and a fixed cost $f$ that is integer-valued up to an affine rescaling, the expected cost per constraint, averaged over the offset, satisfies
$$
\E_{x_0}\E\frac{H_f(x)}m\ge\langle f\rangle_{p_{\beta_d}}-o(1),\qquad H_f(x)=\sum_j f(x_j).
$$
The result allows every normalized complex amplitude function determined by total cost and chosen independently of the offset, over cyclic alphabets and fixed finite fields (Theorem~\ref{thm:common_cost_local} and Corollary~\ref{cor:common_cost_field}). The neighborhood of the local estimate may contain any number $v=o(\sqrt m)$ of qudits, which includes the full logarithmic-radius range of Section~\ref{sec:ov_locality}. For costs that take two values, Theorem~\ref{thm:approx_sym} permits any $v=o(m)$ and does not require $k\ge3$.

Over cyclic alphabets and finite fields, product amplitude functions approach the bound along inverse temperatures without reconstruction tending to $\beta_d$, with the two-stage BPQM decoder (Proposition~\ref{prop:sharp} and Corollary~\ref{cor:field_sharp}). Thus the bound is tight. For Max-XORSAT it takes the familiar form that the expected satisfied fraction is at most
$$
s_d+o(1),\qquad s_d=\frac12+\frac12\tanh(\beta_d/2).
$$

The first step of the proof is the output guarantee. It lets us replace the expected objective under the output distribution by its expectation under $|\Phi(x)|^2$, up to vanishing error. With product amplitudes, this is the single-constraint Gibbs mean. The mean decreases with $\beta$, and reliable local decoding requires non-reconstruction, so it cannot be lower than the value at $\beta_d$.

To extend the sampling bound beyond product amplitudes, we approximate the reduced state of a general symmetric state on a neighborhood of a codeword symbol by a mixture of identical-factor product states. A local measurement's correctness probability is then the average of its correctness probabilities on these components. In the binary case, each factor is, up to a diagonal phase, a Gibbs factor at some nonnegative inverse temperature, or gives a satisfied fraction below one half. The tree bound therefore excludes components that would exceed $s_d$.

For a larger alphabet, an arbitrary product factor need not have the Gibbs weights of the given cost. The additional restriction is that words with exactly the same total cost have equal amplitudes. At any attainable mean cost, Gibbs weights maximize entropy. Integer-cost rounding gives sufficiently many words at that exact cost, and a product vector whose magnitudes are far from Gibbs weights has exponentially small overlap with the subspace of equal-cost amplitudes. Weighting product vectors by their squared overlap with $\ket\Phi$ consequently gives a local product mixture concentrated near Gibbs factors. The quantitative error vanishes for $v=o(\sqrt m)$ (Appendix~\ref{app:common_mixture}).

For $k\ge3$, every reconstructing product component has a uniformly positive tree error, even when the Gibbs channels are not ordered by inverse temperature. Reliable decoding through a retained initial local estimate forces the total weight of components above $\beta_d$ to vanish. The remaining Gibbs factors have mean cost at least $\langle f\rangle_{p_{\beta_d}}$. Their average bounds the cost in $\ket\Phi$, and the output guarantee transfers it to DQI. This proves Theorem~\ref{thm:common_cost_local} without assuming that the non-reconstruction set is an interval.

For the mean-optimal polynomial amplitude functions of Jordan et al., the same reasoning explains their local resemblance to product amplitudes. Their asymptotic semicircle value is attained by a unique single-qubit factor, so the reduced state on any fixed number of qubits converges to that product state (Proposition~\ref{prop:semicircle_local}). Equality of these local marginals and mean objectives does not imply equality of the probabilities of rare optimal assignments; the latter comparison is the subject of Section~\ref{sec:ov_exact}.

With BPQM, DQI reaches a larger satisfied fraction than with classical belief propagation on all fifteen benchmark families, and on nine of them it exceeds the Shannon limit for classical decoding (Table~\ref{tab:benchmark}). Classical optimizers, however, work directly with the primal instance and are not constrained by the decoding problem induced by DQI. Under Conjecture~\ref{conj:classical}, or the corresponding sampling hypothesis in a field model, classical Gibbs sampling approaches the same expected-cost bound.

In the extended experiments, simulated annealing attains a mean largest satisfied fraction above the estimated $s_d$ on all $62$ binary degree pairs and all $105$ larger-alphabet tuples (Table~\ref{tab:sim_summary}). The binary comparison took less than four hours of elapsed time on one MacBook Pro (Section~\ref{sec:approx_bench}). Annealing continues improving the objective after leaving equilibrium, so exceeding the Gibbs satisfied fraction at $\beta_d$ is compatible with slow fixed-temperature dynamics above reconstruction. Appendix~\ref{app:dynamics} reports the full protocols and results.

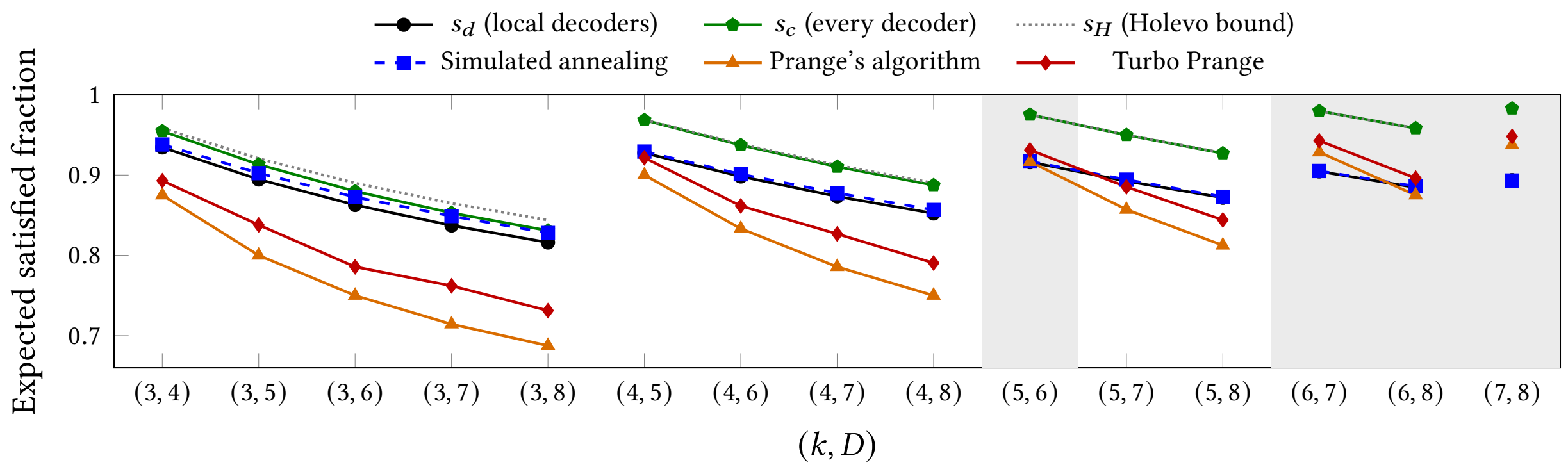
\begin{figure}[t]
\centering
\pgfplotstableread{
i    sd     sa     prange turbo  sh    sc
1    0.9345 0.9381 0.8750 0.8930 0.9583 0.9547
2    0.8946 0.9024 0.8000 0.8379 0.9206 0.9132
3    0.8628 0.8726 0.7500 0.7857 0.8900 0.8798
4    0.8372 0.8489 0.7143 0.7621 0.8648 0.8529
5    0.8161 0.8279 0.6875 0.7312 0.8439 0.8308
5.5  nan    nan    nan    nan    nan   nan
6    0.9275 0.9293 0.9000 0.9216 0.9689 0.9685
7    0.8983 0.9011 0.8333 0.8616 0.9385 0.9373
8    0.8734 0.8776 0.7857 0.8267 0.9124 0.9104
9    0.8523 0.8566 0.7500 0.7905 0.8900 0.8874
9.5  nan    nan    nan    nan    nan   nan
10   0.9162 0.9170 0.9167 0.9312 0.9754 0.9753
11   0.8924 0.8943 0.8571 0.8853 0.9502 0.9499
12   0.8718 0.8730 0.8125 0.8441 0.9276 0.9271
12.5 nan    nan    nan    nan    nan   nan
13   0.9046 0.9051 0.9286 0.9427 0.9797 0.9797
14   0.8845 0.8860 0.8750 0.8962 0.9583 0.9583
14.5 nan    nan    nan    nan    nan   nan
15   0.8936 0.8932 0.9375 0.9481 0.9829 0.9829
}\fracdata
\begin{tikzpicture}
\begin{axis}[
  width=\textwidth, height=4.4cm,
  xmin=0.5, xmax=15.5, ymin=0.66, ymax=1,
  xtick={1,...,15},
  xticklabels={{$(3,4)$},{$(3,5)$},{$(3,6)$},{$(3,7)$},{$(3,8)$},{$(4,5)$},{$(4,6)$},{$(4,7)$},{$(4,8)$},{$(5,6)$},{$(5,7)$},{$(5,8)$},{$(6,7)$},{$(6,8)$},{$(7,8)$}},
  xticklabel style={font=\scriptsize},
  yticklabel style={font=\scriptsize, /pgf/number format/fixed, /pgf/number format/precision=2},
  xlabel={$(k,D)$}, xlabel style={font=\small},
  ylabel={Expected satisfied fraction}, ylabel style={font=\small},
  unbounded coords=jump,
  legend columns=3,
  legend style={font=\scriptsize, draw=none, at={(0.5,1.02)}, anchor=south, /tikz/every even column/.append style={column sep=8pt}},
  every axis plot/.append style={line width=0.8pt, mark size=1.7pt},
]
\fill[black!8] (axis cs:9.5,0.66) rectangle (axis cs:10.5,1);
\fill[black!8] (axis cs:12.5,0.66) rectangle (axis cs:15.5,1);
\addplot[black, mark=*] table[x=i, y=sd] {\fracdata};
\addlegendentry{$s_d$ (local decoders)}
\addplot[green!50!black, mark=pentagon*] table[x=i, y=sc] {\fracdata};
\addlegendentry{$s_c$ (every decoder)}
\addplot[gray, densely dotted] table[x=i, y=sh] {\fracdata};
\addlegendentry{$s_H$ (Holevo bound)}
\addplot[blue, dashed, mark=square*, mark options={solid}] table[x=i, y=sa] {\fracdata};
\addlegendentry{Simulated annealing}
\addplot[orange!85!black, mark=triangle*] table[x=i, y=prange] {\fracdata};
\addlegendentry{Prange's algorithm}
\addplot[red!75!black, mark=diamond*] table[x=i, y=turbo] {\fracdata};
\addlegendentry{Turbo Prange}
\end{axis}
\end{tikzpicture}
\caption{Expected satisfied fractions on the fifteen benchmark families of Table~\ref{tab:benchmark}. With symmetric amplitude functions, DQI reaches at most $s_d$ with local decoders and, on the configuration model, at most $s_c$ with every decoder; $s_H$ is the limit given by the Holevo bound. The values of $s_d$ and $s_c$ are from Table~\ref{tab:condensation}; for $k\ge5$, $s_c$ and $s_H$ differ by less than $10^{-3}$. Simulated annealing is shown for $2520$ variables and $10^6$ sweeps, as in Table~\ref{tab:benchmark}; with about $10080$ variables and the same number of sweeps, its mean exceeds $s_d$ on every pair (Section~\ref{sec:approx_bench}). With FGUM, Regev's reduction~\cite{regev2009lattices}, a procedure closely related to DQI, reaches the values of Turbo Prange, a variant of Prange's algorithm~\cite{shutty2026locallyquantumdecoders}, and DQI with unambiguous discrimination reaches at most those of Prange's algorithm (Section~\ref{sec:ud_prange}). On every degree pair, the best classical value (on Gallager's ensemble) exceeds our estimate of $s_d$ and lies below $s_c$ (on the configuration model), and on the shaded pairs a known efficient nonlocal method exceeds $s_d$.}
\label{fig:fractions}
\end{figure}

\subsection{Exact optimization}\label{sec:ov_exact}

For exact optimization, we must compare optimum probabilities rather than expected costs. These probabilities can be exponentially small, so equality of local marginals does not provide the required control. We extend the classical comparison from Gibbs target states to every complex polynomial target state in a fixed subcritical-degree range.

Quantum minimum finding uses about $\lambda^{-1/2}$ preparations of a state whose optimum probability is $\lambda$, whereas independent sampling from that law requires about $\lambda^{-1}$ samples. We count coherent preparation calls and their inverses, and supply both search algorithms with a lower bound on the relevant optimum probability within a polynomial factor (Section~\ref{sec:extraction}).

For a Gibbs target state, the sampling result already ensures that reliable local decoding requires non-reconstruction. Under Conjecture~\ref{conj:classical}, efficient classical sampling then finds an optimum within a polynomial factor of the square of the quantum preparation count (Corollary~\ref{cor:quadratic}). The classical sampler may run at any larger inverse temperature where it remains efficient, because the probability of an optimum is nondecreasing in inverse temperature.

For binary Max-XORSAT, consider any normalized complex polynomial amplitude function of degree at most $\rho m$, with fixed $\rho<\rho_d$, where $\rho_d=\alpha_d^\perp$ is the error rate at $\beta_d$ and $\alpha_d=p_{\beta_d}(1)$. The matching product amplitude function has independent decoding errors of rate $\rho$, at an inverse temperature $\beta_\rho<\beta_d$. If the polynomial is chosen independently of the offset and its decoding problem can be solved reliably, its target-state optimum probability $\lambda_\Phi$ satisfies
$$
  \lambda_\Phi\le\poly(m)\lambda_{\beta_\rho}
$$
for most uniformly random offsets (Theorem~\ref{thm:poly_target}). Under Conjecture~\ref{conj:classical}, classical sampling at $\beta_\rho$ therefore finds an optimum in time $\poly(m+n)T_\Phi^2$, where $T_\Phi=\lambda_\Phi^{-1/2}$, on both random regular ensembles with $3\le k<D$ (Corollary~\ref{cor:poly_quadratic}). The theorem covers all polynomials in this degree range, not only the mean-optimal choice of Jordan et al.

The proof controls rare optimal patterns directly. A Dicke state of fixed Fourier weight is a phase average of identical-factor product states, with a normalization bounded by a polynomial. Expanding a degree-bounded polynomial in these Dicke states bounds every squared amplitude by a polynomial multiple of a suitable product probability. All optimal patterns have the same weight, so one product law bounds their total mass. Reliable decoding then prevents the target state's coset mass from being too small, while a union bound prevents the product coset masses from being too large for most offsets. This compares the normalized optimum probabilities without replacing a normalization by its expectation.

For the mean-optimal polynomials, reliable local decoding forces the limiting degree ratio to be at most $\rho_d$. At the endpoint, the classical cost is at most $\poly(m+n)e^{\eta m}T_\Phi^2$ for every fixed $\eta>0$, with an algorithm that may depend on $\eta$; polynomial overhead at the endpoint remains open. Appendix~\ref{sec:exact_heavy} proves both statements. It also extends to polynomial target states the unconditional comparison with a sampling version of Prange's algorithm (Section~\ref{sec:prange}), which holds at inverse temperatures where a union bound over the codewords converges, for every offset when the decoding is exact.

These are comparisons of target states. With exact decoding, DQI prepares the target state, so they apply to its output. With an imperfect decoder, an additive error of order $\sqrt\varepsilon$ can exceed an exponentially small optimum probability even when $\varepsilon\to0$. Extending the comparison to that output requires additional control at the scale of the optimum probability (Section~\ref{sec:prepared}).

\subsection{Optimal decoding and the condensation threshold}\label{sec:ov_condensation}\label{sec:ov_nonlocal}

For binary Max-XORSAT on the configuration model with $3\le k<D$, the optimal decoder has the tight threshold $\beta_c$: the PGM succeeds below it, and every decoder has exponentially small block success above it. With arbitrary symmetric amplitude functions, reliable decoding permits an expected satisfied fraction of at most $s_c$, which product amplitudes and the PGM approach (Theorem~\ref{thm:condensation} and Corollary~\ref{cor:condensation_sym}). This answers Question~\ref{q:any} more sharply than the limit derived from the Holevo bound~\cite[Supplementary Information, Section~10]{jordan2024optimization}, because $\beta_c<\beta_{\mathrm H}$.

The decoding criterion concerns the weights of the cosets of $K$. A codeword's phase is constant on each coset, so information about the codeword is encoded in the relative phases of these orthogonal coset components. For the product law with violation probability $\alpha=p_\beta(1)$, let $W_{B,\alpha}$ be the law of $E+K$ when $E$ has independent Bernoulli$(\alpha)$ entries. The PGM's optimal block success is the square of the Bhattacharyya coefficient of $W_{B,\alpha}$ and the uniform law on cosets (Lemma~\ref{lem:pgm_success}). Equal coset weights make the codeword states orthogonal; asymptotic uniformity is exactly what permits reliable decoding.

To locate this change in the coset law, we use its relative entropy $F_B(\alpha)$ from uniform. This entropy equals the difference between $m$ times the capacity of the binary symmetric channel and the mutual information of the low-density generator-matrix code with generator matrix $B$. The known variational formula for the latter quantity determines the limit of $\E_BF_B(\alpha)/n$~\cite{vandenbrand2017mutual}. Above $\beta_c$, the limit is positive and $F_B(\alpha)$ concentrates over the instance. The log-density then separates the typical cosets under the two laws, making their Bhattacharyya coefficient, and optimal decoding success, exponentially small.

Below $\beta_c$, sublinear relative entropy does not by itself imply total-variation closeness to uniform. We begin at a slightly smaller violation probability, truncate the density, and then add independent noise to reach $\alpha$. Entropy controls the mass removed by truncation, while the cutoff bounds the remaining second moment. Every nonconstant Fourier character of the coset group is indexed by a nonzero codeword of $C$. Linear minimum distance therefore makes the added noise damp every such coefficient exponentially, so the remaining variance vanishes. This proves coset uniformity and successful decoding (Lemma~\ref{lem:smoothing}).

For symmetric amplitude functions, the coset law is a mixture over the weight of the violation pattern. Exceeding $s_c$ by a fixed amount requires positive mixture mass on weights below $\alpha_cm$ by a fixed amount. Their coset laws are bounded by polynomial multiples of corresponding product coset laws, and the preceding failure estimate is uniform over these weights. Their positive mass prevents reliable decoding. Thus optimizing the symmetric amplitude function cannot improve the ceiling $s_c$.

The strict Holevo refinement follows from the uncertainty left after finitely many noisy observations. At $\beta_{\mathrm H}$, the capacity term and code-rate term cancel in the variational formula. The trial law whose incoming messages reveal their variables leaves a positive conditional entropy after $D$ observations, so the variational value is still positive. By continuity, the transition occurs at a strictly smaller inverse temperature. A uniform estimate over all trial laws also shows that $\beta_{\mathrm H}-\beta_c$ decreases exponentially at a fixed degree ratio (Proposition~\ref{prop:condensation_holevo}).

Implementing the PGM is a separate computational problem. Its measurement vectors are Fourier superpositions of normalized Gibbs states on the cosets. Coherent preparation of those states and its inverse therefore implement the PGM. Conversely, any coherent implementation of the PGM with access to its inverse prepares the Gibbs state of each specified right-hand side, with heralded success equal to the PGM's block success (Corollary~\ref{cor:pgm_equiv}). Below $\beta_c$, an efficient implementation would consequently prepare these states efficiently for all offsets, including planted right-hand sides.

The cavity method predicts that, at inverse temperatures near $\beta_c$, the Gibbs measure of a sufficiently low-noise planted regular instance is correlated with its planted assignment. Under this prediction, an efficient coherent PGM near $\beta_c$ would yield efficient weak recovery for regular noisy XOR systems. Known lower bounds for restricted classes of classical algorithms concern uniformly random equations, not this regular ensemble; no polynomial-time classical or quantum algorithm is known for the corresponding sparse recovery task. The precise prediction, its first-moment motivation and the scope of this evidence are given in Section~\ref{sec:window}.

\subsection{Larger alphabets and other costs}\label{sec:ov_alphabets}

The local-decoding results hold over every fixed $\Z_q$ and for every fixed cost, on both regular ensembles of zero-one matrices, and for $k\ge3$ the corresponding statements hold over every fixed field $\mathbb F_q$ (Corollary~\ref{cor:field_sharp}). Section~\ref{sec:ov_approx} states the approximation bounds for these alphabets and costs.

For $\Z_q$, a BPQM message is a probability vector rather than one overlap, and a check-node label takes $q$ values~\cite{mandal2026qary,mandal2026abelian}. Fourier magnitudes replace binary overlaps and posterior biases. The Nishimori property identifies the check-node label with the relative translation of two posterior messages. For composite $q$, tree projection is proved with subgroup annihilators, and completion uses linear congruences rather than field Gaussian elimination.

Over $\mathbb F_{p^r}$, trace characters replace cyclic phases. A prime-field basis splits the code constraints and their projections, not the cost or quantum channel. The same Fourier recursion proves tree-error decay, and field Gaussian elimination completes the codeword. The sequential and finite-precision arguments are unchanged. Both output guarantees hold in this field model, as in the original finite-field analysis of Regev's reduction~\cite{shutty2026locallyquantumdecoders}; Appendix~\ref{app:common_field} verifies the complete transfer.

The condensation threshold and the polynomial exact-search comparison remain binary Max-XORSAT results. The coherent PGM equivalence over $\Z_q$ holds for every modulus and every instance, rather than only for the random ensembles (Corollary~\ref{cor:pgm_equiv}). Table~\ref{tab:results} records these distinctions.

Over a field, the classical comparisons use the analogue of Conjecture~\ref{conj:classical}: at each fixed inverse temperature without reconstruction, with probability $1-o(1)$ over a random field instance and a uniformly random offset, some classical algorithm samples from a law within total variation distance $\epsilon$ of the Gibbs measure in time polynomial in $m+n$ and $\log(1/\epsilon)$ (Hypothesis~\ref{hyp:sampler}).

\begingroup
\small
\setlength{\tabcolsep}{3pt}
\setlength{\LTleft}{0pt}
\setlength{\LTright}{0pt}
\begin{longtable}{@{}>{\raggedright\arraybackslash}p{0.43\textwidth}>{\raggedright\arraybackslash}p{0.18\textwidth}>{\centering\arraybackslash}p{0.09\textwidth}>{\centering\arraybackslash}p{0.105\textwidth}>{\raggedright\arraybackslash}p{0.14\textwidth}@{}}
\caption{Domains of applicability for our rigorous results.}\label{tab:results}\\
\toprule
Result & Problem & \shortstack{Config.\\model} & \shortstack{Gallager's\\ensemble} & \shortstack[l]{Formal\\statement}\\
\midrule
\endfirsthead
\multicolumn{5}{@{}l}{Table~\thetable\ (continued)}\\
\toprule
Result & Problem & \shortstack{Config.\\model} & \shortstack{Gallager's\\ensemble} & \shortstack[l]{Formal\\statement}\\
\midrule
\endhead
\midrule
\multicolumn{5}{r@{}}{\emph{Continued on next page}}\\
\endfoot
\bottomrule
\endlastfoot
\multicolumn{5}{@{}l}{\textbf{Local decoders}}\\*
\addlinespace[2pt]
\multicolumn{5}{@{}l}{\hspace{0.5em}\emph{Decoding a single symbol} (Sections~\ref{sec:ov_duality} and~\ref{sec:ov_locality})}\\*
Some measurement on the qudits within distance $2c_0\log m$ of a typical symbol recovers the symbol with probability tending to one if and only if there is no reconstruction on the tree, and BPQM is an optimal measurement of this kind & any $q$, $f$ & \checkmark & \checkmark & Cor.~\ref{cor:typical_nogo}\\
\addlinespace[2pt]
\multicolumn{5}{@{}l}{\hspace{0.5em}\emph{Sampling} (Section~\ref{sec:ov_sampling})}\\*
No local decoder succeeds above $\beta_d$ & any $q$, $f$ & \checkmark & \checkmark & Thm.~\ref{thm:only_nonrecon}\\
Two-stage BPQM succeeds at every finite inverse temperature without reconstruction and approaches the approximation bounds along such temperatures tending to $\beta_d$ & any $q,f$; $k\ge3$ & \checkmark & \checkmark & Prop.~\ref{prop:sharp}\\
The local obstruction, two-stage attainment, and output guarantees also hold over finite fields & any $\mathbb F_q$, $f$; $k\ge3$ & \checkmark & \checkmark & Cor.~\ref{cor:field_sharp}\\
\addlinespace[2pt]
\multicolumn{5}{@{}l}{\hspace{0.5em}\emph{Approximate optimization} (Section~\ref{sec:ov_approx})}\\*
With a symmetric amplitude function and $v=o(m)$, the expected fraction with the smaller cost is at most its Gibbs probability at $\beta_d$, up to $o(1)$ & any $q$; $f$ takes two values & \checkmark & \checkmark & Thm.~\ref{thm:approx_sym}\\
For costs that are integer-valued up to an affine rescaling, a symmetric amplitude function and $v=o(\sqrt m)$ give expected cost at least its Gibbs average at $\beta_d$, up to $o(1)$ & any $q$ over $\Z_q$ or $\mathbb F_q$; $k\ge3$ & \checkmark & \checkmark & Thm.~\ref{thm:common_cost_local}, Cor.~\ref{cor:common_cost_field}\\
With the product amplitude function, the expected cost per constraint is at least the Gibbs average of $f$ at $\beta_d$, up to $o(1)$ & any $q$, $f$ & \checkmark & \checkmark & Thm.~\ref{thm:approx_gibbs}\\
\addlinespace[2pt]
\multicolumn{5}{@{}l}{\hspace{0.5em}\emph{Exact optimization} (Section~\ref{sec:ov_exact})}\\*
Under Conjecture~\ref{conj:classical}, where a local decoder succeeds, quantum minimum finding on the Gibbs state is at most quadratically faster than classical sampling & any $q$, $f$ & \checkmark & \checkmark & Cor.~\ref{cor:quadratic}\\
Under Conjecture~\ref{conj:classical}, the same quadratic comparison holds for every complex polynomial target state of degree at most $\rho m$, with fixed $\rho<\rho_d$ & \mbox{Max-XORSAT}, $k\ge3$ & \checkmark & \checkmark & Cor.~\ref{cor:poly_quadratic}\\
For the Gibbs state, the quadratic comparison holds without Conjecture~\ref{conj:classical} up to an explicit inverse temperature $\beta_U$, through a sampling version of Prange's algorithm & \mbox{Max-XORSAT}, five pairs $(k,D)$ & \checkmark & -- & Cor.~\ref{cor:unconditional}\\
\midrule
\multicolumn{5}{@{}l}{\textbf{Optimal decoding} (Section~\ref{sec:ov_condensation})}\\*
\addlinespace[2pt]
\multicolumn{5}{@{}l}{\hspace{0.5em}\emph{Sampling}}\\*
The pretty-good measurement succeeds below $\beta_c$, every decoder fails above $\beta_c$, and $\beta_d\le\beta_c$ & \mbox{Max-XORSAT}, $k\ge3$ & \checkmark & -- & Cor.~\ref{cor:two_thresholds}\\
\addlinespace[2pt]
\multicolumn{5}{@{}l}{\hspace{0.5em}\emph{Approximate optimization}}\\*
With a symmetric amplitude function and any decoder, DQI reaches a satisfied fraction of at most $s_c+o(1)$, and with the product amplitude function and the pretty-good measurement, it approaches $s_c$ as $\beta\uparrow\beta_c$ & \mbox{Max-XORSAT}, $k\ge3$ & \checkmark & -- & Cor.~\ref{cor:condensation_sym}, Section~\ref{sec:condensation}\\
\addlinespace[2pt]
\multicolumn{5}{@{}l}{\hspace{0.5em}\emph{Comparison with the Holevo bound}}\\*
$\beta_c<\beta_{\mathrm H}$ at every degree pair, and at a fixed ratio $k/D$, $\beta_{\mathrm H}-\beta_c$ vanishes exponentially in the degrees & \mbox{Max-XORSAT}, $k\ge3$ & \checkmark & -- & Prop.~\ref{prop:condensation_holevo}\\
\addlinespace[2pt]
\multicolumn{5}{@{}l}{\hspace{0.5em}\emph{Coherent implementation}}\\*
Coherent access to the PGM and its inverse is equivalent, up to polynomial overhead, to coherent Gibbs-state preparation and its inverse for all offsets at finite $\beta$, when PGM success is at least $1/\poly(m+n)$ & any $q$, $f$ & \multicolumn{2}{c}{every instance} & Cor.~\ref{cor:pgm_equiv}\\
\end{longtable}
\endgroup

\section{Preliminaries}\label{sec:setup}

This section defines the primal problem, the decoding problem induced by DQI, and BPQM, and the rest of the paper relates these three objects. Section~\ref{sec:primal} defines the primal problem and its Gibbs measure. Section~\ref{sec:dqi} derives the decoding problem that DQI solves for a general amplitude function, states the guarantees that turn decoding success into properties of the output, and distinguishes classical from quantum decoders. Section~\ref{sec:bpqm} introduces BPQM, the quantum decoder that our results concern most directly. Section~\ref{sec:strategy} defines local estimates and the two conditions on the neighborhood of a symbol under which computing a local estimate reduces to decoding a tree code, a reduction that is carried out in Sections~\ref{sec:tree} and~\ref{sec:locality}.

\subsection{Sparse constraint problems and their Gibbs measures}\label{sec:primal}

DQI prepares states whose amplitudes are determined by a sparse constraint problem. We write the problem in terms of a matrix $B$, because the code that DQI must decode is determined by $B$, and because the factor graph of $B$ is the graph on which all our recursions run.

Fix an integer $q\ge2$ and write $\omega=e^{2\pi i/q}$. Let $B\in\{0,1\}^{m\times n}$, viewed over $\Z_q$, and let $x_0\in\Z_q^m$. Fix a cost $f:\Z_q\to\mathbb R$ and $\beta\ge0$, and set
$$
  p(z)=\frac{e^{-\beta f(z)}}{\sum_{w\in\Z_q}e^{-\beta f(w)}},\qquad z\in\Z_q .
$$
When the inverse temperature must be shown, we write $p_\beta$ for $p$. We also write $p_\infty=\lim_{\beta\to\infty}p_\beta$, which is the uniform distribution on the minimizers of $f$. The primal problem is to minimize
\begin{equation}\label{eq:primal_cost}
  H(z)=\sum_{j=1}^m f\big((Bz)_j+x_{0,j}\big),\qquad z\in\Z_q^n ,
\end{equation}
and the primal Gibbs measure is $\mu_\beta(z)\propto e^{-\beta H(z)}\propto\prod_jp\big((Bz)_j+x_{0,j}\big)$. The cost depends on $z$ only through the violation pattern $x=Bz+x_0$. With $K=\{Bz:z\in\Z_q^n\}\le\Z_q^m$, the vector $x$ ranges over the coset $x_0+K$, and the image of $\mu_\beta$ under $z\mapsto Bz+x_0$ is the measure $\pi_\beta(x)\propto\prod_jp(x_j)$ on $x_0+K$. Each $x$ has exactly $|\ker B|$ preimages, so a sample from $\pi_\beta$ together with a uniformly random preimage is a sample from $\mu_\beta$.

The factor graph $G_B$ is the bipartite graph on rows $j\in[m]$ and columns $a\in[n]$, with an edge whenever $B_{ja}=1$. We say that $B$ is $(k,D)$-regular if every row has $k$ ones and every column has $D$ ones. Throughout, $k\ge2$ and $D\ge2$. In the primal problem, rows are \emph{clauses} with $k$ variables, and columns are \emph{variables} that occur in $D$ clauses.

\begin{remark}[Binary case and units]\label{rem:units}
For $q=2$ and $f(z)=z$, the cost $H$ counts violated parity constraints, so the primal problem is Max-$k$-XORSAT, and $p=(1,e^{-\beta})/(1+e^{-\beta})$. Zhou et al.~\cite{zhou2026gibbs} and Piveteau et al.~\cite{piveteau2026gibbs} use $\pm1$ spins, for which each clause contributes $\pm1$ to the energy. Up to an additive constant, their Hamiltonians are twice our $H$; for Piveteau et al.\ this holds after the uniformly random right-hand sides are complemented. Their inverse temperature is therefore half of our $\beta$. Our notation follows Shutty et al.~\cite{shutty2026locallyquantumdecoders}, who call the primal problem $D$-regular Max-$k$-XORSAT when $B$ is $(k,D)$-regular.
\end{remark}

\subsection{DQI and the decoding problem}\label{sec:dqi}

All our bounds concern the decoding problem that DQI must solve, so we derive it from the target state and state it precisely. We allow a general amplitude function, because Jordan et al.\ and Shutty et al.\ use amplitude functions that are not products.

An \emph{amplitude function} is a function $\Phi:\Z_q^m\to\mathbb C$ with $\sum_x|\Phi(x)|^2=1$, and we write $\ket\Phi=\sum_x\Phi(x)\ket x$ for the state with these amplitudes on all violation patterns. The \emph{target state} of DQI is the normalization of $\sum_{x\in x_0+K}\Phi(x)\ket x$, which is the projection of $\ket\Phi$ onto the coset $x_0+K$, the set of violation patterns of all assignments. Measuring the target state yields a violation pattern $x$ with probability proportional to $|\Phi(x)|^2$, and a uniformly random preimage under $z\mapsto Bz+x_0$ then yields an assignment. An amplitude function is a \emph{product amplitude function} if it is a product over $j$ of the same function of $x_j$. The product amplitude function \emph{of Gibbs form} at inverse temperature $\beta$ is $\Phi_\beta(x)=\prod_j\sqrt{p(x_j)}$, and we refer to it as the product amplitude function. For it, $\ket{\Phi_\beta}$ is a product state and the target law is $\pi_\beta$. For Max-XORSAT, Jordan et al.~\cite{jordan2024optimization} use amplitude functions that are polynomials in the number of satisfied constraints, which depend only on $|x|$. We call an amplitude function symmetric if $\Phi(x)$ depends only on the total cost $\sum_jf(x_j)$ of the violation pattern; for Max-XORSAT, this total cost is $|x|$. Shutty et al.\ also use an amplitude function that is a product over blocks of symbols (Section~\ref{sec:approx_block}).

Let $C=K^\perp=\{c\in\Z_q^m:B^\top c=0\}$. This is the code that DQI must decode. Its factor graph is $G_B$ with the roles exchanged: rows are code \emph{symbols}, each in $k$ checks, and columns are parity \emph{checks}, each on $D$ symbols. For $c\in\Z_q^m$ let
\begin{equation}\label{eq:states_general}
  \ket{\psi^\Phi_c}=\sum_{x\in\Z_q^m}\Phi(x)\,\omega^{c\cdot x}\ket x=Z^c\ket\Phi,
\end{equation}
where $Z^c=\bigotimes_jZ^{c_j}$ and $Z\ket z=\omega^z\ket z$. By character orthogonality,
$$
  \mathbf 1[x\in x_0+K]=\frac1{|C|}\sum_{c\in C}\omega^{c\cdot(x-x_0)},
$$
and therefore
\begin{equation}\label{eq:dqi_identity}
  \sum_{x\in x_0+K}\Phi(x)\ket x=\frac1{|C|}\sum_{c\in C}\omega^{-c\cdot x_0}\ket{\psi^\Phi_c}.
\end{equation}
DQI prepares the left-hand side as follows~\cite{jordan2024optimization,zhou2026gibbs,shutty2026locallyquantumdecoders}. First, prepare the uniform superposition $|C|^{-1/2}\sum_{c\in C}\ket c$ together with $\ket\Phi$; the superposition can be prepared efficiently because $C$ is linear. Preparing $\ket\Phi$ is efficient for product amplitude functions and for the polynomial amplitude functions of Jordan et al. Second, apply $Z^c$ controlled on the first register and the phase $\omega^{-c\cdot x_0}$, which produces $|C|^{-1/2}\sum_c\omega^{-c\cdot x_0}\ket c\ket{\psi^\Phi_c}$. Third, uncompute the first register by running in reverse a coherent decoder that computes $c$ from $\ket{\psi^\Phi_c}$. If the decoder were exact, the result would be the right-hand side of~\eqref{eq:dqi_identity} up to normalization. In the form used by Jordan et al.\ and Zhou et al., a final Fourier transform maps the result to the computational basis. In our form this transform is not needed, because the states~\eqref{eq:states_general} are already written in the Fourier basis.

For the product amplitude function the states factorize, and we write $\ket{\psi_c}\coloneqq\ket{\psi^{\Phi_\beta}_c}=\bigotimes_j\ket{\psi^p_{c_j}}$, where for a probability vector $\nu$ on $\Z_q$ and $y\in\Z_q$
\begin{equation}\label{eq:family}
  \ket{\psi^\nu_y}=\sum_{z\in\Z_q}\sqrt{\nu(z)}\,\omega^{yz}\ket z .
\end{equation}
We call $\nu$ the \emph{spectrum} of the family $\{\ket{\psi^\nu_y}\}_{y\in\Z_q}$, because the Gram matrix of the family has the eigenvalues $q\nu(z)$, $z\in\Z_q$. The map $y\mapsto\ket{\psi^p_y}$ is the pure-state classical--quantum channel induced by DQI for Gibbs states. Its outputs are states of a qudit, a $q$-level quantum system, which is a qubit when $q=2$. For $q=2$ the two states are $\sqrt{p(0)}\ket0\pm\sqrt{p(1)}\ket1$, which is the binary pure-state channel studied in the BPQM literature~\cite{renes2017bpqm,piveteau2022messagepassing}. Applying the inverse quantum Fourier transform to each symbol maps $\ket{\psi^p_y}$ to $\sum_eg(e)\ket{y+e}$ with $g(e)=q^{-1/2}\sum_z\sqrt{p(z)}\,\omega^{-ez}$. This is the form used by Zhou et al.~\cite{zhou2026gibbs}: the codeword symbol $y$ is subjected to a coherent superposition of errors $e$.

The uncomputation of the first register is a decoding problem in which the codeword is uniformly distributed on $C$.

\begin{definition}[Decoding problem induced by DQI]\label{def:decoding}
Fix an amplitude function $\Phi$. Draw $c$ uniformly from $C$ and give the decoder the state $\ket{\psi^\Phi_c}$. The decoder outputs an estimate $\hat c\in\Z_q^m$. Its \emph{block success} is $\Pr[\hat c=c]$, and its \emph{symbol error at $j$} is $\Pr[\hat c_j\ne c_j]$. Both probabilities are averaged over $c$ and over the outcome of the decoder. When no amplitude function is specified, $\Phi$ is the product amplitude function $\Phi_\beta$.
\end{definition}

To translate bounds on the decoding problem into statements about DQI, we need to know how decoding success affects the output. Zhou et al.~\cite{zhou2026gibbs} prove such a guarantee for Gibbs states with $q=2$, adapting a result of Shutty et al. Shutty et al.~\cite{shutty2026locallyquantumdecoders} prove guarantees for every amplitude function over every finite field, for Regev's reduction, a closely related procedure that originates in lattice cryptography~\cite{regev2009lattices}. With an exact decoder, Regev's reduction prepares the same target state. We use the following form of it, which differs from the procedure above in two steps. After the decoder has been applied, a projective measurement checks whether the decoded word equals the content of the codeword register, and the procedure restarts if it does not. At the end, the violation pattern is measured. If the decoder is not exact, this pattern can lie outside the coset $x_0+K$, and the procedure then outputs $x_0$, which is the violation pattern of the zero assignment. The sampler of Zhou et al.\ also outputs a fixed assignment in this case~\cite{zhou2026gibbs}, whereas Shutty et al.\ condition on $x_0+K$. For a decoder with block success $1-\varepsilon$, the check passes with probability $1-\varepsilon$ for every offset~\cite{shutty2026locallyquantumdecoders}, so for $\varepsilon<1$ the procedure requires $1/(1-\varepsilon)$ attempts in expectation. Statements about the output of DQI in this paper refer to this form. The following theorem collects the statements that we use.

\begin{theorem}[Output guarantees]\label{thm:guarantee}
Let $q\ge2$ and let $\Phi$ be an amplitude function. Suppose that a decoder for the decoding problem with amplitude function $\Phi$ has block success $1-\varepsilon$ with $\varepsilon<1$. Let $x_0$ be uniformly distributed on $\Z_q^m$, and let $\nu_{x_0}$ be the law of the violation pattern that Regev's reduction outputs when it uses this decoder.
\begin{enumerate}
\item Suppose that $\Phi$ has positive mass $\sum_{x\in X}|\Phi(x)|^2$ on every coset $X$ of $K$, and let $\pi^\Phi_{x_0}(x)\propto|\Phi(x)|^2$ on $x_0+K$. Then $\E_{x_0}\|\nu_{x_0}-\pi^\Phi_{x_0}\|_{\mathrm{TV}}\le\sqrt\varepsilon$.
\item For every function $h:\Z_q^m\to[0,1]$,
$$
  \Big|\E_{x_0}\E_{x\sim\nu_{x_0}}h(x)-\sum_{x\in\Z_q^m}|\Phi(x)|^2h(x)\Big|\le2\sqrt\varepsilon .
$$
\end{enumerate}
\end{theorem}

\begin{proof}
Shutty et al.\ prove both parts over every finite field~\cite[Theorems~4 and~6]{shutty2026locallyquantumdecoders}, in which part~1 concerns the law of the pattern before the replacement of patterns outside $x_0+K$, and Chailloux and Hermouet prove a reduction of the same kind over $\Z_q$ for every modulus $q$~\cite[Theorem~4.1]{chailloux2026equivalence}. We give the proof in our notation, which covers codes $C$ over $\Z_q$ that need not be free $\Z_q$-modules. It uses only the identity $|C|^{-1}\sum_{c\in C}\omega^{c\cdot y}=\mathbf 1[y\in K]$, which holds for every $q$ because $C^\perp=K$ (proof of Lemma~\ref{lem:tp}).

Write $s=1-\varepsilon$. Dilate the decoder to a unitary $U$ that retains all work registers, let $\Pi_c$ project onto the outcome $c$ in its outcome register, and set $Q_c=U^\dagger\Pi_cU$. These are mutually orthogonal projections. Write $\ket{\bar\psi_c}=\ket{\psi^\Phi_c}\ket0$ for the received state together with the ancillary input of the decoder, and set $v_c=Q_c\ket{\bar\psi_c}$, so that $s=|C|^{-1}\sum_c\|v_c\|^2$. Regev's reduction applies $U$ to $|C|^{-1/2}\sum_{c\in C}\omega^{-c\cdot x_0}\ket c_A\ket{\bar\psi_c}$, where $A$ is the codeword register. The check passes with probability $|C|^{-1}\sum_c\|\Pi_cU\ket{\bar\psi_c}\|^2=s$ for every offset. After it passes, subtracting the outcome from $A$ and applying $U^\dagger$ leaves, with $A$ in the state $\ket0$, the vector
\begin{equation}\label{eq:actual_ideal}
  \ket{A_{x_0}}=\frac1{\sqrt{|C|s}}\sum_{c\in C}\omega^{-c\cdot x_0}v_c,
  \qquad\text{and we compare it with}\qquad
  \ket{T_{x_0}}=\frac1{\sqrt{|C|}}\sum_{c\in C}\omega^{-c\cdot x_0}\ket{\bar\psi_c}.
\end{equation}
The vectors $v_c$ are orthogonal, so $\|A_{x_0}\|=1$ for every offset. For $c\ne c'$ in $\Z_q^m$, the average of $\omega^{(c-c')\cdot x_0}$ over a uniformly random offset vanishes, and $\langle v_c|\bar\psi_c\rangle=\|v_c\|^2$. Hence
\begin{equation}\label{eq:average_identities}
  \E_{x_0}\|T_{x_0}\|^2=1,\qquad \E_{x_0}\langle A_{x_0}|T_{x_0}\rangle=\frac1{|C|\sqrt s}\sum_{c\in C}\|v_c\|^2=\sqrt s .
\end{equation}
By~\eqref{eq:dqi_identity}, $\ket{T_{x_0}}=\sqrt{|C|}\sum_{x\in x_0+K}\Phi(x)\ket x\ket0$. If the coset $X=x_0+K$ has positive mass $w_X=\sum_{x\in X}|\Phi(x)|^2$, then $\ket{T_{x_0}}=\sqrt{|C|w_X}\,\ket{\Theta_{x_0}}\ket0$, where $\ket{\Theta_{x_0}}$ is the target state, and otherwise $T_{x_0}=0$.

For part~1, every coset has positive mass, and $\E_{x_0}|C|w_{x_0+K}=1$ because each $x$ lies in $x_0+K$ with probability $1/|C|$. The Cauchy--Schwarz inequality applied to~\eqref{eq:average_identities} gives $s\le\E_{x_0}|\langle A_{x_0}|\Theta_{x_0},0\rangle|^2$. The trace distance between two unit vectors $u,v$ is $\sqrt{1-|\langle u|v\rangle|^2}$, so by Jensen's inequality the average trace distance between $\ket{A_{x_0}}$ and $\ket{\Theta_{x_0}}\ket0$ is at most $\sqrt{1-s}=\sqrt\varepsilon$. Measuring the violation pattern does not increase this distance, and it yields the law $\nu'_{x_0}$ of the pattern before the replacement and the target law $\pi^\Phi_{x_0}$, respectively. The law $\nu_{x_0}$ is the image of $\nu'_{x_0}$ under the map that fixes every pattern in $x_0+K$ and sends every other pattern to $x_0$. This map fixes $\pi^\Phi_{x_0}$, and total variation does not increase under a map, which proves part~1. Zhou et al.\ use the same argument for their sampler~\cite[Eq.~(3.18)]{zhou2026gibbs}.

For part~2, let $\mathsf H_{x_0}$ be the operator that is diagonal on the violation register, with entry $h(x)$ for $x\in x_0+K$ and $h(x_0)$ for every other $x$, and acts as the identity on the other registers. Its expectation in $\ket{A_{x_0}}$ is $\E_{x\sim\nu_{x_0}}h(x)$. For two vectors $u,v$ and a Hermitian operator $\mathsf H$ with $\|\mathsf H\|\le1$, the difference $\langle u|\mathsf H|u\rangle-\langle v|\mathsf H|v\rangle$ is the real part of $\langle u-v|\mathsf H|u+v\rangle$, so its absolute value is at most $\|u-v\|\,\|u+v\|$. By~\eqref{eq:average_identities}, $\E_{x_0}\|A_{x_0}-T_{x_0}\|^2=2-2\sqrt s$ and $\E_{x_0}\|A_{x_0}+T_{x_0}\|^2=2+2\sqrt s$, so by the Cauchy--Schwarz inequality over the offset,
$$
  \E_{x_0}\big|\langle A_{x_0}|\mathsf H_{x_0}|A_{x_0}\rangle-\langle T_{x_0}|\mathsf H_{x_0}|T_{x_0}\rangle\big|\le\sqrt{(2-2\sqrt s)(2+2\sqrt s)}=2\sqrt\varepsilon .
$$
Finally, $\E_{x_0}\langle T_{x_0}|\mathsf H_{x_0}|T_{x_0}\rangle=\E_{x_0}|C|\sum_{x\in x_0+K}|\Phi(x)|^2h(x)=\sum_x|\Phi(x)|^2h(x)$, again because each $x$ lies in $x_0+K$ with probability $1/|C|$. This identity holds also on cosets of zero mass, on which $T_{x_0}=0$, so part~2 does not require positive mass. This proves part~2.
\end{proof}

In part~1, the assumption that $\Phi$ has positive mass on every coset makes the target law $\pi^\Phi_{x_0}$ well defined; part~2 does not use it. The assumption holds for product amplitude functions at finite $\beta$, and for the amplitude functions of Proposition~\ref{prop:bounded_distance}, whose mass on every coset is $1/|C|$. Without the assumption, cosets of zero mass are rare when the block success is close to one. The states $\ket{\psi^\Phi_c}$ form a geometrically uniform family, so the optimal block success is $|C|^{-1}\big(\sum_Xw_X^{1/2}\big)^2$, where $X$ runs over the cosets of $K$ and $w_X=\sum_{x\in X}|\Phi(x)|^2$~\cite{eldar2001square}. By the Cauchy--Schwarz inequality this is at most one minus the fraction of cosets of zero mass, so a uniformly random offset lies in such a coset with probability at most $\varepsilon$. The assumption does not ensure that the pattern measured at the end of Regev's reduction lies in $x_0+K$ with positive probability, even when the block success is close to one, so the law of this pattern conditioned on $x_0+K$ need not be defined.

Theorem~\ref{thm:guarantee} holds on average over the offset $x_0$, and not for every offset: Chailloux and Tillich give an example in which, for a fixed offset, the decoding errors interfere constructively~\cite{chailloux2024quantumdecoding}.

We prove our bounds for local decoders under the assumption that the block success tends to one. The proofs use a local estimate at one symbol, and for decoders that make such an estimate, this assumption is essential, as we explain after Definition~\ref{def:local}. Franco Garrido and Chailloux note that Regev's reduction can sometimes succeed when the decoding problem cannot be solved reliably~\cite{garrido2026regev}. Similarly, a coherent implementation of the pretty-good measurement, with access to its inverse, prepares the target state of the product amplitude function exactly with probability equal to its block success, even when this block success does not tend to one (Proposition~\ref{prop:pgm_gibbs}).

\paragraph{Error rates for $q=2$.} For $q=2$ and $p(1)=\alpha\le1/2$, the coefficients above are $g(0)=\sqrt{1-\alpha^\perp}$ and $g(1)=\sqrt{\alpha^\perp}$, so the inverse Fourier transform maps $\ket{\psi^p_y}$ to $\sqrt{1-\alpha^\perp}\ket y+\sqrt{\alpha^\perp}\ket{y+1}$, where
\begin{equation}\label{eq:perp}
  \alpha^\perp=\tfrac12-\sqrt{\alpha(1-\alpha)} .
\end{equation}
The map $\alpha\mapsto\alpha^\perp$ is a decreasing bijection of $[0,1/2]$ onto itself, and $(\alpha^\perp)^\perp=\alpha$. Measuring each symbol therefore produces bit flips with probability $\alpha^\perp$, which equals $\tfrac12-\tfrac12\operatorname{sech}(\beta/2)$ and agrees with the error rate of Zhou et al.\ after the change of units in Remark~\ref{rem:units}. Conversely, the product amplitude function whose decoding problem has error rate $\rho$ leads to the satisfied fraction $1-\rho^\perp=\tfrac12+\sqrt{\rho(1-\rho)}$ per constraint.

\paragraph{The limit for every decoder.} To place the thresholds of particular decoders in context, we record the limit that applies to every decoder for product amplitude functions. The overlaps of the single-symbol states are $\langle\psi^p_{y'}|\psi^p_y\rangle=\sum_zp(z)\omega^{(y-y')z}$. For $q=2$ and $f(z)=z$,
$$
  \sigma_\beta\coloneqq\langle\psi^p_0|\psi^p_1\rangle=p(0)-p(1)=\tanh(\beta/2).
$$
The two states are orthogonal at $\beta=0$ and become identical as $\beta\to\infty$. The average of $\ket{\psi^p_y}\!\bra{\psi^p_y}$ over uniform $y$ is $\sum_zp(z)\ket z\!\bra z$, so the Holevo information of the uniform input ensemble is the entropy $H(p)$. By the converse of the Holevo--Schumacher--Westmoreland theorem, no decoder achieves vanishing block error at rate $R$ unless $H(p)/\ln q\ge R$. For the $(k,D)$-regular codes below, the rate of $C$ is at least the design rate $1-k/D$. The entropy $H(p_\beta)$ is nonincreasing in $\beta$, because its derivative is $-\beta\operatorname{Var}_{p_\beta}(f)$. We write $\beta_{\mathrm H}=\sup\{\beta\ge0:H(p_\beta)/\ln q\ge1-k/D\}$. No decoder achieves vanishing block error above $\beta_{\mathrm H}$. The value $\beta_{\mathrm H}$ is finite if and only if $H(p_\infty)/\ln q<1-k/D$, for example when $k<D$ and $f$ has a unique minimizer. For $q=2$ and $f(z)=z$, we write $\alpha_H=p_{\beta_{\mathrm H}}(1)$, which solves $H_2(\alpha_H)=1-k/D$ for the binary entropy $H_2$ in bits. We call $\rho_H=\alpha_H^\perp$ the Holevo rate, and $s_H=1-\alpha_H$ is the corresponding satisfied fraction. Jordan et al.\ derived this limit on the satisfied fraction that DQI reaches with any quantum decoder from the Holevo bound, after approximating the coherent errors of the decoding problem induced by DQI by a product pure-state channel, and they found no degree pair on Gallager's ensemble at which simulated annealing or Prange's algorithm exceeds it~\cite[Supplementary Information, Section~10]{jordan2024optimization}. This limit depends on the code only through its rate. For $q=2$, $f(z)=z$, and random $(k,D)$-regular instances from the configuration model with $3\le k<D$, the exact limit of every decoder is the condensation threshold $\beta_c$ of Theorem~\ref{thm:condensation}, which is strictly smaller than $\beta_{\mathrm H}$ (Proposition~\ref{prop:condensation_holevo}); numerically, $\beta_c\approx1.99$ and $\beta_{\mathrm H}\approx2.09$ for $(k,D)=(3,6)$.

\paragraph{Classical and quantum decoders.} A \emph{classical decoder} applies a classical algorithm, run reversibly, to the computational-basis values $y+e$ obtained after the inverse Fourier transform. It therefore uses only the classical channel $y\mapsto y+e$, which for $q=2$ is the binary symmetric channel with crossover probability $\alpha^\perp$. Belief propagation, Gaussian elimination, and the decoders used by Jordan et al.\ are classical decoders. A \emph{quantum decoder} may apply any quantum operation to the qudits. Unambiguous discrimination followed by Gaussian elimination~\cite{chailloux2024quantumdecoding,zhou2026gibbs}, FGUM followed by Gaussian elimination~\cite{shutty2026locallyquantumdecoders}, BPQM~\cite{renes2017bpqm}, and the pretty-good measurement are quantum decoders. Section~\ref{sec:strategy} defines the property that our theorems use: whether the estimate of a symbol depends only on the qudits near that symbol.

\subsection{Belief propagation with quantum messages}\label{sec:bpqm}

BPQM combines quantum channel outputs along the factor graph before estimating a codeword symbol~\cite{renes2017bpqm}. On a tree, its node operations preserve the information required to decode the root, so the final measurement is optimal. We describe these operations for binary pure-state channels, then explain their coherent and sequential implementation. Section~\ref{sec:tree} gives the corresponding operations over $\Z_q$ and relates their recursion to primal reconstruction.

\emph{Binary messages.} Classical belief propagation passes likelihoods obtained from classical channel outputs. BPQM instead retains the received quantum states. Write
$$
  \ket{\psi_y(a)}=\sqrt{\frac{1+a}{2}}\,\ket0+(-1)^y\sqrt{\frac{1-a}{2}}\,\ket1,\qquad a\in[-1,1],\quad y\in\{0,1\}.
$$
The two states have overlap $a$. For the channel induced by the product amplitude function $\Phi_\beta$, this overlap is $\sigma_\beta=\tanh(\beta/2)$. Two equally likely pure states can be distinguished with optimal probability $\tfrac12\big(1+\sqrt{1-a^2}\big)$, so the optimal success probability increases as $|a|$ decreases. BPQM combines messages using two operations~\cite{renes2017bpqm,piveteau2022messagepassing}.

\emph{Symbol nodes.} Suppose two incoming messages about the same bit $y$ have overlaps $a$ and $b$. The two possible joint states have overlap $ab$, so a unitary depending on $a$ and $b$, but not on $y$, maps
$$
  \ket{\psi_y(a)}\ket{\psi_y(b)}\longmapsto\ket{\psi_y(ab)}\ket0.
$$
The information about $y$ is therefore preserved in one outgoing qubit, while the other qubit is left in a fixed state.

\emph{Check nodes.} A check combines messages about two bits into a message about their parity. BPQM applies a CNOT controlled by the second qubit and reads the first qubit in the computational basis. The outcome is a label $\Delta$, called a herald in the BPQM literature. It equals zero with probability $(1+ab)/2$ and one with probability $(1-ab)/2$, independently of the parity. Conditional on an outcome of positive probability, the remaining qubit is
$$
  \begin{cases}
  \displaystyle\ket{\psi_{y_1\oplus y_2}\left(\frac{a+b}{1+ab}\right)},&\Delta=0,\\[6pt]
  \displaystyle\ket{\psi_{y_1\oplus y_2}\left(\frac{b-a}{1-ab}\right)},&\Delta=1.
  \end{cases}
$$
In a uniformly random tree codeword, $y_1$ is uniform given the parity. Averaging over this bit removes the coherence between the two herald values, so reading the herald loses no information about the parity. The herald must nevertheless be retained, because it determines the overlap of the outgoing message.

\emph{Decoding a tree.} To estimate the root symbol, BPQM applies the node operations from the leaves towards the root. At a check node, it combines the messages from the child symbols into a message about the parent symbol, whose value is fixed by the parity check. At a symbol node, it combines the received qubit with the messages from its child checks. Given the heralds, the final qubit is in one of two pure states, according to the root value. Applying the Helstrom measurement for these states attains the largest success probability of any measurement on the original tree input~\cite{renes2017bpqm,rengaswamy2021bpqm,piveteau2022messagepassing}.

\emph{Coherent and sequential implementation.} In a coherent implementation, heralds are computed into registers rather than measured. These registers control the later node operations. The overlaps can likewise be represented in quantum registers and computed to finite precision; efficient circuits for this form of message passing were developed in~\cite{piveteau2022messagepassing}.

Sequential receivers estimate codeword symbols one at a time~\cite{rengaswamy2021bpqm,piveteau2022messagepassing}. For each symbol, the decoder applies its BPQM circuit, records the estimate, and reverses the other operations before proceeding. In the coherent version, the final estimate is copied to a fresh register instead of being measured. Later operations preserve this register.

Reversing the circuit does not undo the disturbance caused by recording the estimate. Nor does it automatically return every work register to its initial state. We therefore realize each local call with work registers initialized independently of the received input; a register may be reused when it is known to return exactly to a fixed state. At the prescribed logarithmic depths, the computation trees and the number of calls have polynomial size, so retaining separate work registers preserves polynomial resource bounds.

Local heralds may control operations within a call. Dependence on estimates from earlier calls is a different matter: it changes the instrument applied to the next neighborhood. Section~\ref{sec:strategy} distinguishes these cases when establishing locality. When the individual estimates are sufficiently accurate, the disturbance accumulated during sequential decoding can be bounded; this is the argument used for reliable block recovery in Proposition~\ref{prop:sharp} and Appendix~\ref{app:sharp}~\cite{mandal2026blockerror}.

\emph{Density evolution.} For product channel inputs on a regular tree, disjoint subtrees produce independent heralds. The message overlaps are consequently random, and their distribution evolves recursively~\cite{brandsen2022bpqm,mandal2026blockerror}. Let $V_t$ be the overlap of a message from a symbol after $t$ completed symbol--check levels, starting with $V_0=\sigma_\beta$. If $\Gamma$ combines its inputs two at a time using the random check-node rule, then
$$
  W_t=\Gamma\!\left(V_t^{(1)},\ldots,V_t^{(D-1)}\right),\qquad V_{t+1}=\sigma_\beta\prod_{s=1}^{k-1}W_t^{(s)},
$$
where superscripts denote independent copies. At the root, the product contains $k$ check messages rather than $k-1$. Reliable root decoding is equivalent to the resulting overlap tending to zero in probability. Section~\ref{sec:tree} derives this recursion over every $\Z_q$ and proves that this convergence is equivalent to non-reconstruction for the primal Gibbs measure.

\emph{Codes with cycles.} Unrolling classical belief propagation around a symbol can place the same channel output in several branches of the computation tree. Unknown quantum states cannot be copied exactly. BPQM with approximate cloning supplies approximate copies for these repeated occurrences~\cite{piveteau2022messagepassing}. A cloning operation can be implemented through a unitary dilation, with its auxiliary registers retained so that the operation can subsequently be reversed.

A BPQM call unrolled for $t$ iterations still uses original channel outputs only from within distance $2t$ of its target symbol. Cloning therefore remains part of a measurement on those original inputs; it does not provide information from outside the neighborhood. Paired-measurement BPQM, introduced for mixed-state channels, similarly adds local operations and reduces to standard BPQM on pure-state channels~\cite{brandsen2022bpqm}.

The two-stage decoder takes a different route~\cite{mandal2026blockerror}. It applies BPQM at depth $\Theta(\log\log m)$ to symbols with suitable tree neighborhoods, then treats the remaining symbols as erasures and recovers them by Gaussian elimination. Proposition~\ref{prop:sharp} verifies the required tree-error decay, distance, and completion properties for our two regular ensembles.

\emph{Larger alphabets and locality.} BPQM was extended to prime alphabets and finite abelian groups in~\cite{mandal2026qary,mandal2026abelian}. A message is then described by a probability vector rather than a single overlap. The same two node operations combine repeated-symbol messages and parity-check messages; Lemma~\ref{lem:nodes} states them over every $\Z_q$.

For both binary and larger-alphabet implementations, the relevant locality is the support on the original received qudits. A single BPQM call with $t$ iterations defines a measurement on the neighborhood of its target symbol, including any local reliability, herald, and cloning registers. Whether a sequence of such calls satisfies Definition~\ref{def:local_decoder} also depends on its schedule and on how recorded estimates are used. The joint-locality and first-estimate arguments in the following subsection and Section~\ref{sec:only_nonrecon} establish the two kinds of coverage described in Section~\ref{sec:ov_setting}.

\subsection{Local estimates and tree codes}\label{sec:strategy}

The code $C$ is long and its factor graph has many cycles, so optimal decoding for the whole code is difficult to analyze. We therefore study the largest probability of correctly estimating one symbol using only the received qudits in a neighborhood of that symbol. We first define these estimates and the conditions under which their input is a tree decoding problem. We then give operational criteria for joint locality and explain how a retained initial estimate bounds block success.

Let $N_t(i)$ be the symbols at graph distance at most $2t$ from symbol $i$ in $G_B$, let $F_t(i)$ be the checks at distance at most $2t-1$, and let $L_t(i)$ be the symbols at distance exactly $2t$. Every check in $F_t(i)$ has all its symbols in $N_t(i)$.

\begin{definition}[Local estimates]\label{def:local}
A decoder makes a \emph{$t$-local estimate at $i$} if there is a POVM $\{E_y\}_{y\in\Z_q}$ on the qudits of $N_t(i)$ such that $\Pr[\hat c_i=y\mid c]=\tr\big((E_y\otimes I)\ket{\psi^\Phi_c}\!\bra{\psi^\Phi_c}\big)$ for every $c\in C$.
\end{definition}

The measurement may be collective and may use ancillas. This definition concerns a measurement of the original received state. In sequential decoding, later estimates may be computed after previous operations have disturbed their input, so the support of their circuits alone does not establish locality in this sense.

\begin{definition}[Local conditions]\label{def:conditions}
The condition $\mathrm{LT}_t(i)$ (local tree) holds if the ball of radius $2t$ around $i$ in $G_B$ is a tree. The condition $\mathrm{TP}_t(i)$ (tree projection) holds if the restriction map $c\mapsto c|_{N_t(i)}$ sends $C$ onto the code $\mathcal C_t(i)$ of all assignments on $N_t(i)$ that satisfy the checks in $F_t(i)$.
\end{definition}

The inclusion $C|_{N_t(i)}\subseteq\mathcal C_t(i)$ always holds. Tree projection states that checks outside the ball impose no additional constraints inside it. Since restriction is a group homomorphism, under $\mathrm{TP}_t(i)$ a uniformly random codeword restricts to a uniformly random element of $\mathcal C_t(i)$. Under $\mathrm{LT}_t(i)$, this is a tree code. For the product amplitude function, the received qudits therefore have exactly the distribution of the tree decoding problem studied in Section~\ref{sec:tree}.

This reduction applies at any depth at which both conditions hold. The logarithmic restriction below serves a different purpose: it ensures that the conditions hold at a typical symbol of a random instance. For fixed $2\le k<D$, choose any fixed constant
$$
  0<c_0<\frac{1}{2\log((k-1)(D-1))}.
$$
Corollary~\ref{cor:typical_nogo} establishes typicality for $t\le c_0\log m$ in both regular ensembles.

\begin{definition}[Local decoders]\label{def:local_decoder}
Let $S$ be a set of symbols at which $\mathrm{LT}_t$ and $\mathrm{TP}_t$ hold and whose sets $N_t(i)$, $i\in S$, are pairwise disjoint. A decoder makes \emph{$t$-local estimates on $S$} if there are POVMs $\{E^{(i)}_y\}_{y\in\Z_q}$ on the qudits of $N_t(i)$, $i\in S$, such that for every $c\in C$ and every $(y_i)_{i\in S}$,
$$
  \Pr\big[\hat c_i=y_i\text{ for all }i\in S\mid c\big]=\tr\Big(\Big(\bigotimes_{i\in S}E^{(i)}_{y_i}\otimes I\Big)\ket{\psi^\Phi_c}\!\bra{\psi^\Phi_c}\Big).
$$
A sequence of decoders for matrices with $m\to\infty$ rows is \emph{local} if there are depths $t_m\le c_0\log m$, where $c_0$ is as in Corollary~\ref{cor:typical_nogo}, and numbers $s_m$ with $s_m/\log m\to\infty$ such that the following holds. The $m$-th decoder first makes a random choice that does not depend on its input, and except for choices of total probability at most $2^{-s_m}$, the decoder that it then applies makes $t_m$-local estimates on a set of at least $s_m$ symbols.
\end{definition}

The joint law specifies separate measurements, not independent outcomes: the received state may be entangled. Nor does individual tree projection imply that the code projects onto a product of several tree codes. Requiring many estimates also makes artificial randomization costly. Adding independent uniform offsets to $s$ output coordinates makes those coordinates jointly uniform, but reduces block success to at most $q^{-s}$. Regev's reduction then requires at least $q^s$ attempts in expectation, which is superpolynomial when $s/\log m\to\infty$.

\begin{lemma}[Measurements on disjoint neighborhoods]\label{lem:disjoint_measurements}
Let $\rho$ be an arbitrary received state and let the neighborhoods $N_t(i)$, $i\in S$, be disjoint. A decoder applies a fixed sequence of instruments, which may be controlled by earlier records. Each selected call uses its own unused ancillas, with the received qudits and all work registers initially in the state $\rho\otimes\bigotimes_{i\in S}\sigma_i\otimes\sigma_W$, where the ancillary states are independent of the input. It applies a fixed instrument on its neighborhood and ancillas, before any other operation touches these registers, and the instrument is not controlled by earlier records. Its outcome is stored in a fresh record $Y_i$ that later operations do not change. Let $\{E_y^{(i)}\}$ be the induced POVM on $N_t(i)$. Then, for every tuple $(y_i)_{i\in S}$,
$$
  \Pr[Y_i=y_i\text{ for all }i\in S]=\tr\Big(\Big(\bigotimes_{i\in S}E_{y_i}^{(i)}\otimes I\Big)\rho\Big).
$$
If both local conditions hold and the decoder outputs $\hat c_i=Y_i$ on every branch, it makes $t$-local estimates on $S$.
\end{lemma}

\begin{proof}
Each selected instrument commutes with the operations preceding it, since they act on disjoint registers and do not control that instrument. Move the selected instruments to the beginning in their original order. With the ancillas initially in a product state, tracing them out yields the tensor-product POVM law. Subsequent operations, summed over their outcomes, are trace preserving and leave the records unchanged, so they preserve this joint law.
\end{proof}

There are sufficiently many disjoint neighborhoods throughout the permitted radius range. With probability tending to one, all but $o(m)$ symbols satisfy both local conditions. For $\delta=\tfrac12-c_0\log((k-1)(D-1))>0$, each ball meets at most $O(m^{1-2\delta})$ other balls. Greedily selecting disjoint good balls therefore gives $\Omega(m^{2\delta})$ symbols. At depth $O(\log\log m)$, it gives $m^{1-o(1)}$ symbols.

\begin{corollary}[Uniformly random sequential orders]\label{cor:random_order_local}
Fix $q\ge2$, $2\le k<D$, and $c_0$ as above, and let $t_m\le c_0\log m$ be nonnegative integers. Let $J_m$ be determined by the instance, with $|J_m|\ge\lambda m$ with probability tending to one for some fixed $\lambda>0$. Each symbol in $J_m$ has a fixed instrument on its neighborhood and fresh independent ancillas. The instrument may include local operations after recording its estimate, but does not depend on estimates at other symbols. Process these instruments in a uniformly random order and retain every recorded estimate in the output on every branch. With probability tending to one over either regular ensemble, this decoder is local, for every amplitude function, with
$$
  s_m=\lceil(\log m)^2\rceil,\qquad\Pr[\text{exceptional order}\mid B]\le4^{-s_m}\le2^{-s_m}.
$$
\end{corollary}

The proof in Appendix~\ref{app:random_order_local} examines the first $2s_m$ calls and keeps good symbols whose balls meet no earlier processed ball, including balls of rejected symbols. Lemma~\ref{lem:disjoint_measurements} gives their joint law. Typicality and the overlap bound imply that fewer than $s_m$ symbols are kept with probability at most $4^{-s_m}$.

A fixed instrument may contain operations controlled by heralds generated within that call; it excludes control by earlier symbol estimates. Thus Corollary~\ref{cor:random_order_local} applies to the sequential BPQM calls described in Section~\ref{sec:bpqm} when implemented with fresh work registers and the stated output convention~\cite{piveteau2022messagepassing,mandal2026blockerror}. Classical belief propagation with a fixed number of flooding iterations after separate qudit measurements also gives joint local estimates on disjoint balls, because each hard decision is a function of the measurement outcomes in its ball.

\paragraph{An initial local estimate.} Fixed sequential orders need not satisfy Definition~\ref{def:local_decoder}, but their block success can still be bounded through an estimate obtained before its input is disturbed. To formulate this condition, fix $B$ and $\Phi$. Let the decoder draw a seed $R$ from a finite set, with a law that may depend on $B$ and $\Phi$ but not on the uniformly distributed codeword $c$. For a set $\mathcal R_0$ of seeds of probability $g>0$, suppose that each $r\in\mathcal R_0$ determines a symbol $i_r$, a radius $t_r$ satisfying both local conditions, and an estimate $Y_r$ whose law is that of a POVM on the original received qudits of $N_{t_r}(i_r)$:
\begin{equation}\label{eq:first_estimate_law}
  \Pr[Y_r=y\mid c,R=r]=\tr\big((E_y^{(r)}\otimes I)\ket{\psi_c^\Phi}\!\bra{\psi_c^\Phi}\big)\qquad\text{for every }c\in C\text{ and }y\in\Z_q.
\end{equation}
Later operations may be arbitrary. To account for the possibility that they correct an initially wrong estimate, define
\begin{equation}\label{eq:first_estimate_repair}
  a=\Pr[\hat c=c,\ R\in\mathcal R_0,\ Y_R\ne c_{i_R}],
\end{equation}
the probability of correct block recovery despite an incorrect initial estimate. The bounds in Section~\ref{sec:only_nonrecon} use $g$ and $a$ without requiring joint locality at many symbols.

\begin{lemma}[Output conventions]\label{lem:first_estimate_output}
If, for every $r\in\mathcal R_0$, the decoder retains $Y_r$ at coordinate $i_r$ whenever it outputs a codeword, then $a=0$. Alternatively, suppose that, for every $r\in\mathcal R_0$, whenever the output differs from $Y_r$ at that coordinate, it is a fixed vector $d_r$ determined by the seed alone. Then $a\le1/|C|\le q^{-(m-n)}$.
\end{lemma}

\begin{proof}
In the first case, correct recovery with an incorrect retained estimate is impossible. In the second, that event requires $c=d_R$. Since $c$ is uniform and independent of $R$, its probability is at most $1/|C|$. Finally, $|C|=q^m/|\operatorname{im}B^\top|\ge q^{m-n}$.
\end{proof}

For a first coordinate fixed independently of the instance, Corollary~\ref{cor:typical_nogo} makes its neighborhood good with probability tending to one. The same applies to a fixed-order scan that includes this coordinate whenever its neighborhood is good. For binary codes, greedy information-set selection admits a good first coordinate because its value is not constant on $C$. If the first coordinate is uniform in an instance-determined linear-sized set, its probability of being good tends to one because only $o(m)$ coordinates are bad. These facts justify the standard orders in Section~\ref{sec:ov_setting}; arbitrary instance-dependent choices require a separate check.

The reconstruction correspondence in Section~\ref{sec:tree} holds at every tree depth and in the infinite-depth limit. The finite-instance bound in Section~\ref{sec:locality} requires the local conditions, not a particular growth rate for the radius. Our typicality theorem guarantees those conditions for $t\le c_0\log m$, but does not determine the reliable-decoding range of unrestricted-depth BPQM on finite cyclic graphs. A retained estimate from a neighborhood satisfying both local conditions still constrains block success, regardless of later depth.

\section{Optimal local decoding and tree reconstruction}\label{sec:tree}

The local input studied in Section~\ref{sec:strategy} is a tree code whenever the neighborhood is a tree and the rest of the code imposes no additional constraints inside it. We therefore determine the largest probability $P_t$ of correctly decoding its root symbol. The main conclusion is that $P_t\to1$ precisely when the primal Gibbs measure has no reconstruction (Theorem~\ref{thm:duality}). BPQM attains $P_t$, so the correspondence bounds every measurement on the same tree input.

The proof has three steps. The lossless node operations in Section~\ref{sec:nodes} reduce root decoding to a random recursion on message spectra. Section~\ref{sec:spectral} defines this recursion and shows that its output determines $P_t$. In Sections~\ref{sec:primal_recon} and~\ref{sec:duality}, we match it to the recursion for the posterior of a primal variable given distant observations. This comparison holds at every tree depth; the logarithmic radius used for finite random instances is not a restriction on the tree theorem.

The binary operations and their tree optimality are established in~\cite{renes2017bpqm,rengaswamy2021bpqm,piveteau2022messagepassing}. Their extensions to prime alphabets and finite abelian groups are given in~\cite{mandal2026qary,mandal2026abelian,mandal2026blockerror}. We state the operations over $\Z_q$ and include their proofs to make the correspondence self-contained. Finally, Section~\ref{sec:monotone} explains when non-reconstruction holds on an interval of inverse temperatures, and Section~\ref{sec:binary_bound} gives an explicit lower bound on the binary decoding error.

\subsection{Two lossless node operations}\label{sec:nodes}

We process the tree from the leaves to the root. The following lemma describes the operation at each type of node and shows that neither operation loses information about the symbol values.

For probability vectors $\lambda,\mu$ on $\Z_q$, let $(\lambda*\mu)(z)=\sum_w\lambda(w)\mu(z-w)$ and $R\lambda(z)=\lambda(-z)$. Write $\unif$ for the uniform vector.

\begin{lemma}[Node operations]\label{lem:nodes}
Let $\lambda,\mu$ be spectra.
\begin{enumerate}
\item \emph{Symbol node.} There is an isometry $V$, independent of $y$, with $V\ket{\psi^{\lambda*\mu}_y}=\ket{\psi^\lambda_y}\ket{\psi^\mu_y}$ for all $y$.
\item \emph{Check node.} For $y\in\Z_q$ let
$$
  \rho_y=\frac1q\sum_{y_1\in\Z_q}\ket{\psi^\lambda_{y_1}}\!\bra{\psi^\lambda_{y_1}}\otimes\ket{\psi^\mu_{y-y_1}}\!\bra{\psi^\mu_{y-y_1}}
$$
be the state of two symbols with uniformly random values $y_1,y_2$ conditioned on $y_1+y_2=y$. For $\delta\in\Z_q$ let $\Pi_\delta=\sum_{s-l=\delta}\ket{s,l}\!\bra{s,l}$. Then $\rho_y=\sum_\delta\Pi_\delta\rho_y\Pi_\delta$, the probability $\tr(\Pi_\delta\rho_y)=P_{\lambda,\mu}(\delta)\coloneqq\sum_l\lambda(l+\delta)\mu(l)$ does not depend on $y$, and, for every $\delta$ with $P_{\lambda,\mu}(\delta)>0$,
$$
  \frac{\Pi_\delta\rho_y\Pi_\delta}{P_{\lambda,\mu}(\delta)}=W_\delta\ket{\psi^{\nu_\delta}_y}\!\bra{\psi^{\nu_\delta}_y}W_\delta^\dagger,
  \qquad
  \nu_\delta(l)=\frac{\lambda(l+\delta)\mu(l)}{P_{\lambda,\mu}(\delta)},
$$
for an isometry $W_\delta$ that does not depend on $y$.
\end{enumerate}
\end{lemma}

\begin{proof}
\emph{Symbol node.} Both families have Gram matrix $G_{y'y}=\sum_z(\lambda*\mu)(z)\,\omega^{(y-y')z}$. Two families of vectors with the same Gram matrix are related by an isometry from the span of one onto the span of the other, and this isometry extends to the whole space.

\emph{Check node.} In the computational basis, $\ket{\psi^\lambda_{y_1}}\ket{\psi^\mu_{y-y_1}}=\sum_{s,l}\sqrt{\lambda(s)\mu(l)}\,\omega^{y_1(s-l)+yl}\ket{s,l}$. Averaging the projector over $y_1$ removes every matrix element between $\ket{s,l}$ and $\ket{s',l'}$ with $s-l\ne s'-l'$, so the averaged state is block diagonal.

Inside the block $s-l=\delta$, the factor $\omega^{y_1\delta}$ is a global phase. The block therefore contains the single pure state $\sum_l\sqrt{\lambda(l+\delta)\mu(l)}\,\omega^{yl}\ket{l+\delta,l}$, whose squared norm is $P_{\lambda,\mu}(\delta)$. The isometry $W_\delta:\ket l\mapsto\ket{l+\delta,l}$ sends $\sqrt{P_{\lambda,\mu}(\delta)}\ket{\psi^{\nu_\delta}_y}$ to this vector.
\end{proof}

In a coherent implementation of BPQM, the label $\delta$ is computed into an ancilla rather than measured. Since $\rho_y$ is block diagonal in $\delta$, the success probability is the same in both cases, and we treat $\delta$ as a measured classical outcome.

Define the random \emph{check-node operation} $\Gamma(\lambda,\mu)=\nu_\Delta$, where $\Delta$ has law $P_{\lambda,\mu}$. Thus $\Gamma(\lambda,\mu)$ is the spectrum of the outgoing message after the label has been measured. For several arguments, set $\Gamma(\lambda_1)=\lambda_1$ and $\Gamma(\lambda_1,\dots,\lambda_h)=\Gamma(\Gamma(\lambda_1,\dots,\lambda_{h-1}),\lambda_h)$, with fresh randomness at each step. We use two facts about $\Gamma$, the first in Section~\ref{sec:duality} and the second in the proof of Proposition~\ref{prop:tree_optimal}. First, if $\lambda$ is replaced by its translate $\lambda(\cdot+g)$, then $P_{\lambda,\mu}$ becomes $P_{\lambda,\mu}(\cdot+g)$ and $\nu_\delta$ becomes $\nu_{\delta+g}$. If $\mu$ is replaced by $\mu(\cdot+g)$, then $P_{\lambda,\mu}$ becomes $P_{\lambda,\mu}(\cdot-g)$ and $\nu_\delta$ becomes the translate $\nu_{\delta-g}(\cdot+g)$. Hence the law of the translation orbit of $\Gamma(\lambda,\mu)$ depends only on the translation orbits of $\lambda$ and $\mu$. Second, the unitary $\ket z\mapsto\ket{-z}$ maps $\ket{\psi^\nu_{-y}}$ to $\ket{\psi^{R\nu}_y}$ for every $y$.

\subsection{The spectral recursion}\label{sec:spectral}

Applying Lemma~\ref{lem:nodes} at every node of the tree turns the decoding problem for the root into a random recursion on spectra. We first define the tree and the recursion, and then prove that the recursion determines the optimal success probability.

Let $\mathcal T_t$ be the $(k,D)$ tree of depth $t$ rooted at a symbol. The root has $k$ child checks, every check has $D-1$ child symbols, every non-root symbol at depth less than $t$ has $k-1$ child checks, and the symbols at depth $t$, measured in check layers, are the leaves. The \emph{tree code} $\mathcal C_t$ is the set of assignments of values in $\Z_q$ to the symbols of $\mathcal T_t$ that satisfy every check. No other constraint acts on the leaves.

\begin{definition}[Spectral recursion]\label{def:spectral}
Let $V^{(0)}=p$ and, for $t\ge0$,
$$
  C^{(t)}=R\,\Gamma\big(V^{(t)}_1,\dots,V^{(t)}_{D-1}\big),
  \qquad
  V^{(t+1)}=p*C^{(t)}_1*\cdots*C^{(t)}_{k-1},
$$
where the arguments on each right-hand side are independent copies. The root spectrum is $N^{(0)}=p$ and $N^{(t)}=p*C^{(t-1)}_1*\cdots*C^{(t-1)}_k$ for $t\ge1$. Define
$$
  S(\nu)=\frac1q\Big(\sum_{z\in\Z_q}\sqrt{\nu(z)}\Big)^2,
  \qquad
  P_t=\E\,S\big(N^{(t)}\big).
$$
\end{definition}

\begin{proposition}[Optimal tree decoding]\label{prop:tree_optimal}
Draw $y$ uniformly from $\mathcal C_t$ and give a decoder the state $\bigotimes_{v\in\mathcal T_t}\ket{\psi^p_{y_v}}$. The largest probability, over all measurements, of outputting the root value $y_{\mathrm{root}}$ is $P_t$, and depth-$t$ BPQM over $\Z_q$, described in Section~\ref{sec:bpqm}, attains it. Moreover $P_t\le P_{t+1}\le1$, and $S(\nu)=1$ if and only if $\nu=\unif$.
\end{proposition}

\begin{proof}
Every check imposes one linear equation, and the tree has no cycles. Hence the root value is uniform on $\Z_q$. Conditional on the value of a symbol, the subtrees below its child checks are independent. Each child check fixes only the sum of its $D-1$ child symbols, and these child symbols are otherwise uniform.

We claim, by induction on the height $h$ of a non-root symbol, with leaves at height $0$, that conditional on the value $y$ of the symbol, the state of its subtree is, up to an isometry independent of $y$,
\begin{equation}\label{eq:flagged}
  \sum_{\mathbf a}P(\mathbf a)\,\ket{\mathbf a}\!\bra{\mathbf a}\otimes\ket{\psi^{\nu_{\mathbf a}}_y}\!\bra{\psi^{\nu_{\mathbf a}}_y},
\end{equation}
where $\mathbf a$ is a classical record whose law does not depend on $y$, and $\nu_{\mathbf a}$ has the law of $V^{(h)}$.

\emph{Leaves.} The claim holds with $\nu=p$ and an empty record.

\emph{Check nodes.} Consider a check whose $D-1$ children have independent states of the form~\eqref{eq:flagged}. If the parent symbol has value $y$, the children sum to $-y$. Apply Lemma~\ref{lem:nodes}(2), with $-y$ in place of $y$, to the children one at a time, conditional on their records. The projections $\Pi_\delta$ are block diagonal for every value of the sum, and their probabilities do not depend on it. After the outcomes $\delta$ are added to the record, the state has the form~\eqref{eq:flagged} with $-y$ in place of $y$ and spectrum $\Gamma(V_1,\dots,V_{D-1})$.

The unitary $\ket z\mapsto\ket{-z}$ changes the sign of the encoded symbol and reflects the spectrum. The state therefore has the form~\eqref{eq:flagged} with value $y$ and spectrum $R\,\Gamma(V_1,\dots,V_{D-1})$, which has the law of $C^{(h)}$.

\emph{Symbol nodes and the root.} At a symbol with own spectrum $p$ and $k-1$ child checks, Lemma~\ref{lem:nodes}(1) combines the pure parts into the spectrum $p*C_1*\cdots*C_{k-1}$, which has the law of $V^{(h+1)}$. The same step at the root, which has $k$ child checks, produces the spectrum $N^{(t)}$.

The root state is block diagonal in the record basis, with weights that do not depend on $y$. For any POVM $\{E_y\}$, the compressions $E_y^{\mathbf a}=(\bra{\mathbf a}\otimes I)E_y(\ket{\mathbf a}\otimes I)$ form a POVM for each $\mathbf a$, and the success probability equals $\sum_{\mathbf a}P(\mathbf a)\,q^{-1}\sum_y\tr\big(E^{\mathbf a}_y\ket{\psi^{\nu_{\mathbf a}}_y}\!\bra{\psi^{\nu_{\mathbf a}}_y}\big)$. The optimum is therefore the average over the record of the optimal success probability for the family $\{\ket{\psi^\nu_y}\}_y$ with uniform prior. This family is geometrically uniform under the cyclic group generated by $\ket z\mapsto\omega^z\ket z$, so the square-root measurement is optimal~\cite{eldar2001square}. Its success probability is $q^{-1}\sum_y|\langle\mu_y|\psi^\nu_y\rangle|^2$ with $\ket{\mu_y}=q^{-1/2}\sum_z\omega^{yz}\ket z$, which equals $S(\nu)$. The measurement in the orthonormal basis $\{\ket{\mu_y}\}$, which applies the inverse quantum Fourier transform and measures in the computational basis, has the same success probability. Depth-$t$ BPQM applies the node operations above and then this measurement at the root, so it attains $P_t$.

For monotonicity in $t$, the restriction of a uniform element of $\mathcal C_{t+1}$ to $\mathcal T_t$ is a uniform element of $\mathcal C_t$, because every element of $\mathcal C_t$ has the same number of extensions. The depth-$t$ problem is therefore obtained from the depth-$(t+1)$ problem by discarding qudits, which cannot increase the optimal success probability. Finally, by the Cauchy--Schwarz inequality, $(\sum_z\sqrt{\nu(z)})^2\le q$, with equality if and only if $\nu$ is uniform.
\end{proof}

\paragraph{The binary case.} For $q=2$, encode a spectrum $\nu$ by its overlap $v=\nu(0)-\nu(1)=\langle\psi^\nu_0|\psi^\nu_1\rangle$. Then $R$ is the identity, convolution multiplies overlaps, and $\Gamma$ sends overlaps $(a,b)$ to $(a+b)/(1+ab)$ with probability $(1+ab)/2$ and to $(b-a)/(1-ab)$ with probability $(1-ab)/2$. These are the check-node rule and the label probabilities of BPQM described in Section~\ref{sec:bpqm}, and the multiplication of overlaps is its symbol-node rule. The success function is the Helstrom value $S=\tfrac12(1+\sqrt{1-v^2})$. Proposition~\ref{prop:tree_optimal} therefore recovers the optimality of binary BPQM on trees~\cite{renes2017bpqm,rengaswamy2021bpqm,piveteau2022messagepassing}. For prime $q$, the $q$-ary BPQM of Mandal and Pfister~\cite{mandal2026qary} performs the node operations of Lemma~\ref{lem:nodes} followed by the square-root measurement, and Mandal et al.\ show that it is optimal on trees~\cite{mandal2026blockerror}.

\subsection{The primal reconstruction problem}\label{sec:primal_recon}

To determine when $P_t\to1$, we compare the spectral recursion with the primal recursion, which we define now. The primal recursion runs on trees with the same branching numbers as the trees of the spectral recursion, with the roles of the two node types exchanged: code symbols become clauses, and checks become variables. Let $\mathcal T'_t$ be the tree rooted at a primal variable $a_0$ that has $D-1$ child clauses. Every clause has $k-1$ child variables, and every non-root variable above the leaves has $D-1$ child clauses. The leaves are the variables after $t$ clause layers. Draw the variable values $z^\star$ uniformly and, independently for each clause $j$, draw $u_j\sim p$. Reveal $w_j=u_j-\sum_{a\in\partial j}z^\star_a$ for each clause and the values $z^\star_a$ of all leaves. Given these observations, the posterior law of the variable values is the primal Gibbs measure on the tree with clause weights $p\big(\sum_{a\in\partial j}z_a+w_j\big)$ and with the leaves fixed. Let
$$
  M^{(t)}(x)=\Pr\big[z_{a_0}=z^\star_{a_0}+x\ \big|\ \text{observations}\big],\qquad x\in\Z_q,
$$
be the posterior of the root, measured relative to the true value.

\begin{definition}[Reconstruction]
The primal model at $(k,D,q,f,\beta)$ has \emph{non-reconstruction} if $M^{(t)}\to\unif$ in probability as $t\to\infty$, and \emph{reconstruction} otherwise.
\end{definition}

This is the standard notion of tree reconstruction for the planted model~\cite{mezard2006reconstruction}, in which the observations are generated from a hidden assignment $z^\star$. The cavity method predicts that the reconstruction threshold, defined in Section~\ref{sec:monotone}, equals the dynamical threshold of the corresponding random sparse model~\cite{mezard2006reconstruction}. The standard definition roots the tree at a variable with $D$ child clauses. The next lemma shows that the two choices of root degree define the same notion.

\begin{lemma}\label{lem:root_degree}
Let $\widetilde M^{(t)}$ be the root posterior, relative to the truth, when the root variable has $D$ child clauses and the rest of the tree is as above. Then $M^{(t)}\to\unif$ in probability if and only if $\widetilde M^{(t)}\to\unif$ in probability.
\end{lemma}

\begin{proof}
For $s\in\Z_q$, write $\hat\nu(s)=\sum_x\nu(x)\omega^{sx}$. Since $M^{(t)}$ is a posterior measured relative to the truth, $\E\,\hat M^{(t)}(s)=\E\big|\E[\omega^{sz^\star_{a_0}}\mid\text{observations}]\big|^2$. This quantity is nonnegative and does not decrease when observations are added. Also $M\to\unif$ in probability if and only if $\E|\hat M(s)|^2\to0$ for each $s\ne0$, and $|\hat M(s)|^2$ equals the integrand above. The tree for $\widetilde M^{(t)}$ contains the tree for $M^{(t)}$ together with one more subtree, so $\widetilde M^{(t)}\to\unif$ implies $M^{(t)}\to\unif$. Conversely, the root posterior with $D$ child clauses is the normalized product of $D$ independent clause messages. Each clause message is a translate of $p*RM_1*\cdots*RM_{k-1}$, as computed in the proof of Theorem~\ref{thm:duality} below, with $M_b$ distributed as $M^{(t-1)}$. If $M^{(t-1)}\to\unif$ in probability, each clause message tends to $p*\unif=\unif$ in probability, and so does the normalized product of $D$ such messages.
\end{proof}

\paragraph{Nishimori property.} To match the random check-node label to the primal recursion, we need the law of the unknown translation of a posterior message. Write $M=M^{(t)}$ and let $[M]=\{M(\cdot+g):g\in\Z_q\}$ be its translation orbit. Fix a representative $o\in[M]$ as a function of the orbit. Conditional on $M$, draw $G$ uniformly from $\{g:M=o(\cdot+g)\}$ using independent randomness. This set is a coset of the stabilizer of $o$, and $o$ is constant on cosets of its stabilizer.

Given the observations, $z^\star_{a_0}$ is distributed according to the unshifted posterior, which we write as $o(\cdot+h)$. Thus $G=h+z^\star_{a_0}+G'$ with $G'$ uniform on the stabilizer $\Sigma$ of $o$, and $\Pr[G=g\mid\text{observations}]=\sum_zo(z+h)\Pr[G'=g-h-z]=|\Sigma|^{-1}\sum_{g'\in\Sigma}o(g-g')=o(g)$. Since $[M]$ is a function of the observations,
\begin{equation}\label{eq:nishimori}
  \Pr\big[G=g\mid [M]\big]=o(g).
\end{equation}
In words, the unknown translation of a posterior message is distributed according to the message itself. The same holds for every posterior message on the tree, and messages from disjoint subtrees are independent.

\subsection{The duality theorem}\label{sec:duality}

With both recursions defined, we can compare them. We compare their translation orbits and their Fourier magnitudes, which are the quantities that determine whether $P_t\to1$ and whether the primal model has reconstruction. For a probability vector $\nu$ on $\Z_q$, the magnitudes $|\hat\nu(s)|$ do not change under translation and reflection, $\widehat{\lambda*\mu}=\hat\lambda\hat\mu$, and $\nu=\unif$ if and only if $\hat\nu(s)=0$ for every $s\ne0$.

To prove the duality, we match the two recursions node by node. The quantum symbol node convolves spectra, and the primal clause node also convolves messages. The quantum check node produces a label $\delta$ with probability $P_{\lambda,\mu}(\delta)$ and returns $\nu_\delta$. The primal variable node multiplies two posterior messages. By~\eqref{eq:nishimori}, each message is a translate of a representative of its translation orbit, and the translation is distributed according to that representative. The product depends only on the relative translation of the two messages, and the law of the relative translation is exactly $P_{\lambda,\mu}$. Measuring the label at a quantum check node and multiplying posteriors at a primal variable node are therefore the same random operation. This is the exchange of node types familiar from the duality of codes. The equality of laws then implies that the two thresholds coincide, because Fourier magnitudes do not change under translation and multiply under convolution.

\begin{theorem}[Decoder--reconstruction duality]\label{thm:duality}
For every $q\ge2$, every cost $f$, every $\beta\ge0$, and every $t\ge0$:
\begin{enumerate}
\item The translation orbits satisfy $[C^{(t)}]\overset d=[R\,M^{(t+1)}]$.
\item $P_t\to1$ if and only if the primal model has non-reconstruction.
\end{enumerate}
\end{theorem}

\begin{proof}
Let $U^{(t)}$ denote the message, relative to the truth, that a clause sends to its parent variable when its $k-1$ child variables send $M^{(t)}_1,\dots,M^{(t)}_{k-1}$. We prove $[U^{(t)}]\overset d=[V^{(t)}]$ and $[M^{(t+1)}]\overset d=[\Gamma(V^{(t)}_1,\dots,V^{(t)}_{D-1})]$ together by induction on $t$.

\emph{Clause step.} Write the parent variable as $z^\star_a+x$ and each child as $z^\star_b+y_b$. Since $w_j+z^\star_a+\sum_bz^\star_b=u_j$,
$$
  U(x)\propto\sum_{y}p\Big(x+\sum_by_b+u_j\Big)\prod_bM_b(y_b)=(p*RQ)(x+u_j),
  \qquad Q=M_1*\cdots*M_{k-1}.
$$
Hence $[U]=[p*RM_1*\cdots*RM_{k-1}]$, and the orbit of a convolution depends only on the orbits of its factors. In the base case the leaves are revealed, so $M^{(0)}=\delta_0$ and $[U^{(0)}]=[p]=[V^{(0)}]$.

\emph{Variable step.} The message that a variable sends to its parent clause is $M\propto\prod_jU_j$, where the product runs over its $D-1$ child clauses and the factors are independent. Consider two factors and write $U_i=o_i(\cdot+G_i)$ with $G_i$ distributed as in~\eqref{eq:nishimori}, independently. Substituting $x=l-G_2$,
$$
  o_1(x+G_1)\,o_2(x+G_2)=o_1(l+\Delta)\,o_2(l),
  \qquad \Delta=G_1-G_2,
$$
and $\Pr[\Delta=\delta]=\sum_lo_1(l+\delta)o_2(l)=P_{o_1,o_2}(\delta)$.
So $[U_1U_2/\|U_1U_2\|_1]=[\nu_\Delta]$ with $\Delta\sim P_{o_1,o_2}$, which is the orbit of $\Gamma(o_1,o_2)$. The normalized product is again a posterior message, so~\eqref{eq:nishimori} holds for it, and the step can be repeated for all $D-1$ factors. By the induction hypothesis $[U^{(t)}]\overset d=[V^{(t)}]$, and the law of the orbit of $\Gamma$ depends only on the orbits of its arguments. Therefore $[M^{(t+1)}]\overset d=[\Gamma(V^{(t)}_1,\dots,V^{(t)}_{D-1})]$, and so $[C^{(t)}]\overset d=[RM^{(t+1)}]$. This proves part 1. By part 1 and the clause step, $[U^{(t+1)}]=[p*RM^{(t+1)}_1*\cdots*RM^{(t+1)}_{k-1}]\overset d=[p*C^{(t)}_1*\cdots*C^{(t)}_{k-1}]=[V^{(t+1)}]$, which completes the induction.

For part 2, $S$ is continuous on the simplex and equals one only at $\unif$. By compactness, $P_t=\E S(N^{(t)})\to1$ if and only if $N^{(t)}\to\unif$ in probability, which holds if and only if $\E|\hat N^{(t)}(s)|\to0$ for each $s\ne0$. By independence and part 1,
$$
  \E\big|\hat N^{(t)}(s)\big|=|\hat p(s)|\,\Big(\E\big|\hat C^{(t-1)}(s)\big|\Big)^{k}=|\hat p(s)|\,\Big(\E\big|\hat M^{(t)}(s)\big|\Big)^{k}.
$$
\emph{Non-reconstruction implies reliable decoding.} If the primal model has non-reconstruction, the right-hand side tends to zero for every $s\ne0$, so $P_t\to1$.

\emph{Reliable decoding implies non-reconstruction.} Suppose that $P_t\to1$, and let $K'$ be the set of $s\ne0$ with $\hat p(s)\ne0$. Then $\E|\hat M^{(t)}(s)|\to0$ for every $s\in K'$. By the clause step, the message $U^{(t)}$ that a clause sends to its parent is a translate of $p*RM_1*\cdots*RM_{k-1}$ with $M_b$ independent copies of $M^{(t)}$, so $|\hat U^{(t)}(s)|=|\hat p(s)|\prod_b|\hat M_b(s)|$. This is zero for $s\notin K'\cup\{0\}$ and tends to zero in probability for $s\in K'$. Hence $U^{(t)}\to\unif$ in probability.

The posterior $M^{(t+1)}$ is the normalized product of $D-1$ independent copies of $U^{(t)}$. The normalized product of finitely many probability vectors converging to $\unif$ also converges to $\unif$, so $M^{(t)}\to\unif$ in probability. This is non-reconstruction.
\end{proof}

\begin{remark}
Mandal et al.~\cite{mandal2026blockerror} draw the nonzero entries of $B$ uniformly at random from the units of $\Z_q$. For such matrices, the spectral recursion can be adapted by relabeling spectra through the corresponding automorphisms of $\Z_q$. Our matrices have all nonzero entries equal to one, and for $q>2$ these matrices have vanishing probability under random units, so Proposition~\ref{prop:sharp} does not follow from their results; it follows their two-stage decoder, and we prove the required properties of these codes in Appendix~\ref{app:sharp}.
\end{remark}

\subsection{Monotonicity in the inverse temperature}\label{sec:monotone}

Theorem~\ref{thm:duality} characterizes success at a fixed inverse temperature. To speak of a single threshold, we also need to know how success depends on $\beta$. We first define the threshold in a way that needs no monotonicity, and then give a condition under which the set of inverse temperatures with non-reconstruction is an interval. In this subsection we write $p_\beta$ for the weights $p$ at inverse temperature $\beta$, and $P_t(\beta)$ for the corresponding optimal success probability.

\begin{definition}[Reconstruction threshold]\label{def:beta_d}
For fixed $(k,D,q,f)$, let $\mathcal N$ be the set of $\beta\ge0$ at which the primal model has non-reconstruction. The \emph{reconstruction threshold} is $\beta_d=\sup\mathcal N\in[0,\infty]$.
\end{definition}

At $\beta=0$ the weights are uniform, every clause message is uniform, and the model has non-reconstruction, so $\mathcal N$ is not empty. Every $\beta>\beta_d$ lies outside $\mathcal N$, so the primal model has reconstruction there. This is the only property of $\beta_d$ that our results use, and it needs no monotonicity in $\beta$.

The set of inverse temperatures without reconstruction is an interval whenever the channels at different inverse temperatures can be compared in the following sense. Suppose that the spectrum $p_{\beta'}$ at a smaller inverse temperature $\beta'$ is obtained from $p_\beta$ by convolution with a probability vector. Then the symbol-node identity of Lemma~\ref{lem:nodes} turns each state at $\beta'$ into a state at $\beta$ together with an extra register, which can be discarded. Every success probability that a decoder reaches at $\beta$ is therefore also reached at $\beta'$.

\begin{proposition}[Degradation]\label{prop:degrade}
Let $0\le\beta'<\beta$, and suppose that there is a probability vector $\kappa$ on $\Z_q$ with $p_{\beta'}=p_\beta*\kappa$. Then $P_t(\beta')\ge P_t(\beta)$ for every $t$. If such a $\kappa$ exists for every pair $\beta'<\beta$, then the set of $\beta$ with non-reconstruction is $[0,\beta_d)$ or $[0,\beta_d]$.
\end{proposition}

\begin{proof}
By Lemma~\ref{lem:nodes}(1) there is an isometry $V$, independent of $y$, with $V\ket{\psi^{p_{\beta'}}_y}=\ket{\psi^{p_\beta}_y}\ket{\psi^\kappa_y}$. Applying $V$ to every qudit of the depth-$t$ tree problem at $\beta'$ and discarding the second registers produces the input of the depth-$t$ tree problem at $\beta$, with the same codeword. Any measurement for the problem at $\beta$ therefore succeeds at $\beta'$ with the same probability, so $P_t(\beta')\ge P_t(\beta)$. If such a $\kappa$ exists for all pairs, then $P_\infty(\beta)=\lim_tP_t(\beta)$ is nonincreasing in $\beta$, so the set of $\beta$ with $P_\infty(\beta)=1$ is an interval that contains $0$. By Theorem~\ref{thm:duality}, this set is the set of $\beta$ with non-reconstruction, and its supremum is $\beta_d$.
\end{proof}

The hypothesis of Proposition~\ref{prop:degrade} can be checked through an infinitesimal condition. For a probability vector $\lambda$ write $\langle f\rangle_\lambda=\sum_z\lambda(z)f(z)$.

\begin{lemma}[A criterion for degradation]\label{lem:degrade_criterion}
Let $I$ be an interval on which $\hat p_b(s)\ne0$ for every $b\in I$ and every $s$. For $b\in I$, let $\nu_b$ be the function on $\Z_q$ determined by $p_b*\nu_b=p_b\,(f-\langle f\rangle_{p_b})$. If $\nu_b(z)\ge0$ for every $z\ne0$ and every $b\in I$, then for all $\beta'<\beta$ in $I$ there is a probability vector $\kappa$ with $p_{\beta'}=p_\beta*\kappa$. Conversely, if such a $\kappa$ exists for all $\beta'<\beta$ in $I$, then $\nu_b(z)\ge0$ for every $z\ne0$ and every $b$ in the interior of $I$.
\end{lemma}

\begin{proof}
Since $p_b(z)\propto e^{-bf(z)}$, we have $\partial_bp_b=-p_b(f-\langle f\rangle_{p_b})=-p_b*\nu_b$, so $\partial_b\hat p_b(s)=-\hat p_b(s)\hat\nu_b(s)$. Since $\hat p_b(0)=1$ for every $b$, we have $\hat\nu_b(0)=0$, so the entries of $\nu_b$ sum to zero. Integrating, $\hat p_{\beta'}(s)=\hat p_\beta(s)\exp\big(\int_{\beta'}^\beta\hat\nu_b(s)\,db\big)$. Divide $[\beta',\beta]$ into $N$ intervals of length $h=(\beta-\beta')/N$ with left endpoints $b_1,\dots,b_N$. The function $\nu_b$ is continuous in $b$, so for large $N$ each vector $\delta_0+h\nu_{b_i}$ is a probability vector: its entries off $0$ are nonnegative by hypothesis, its entry at $0$ is $1+h\nu_{b_i}(0)\ge0$, and its entries sum to one. Their convolution $\kappa_N$ is a probability vector with $\hat\kappa_N(s)=\prod_i(1+h\hat\nu_{b_i}(s))$, which tends to $\exp\big(\int_{\beta'}^\beta\hat\nu_b(s)\,db\big)=\hat p_{\beta'}(s)/\hat p_\beta(s)$ as $N\to\infty$. A limit $\kappa$ of probability vectors is a probability vector, and it satisfies $p_{\beta'}=p_\beta*\kappa$. Conversely, if $p_{b-h}=p_b*\kappa_h$, then $\kappa_h$ is determined by $\hat\kappa_h(s)=\hat p_{b-h}(s)/\hat p_b(s)=1+h\hat\nu_b(s)+o(h)$. Hence $\kappa_h(z)=h\nu_b(z)+o(h)$ for $z\ne0$, and $\kappa_h(z)\ge0$ forces $\nu_b(z)\ge0$.
\end{proof}

Every cost that takes only two values satisfies the hypothesis of Proposition~\ref{prop:degrade} for all pairs $\beta'<\beta$. Write such a cost as $f(z)=c+\Delta\,\mathbf 1[z\notin A]$, where $A$ is a nonempty proper subset of $\Z_q$, $c$ is a constant, and $\Delta>0$. The Potts cost $f(z)=\mathbf 1[z\ne0]$ is the case $A=\{0\}$, and so is Max-XORSAT, with $q=2$ and $f(z)=z$. Let $a=|A|$, $b=q-a$, and let $U_A$ be the uniform probability vector on $A$. Then
\begin{equation}\label{eq:uniformmixture}
  p_\beta=\theta_\beta\,U_A+(1-\theta_\beta)\,\unif,\qquad \theta_\beta=\frac{a\big(1-e^{-\beta\Delta}\big)}{a+b\,e^{-\beta\Delta}},
\end{equation}
as one checks on $A$, where both sides equal $1/(a+be^{-\beta\Delta})$, and on its complement, where both sides equal $e^{-\beta\Delta}/(a+be^{-\beta\Delta})$. The coefficient $\theta_\beta$ increases from $0$ at $\beta=0$ to $1$ as $\beta\to\infty$, and no subgroup property of $A$ is needed. Convolution with $\delta_0$ fixes a probability vector, and convolution with $\unif$ sends it to $\unif$. Hence, for $\beta'<\beta$, the probability vector
$$
  \kappa=\frac{\theta_{\beta'}}{\theta_\beta}\,\delta_0+\Big(1-\frac{\theta_{\beta'}}{\theta_\beta}\Big)\unif
$$
satisfies $p_{\beta'}=p_\beta*\kappa$, and the set of $\beta$ with non-reconstruction is $[0,\beta_d)$ or $[0,\beta_d]$. The same comparison holds with $p_\infty=U_A$ in place of $p_\beta$: since $p_{\beta'}=U_A*\big(\theta_{\beta'}\delta_0+(1-\theta_{\beta'})\unif\big)$, the proof of Proposition~\ref{prop:degrade} gives $P_t(p_{\beta'})\ge P_t(U_A)$ for every finite $\beta'$, where $P_t(U_A)$ is the optimal success probability for the spectrum $U_A$. For the Potts cost, $\hat p_\beta(s)=\theta_\beta=(1-e^{-\beta})/(1+(q-1)e^{-\beta})$ for $s\ne0$, which equals $\sigma_\beta=\tanh(\beta/2)$ when $q=2$.

Other costs need not satisfy the hypothesis. For the clock cost $f(z)=-\cos(2\pi z/q)$ with $q\in\{5,7,8,11,13\}$, a numerical evaluation of $\nu_b$ on a grid of $b\in(0,12]$ finds negative entries off $0$, so by Lemma~\ref{lem:degrade_criterion} the channels are not ordered by degradation. Negative entries off $0$ also occur for about three quarters of the costs that we drew at random, with $q\in\{3,5,7\}$, $f(0)=0$, and the other values of $f$ uniform in $[0,3]$. For such costs we do not know whether the set of $\beta$ with non-reconstruction is an interval. For the clock cost with $q=5$ and $q=8$ on the $(3,6)$ tree, we evaluated the spectral recursion by population dynamics (Appendix~\ref{app:numerics}) with $10^4$ messages and $120$ levels at the grid points $\beta\in\{0.2,0.3,\dots,6.0\}$. The population mean of $1-S(V^{(120)})$, with $V^{(120)}$ as in Definition~\ref{def:spectral}, is below $10^{-6}$ up to $\beta=2.1$ for $q=5$ and up to $\beta=2.3$ for $q=8$, and it is positive and increasing in $\beta$ at every larger grid point, for example $0.19$ at $\beta=2.2$ for $q=5$ and $0.18$ at $\beta=2.4$ for $q=8$. A repetition at the seven grid points around each change, with $2\times10^4$ messages and $240$ levels, gives the same change. These results agree with the thresholds $2.165$ and $2.358$ of Table~\ref{tab:qary}, and they suggest that the inverse temperatures without reconstruction form an interval in these cases as well.

\subsection{The binary case: an explicit bound}\label{sec:binary_bound}

Theorem~\ref{thm:duality} shows that $1-P_t$ is bounded away from zero when the primal model has reconstruction, but it does not bound $1-P_t$ explicitly. For $q=2$ we can give an explicit lower bound in terms of the mean posterior bias of the root, which is the usual order parameter for reconstruction. Write ${\mathfrak m}_t=M^{(t)}(0)-M^{(t)}(1)\in[-1,1]$ and $\bar {\mathfrak m}_t=\E {\mathfrak m}_t$. By the Nishimori property, $\E {\mathfrak m}_t=\E {\mathfrak m}_t^2$. The sequence $\bar {\mathfrak m}_t$ is nonincreasing in $t$, because the leaves of a deeper tree carry less information about the root. Non-reconstruction is the statement $\bar {\mathfrak m}_t\to0$.

\begin{corollary}[Binary duality]\label{cor:binary}
For $q=2$ and $f(z)=z$, the optimal error for the root of $\mathcal T_t$ satisfies
$$
  1-P_t\ \ge\ \tfrac14\,\sigma_\beta^2\,\bar {\mathfrak m}_t^{2k}\ \ge\ \tfrac14\,\sigma_\beta^2\,\bar {\mathfrak m}_\infty^{2k},
  \qquad \bar {\mathfrak m}_\infty=\lim_{t\to\infty}\bar {\mathfrak m}_t .
$$
Moreover, the set of $\beta$ with non-reconstruction is an interval $[0,\beta_d)$ or $[0,\beta_d]$, and the tree threshold of BPQM equals $\beta_d$.
\end{corollary}

\begin{proof}
Let $v$ be the overlap of $N^{(t)}$. The error is $\E\tfrac12(1-\sqrt{1-v^2})\ge\tfrac14\E v^2\ge\tfrac14(\E|v|)^2$. Overlaps multiply under convolution, so by independence and Theorem~\ref{thm:duality}, $\E|v|=\sigma_\beta\,(\E|{\mathfrak m}_t|)^k\ge\sigma_\beta\,\bar {\mathfrak m}_t^k$, using $|{\mathfrak m}|\ge {\mathfrak m}^2$. The second claim is Proposition~\ref{prop:degrade} applied to the binary family. The last claim follows because $P_t$ is the success probability of depth-$t$ BPQM.
\end{proof}

\section{Local estimates in random instances}\label{sec:locality}

An acyclic neighborhood need not contain a uniformly random tree codeword: checks outside the neighborhood can constrain the values inside it. We therefore establish the tree-projection condition before applying the optimal decoding result of Section~\ref{sec:tree}. Whenever $\mathrm{LT}_t(i)$ and $\mathrm{TP}_t(i)$ hold, every $t$-local estimate at $i$ has success at most $P_t$ (Theorem~\ref{thm:locality}). This statement has no radius restriction beyond the two conditions.

For random instances, Theorem~\ref{thm:tp_typical} proves that both conditions hold at a typical symbol throughout $t\le c_0\log m$, for any fixed admissible $c_0$. Thus the infinite-tree obstruction becomes a typical-instance obstruction in this range. The notation $N_t(i)$, $F_t(i)$, and $L_t(i)$ is as in Section~\ref{sec:strategy}.

We first characterize projection through the vectors of $K$ supported in the ball and prove the local measurement bound. Section~\ref{sec:typical} then reduces projection to the kernel of the matrix outside the ball. A first-moment argument excludes additional kernel vectors: expansion controls small supports, and the zero-sum constraints reduce the entropy of large supports. We verify these steps separately for the configuration model and Gallager's ensemble.

\begin{lemma}\label{lem:tp}
The condition $\mathrm{TP}_t(i)$ holds if and only if every vector of $K$ supported in $N_t(i)$ is a $\Z_q$-linear combination of the columns of $B$ indexed by $F_t(i)$. In particular, $\mathrm{TP}_t(i)$ holds if no nonzero combination of the columns indexed by checks outside $F_t(i)$ is supported in $L_t(i)$.
\end{lemma}

\begin{proof}
Write $S=N_t(i)$. For a subgroup $H\le\Z_q^S$ let $H^\perp=\{w\in\Z_q^S:w\cdot y=0\text{ for all }y\in H\}$. By finite-group character theory, $|H^\perp|=q^{|S|}/|H|$ and $(H^\perp)^\perp=H$, so two subgroups $H_1\subseteq H_2$ agree precisely when their annihilators agree.

The annihilator of $C|_S$ consists of the vectors whose extension by zero lies in $C^\perp=K$. The annihilator of $\mathcal C_t(i)$ is the image of the transposed check map, namely the $\Z_q$-linear combinations of columns indexed by $F_t(i)$, restricted to $S$. Comparing these annihilators proves the first claim.

For the second claim, let $\kappa=\sum_az_ab_a\in K$ be supported in $S$, with $b_a$ the column indexed by check $a$. Subtract its contribution from checks in $F_t(i)$ to obtain $w=\sum_{a\notin F_t(i)}z_ab_a$, also supported in $S$. An external check meets $S$ only in $L_t(i)$, so $w$ is supported there. The hypothesis gives $w=0$, and $\kappa$ is a combination of internal columns.
\end{proof}

The next theorem makes the reduction of Section~\ref{sec:strategy} precise. The outcome of a measurement on the qudits of $N_t(i)$ depends on the codeword only through its restriction to $N_t(i)$. Under $\mathrm{LT}_t(i)$ and $\mathrm{TP}_t(i)$, this restriction is a uniformly random codeword of the tree code, so the measurement is a decoder for the problem of Proposition~\ref{prop:tree_optimal}.

\begin{theorem}[Locality bound]\label{thm:locality}
Let $B$ be $(k,D)$-regular, and suppose that $\mathrm{LT}_t(i)$ and $\mathrm{TP}_t(i)$ hold. Then every decoder for the decoding problem with the product amplitude function that makes a $t$-local estimate at $i$ has symbol error at $i$ at least
$$
  1-P_t\ \ge\ 1-P_\infty,\qquad P_\infty=\lim_{t\to\infty}P_t .
$$
If the primal model has reconstruction, then $1-P_\infty>0$. For $q=2$ and $f(z)=z$, the bound is at least $\tfrac14\sigma_\beta^2\,\bar {\mathfrak m}_\infty^{2k}$. In every case the block success of the decoder is at most $P_t$.
\end{theorem}

\begin{proof}
The success probability at $i$ is $|C|^{-1}\sum_{c\in C}\tr\big(E_{c_i}\bigotimes_{j\in N_t(i)}\ket{\psi^p_{c_j}}\!\bra{\psi^p_{c_j}}\big)$, which depends on $c$ only through $c|_{N_t(i)}$. Restriction maps the uniform law on $C$ to the uniform law on $C|_{N_t(i)}$, which equals $\mathcal C_t(i)$ by $\mathrm{TP}_t(i)$. By $\mathrm{LT}_t(i)$, this is the tree code $\mathcal C_t$, so Proposition~\ref{prop:tree_optimal} bounds success by $P_t\le P_\infty$.

Under reconstruction, Theorem~\ref{thm:duality} gives $P_t\not\to1$. Since $P_t$ is nondecreasing, $P_\infty<1$. Corollary~\ref{cor:binary} gives the explicit binary error bound. Correct block recovery requires a correct estimate at $i$, so block success is at most $P_t$.
\end{proof}

Theorem~\ref{thm:locality} applies to any instance and to any symbol at which the two local conditions hold. We need $\mathrm{TP}_t(i)$ because a local decoder could use any constraint that the rest of the code imposes on the values inside $N_t(i)$. We now show that both conditions hold at a typical symbol of a random regular instance over $\Z_q$.

\subsection{The local conditions are typical}\label{sec:typical}

Theorem~\ref{thm:locality} constrains a decoder on a random instance only if $\mathrm{LT}_t(i)$ and $\mathrm{TP}_t(i)$ hold at a typical symbol. We prove this for two ensembles of random $(k,D)$-regular instances, for every alphabet $\Z_q$. The first is the configuration model, which is standard in coding theory. The second is Gallager's ensemble, which is the ensemble used for Max-XORSAT by Jordan et al., by Anschuetz et al., and by Shutty et al.~\cite{jordan2024optimization,anschuetz2025decoded,shutty2026locallyquantumdecoders}. Throughout we assume $k<D$. This says that the design rate $1-k/D$ of $C$ is positive, or equivalently that the primal problem has more clauses than variables, which is the regime in which DQI is used.

\paragraph{The configuration model.} Give each of the $m$ symbols $k$ half-edges and each of the $n=mk/D$ checks $D$ half-edges, pair the two sets of half-edges by a uniformly random bijection, and condition on the resulting bipartite graph having no multiple edges. For fixed $(k,D)$, the probability of no multiple edges is bounded below by a positive constant~\cite{richardson2008modern}. Hence any event of probability $o(1)$ in the unconditioned model also has probability $o(1)$ in the ensemble. For $\mathrm{LT}_t$, the standard exploration argument shows that $\Pr[\mathrm{LT}_t(i)]=1-O\big(((k-1)(D-1))^{2t}/m\big)$ for a uniformly random symbol $i$~\cite{richardson2008modern}.

\paragraph{Gallager's ensemble.} Assume that $D$ divides $m$. Gallager's ensemble~\cite{gallager1962ldpc} builds the checks in $k$ layers. Each layer is a partition of the $m$ symbols into $m/D$ blocks of size $D$, and each block is a check. The first layer is fixed, and each other layer is the image of the first under an independent uniformly random permutation of the symbols. Every symbol lies in exactly one check of each layer, so $B$ is $(k,D)$-regular and its factor graph has no multiple edges. After a uniformly random relabeling of the symbols, the $k$ layers become independent uniformly random partitions. The events $\mathrm{LT}_t(i)$ and $\mathrm{TP}_t(i)$ for a uniformly random symbol $i$ do not depend on the labels, so for these events we may work with independent uniformly random partitions.

We first reduce $\mathrm{TP}_t(i)$ to a condition on a kernel. Write $E$ for the set of checks not in $F_t(i)$, and let $M$ be the submatrix of $B$ with rows $[m]\setminus N_t(i)$ and columns $E$.

\begin{lemma}\label{lem:tp_kernel}
If every $z\in\ker M$ is the restriction to $E$ of some $\kappa\in\ker B$, then $\mathrm{TP}_t(i)$ holds.
\end{lemma}

\begin{proof}
Write $S=N_t(i)$ and $F=F_t(i)$. By Lemma~\ref{lem:tp}, it suffices to show that every $w=Bz\in K$ supported in $S$ is a combination of the columns indexed by $F$. Split $z=z_F+z_E$. The columns indexed by $F$ are supported in $S$, so $Bz_E=w-Bz_F$ is supported in $S$, and hence $Mz_E=0$. By hypothesis $z_E=\kappa|_E$ for some $\kappa\in\ker B$. Then $Bz_E=-B\kappa_F$, where $\kappa_F$ is the part of $\kappa$ on $F$, and so $w=B(z_F-\kappa_F)$ is a combination of the columns indexed by $F$.
\end{proof}

A vector $z\in\ker M$ assigns a value in $\Z_q$ to each check in $E$, which we call an external check, so that at every symbol outside $N_t(i)$ the values of its checks sum to zero. The leaves are exempt from this condition. Some vectors of $\ker M$ are always present, and they satisfy the hypothesis of Lemma~\ref{lem:tp_kernel} because they are restrictions of vectors of $\ker B$. In the configuration model, let $\Lambda_k=\{\gamma\in\Z_q:k\gamma=0\}$. Every symbol outside $N_t(i)$ has all its $k$ checks in $E$, so $\gamma\mathbf 1_E\in\ker M$ for $\gamma\in\Lambda_k$, and $\gamma\mathbf 1_E$ is the restriction of $\gamma\mathbf 1\in\ker B$. For $q=2$, the set $\Lambda_k$ is $\{0,1\}$ if $k$ is even and $\{0\}$ if $k$ is odd. In Gallager's ensemble, let $\mathbf 1_j\in\Z_q^n$ be the indicator of the checks of layer $j$. Every symbol lies in exactly one check of each layer, so $\sum_ja_j\mathbf 1_j\in\ker B$ whenever $\sum_ja_j=0$, and its restriction to $E$ lies in $\ker M$. The next theorem shows that, typically, $\ker M$ contains no vector that violates the hypothesis of Lemma~\ref{lem:tp_kernel}.

The proof counts the vectors of $\ker M$ by the first-moment method. Vectors with small support are rare because random sparse graphs expand. For vectors with support of linear size we use the following counting lemma. It says that the requirement that the values at every symbol sum to zero reduces the entropy by at least the entropy of one coordinate per symbol, because one coordinate is determined by the others. Because there are more symbols than checks, this reduction exceeds the entropy of the values on the checks. We write $H(\pi)=-\sum_\gamma\pi(\gamma)\ln\pi(\gamma)$ for the entropy of a probability vector $\pi$ on $\Z_q$, and $\mathcal A_{q,k}=\{b\in\Z_q^k:\sum_jb_j=0\}$, a set of size $q^{k-1}$.

\begin{lemma}[Counting zero-sum labelings]\label{lem:even_count}
Let $q\ge2$, $k\ge2$, and $s\ge1$.
\begin{enumerate}
\item For each $j\in[k]$ let $\pi_j$ be a probability vector on $\Z_q$ with $s\pi_j(\gamma)\in\mathbb Z$. The number of sequences $(b(1),\dots,b(s))\in\mathcal A_{q,k}^s$ in which, for every $j$ and $\gamma$, exactly $s\pi_j(\gamma)$ entries have $j$-th coordinate $\gamma$ is at most
$$
  (s+1)^{q^{k-1}}\exp\Big(s\Big[\sum_jH(\pi_j)-\max_jH(\pi_j)\Big]\Big).
$$
It is zero unless $\pi_j(\gamma)\le\sum_{i\ne j}(1-\pi_i(0))$ for every $j$ and every $\gamma\ne0$.
\item Let $\pi$ be a probability vector on $\Z_q$ with $k s\pi(\gamma)\in\mathbb Z$. The number of sequences in $\mathcal A_{q,k}^s$ in which, for every $\gamma$, exactly $k s\pi(\gamma)$ of the $k s$ coordinates equal $\gamma$ is at most $(s+1)^{q^{k-1}}\exp\big(s(k-1)H(\pi)\big)$. It is zero if $\pi(\gamma)>1-1/k$ for some $\gamma$ with $k\gamma\ne0$.
\end{enumerate}
\end{lemma}

\begin{proof}
A sequence in $\mathcal A_{q,k}^s$ has a type, the empirical distribution $P$ of its entries on $\mathcal A_{q,k}$. There are at most $(s+1)^{|\mathcal A_{q,k}|}$ types, and at most $e^{sH(P)}$ sequences of type $P$~\cite{cover2006elements}. If $b$ has law $P$ and $P_i$ is the law of $b_i$, each coordinate is determined by the others, so $H(P)\le\sum_{i\ne j}H(P_i)$ for every $j$.

In part 1, $P_j=\pi_j$. Choosing $j$ with largest $H(\pi_j)$ gives the claimed entropy reduction. The event $b_j=\gamma\ne0$ requires $b_i\ne0$ for some $i\ne j$, which gives the compatibility inequality.

In part 2, average $P$ over coordinate permutations to obtain $\bar P$. These permutations preserve $\mathcal A_{q,k}$, and entropy concavity gives $H(\bar P)\ge H(P)$. Every coordinate of $\bar P$ has law $\pi$, so $H(\bar P)\le(k-1)H(\pi)$. If $k\gamma\ne0$, the constant vector $(\gamma,\dots,\gamma)$ is absent from $\mathcal A_{q,k}$. Each entry then has at most $k-1$ coordinates equal to $\gamma$, and $k s\pi(\gamma)\le(k-1)s$.
\end{proof}

For vectors with small support we use the generating function $E_k(x)=(1+x)^k-k x$, which counts the subsets of $[k]$ of size different from one by their size, and the constants
$$
  A=(q-1)(k+1)^{D-1}e^{\,k+1+2^{k}D/k},
  \qquad
  \epsilon=(2A)^{-1/(D/2-1)} .
$$

\begin{theorem}[Typical tree projection]\label{thm:tp_typical}
Let $q\ge2$, $k\ge2$, $D\ge3$, and $k<D$. Let $\delta>0$ and $t\le(\tfrac12-\delta)\log_{(k-1)(D-1)}m$. Then, for random $(k,D)$-regular instances over $\Z_q$ from the configuration model or from Gallager's ensemble, and for a uniformly random symbol $i$,
$$
  \Pr\big[\mathrm{LT}_t(i)\text{ and }\mathrm{TP}_t(i)\big]=1-o(1).
$$
\end{theorem}

\begin{proof}
For $t=0$ the conclusion follows from the case $t=1$. Indeed, $\mathrm{LT}_0(i)$ always holds, and $\mathcal C_0(i)=\Z_q$ because $F_0(i)$ is empty. On $\mathrm{LT}_1(i)$, every value of the root extends to an assignment that satisfies the checks in $F_1(i)$, so the restriction of $\mathcal C_1(i)$ to $\{i\}$ is $\Z_q$, and $\mathrm{TP}_1(i)$ implies $\mathrm{TP}_0(i)$. We therefore assume $t\ge1$, and we treat the two ensembles in turn. In both, we expose the radius-$2t$ ball around $i$ and count, by the first-moment method, the vectors of $\ker M$ that would violate the hypothesis of Lemma~\ref{lem:tp_kernel}. We treat vectors with small support and vectors with support of linear size separately. For a vector $z$ on the external checks, $|z|$ denotes the size of its support.

\emph{Configuration model.} Work in the unconditioned configuration model. Expose the radius-$2t$ ball around $i$ by revealing the partners of all half-edges met while exploring from $i$, and stop at the leaves $L=L_t(i)$. On the event $\mathrm{LT}_t(i)$, the checks in $F=F_t(i)$ have all their half-edges paired inside the ball, each interior symbol has all $k$ half-edges paired inside the ball, and each leaf has exactly $k-1$ unpaired half-edges. Conditional on the exposed part, the unpaired half-edges are paired by a uniformly random bijection. The unpaired half-edges on the symbol side are the $k m'$ half-edges of the $m'=m-|N_t(i)|$ symbols outside the ball, together with $\Delta=(k-1)|L|$ half-edges of the leaves. The unpaired half-edges on the check side are the $Dn'$ half-edges of the $n'=n-|F|$ external checks, and $Dn'=k m'+\Delta$. The number of leaves is at most $k((k-1)(D-1))^{t}$, so $\Delta=O(m^{1/2-\delta})$, and in particular $\Delta\le\sqrt{n'}$ for large $m$.

Suppose that $\mathrm{LT}_t(i)$ holds and $\mathrm{TP}_t(i)$ fails. By Lemma~\ref{lem:tp_kernel}, there is $z\in\ker M$ that is not the restriction of an element of $\ker B$. Let $\gamma^\ast$ be a value in $\Lambda_k$ that $z$ takes most often among the values in $\Lambda_k$. The vector $z-\gamma^\ast\mathbf 1_E$ lies in $\ker M$ and is not the restriction of an element of $\ker B$, because $\gamma^\ast\mathbf 1_E$ is the restriction of $\gamma^\ast\mathbf 1\in\ker B$. Replacing $z$ by this vector, we may assume that $z$ takes the value $0$ at least as often as any other value in $\Lambda_k$. We call such a vector reduced. A reduced vector of this kind is nonzero, because $0$ is the restriction of $0\in\ker B$. It therefore suffices to show that the expected number of nonzero reduced vectors in $\ker M$ is $o(1)$, conditional on the exposed ball and uniformly over exposed balls on which $\mathrm{LT}_t(i)$ holds.

Let $N_h$ be the number of $z\in\ker M$ with $|z|=h$. If $z\in\ker M$ has support $Y$, then no symbol outside the ball is joined to $Y$ by exactly one edge, since at such a symbol the values of its checks would sum to a single nonzero value. Let $N'_h$ be the number of sets $Y$ of $h$ external checks with this property, so that $N_h\le(q-1)^hN'_h$. Fix $Y$. The $Dh$ half-edges of $Y$ are paired with a uniformly random subset of size $Dh$ of the $Dn'$ unpaired half-edges on the symbol side. The number of such subsets that meet no constrained symbol in exactly one half-edge is the coefficient $[x^{Dh}]\,E_k(x)^{m'}(1+x)^{\Delta}$. Since this polynomial has nonnegative coefficients, the coefficient is at most $E_k(x)^{m'}(1+x)^\Delta x^{-Dh}$ for every $x>0$. Hence
\begin{equation}\label{eq:first_moment}
  \E N_h\le(q-1)^h\binom{n'}{h}\frac{E_k(x)^{m'}(1+x)^\Delta x^{-Dh}}{\binom{Dn'}{Dh}}\qquad\text{for all }x>0 .
\end{equation}
Note that $m'/n'\le D/k$.

\emph{Small sets, $1\le h\le\epsilon n'$.} Take $x=\sqrt{h/n'}\le1$. Every term of $E_k(x)$ other than the constant term has degree at least two, so $E_k(x)\le1+2^{k}x^2\le\exp(2^{k}h/n')$, and $E_k(x)^{m'}\le\exp(2^{k}Dh/k)$. Also $(1+x)^\Delta\le e^{\Delta x}\le e^{\sqrt h}\le e^h$, using $\Delta\le\sqrt{n'}$. With $\binom{n'}h\le(en'/h)^h$ and $\binom{Dn'}{Dh}\ge(n'/h)^{Dh}$, the bound~\eqref{eq:first_moment} becomes
$$
  \E N_h\le\Big[(q-1)e^{2+2^{k}D/k}(h/n')^{D/2-1}\Big]^h\le\Big[A\,(h/n')^{D/2-1}\Big]^h .
$$
For $h\le\epsilon n'$ the bracket is at most $A\epsilon^{D/2-1}=1/2$. Let $h_0=\lceil\ln^2n'\rceil$. For $h\le h_0$ the bracket is at most $b_0=A(h_0/n')^{1/2}$, since $D/2-1\ge1/2$, and $b_0\to0$. Therefore $\sum_{h=1}^{\epsilon n'}\E N_h\le\sum_{h\ge1}b_0^h+\sum_{h>h_0}2^{-h}=o(1)$.

\emph{Large sets, $h\ge\epsilon n'$.} We count vectors by type instead of using~\eqref{eq:first_moment}. For a probability vector $\chi$ with $n'\chi(\gamma)\in\mathbb Z$, let $N_\chi$ count the vectors in $\ker M$ taking each value $\gamma$ on exactly $n'\chi(\gamma)$ external checks. There are at most $e^{n'H(\chi)}$ candidate vectors~\cite{cover2006elements}.

Fix a candidate. Label each unpaired half-edge on the symbol side by the value at its partner check. Under the random pairing this is uniform among the labelings with $Dn'\chi(\gamma)$ occurrences of each $\gamma$, of which there are at least $e^{Dn'H(\chi)}/(Dn'+1)^q$.

The candidate lies in $\ker M$ exactly when the labels at each symbol outside the ball sum to zero. The leaf half-edges contribute at most $q^\Delta$ labelings. The other symbols form a sequence in $\mathcal A_{q,k}^{m'}$. If $\pi$ is their empirical label distribution, Lemma~\ref{lem:even_count}(2) with $s=m'$ gives
$$
  \E N_\chi\le\frac{(Dn'+1)^q\,q^\Delta(m'+1)^{q^{k-1}}}{e^{(D-1)n'H(\chi)}}\max_\pi e^{m'(k-1)H(\pi)},
$$
where the term with $\pi$ vanishes if $\pi(\gamma)>1-1/k$ for some $\gamma$ with $k\gamma\ne0$. Since $Dn'=k m'+\Delta$ and the leaves carry $\Delta$ labels, $|\pi(\gamma)-\chi(\gamma)|\le2\Delta/(k m')$ for every $\gamma$. Also $m'(k-1)\le Dn'(1-1/k)$. Hence
$$
  \ln\E N_\chi\le-n'\Big(\frac Dk-1\Big)H(\chi)+Dn'\Big(1-\frac1k\Big)\big(H(\pi)-H(\chi)\big)+O(\Delta+\log m).
$$
The entropy $H$ is uniformly continuous on the set of probability vectors on $\Z_q$, and $\Delta=o(m)$, so the last two terms are $o(n')$ uniformly in $\chi$.

It remains to bound $H(\chi)$ from below when $\chi$ is the type of a reduced vector with $|z|\ge\epsilon n'$. Whether $z$ is reduced depends only on its type. If a most frequent value of $z$ lies in $\Lambda_k$, then $0$ is a most frequent value, and $\max_\gamma\chi(\gamma)=\chi(0)=1-|z|/n'\le1-\epsilon$. Otherwise a most frequent value $\gamma$ satisfies $k\gamma\ne0$, and the bound on $\E N_\chi$ vanishes unless $\chi(\gamma)\le1-1/k+2\Delta/(k m')$, which is at most $1-1/(2k)$ for large $m$. Since $H(\chi)\ge-\ln\max_\gamma\chi(\gamma)$, in both cases $H(\chi)\ge c_1\coloneqq-\ln\big(1-\min\{\epsilon,1/(2k)\}\big)>0$. Since $k<D$, we obtain $\E N_\chi\le\exp(-n'(D/k-1)c_1/2)$ for large $m$, for every such type. There are at most $(n'+1)^q$ types, so the expected number of reduced vectors $z\in\ker M$ with $|z|\ge\epsilon n'$ is $o(1)$.

Together with the small sets, the expected number of nonzero reduced vectors in $\ker M$ is $o(1)$, uniformly over exposed balls on which $\mathrm{LT}_t(i)$ holds. By Markov's inequality, the conditional probability that $\mathrm{TP}_t(i)$ fails is $o(1)$. Together with $\Pr[\mathrm{LT}_t(i)]=1-o(1)$ and the transfer to the ensemble without multiple edges, this proves the theorem for the configuration model.

\emph{Gallager's ensemble.} Work with independent uniformly random partitions. Explore the ball around $i$ by revealing, for each interior symbol and each layer, the block of that layer that contains the symbol. When such a block is revealed, its other $D-1$ symbols form a uniformly random subset of the symbols not yet covered by revealed blocks of that layer. The ball fails to be a tree only if one of these symbols has already been explored. The ball contains at most $v_t=O(((k-1)(D-1))^t)$ symbols, so by a union bound $\Pr[\mathrm{LT}_t(i)\text{ fails}]\le Dv_t^2/(m-Dv_t)=O\big(((k-1)(D-1))^{2t}/m\big)$; Shutty et al.\ prove a similar statement~\cite{shutty2026locallyquantumdecoders}.

On the event $\mathrm{LT}_t(i)$, condition on the revealed blocks. For each layer $j$, the unrevealed blocks of layer $j$ then form a uniformly random partition of the set $R_j$ of symbols not covered by revealed blocks of layer $j$, independently over $j$. The set $R_j$ consists of the $m'=m-|N_t(i)|$ symbols outside the ball and of the leaves whose parent check is not in layer $j$. Write $\Delta_j$ for the number of these leaves and $\Delta=\sum_j\Delta_j=(k-1)|L|=O(m^{1/2-\delta})$. Layer $j$ has $n'_j=|R_j|/D=(m'+\Delta_j)/D$ external checks, and $n'=\sum_jn'_j$ satisfies $Dn'=k m'+\Delta$.

Suppose that $\mathrm{TP}_t(i)$ fails. By Lemma~\ref{lem:tp_kernel} there is $z\in\ker M$ that is not the restriction of an element of $\ker B$. For $a\in\Z_q^k$ with $\sum_ja_j=0$, the vector $z-(\sum_ja_j\mathbf 1_j)|_E$ has the same property, and it subtracts $a_j$ from the values of $z$ on the external checks of layer $j$. For $j<k$, let $a_j$ be a most frequent value of $z$ on the external checks of layer $j$, and let $a_k=-\sum_{j<k}a_j$. After this replacement, $0$ is a most frequent value of $z$ on the external checks of every layer $j<k$. We call such a vector reduced. It is nonzero.

For $\vec h=(h_1,\dots,h_k)$, let $N_{\vec h}$ be the number of $z\in\ker M$ with $h_j$ nonzero values on the external checks of layer $j$, and write $h=\sum_jh_j$. Label the external checks of each layer uniformly at random, and fix $h_j$ labels in each layer. The union $U_j$ of the chosen checks of layer $j$ is a uniformly random subset of $R_j$ of size $Dh_j$, independently over $j$. As in the configuration model, if the chosen checks support a vector of $\ker M$, then no symbol outside the ball lies in exactly one of the sets $U_j$. Hence
$$
  \E N_{\vec h}\le(q-1)^hQ(\vec h)\prod_j\binom{n'_j}{h_j}\binom{Dn'_j}{Dh_j}^{-1},
$$
where $Q(\vec h)$ is the number of tuples $(U_1,\dots,U_k)$ with $U_j\subseteq R_j$, $|U_j|=Dh_j$, and no symbol outside the ball in exactly one of them.

\emph{Small sets, $1\le h\le\epsilon n'$.} The number $Q(\vec h)$ is the coefficient of $\prod_jx_j^{Dh_j}$ in $\big(\prod_j(1+x_j)-\sum_jx_j\big)^{m'}\prod_j(1+x_j)^{\Delta_j}$. This polynomial has nonnegative coefficients, so, setting every $x_j$ equal to $x>0$, we get $Q(\vec h)\le E_k(x)^{m'}(1+x)^\Delta x^{-Dh}$. Take $x=\sqrt{h/n'}$. As in the configuration model, $E_k(x)^{m'}\le\exp(2^{k}Dh/k)$, $(1+x)^\Delta\le e^h$, and $x^{-Dh}=(n'/h)^{Dh/2}$. For the binomial coefficients, $\binom{n'_j}{h_j}\binom{Dn'_j}{Dh_j}^{-1}\le e^{h_j}(h_j/n'_j)^{(D-1)h_j}\le e^{h_j}((k+1)h/n')^{(D-1)h_j}$, because $n'_j\ge m'/D\ge n'/(k+1)$ for large $m$. There are at most $(h+1)^{k-1}\le e^{(k-1)h}$ vectors $\vec h$ with $\sum_jh_j=h$. Altogether,
$$
  \sum_{\vec h:\,\sum_jh_j=h}\E N_{\vec h}\le\Big[A\,(h/n')^{D/2-1}\Big]^h ,
$$
and the sum over $1\le h\le\epsilon n'$ is $o(1)$, exactly as for the configuration model.

\emph{Large reduced sets, $h\ge\epsilon n'$.} We count by types, as for the configuration model. For probability vectors $\chi_1,\dots,\chi_k$ on $\Z_q$ with $n'_j\chi_j(\gamma)\in\mathbb Z$, let $N_{\vec\chi}$ count the vectors in $\ker M$ taking each value $\gamma$ on exactly $n'_j\chi_j(\gamma)$ external checks of layer $j$. There are at most $\prod_je^{n'_jH(\chi_j)}$ candidate vectors.

Fix a candidate and label each symbol of $R_j$ by its value at the check of layer $j$. The labelings are independent over layers, and each is uniform among the at least $e^{Dn'_jH(\chi_j)}/(Dn'_j+1)^q$ labelings with $Dn'_j\chi_j(\gamma)$ occurrences of each $\gamma$.

Membership in $\ker M$ requires the $k$ labels at every symbol outside the ball to sum to zero. Let $\pi_j$ be the empirical label distribution of layer $j$ on these $m'$ symbols, so that $|\pi_j(\gamma)-\chi_j(\gamma)|\le2\Delta/m'$. The leaf labels have at most $q^\Delta$ choices. Lemma~\ref{lem:even_count}(1) with $s=m'$ gives
$$
  \E N_{\vec\chi}\le\frac{q^\Delta(m'+1)^{q^{k-1}}\prod_j(Dn'_j+1)^q}{\prod_je^{(D-1)n'_jH(\chi_j)}}\max_{\vec\pi}\exp\Big(m'\Big[\sum_jH(\pi_j)-\max_jH(\pi_j)\Big]\Big),
$$
where the term with $\vec\pi$ vanishes unless $\pi_j(\gamma)\le\sum_{i\ne j}(1-\pi_i(0))$ for every $j$ and every $\gamma\ne0$. Since $Dn'_j=m'+\Delta_j$ and $H$ is uniformly continuous,
$$
  \ln\E N_{\vec\chi}\le m'\Big[\frac1D\sum_jH(\chi_j)-\max_jH(\chi_j)\Big]+o(m)\le-m'\Big(1-\frac k D\Big)\max_jH(\chi_j)+o(m),
$$
uniformly in $\vec\chi$. It remains to bound $\max_jH(\chi_j)$ from below when $\vec\chi$ is the type of a reduced vector with $h\ge\epsilon n'$. Let $\eta=\min\{\epsilon,1/(4k)\}$. Since $\sum_jn'_j(1-\chi_j(0))=h\ge\epsilon n'$ and $\sum_jn'_j=n'$, some layer $j$ has $\chi_j(0)\le1-\epsilon$. If $\max_\gamma\chi_j(\gamma)\le1-\eta$, then $H(\chi_j)\ge-\ln(1-\eta)$. Otherwise $\chi_j(\gamma)>1-\eta\ge3/4$ for some $\gamma$, and $\gamma\ne0$ because $\chi_j(0)\le1-\epsilon$. Then $0$ is not a most frequent value in layer $j$, so $j=k$ because $z$ is reduced. The bound on $\E N_{\vec\chi}$ vanishes unless $\sum_{i<k}(1-\pi_i(0))\ge\pi_k(\gamma)\ge3/4-2\Delta/m'$. So some layer $i<k$ has $1-\chi_i(0)\ge(3/4-2\Delta/m')/(k-1)-2\Delta/m'$, which is at least $\eta$ for large $m$. Since $0$ is a most frequent value in layer $i$, we get $\max_\gamma\chi_i(\gamma)=\chi_i(0)\le1-\eta$ and $H(\chi_i)\ge-\ln(1-\eta)$. Hence $\max_jH(\chi_j)\ge c_1\coloneqq-\ln(1-\eta)$, and, since $k<D$, $\E N_{\vec\chi}\le\exp(-m'(1-k/D)c_1/2)$ for large $m$. The sum over the at most $(n'+1)^{qk}$ type vectors is therefore $o(1)$.

By Markov's inequality, the conditional probability that a nonzero reduced vector exists is $o(1)$, uniformly over exposed balls on which $\mathrm{LT}_t(i)$ holds. Together with $\Pr[\mathrm{LT}_t(i)]=1-o(1)$, this proves the theorem for Gallager's ensemble.
\end{proof}

\begin{corollary}[Local estimates in random instances]\label{cor:typical_nogo}
Let $q\ge2$, let $f$ be any cost, let $k\ge2$, $D\ge3$, and $k<D$, and let $0<c_0<1/(2\log((k-1)(D-1)))$, with the natural logarithm. For $t\le c_0\log m$ and a uniformly random symbol $i$, with probability $1-o(1)$ over random $(k,D)$-regular instances over $\Z_q$ from the configuration model or from Gallager's ensemble, $\mathrm{LT}_t(i)$ and $\mathrm{TP}_t(i)$ hold, so the conclusion of Theorem~\ref{thm:locality} holds at $i$. On this event, the bound of Theorem~\ref{thm:locality} on the symbol error is attained: some $t$-local estimate at $i$ equals $c_i$ with probability exactly $P_t$, and depth-$t$ BPQM for symbol $i$ computes such an estimate. In particular, for $t=\lfloor c_0\log m\rfloor$, the largest probability with which a $t$-local estimate at $i$ equals $c_i$ is $P_t$, and $P_t$ tends to one as $m\to\infty$ if and only if the primal model has non-reconstruction. The same holds for any fixed symbol $i$.
\end{corollary}

\begin{proof}
Let $\delta=\tfrac12-c_0\log((k-1)(D-1))$, which is positive. Then $t\le c_0\log m$ implies $t\le(\tfrac12-\delta)\log_{(k-1)(D-1)}m$, and Theorem~\ref{thm:tp_typical} shows that $\mathrm{LT}_t(i)$ and $\mathrm{TP}_t(i)$ hold with probability $1-o(1)$, and Theorem~\ref{thm:locality} then applies at $i$. On this event, as in the proof of Theorem~\ref{thm:locality}, the restriction of a uniformly random codeword $c$ of $C$ to $N_t(i)$ is a uniformly random codeword of the tree code $\mathcal C_t$, with $i$ as the root. Given $c$, the qudits of $N_t(i)$ are in the state $\bigotimes_{j\in N_t(i)}\ket{\psi^p_{c_j}}$, so they carry exactly the input of the problem of Proposition~\ref{prop:tree_optimal}. A measurement that is optimal for that problem, applied to these qudits, therefore computes a $t$-local estimate at $i$ that equals $c_i$ with probability $P_t$, and depth-$t$ BPQM is such a measurement by Proposition~\ref{prop:tree_optimal}. Since $P_t$ is nondecreasing in $t$ and $\lfloor c_0\log m\rfloor\to\infty$, Theorem~\ref{thm:duality} shows that $P_{\lfloor c_0\log m\rfloor}\to1$ if and only if the primal model has non-reconstruction. For a fixed symbol, note that the configuration model is invariant under relabeling the symbols, and that Gallager's ensemble is invariant under relabelings that preserve the first layer, which act transitively on the symbols. Hence the probability is the same for every symbol.
\end{proof}

\section{Consequences for DQI with local decoders}\label{sec:consequences}

In this section we derive the consequences of Sections~\ref{sec:tree} and~\ref{sec:locality} for the three tasks for which DQI is used on random $(k,D)$-regular instances, and we compare them with classical algorithms. For sampling, a local decoder has block success tending to one only at inverse temperatures without reconstruction, and for $k\ge3$, over every $\Z_q$, for every cost, and for both ensembles, the two-stage decoder of Mandal et al.\ with the node operations of Lemma~\ref{lem:nodes} has block success tending to one at all finite inverse temperatures without reconstruction (Section~\ref{sec:sampling}). For approximate optimization with a cost that takes two values, we call a constraint satisfied if its cost takes the smaller value. With every amplitude function that depends only on the total cost and a local decoder whose block success tends to one, the expected satisfied fraction that DQI reaches is at most the probability that a single constraint is satisfied under its Gibbs weights at $\beta_d$, up to $o(1)$ (Section~\ref{sec:approx}). For Max-XORSAT, these are the amplitude functions that depend only on the number of violated constraints, and the bound is $s_d+o(1)$, where $s_d=p_{\beta_d}(0)$. We compare the block amplitude function of Shutty et al., which is not symmetric, with product amplitude functions on four degree pairs (Section~\ref{sec:approx_block}). For exact optimization, quantum minimum finding on the ideal Gibbs state, at an inverse temperature at which a local decoder succeeds, is at most quadratically faster than classical sampling if Conjecture~\ref{conj:classical} holds and both algorithms are given a lower bound within a polynomial factor on the probability of an optimum. For Max-XORSAT on the configuration model with several degree pairs, the same conclusion holds without the conjecture up to an explicit inverse temperature $\beta_U$ (Section~\ref{sec:exact}). Nonlocal decoders, and the nonlocal classical algorithms that reach the same values, are discussed in Section~\ref{sec:nonlocal}.

\subsection{Sampling}\label{sec:sampling}

For sampling, the target is the primal Gibbs measure $\mu_\beta$. The output guarantee reduces the task to reliable recovery of the codeword. Section~\ref{sec:only_nonrecon} proves that a local or retained initial estimate permits this only without reconstruction. Section~\ref{sec:sharp} proves the converse for $k\ge3$: tree BPQM followed by erasure completion recovers the entire word at every finite inverse temperature without reconstruction. Section~\ref{sec:status} discusses the separate question of efficient classical sampling.

\subsubsection{Local decoders succeed only without reconstruction}\label{sec:only_nonrecon}

A local estimate limits block success because correct recovery of the whole codeword requires that estimate to be correct. Theorems~\ref{thm:duality} and~\ref{thm:locality} therefore imply that reliable local decoding requires non-reconstruction. We first state this consequence for a local estimate at a fixed symbol, then extend it to an initial estimate selected using input-independent randomness.

\begin{theorem}[Local decoders succeed only without reconstruction]\label{thm:only_nonrecon}
Fix $(k,D,q,f)$ and $\beta\ge0$. Let $(B_n)$ be a sequence of $(k,D)$-regular matrices, and for each $n$ let a decoder for the decoding problem of $B_n$ with the product amplitude function $\Phi_\beta$ be given. Suppose that for each $n$ there are a radius $t_n$ and a symbol $i_n$ such that $\mathrm{LT}_{t_n}(i_n)$ and $\mathrm{TP}_{t_n}(i_n)$ hold and the decoder makes a $t_n$-local estimate at $i_n$. If the block success of the decoders tends to one, then the primal model at $\beta$ has non-reconstruction.
\end{theorem}

\begin{proof}
By Theorem~\ref{thm:locality}, the block success of the $n$-th decoder is at most $P_{t_n}\le P_\infty$. If the block success tends to one, then $P_\infty=1$, so $P_t\to1$. By Theorem~\ref{thm:duality}, the primal model has non-reconstruction.
\end{proof}

No condition on the growth of $t_n$ is needed, as long as the two local conditions hold. For random instances, Corollary~\ref{cor:typical_nogo} supplies them for $t_n\le c_0\log m$. An estimate chosen using a random processing order need not be a local estimate at a fixed symbol in the sense of Definition~\ref{def:local}. The initial-estimate conditions of Section~\ref{sec:strategy} allow us to average over this choice directly.

\begin{proposition}[Block success from an initial local estimate]\label{prop:first_local_block}
Let $q\ge2$, let $f$ be a cost, let $\beta\ge0$, and let $B$ be $(k,D)$-regular. Consider the decoding problem with product amplitude function $\Phi_\beta$. Suppose that the decoder has an initial estimate satisfying~\eqref{eq:first_estimate_law} on a set $\mathcal R_0$ of input-independent seeds with probability $g>0$, and write $a$ for~\eqref{eq:first_estimate_repair}. If its block success is $1-\varepsilon$, then
\begin{equation}\label{eq:first_local_block}
  1-\varepsilon\le1-g+\sum_{r\in\mathcal R_0}\Pr[R=r]P_{t_r}(p_\beta)+a\le1-g+gP_\infty(p_\beta)+a,
\end{equation}
where $P_t(p_\beta)$ is the optimal success probability for the root of the tree code, and $P_\infty(p_\beta)=\lim_tP_t(p_\beta)$. Equivalently,
$$
  g\big(1-P_\infty(p_\beta)\big)\le\varepsilon+a.
$$
\end{proposition}

\begin{proof}
Split the event $\hat c=c$ according to whether $R\in\mathcal R_0$ and whether the initial estimate is correct:
$$
  1-\varepsilon\le1-g+\Pr[R\in\mathcal R_0,\ Y_R=c_{i_R}]+a.
$$
Given $R=r\in\mathcal R_0$, the codeword remains uniform on $C$ because $R$ is independent of $c$. The estimate is the outcome of a POVM on the original received qudits of $N_{t_r}(i_r)$, and both local conditions hold there. Theorem~\ref{thm:locality} therefore gives $\Pr[Y_r=c_{i_r}\mid R=r]\le P_{t_r}(p_\beta)\le P_\infty(p_\beta)$. Multiply by $\Pr[R=r]$ and sum over $r\in\mathcal R_0$ to obtain~\eqref{eq:first_local_block}.
\end{proof}

\begin{corollary}[Reliable decoding with an initial local estimate]\label{cor:first_local_nonrecon}
Fix $(k,D,q,f)$ and $\beta\ge0$. For a sequence of $(k,D)$-regular matrices $B_n$ and decoders as in Proposition~\ref{prop:first_local_block}, let $g_n$, $a_n$, and $1-\varepsilon_n$ be their initial-estimate parameters and block success. If $(\varepsilon_n+a_n)/g_n\to0$, then the primal model at $\beta$ has non-reconstruction, and in particular $\beta\le\beta_d$.
\end{corollary}

\begin{proof}
Proposition~\ref{prop:first_local_block} gives $1-P_\infty(p_\beta)\le(\varepsilon_n+a_n)/g_n\to0$. Hence $P_\infty(p_\beta)=1$, and Theorem~\ref{thm:duality} implies non-reconstruction.
\end{proof}

For an order fixed independently of the instance, or a uniformly random order of an instance-determined set of linear size, Section~\ref{sec:strategy} establishes $g_n\to1$ on typical instances under the stated first-coordinate conditions. If successful recovery retains the initial estimate, then $a_n=0$; a fixed default output on failure gives $a_n\le1/|C_n|$ by Lemma~\ref{lem:first_estimate_output}, which is exponentially small when $k<D$. Thus reliable block decoding implies non-reconstruction for these sequential implementations as well. In particular, this applies to the two-stage BPQM decoder~\cite{mandal2026blockerror} and the bounded-radius sequential and approximate-cloning implementations~\cite{piveteau2022messagepassing} described in Section~\ref{sec:ov_setting}.

The bound on the initial estimate requires its neighborhood to satisfy both local conditions, but places no restriction on later computation. Above $\beta_d$, retaining such an estimate prevents reliable block recovery even if later calls use larger neighborhoods or process the whole code. A decoder that corrects its initially wrong estimate need not have $a_n\to0$ and is not excluded by this implication. Whether unrestricted-depth BPQM on finite cyclic graphs can succeed above $\beta_d$ remains open. The decoder-independent limits of Section~\ref{sec:condensation} still apply in their stated setting.

\subsubsection{The threshold is attained}\label{sec:sharp}

By Theorem~\ref{thm:only_nonrecon}, non-reconstruction is necessary for a local decoder to succeed. For $k\ge3$, it is also sufficient, over every alphabet, for every cost, and for both ensembles. The decoder that attains it is the two-stage decoder of Mandal et al.~\cite{mandal2026blockerror}, described in Section~\ref{sec:bpqm}, with the node operations of Lemma~\ref{lem:nodes}. It measures the symbols whose neighborhoods satisfy the local conditions one after another, each with the optimal measurement for the root of its tree code, and then recovers the remaining symbols by solving linear congruences. The reliability bound holds for every processing order chosen independently of the received input. An order that starts with a maximal set of disjoint good balls gives joint locality directly; a uniformly random order gives it by Corollary~\ref{cor:random_order_local}.

\begin{proposition}[The reconstruction threshold is attained]\label{prop:sharp}
Let $q\ge2$, let $f$ be a cost, let $3\le k<D$, and let $\beta\ge0$ be finite and such that the primal model has non-reconstruction. Let $t_m=2\lceil\log_{k-1}(1+\ln m)\rceil$, so that $t_m=O(\log\log m)$. For random $(k,D)$-regular instances from the configuration model or from Gallager's ensemble, the following holds with probability $1-o(1)$ over the instance.

Consider the two-stage decoder of Mandal et al.~\cite{mandal2026blockerror} with the node operations of Lemma~\ref{lem:nodes}. It applies the optimal measurement for the root of the tree code of depth $t_m$ at every symbol satisfying $\mathrm{LT}_{t_m}$ and $\mathrm{TP}_{t_m}$, one symbol after another, in any order chosen independently of the received input, and completes the remaining symbols by solving linear congruences. For every such order, its block success at $\beta$ is at least $1-2m^{-3}$.

If the order starts with a maximal set of good symbols whose balls are pairwise disjoint, the decoder makes $t_m$-local estimates on $m^{1-o(1)}$ symbols. If the order is uniformly random, it satisfies Definition~\ref{def:local_decoder} with $s_m=\lceil(\log m)^2\rceil$ and exceptional probability at most $4^{-s_m}$. Thus both schedules are local. If $p_\beta$ can be computed to $b$ bits of precision in time polynomial in $b$, the decoder can be implemented in polynomial time with block success $1-o(1)$ and the same locality property.

Consequently, with probability $1-o(1)$ over the instance, DQI with this decoder outputs a violation pattern whose law is within $o(1)$ of $\pi_\beta$ in expected total variation over $x_0$, and whose expected cost per constraint differs from $\langle f\rangle_{p_\beta}$ by $o(1)$. For Max-XORSAT, the expected satisfied fraction is $p_\beta(0)-o(1)$.
\end{proposition}

The proof is given in Appendix~\ref{app:sharp}. For $q=2$ and the configuration model, Proposition~\ref{prop:sharp} also follows from the results of Mandal et al.~\cite[Theorem~30 and Corollary~31]{mandal2026blockerror}, which hold for prime $q$, for matrices whose nonzero entries are independent uniformly random units, and for every channel at which BPQM succeeds on the tree. Our matrices have all nonzero entries equal to one, and for $q>2$ such matrices have vanishing probability under random units, so for $q>2$ the proposition does not follow from their theorem by conditioning. We follow their two-stage decoder and verify its two ingredients for our codes. The first is that the optimal error for the root of the tree code converges to zero doubly exponentially in the depth, which Mandal et al.\ prove for prime $q$ throughout the region in which BPQM succeeds on the tree~\cite[Theorem~26]{mandal2026blockerror}; we prove it for every $q$ from non-reconstruction (Lemma~\ref{lem:decay}). The second is that the minimum distance of $C$ is linear in $m$ for both ensembles (Lemma~\ref{lem:distance}). Linear minimum distance is known for related ensembles, by Gallager for codes over $\Z_M$ with zero-one parity-check matrices~\cite{gallager1962ldpc}, by Como and Fagnani for codes over finite abelian groups~\cite{como2008spectra}, and by Yang et al.\ for codes over finite fields with random nonzero entries~\cite{yang2011weights}; we include a short proof for the two ensembles used here.

Together, Theorem~\ref{thm:only_nonrecon} and Proposition~\ref{prop:sharp} show that, for $3\le k<D$, over every $\Z_q$, for every cost, and for both ensembles, DQI with local decoders samples from the Gibbs measure exactly at the finite inverse temperatures without reconstruction. For costs that take two values, which include Max-XORSAT and the Potts cost, these form the interval $[0,\beta_d)$ or $[0,\beta_d]$ (Section~\ref{sec:monotone}).

\begin{corollary}[Local decoding and attainment over finite fields]\label{cor:field_sharp}
Fix a field $\mathbb F_q$ with $q=p^r$, a cost $f:\mathbb F_q\to\mathbb R$, $3\le k<D$, and a finite $\beta\ge0$. Use zero-one matrices over this field and the received states
$$
  \ket{\psi_c}=\sum_x\Phi_\beta(x)\exp\!\left(\frac{2\pi i}{p}\operatorname{Tr}_{\mathbb F_q/\mathbb F_p}(c\cdot x)\right)\ket x,
  \qquad c\in\ker B^\top.
$$
The tree and local-projection statements hold in this field model. In particular, a $t$-local estimate at a symbol satisfying LT and TP has correctness probability at most the field tree optimum $P_t\le P_\infty$, and reliable block decoding through such an estimate requires non-reconstruction.

For a cost that takes two values, the set of finite inverse temperatures without reconstruction is an interval starting at zero, with supremum $\beta_d$.

If the field model at $\beta$ has non-reconstruction, all the attainment conclusions of Proposition~\ref{prop:sharp} hold on either regular ensemble: depth $t_m$ gives exact-arithmetic block success at least $1-2m^{-3}$, uniformly over input-independent processing orders, with Gaussian elimination for completion. The disjoint-first and uniformly random schedules have the same joint-locality guarantees. If the entries of $p_\beta$ are computable to $b$ bits in time polynomial in $b$, a polynomial-time implementation has block success $1-o(1)$. Its DQI output is within $o(1)$ of $\pi_\beta$ in expected total variation over a uniform field offset, and its expected cost per constraint is $\langle f\rangle_{p_\beta}+o(1)$.
\end{corollary}

The proof is given in Appendix~\ref{app:common_field}. The basis decomposition there concerns the linear constraints only; it does not require the field cost or the single-symbol channel to factor over the prime-field coordinates.

\subsubsection{Classical sampling below the reconstruction threshold}\label{sec:status}

Whether Theorem~\ref{thm:only_nonrecon} rules out a speedup of DQI for sampling depends on whether classical algorithms sample efficiently from $\mu_\beta$ at the same inverse temperatures. We use the following notion of efficient classical sampling, which is also the assumption of Section~\ref{sec:conditional}.

\begin{hypothesis}[Efficient classical sampling at $\beta$]\label{hyp:sampler}
There is a classical algorithm that, on input an instance and $\epsilon>0$, outputs a sample from a law $\hat\mu$ with $\|\hat\mu-\mu_\beta\|_{\mathrm{TV}}\le\epsilon$, in time $\poly(m+n,\log(1/\epsilon))$.
\end{hypothesis}

The cavity method identifies $\beta_d$ with the dynamical transition. M\'ezard and Montanari showed that the 1RSB cavity equations with Parisi parameter one, whose nontrivial solution defines the dynamical transition, have a nontrivial solution exactly when tree reconstruction is solvable~\cite[Proposition~1]{mezard2006reconstruction}. Below $\beta_d$, the Gibbs measure is predicted not to split into well-separated clusters, and the equilibrium correlation time of Glauber dynamics is predicted not to grow with the number of variables. Above $\beta_d$, it is predicted to grow exponentially~\cite{montanari2006rigorous,krzakala2011melting}. Montanari and Semerjian proved that this correlation time is at least proportional to the point-to-set correlation length~\cite[Theorem~1]{montanari2006rigorous}.

The prediction concerns the law of Glauber dynamics started from a uniformly random assignment, or from an equilibrium assignment, after polynomially many steps, and not its spectral gap. Montanari and Semerjian note that the spectral gap is expected to become exponentially small already at some inverse temperatures below $\beta_d$, because of metastable states that the dynamics does not reach from such initial assignments~\cite{montanari2006rigorous}. Hypothesis~\ref{hyp:sampler} requires only some polynomial-time sampler, and we state the prediction in this form as a conjecture.

\begin{conjecture}[Classical sampling without reconstruction]\label{conj:classical}
Fix $(k,D,q,f)$. For random $(k,D)$-regular instances over $\Z_q$ from the configuration model or from Gallager's ensemble, with uniformly random $x_0$, and for every $\beta$ at which the primal tree model has non-reconstruction, Hypothesis~\ref{hyp:sampler} holds with probability $1-o(1)$.
\end{conjecture}

Rigorous mixing results are known only for part of the phase without reconstruction. They bound the mixing time of Glauber dynamics from every initial assignment, either directly or through the spectral gap, and so they imply Hypothesis~\ref{hyp:sampler}. For a bound on the spectral gap, this holds because the mixing time to accuracy $\epsilon$ is at most the relaxation time multiplied by $\ln\big(1/(\epsilon\min_z\mu_\beta(z))\big)$, and $\ln(1/\min_z\mu_\beta(z))=O(m+n)$ for fixed $\beta$ and bounded $f$.
\begin{itemize}[leftmargin=*]
\item For binary Max-XORSAT ($q=2$, $f(z)=z$), fixed $(k,D)$, and small enough $\beta$, Dobrushin's uniqueness condition holds and Glauber dynamics mixes in $O(n\log n)$ steps. Each variable interacts with $D(k-1)$ others, and the influence through one clause is at most $\tanh(\beta/2)$. For $(3,6)$ the condition $12\tanh(\beta/2)<1$ holds only for $\beta<0.167$.
\item For binary Max-XORSAT with $k=2$, the primal model is the $\pm J$ Ising spin glass on a random $D$-regular graph with couplings $\beta/2$. The tree reconstruction threshold satisfies $\tanh^2(\beta_d/2)=1/(D-1)$, so $\beta_d\approx2/\sqrt{D-1}$ for large $D$. Eldan, Koehler, and Zeitouni~\cite{eldan2020spectral} prove that Glauber dynamics mixes fast when the spread of the eigenvalues of the coupling matrix is below one. With the norm bound $2\sqrt{D-1}+o(1)$ for signed random regular adjacency matrices~\cite{deshpande2019threshold,mohanty2022explicit}, this holds for $\beta<1/(2\sqrt{D-1})$. Anari, Koehler, and Vuong~\cite{anari2024trickle} report an extension to about $0.59/\sqrt{D-1}$, which is about $30\%$ of $[0,\beta_d)$ for large $D$. Related results for Erd\H{o}s--R\'enyi graphs~\cite{liu2024sparse,efthymiou2026unbounded} hold for couplings up to order one over the square root of the average degree, which again covers a constant fraction of the phase.
\item For $k\ge3$, we know of no proof of rapid mixing that holds up to $\beta_d$.
\end{itemize}
Section~\ref{sec:prange} proves a property weaker than Hypothesis~\ref{hyp:sampler}, which suffices for exact optimization, up to an explicit inverse temperature $\beta_U$. Sampling beyond $\beta_d$ with nonlocal classical algorithms, such as Prange's algorithm, is discussed in Section~\ref{sec:larger_degrees}.

\paragraph{Numerical evidence.} We simulated fixed-temperature Glauber dynamics for Max-XORSAT on $62$ degree pairs from Gallager's ensemble, with $3\le k<D\le36$. To measure equilibrium relaxation, we generated planted instances by violating each equation of a uniformly random assignment independently with probability $p_\beta(1)$. Conditional on the instance, this assignment is an exact sample from $\mu_\beta$. We measured the first time $t_{\mathrm{rel}}$ at which its overlap with the dynamics falls to $0.1$, counting $n$ single-variable updates as one sweep.

Table~\ref{tab:dynamics} gives the fifteen benchmark pairs; Appendix~\ref{app:sim_glauber} reports the other $47$ pairs and the full protocol. Every run at $\beta\le0.95\beta_d$ decorrelates within $6\times10^4$ sweeps on instances with about $30240$ variables. At $0.85\beta_d$ and $0.9\beta_d$, multiplying the size by ten changes the measured relaxation time by at most $21\%$ across all $62$ pairs. The relaxation time grows rapidly near $\beta_d$: power-law fits to the geometric means over three instances place the divergence between $0.998\beta_d$ and $1.039\beta_d$. Above $\beta_d$, the dynamics remains close to its initial planted assignment on the simulated timescales.

From a uniformly random start on instances with uniformly random right-hand sides, Glauber dynamics reaches a mean violated fraction within $0.0040$ of $p_\beta(1)$ at $0.9\beta_d$ and $0.95\beta_d$. At $1.05\beta_d$ and $1.1\beta_d$, it remains above $p_\beta(1)$ by $0.010$ to $0.026$ after $5\times10^4$ sweeps. These observations support the predicted change in relaxation near reconstruction. They do not establish polynomial-time mixing: decorrelation from equilibrium and agreement of the energy with its equilibrium mean are not total-variation mixing guarantees. Annealing's practical optimization performance is a separate comparison, developed in Section~\ref{sec:approx_bench}.

\begin{table}[!htbp]
\centering
\small
\begin{tabular}{@{}cc cccc c cc@{}}
\toprule
& & \multicolumn{4}{c}{$t_{\mathrm{rel}}$, $n=30240$} & $t_{\mathrm{rel}}$, $n=302400$ & & \\
\cmidrule(lr){3-6}\cmidrule(lr){7-7}
$(k,D)$ & $\beta_d$ & $0.6\beta_d$ & $0.8\beta_d$ & $0.9\beta_d$ & $0.95\beta_d$ & $0.9\beta_d$ & $\beta_{\mathrm{div}}/\beta_d$ & $Q$ at $1.1\beta_d$ \\
\midrule
$(3,4)$ & $2.656$ & $48$ & $504$ & $3877$ & $27858$ & $4113$ & $1.016\text{--}1.041$ & $0.96$ \\
$(3,5)$ & $2.139$ & $33$ & $325$ & $2687$ & $15530$ & $2436$ & $1.026\text{--}1.035$ & $0.95$ \\
$(3,6)$ & $1.839$ & $27$ & $222$ & $1677$ & $10627$ & $1692$ & $1.016\text{--}1.024$ & $0.94$ \\
$(3,7)$ & $1.638$ & $24$ & $175$ & $1381$ & $7509$ & $1442$ & $1.008\text{--}1.022$ & $0.93$ \\
$(3,8)$ & $1.490$ & $21$ & $174$ & $1030$ & $6388$ & $1179$ & $0.999\text{--}1.018$ & $0.92$ \\
$(4,5)$ & $2.549$ & $47$ & $565$ & $5317$ & $38884$ & $5805$ & $1.005\text{--}1.047$ & $0.98$ \\
$(4,6)$ & $2.177$ & $36$ & $342$ & $2620$ & $14728$ & $2673$ & $1.008\text{--}1.039$ & $0.98$ \\
$(4,7)$ & $1.931$ & $31$ & $308$ & $2318$ & $15456$ & $2319$ & $0.998\text{--}1.013$ & $0.97$ \\
$(4,8)$ & $1.754$ & $29$ & $244$ & $1621$ & $11968$ & $1731$ & $0.999\text{--}1.019$ & $0.97$ \\
$(5,6)$ & $2.392$ & $44$ & $429$ & $3662$ & $19908$ & $4142$ & $1.021\text{--}1.055$ & $0.99$ \\
$(5,7)$ & $2.115$ & $37$ & $379$ & $3397$ & $24688$ & $3238$ & $1.000\text{--}1.016$ & $0.99$ \\
$(5,8)$ & $1.917$ & $32$ & $284$ & $2199$ & $15336$ & $2475$ & $1.017\text{--}1.046$ & $0.99$ \\
$(6,7)$ & $2.249$ & $43$ & $445$ & $4490$ & $28407$ & $4154$ & $1.012\text{--}1.096$ & $0.99$ \\
$(6,8)$ & $2.036$ & $35$ & $338$ & $2915$ & $17298$ & $2932$ & $1.024\text{--}1.045$ & $0.99$ \\
$(7,8)$ & $2.128$ & $41$ & $379$ & $3183$ & $17555$ & $3273$ & $1.004\text{--}1.043$ & $0.99$ \\
\bottomrule
\end{tabular}
\caption{Glauber dynamics on planted instances of $(k,D)$-regular Max-XORSAT from Gallager's ensemble, for the fifteen benchmark pairs. The relaxation time $t_{\mathrm{rel}}$, in sweeps, is the first time at which the overlap with the planted assignment falls to $0.1$. For $n=30240$ variables it is the geometric mean over three instances, and for $n=302400$ variables it is measured on one instance. The column $\beta_{\mathrm{div}}/\beta_d$ gives the range over the three instances of the fitted divergence point, and the last column gives the smallest overlap after $2\times10^4$ sweeps at $1.1\beta_d$. The full protocol and the relation between planted and uniformly random right-hand sides are given in Appendix~\ref{app:sim_glauber}.}
\label{tab:dynamics}
\end{table}

If Conjecture~\ref{conj:classical} holds, then for random instances DQI with local decoders samples only at inverse temperatures where classical sampling is also efficient, so it provides no exponential speedup for sampling from these Gibbs measures.

\subsection{Approximate optimization}\label{sec:approx}

For approximate optimization, the quantity of interest is the expected objective of the output of DQI. This subsection bounds this quantity when the decoder is local. Section~\ref{sec:approx_gibbs} treats product amplitude functions for every $q$ and every cost. Section~\ref{sec:approx_sym} treats costs that take two values, such as Max-XORSAT and the Potts cost, with every amplitude function that depends only on the total cost, which includes the polynomial amplitude functions of Jordan et al. The block amplitude function of Shutty et al., which is not symmetric, is compared with product amplitude functions in Section~\ref{sec:approx_block}. Section~\ref{sec:approx_bench} compares the resulting bound with the values of known algorithms.

\subsubsection{Product amplitude functions}\label{sec:approx_gibbs}

With a product amplitude function, the satisfied fraction that DQI reaches is determined by the error rate at which the decoder succeeds. For $q=2$, the decoding problem for $\Phi_\beta$ has error rate $\alpha^\perp$ with $\alpha=p_\beta(1)$, and the satisfied fraction is $1-\alpha=\tfrac12+\sqrt{\alpha^\perp(1-\alpha^\perp)}$ by~\eqref{eq:perp}, as Shutty et al.\ observe~\cite{shutty2026locallyquantumdecoders}. A decoder that succeeds at a larger error rate therefore permits a larger $\beta$ and a larger satisfied fraction. The next theorem bounds the satisfied fraction, and more generally the expected cost for every $q$, when the decoder is local.

\begin{theorem}[Approximate optimization with product amplitude functions]\label{thm:approx_gibbs}
Let $q\ge2$, let $f$ be a cost that is not constant, and let $\beta\ge0$. Let $(B_n)$ be a sequence of $(k,D)$-regular matrices with $m_n$ rows. For each $n$, let a decoder for the decoding problem of $B_n$ with the product amplitude function $\Phi_\beta$ have block success at least $1-\varepsilon_n$. Suppose that for each $n$ there are $t_n$ and $i_n$ such that $\mathrm{LT}_{t_n}(i_n)$ and $\mathrm{TP}_{t_n}(i_n)$ hold and the decoder makes a $t_n$-local estimate at $i_n$. If $\varepsilon_n\to0$, then $\beta\le\beta_d$, and the violation pattern $x$ output by DQI satisfies, for every $n$ with $\varepsilon_n<1$,
$$
  \E_{x_0}\E\Big[\frac1{m_n}\sum_{j=1}^{m_n}f(x_j)\Big]\ \ge\ \langle f\rangle_{p_{\beta_d}}-2\big(\max f-\min f\big)\sqrt{\varepsilon_n}.
$$
For $q=2$ and $f(z)=z$, the expected satisfied fraction is at most $s_d+2\sqrt{\varepsilon_n}$, where $s_d=p_{\beta_d}(0)=\tfrac12+\tfrac12\tanh(\beta_d/2)$. If $\beta_d=\infty$, then $p_{\beta_d}$ means $p_\infty$.
\end{theorem}

\begin{proof}
By Theorem~\ref{thm:only_nonrecon}, the primal model at $\beta$ has non-reconstruction, so $\beta\le\beta_d$ by Definition~\ref{def:beta_d}. Write $\Delta=\max f-\min f$, and apply part~2 of Theorem~\ref{thm:guarantee} to $h(x)=\frac1{m_n}\sum_j(f(x_j)-\min f)/\Delta$, which takes values in $[0,1]$. The expectation $\sum_x|\Phi_\beta(x)|^2h(x)$ equals $(\langle f\rangle_{p_\beta}-\min f)/\Delta$, because $\Phi_\beta$ is a product amplitude function. Hence the expected cost per constraint is at least $\langle f\rangle_{p_\beta}-2\Delta\sqrt{\varepsilon_n}$. Finally, $\frac{d}{d\beta}\langle f\rangle_{p_\beta}=-\mathrm{Var}_{p_\beta}(f)\le0$, so $\langle f\rangle_{p_\beta}\ge\langle f\rangle_{p_{\beta_d}}$. For $q=2$ and $f(z)=z$, the satisfied fraction is $1-\frac1{m_n}\sum_jx_j$, and $\langle f\rangle_{p_\beta}=p_\beta(1)$.
\end{proof}

For random instances, Corollary~\ref{cor:typical_nogo} supplies the hypotheses on $\mathrm{LT}$ and $\mathrm{TP}$ with probability $1-o(1)$ for $t_n\le c_0\log m_n$. The bound is approached for $k\ge3$ on both ensembles. By Proposition~\ref{prop:sharp}, DQI with the two-stage decoder reaches the expected cost per constraint $\langle f\rangle_{p_\beta}+o(1)$ at every finite $\beta$ without reconstruction. Since $\beta_d$ is the supremum of these inverse temperatures, there are such $\beta$ that tend to $\beta_d$, and along them $\langle f\rangle_{p_\beta}\to\langle f\rangle_{p_{\beta_d}}$.

A classical algorithm that samples from the Gibbs measure reaches the same values wherever some decoder succeeds, because the Gibbs measure of the instance then has, on average over the offset, the same expected cost per constraint as the product law.

\begin{lemma}[Expected cost under the Gibbs measure]\label{lem:gibbs_cost}
Let $q\ge2$, let $f$ be a cost, let $\beta\ge0$ be finite, and let $B$ be a matrix for which some decoder for the decoding problem with the product amplitude function $\Phi_\beta$ has block success $1-\varepsilon$. Let $x_0$ be uniformly distributed on $\Z_q^m$, and let $\mu_\beta$ be the Gibbs measure of the instance with offset $x_0$. Then
$$
  \Big|\E_{x_0}\E_{z\sim\mu_\beta}\Big[\frac1mH(z)\Big]-\langle f\rangle_{p_\beta}\Big|\ \le\ 3\big(\max f-\min f\big)\sqrt\varepsilon .
$$
\end{lemma}

\begin{proof}
If $f$ is constant, both sides vanish. Otherwise, write $\Delta=\max f-\min f$ and $h(x)=\frac1m\sum_j(f(x_j)-\min f)/\Delta$, which takes values in $[0,1]$, so that $\frac1mH(z)=\min f+\Delta\,h(Bz+x_0)$. The image of $\mu_\beta$ under $z\mapsto Bz+x_0$ is $\pi^{\Phi_\beta}_{x_0}$ (Section~\ref{sec:primal}), and $\Phi_\beta$ has positive mass on every coset of $K$. Let $\nu_{x_0}$ be the law of the violation pattern that Regev's reduction outputs with the given decoder. Since $h$ takes values in $[0,1]$, part~1 of Theorem~\ref{thm:guarantee} gives
$$
  \Big|\E_{x_0}\E_{x\sim\pi^{\Phi_\beta}_{x_0}}h(x)-\E_{x_0}\E_{x\sim\nu_{x_0}}h(x)\Big|\ \le\ \E_{x_0}\|\nu_{x_0}-\pi^{\Phi_\beta}_{x_0}\|_{\mathrm{TV}}\ \le\ \sqrt\varepsilon,
$$
and part~2 bounds the distance between $\E_{x_0}\E_{x\sim\nu_{x_0}}h(x)$ and $\sum_x|\Phi_\beta(x)|^2h(x)=(\langle f\rangle_{p_\beta}-\min f)/\Delta$ by $2\sqrt\varepsilon$, as in the proof of Theorem~\ref{thm:approx_gibbs}. Multiplying by $\Delta$ gives the claim.
\end{proof}

By Proposition~\ref{prop:sharp}, for $k\ge3$ and every finite $\beta$ without reconstruction, a decoder with $\varepsilon=o(1)$ exists with probability $1-o(1)$ over $B$, on both ensembles, and for Max-XORSAT on the configuration model, Theorem~\ref{thm:condensation} supplies one at every $\beta<\beta_c$. At these inverse temperatures, a classical algorithm that samples from $\mu_\beta$ as in Hypothesis~\ref{hyp:sampler} therefore reaches, on average over the offset, the expected cost per constraint $\langle f\rangle_{p_\beta}+o(1)$, which is also the value that DQI reaches with $\Phi_\beta$ and a decoder that succeeds. For Max-XORSAT and $k\ge3$, if Conjecture~\ref{conj:classical} holds, efficient classical sampling thus reaches satisfied fractions arbitrarily close to $s_d$.

\subsubsection{Amplitude functions that depend only on the cost}\label{sec:approx_sym}

Jordan et al.~\cite{jordan2024optimization} prepare target states whose amplitudes are polynomials in the objective, so these amplitude functions depend only on the total cost $\sum_jf(x_j)$; for Max-XORSAT, this is the number of violated constraints. Such an amplitude function is not a product, so the tree analysis of Sections~\ref{sec:tree}--\ref{sec:locality} does not apply to it directly. This subsection extends Theorem~\ref{thm:approx_gibbs} to all such amplitude functions when the cost takes two values, as for Max-XORSAT and the Potts cost. The quantum de Finetti theorem shows that the reduced state of $\ket\Phi$ on a set of $o(m)$ qudits is close to a mixture of product states, and the tree analysis applies to each product component. For a two-level cost, the factors of these product states are uniform within each of the two cost levels. If the spectrum of a factor puts at least as much mass on the lower cost level as the uniform vector does, then it is the Gibbs weights of the same cost at some $\beta\in[0,\infty]$, and these spectra are ordered by degradation (Section~\ref{sec:monotone}). Every other spectrum puts less mass on the lower cost level than the Gibbs weights at every $\beta\ge0$, so the corresponding components cannot lower the expected cost per constraint below its value under the Gibbs weights at $\beta_d$.

Until the paragraph on integer-valued costs below, $f(z)=c+\Delta\,\mathbf 1[z\notin A]$ takes two values, with $A$ a nonempty proper subset of $\Z_q$ and $\Delta>0$, and we use $a=|A|$, $b=q-a$, $U_A$, and $\theta_\beta$ from~\eqref{eq:uniformmixture}. We say that $x$ satisfies constraint $j$ if $x_j\in A$, and write $H_A(x)$ for the number of violated constraints, so $\sum_jf(x_j)=mc+\Delta H_A(x)$. For Max-XORSAT, $A=\{0\}$ and $H_A(x)=|x|$. Symmetry means that $\Phi(x)$ depends only on this total cost. The reduced state on $v$ qudits is therefore independent of the chosen set. The decoding states are $Z^c\ket\Phi$ by~\eqref{eq:states_general}.

A symmetric amplitude function lies in the symmetric subspace of a two-dimensional subspace. Let
\begin{equation}\label{eq:two_vectors}
  \ket{u_A}=a^{-1/2}\sum_{z\in A}\ket z,\qquad \ket{u_{A^c}}=b^{-1/2}\sum_{z\notin A}\ket z,\qquad W=\operatorname{span}\{\ket{u_A},\ket{u_{A^c}}\}\subseteq\mathbb C^q ,
\end{equation}
and let $\ket{D_w}$ be the normalized symmetric tensor in $W^{\otimes m}$ with $w$ factors $\ket{u_{A^c}}$ and $m-w$ factors $\ket{u_A}$. Its coefficient on a word $x$ with $H_A(x)=w$ is $\big[\binom mw a^{m-w}b^w\big]^{-1/2}$, and its coefficient on every other word is zero. Hence, if $\Phi(x)=g(H_A(x))$, then
\begin{equation}\label{eq:weighted_dicke}
  \ket\Phi=\sum_{w=0}^mg(w)\sqrt{\tbinom mw a^{m-w}b^w}\,\ket{D_w},
\end{equation}
which lies in the symmetric subspace of $W^{\otimes m}$. We use the following form of the quantum de Finetti theorem.

\begin{lemma}[Quantum de Finetti theorem~{\cite[Corollary~II.3]{christandl2007oneandahalf}}]\label{lem:definetti}
Let $\ket\Phi$ be a unit vector in the symmetric subspace of $(\mathbb C^d)^{\otimes m}$, and let $\rho$ be its reduced state on $v\le m$ factors. There is a probability measure $\mu$ on the unit vectors of $\mathbb C^d$ such that
$$
  \frac12\Big\|\rho-\int\big(\ket\xi\!\bra\xi\big)^{\otimes v}\,d\mu(\xi)\Big\|_1\le\frac{2dv}m .
$$
\end{lemma}

We apply Lemma~\ref{lem:definetti} to $W^{\otimes m}$ through an isometry of $W$ onto $\mathbb C^2$, which maps the symmetric subspace of $W^{\otimes m}$ onto that of $(\mathbb C^2)^{\otimes m}$ and preserves trace distances. With $d=2$, the reduced state of $\ket\Phi$ on $v$ qudits is within $4v/m$ of a mixture of states $(\ket\xi\!\bra\xi)^{\otimes v}$ with unit vectors $\xi\in W$, for every $q$.

Each product component $\ket\xi^{\otimes v}$ leads to a tree problem with spectrum $p_\xi(z)=|\xi_z|^2$. Indeed, $Z^y\ket\xi=V_\xi\ket{\psi^{p_\xi}_y}$, where $V_\xi$ is the diagonal unitary with entries $\xi_z/|\xi_z|$, or $1$ if $\xi_z=0$, and $V_\xi$ does not depend on $y$. For $\xi\in W$, the spectrum $p_\xi$ is constant on $A$ and on its complement, and we call such a probability vector level-uniform. The following lemma compares the optimal tree success for level-uniform spectra with that for the Gibbs weights $p_\beta$ of $f$.

\begin{lemma}[Level-uniform spectra]\label{lem:binary_spectrum}
For a probability vector $\nu$ on $\Z_q$, let $P_t(\nu)$ be the optimal success probability of Proposition~\ref{prop:tree_optimal} for the family $\{\ket{\psi^\nu_y}\}$. Let $\nu$ be level-uniform, with $r_\nu=\nu(A)$.
\begin{enumerate}
\item If $a/q\le r_\nu<1$, then $\nu=p_{\beta(\nu)}$ with $\beta(\nu)=\Delta^{-1}\ln\frac{b\,r_\nu}{a(1-r_\nu)}\ge0$. If $r_\nu=1$, then $\nu=U_A$.
\item For every $\beta\ge0$ with $r_\nu\ge p_\beta(A)$ and every $t$, $P_t(\nu)\le P_t(p_\beta)$.
\end{enumerate}
\end{lemma}

\begin{proof}
A level-uniform vector is determined by its mass on $A$. The Gibbs weights satisfy $p_\beta(A)=a/(a+be^{-\beta\Delta})$, which increases from $a/q$ at $\beta=0$ to $1$ as $\beta\to\infty$. Solving $p_\beta(A)=r_\nu$ gives $e^{-\beta\Delta}=a(1-r_\nu)/(b\,r_\nu)$, which is at most one exactly when $r_\nu\ge a/q$, and this proves part~1. For part~2, if $r_\nu<1$, then $r_\nu\ge p_\beta(A)\ge a/q$, so $\nu=p_{\beta(\nu)}$ by part~1, and $\beta(\nu)\ge\beta$ because $p_\beta(A)$ increases with $\beta$. If $\beta(\nu)>\beta$, the vector $\kappa$ after~\eqref{eq:uniformmixture} satisfies $p_\beta=p_{\beta(\nu)}*\kappa$, and Proposition~\ref{prop:degrade} gives $P_t(p_{\beta(\nu)})\le P_t(p_\beta)$. If $r_\nu=1$, then $\nu=U_A$, and $P_t(U_A)\le P_t(p_\beta)$ by the comparison with $p_\infty=U_A$ in Section~\ref{sec:monotone}.
\end{proof}

\begin{theorem}[Approximate optimization with symmetric amplitude functions]\label{thm:approx_sym}
Let $q\ge2$, let $f(z)=c+\Delta\,\mathbf 1[z\notin A]$ with $\Delta>0$ be a cost that takes two values, and assume that its reconstruction threshold $\beta_d$ is finite. Let $(B_n)$ be a sequence of $(k,D)$-regular matrices with $m_n$ rows. For each $n$, let $\Phi_n$ be a symmetric amplitude function, and let a decoder for the decoding problem of $B_n$ with amplitude function $\Phi_n$ have block success at least $1-\varepsilon_n$. Suppose that for each $n$ there are $t_n$ and $i_n$ such that $\mathrm{LT}_{t_n}(i_n)$ and $\mathrm{TP}_{t_n}(i_n)$ hold and the decoder makes a $t_n$-local estimate at $i_n$, and write $v_n=|N_{t_n}(i_n)|$. For $\eta>0$, let $s_\eta=p_{\beta_d+\eta}(A)$ and $\kappa_\eta=1-P_\infty(p_{\beta_d+\eta})$, where $P_\infty=\lim_tP_t$; then $\kappa_\eta>0$, and it depends only on $(k,D,q,f,\eta)$. The violation pattern $x$ output by DQI satisfies, for every $n$ with $\varepsilon_n<1$,
\begin{equation}\label{eq:approx_sym}
  \E_{x_0}\E\Big[1-\frac{H_A(x)}{m_n}\Big]\le s_\eta+\frac{\varepsilon_n+4v_n/m_n}{\kappa_\eta}+\frac{4v_n}{m_n}+2\sqrt{\varepsilon_n}.
\end{equation}
In particular, if $\varepsilon_n\to0$ and $v_n/m_n\to0$, the expected satisfied fraction is at most $p_{\beta_d}(A)+o(1)$, and the expected cost per constraint is at least $\langle f\rangle_{p_{\beta_d}}-o(1)$. For Max-XORSAT, the expected satisfied fraction is at most $s_d+o(1)$.
\end{theorem}

\begin{proof}
Fix $n$ and drop the index $n$. Write $N=N_t(i)$ and $v=|N|$. Let $\rho_N$ be the reduced state of $\ket\Phi$ on the qudits in $N$. By~\eqref{eq:weighted_dicke} and Lemma~\ref{lem:definetti}, applied as explained after it, there is a probability measure $\mu$ on the unit vectors of $W$ such that $\tau=\int(\ket\xi\!\bra\xi)^{\otimes v}d\mu(\xi)$ satisfies $\frac12\|\rho_N-\tau\|_1\le4v/m$.

\emph{Step 1: the local estimate.} Let $\{E_y\}$ be the POVM on the qudits of $N$ that produces the estimate $\hat c_i$. Given $c$, the reduced state of $\ket{\psi^\Phi_c}=Z^c\ket\Phi$ on $N$ is $Z^{c_N}\rho_N(Z^{c_N})^\dagger$, where $Z^{c_N}=\bigotimes_{j\in N}Z^{c_j}$. As in the proof of Theorem~\ref{thm:locality}, $\mathrm{LT}_t(i)$ and $\mathrm{TP}_t(i)$ imply that $c_N$ is uniformly distributed on the tree code. Since $0\le E_y\le I$ and conjugation by $Z^{c_N}$ preserves trace distance, replacing $\rho_N$ by $\tau$ changes the success probability at $i$ by at most $4v/m$. For each $\xi$, the states $Z^{c_N}\ket\xi^{\otimes v}=V_\xi^{\otimes v}\bigotimes_{j\in N}\ket{\psi^{p_\xi}_{c_j}}$ are, up to a unitary that does not depend on $c$, the states of the tree problem with spectrum $p_\xi$. Hence
$$
  1-\varepsilon\le\Pr[\hat c=c]\le\Pr[\hat c_i=c_i]\le\int P_t(p_\xi)\,d\mu(\xi)+\frac{4v}m .
$$

\emph{Step 2: components above the threshold.} Since $\beta_d+\eta>\beta_d$, the primal model has reconstruction at $\beta_d+\eta$, so $P_t(p_{\beta_d+\eta})\not\to1$ by Theorem~\ref{thm:duality}, and $\kappa_\eta>0$ because $P_t$ is nondecreasing in $t$. Let $r_\xi=p_\xi(A)=|\langle u_A|\xi\rangle|^2$ and $\mathcal A_\eta=\{\xi:r_\xi\ge s_\eta\}$. For $\xi\in\mathcal A_\eta$, the spectrum $p_\xi$ is level-uniform, so Lemma~\ref{lem:binary_spectrum} gives $P_t(p_\xi)\le P_t(p_{\beta_d+\eta})\le1-\kappa_\eta$. Hence $\int P_t(p_\xi)\,d\mu\le1-\kappa_\eta\,\mu(\mathcal A_\eta)$, and Step~1 yields
$$
  \mu(\mathcal A_\eta)\le\frac{\varepsilon+4v/m}{\kappa_\eta}.
$$

\emph{Step 3: the satisfied fraction in the state $\ket\Phi$.} By symmetry, $\sum_x|\Phi(x)|^2(1-H_A(x)/m)$ is the probability that measuring one qudit of $\ket\Phi$ yields a value in $A$. The one-qudit marginal of $\tau$ is $\int\ket\xi\!\bra\xi\,d\mu$, and tracing out qudits does not increase the trace distance, so
$$
  \sum_x|\Phi(x)|^2\Big(1-\frac{H_A(x)}m\Big)\le\int r_\xi\,d\mu(\xi)+\frac{4v}m\le s_\eta+\mu(\mathcal A_\eta)+\frac{4v}m ,
$$
because $r_\xi<s_\eta$ for $\xi\notin\mathcal A_\eta$ and $r_\xi\le1$ for $\xi\in\mathcal A_\eta$.

\emph{Step 4: the output.} By part~2 of Theorem~\ref{thm:guarantee} with $h(x)=1-H_A(x)/m$, the expected satisfied fraction of the output is at most $\sum_x|\Phi(x)|^2(1-H_A(x)/m)+2\sqrt\varepsilon$. Combining Steps~2--4 proves~\eqref{eq:approx_sym}. For the limit, let $n\to\infty$ in~\eqref{eq:approx_sym} and then let $\eta\to0$; the mass $p_\beta(A)$ is continuous in $\beta$. Finally, the cost per constraint is $\frac1m\sum_jf(x_j)=c+\Delta H_A(x)/m$, and $\langle f\rangle_{p_\beta}=c+\Delta(1-p_\beta(A))$, so the bound on the expected satisfied fraction is equivalent to the bound on the expected cost per constraint.
\end{proof}

For random $(k,D)$-regular instances with $k<D$ and $D\ge3$, Theorem~\ref{thm:tp_typical} and Corollary~\ref{cor:typical_nogo} supply the hypotheses at any fixed symbol with probability $1-o(1)$, for $t_n\le(\tfrac12-\delta)\log_{(k-1)(D-1)}m_n$. The ball then has $v_n=O(m_n^{1/2-\delta})$ symbols, so $v_n/m_n\to0$. Hence, with probability $1-o(1)$ over the instance, DQI with a symmetric amplitude function and a local decoder whose block success tends to one has expected satisfied fraction at most $p_{\beta_d}(A)+o(1)$, which is $s_d+o(1)$ for Max-XORSAT. By Proposition~\ref{prop:sharp} and Section~\ref{sec:monotone}, product amplitude functions at $\beta<\beta_d$ approach this value for $k\ge3$ on both ensembles, so, when the decoder is local, no symmetric amplitude function gives an asymptotically larger expected satisfied fraction than the product amplitude functions. Jordan et al.\ analyze costs that take two values over prime fields with polynomial amplitude functions of bounded degree, under a condition on the dual distance of the code, and determine the optimal expected satisfied fraction within this family~\cite[Theorem~4.1]{jordan2024optimization}. Theorem~\ref{thm:approx_sym} allows every amplitude function of the total cost of a two-level cost over every $\Z_q$, and it bounds the expected satisfied fraction under the hypothesis that the decoder is local.

\paragraph{Costs that are integer-valued up to an affine rescaling.} We first prove the bound for a nonconstant fixed integer-valued cost $f$ and then extend it by affine rescaling. Write $H_f(x)=\sum_j f(x_j)$. Dependence on this total cost is more restrictive than permutation invariance alone. The exact-cost count in Appendix~\ref{app:common_cost} makes a local marginal close to a mixture of product states whose magnitudes are Gibbs weights of $f$. This is the additional fact needed to apply the reconstruction obstruction beyond two-valued costs.

\begin{theorem}[Approximation for integer-valued costs]\label{thm:common_cost_local}
Fix $q\ge2$, a nonconstant integer-valued cost $f:\Z_q\to\mathbb Z$, and degrees $3\le k<D$. Let $L_f=\max f-\min f$, suppose $\beta_d<\infty$, and put $e_d=\langle f\rangle_{p_{\beta_d}}$. Let $B$ be a $(k,D)$-regular zero-one matrix and let $\Phi$ be a normalized amplitude function depending only on $H_f(x)$, chosen independently of the offset. Suppose a decoder has block success $1-\varepsilon>0$ and an initial estimate satisfying~\eqref{eq:first_estimate_law}, with admissible-seed probability $g_{\mathrm{init}}>0$ and recovery probability $a_{\mathrm{rec}}$ from~\eqref{eq:first_estimate_repair}. Set $v=\max_{r\in\mathcal R_0}|N_{t_r}(i_r)|$. Then
\begin{equation}\label{eq:common_cost_local}
\E_{x_0}\E\frac{H_f(x)}m\ge e_d-L_f\left[\frac{(\varepsilon+a_{\mathrm{rec}})/g_{\mathrm{init}}+\gamma_m(v)}{\kappa_{q,k,D}}+\gamma_m(v)+2\sqrt\varepsilon\right],
\end{equation}
where $x_0$ is uniform and $\gamma_m(v)$ and $\kappa_{q,k,D}>0$ are defined in~\eqref{eq:common_marginal_error} and~\eqref{eq:common_tree_gap}. In particular, for fixed parameters, if $v_n^2/m_n\to0$ and $(\varepsilon_n+a_{\mathrm{rec},n})/g_{\mathrm{init},n}\to0$, the expected cost per constraint is at least $e_d-o(1)$.
\end{theorem}

The proof is given in Appendix~\ref{app:common_cost_bound}. A fixed local estimate is the case $g_{\mathrm{init}}=1$, $a_{\mathrm{rec}}=0$. The same theorem applies to the standard sequential orders when their initial estimate satisfies the conditions of Section~\ref{sec:strategy}. Its neighborhood requirement covers the full logarithmic-radius range of Corollary~\ref{cor:typical_nogo}, since the balls there have size $O(m^{1/2-\delta})$ for a fixed $\delta>0$. If $\beta_d=\infty$, the corresponding lower bound is simply $\min f$.

The conclusion also holds for costs that are integer-valued up to an affine rescaling. If $f=c+\lambda h$ with $\lambda>0$ and integer-valued $h$, then $p^f_\beta=p^h_{\lambda\beta}$, $\beta_d(f)=\beta_d(h)/\lambda$, and the expected costs transform by the same affine relation. This includes every fixed rational-valued cost. It does not change the hypotheses of the stronger two-valued theorem above.

\begin{corollary}[Integer-valued costs over finite fields]\label{cor:common_cost_field}
Theorem~\ref{thm:common_cost_local} also holds over every fixed finite field $\mathbb F_q$, for a fixed integer-valued cost on that field, zero-one matrices, and the corresponding additive-character decoding model. If $q=p^r$, the phase encoding is $Z^y\ket z=\exp(2\pi i\operatorname{Tr}_{\mathbb F_q/\mathbb F_p}(yz)/p)\ket z$. Here $\beta_d$ is the reconstruction threshold of the primal Gibbs measure over the field. The local conditions, seed law, and offset independence are interpreted in that model. For either regular ensemble and every prescribed radius sequence in the range of Corollary~\ref{cor:typical_nogo}, both local conditions hold at a typical symbol with probability tending to one, and $v^2/m\to0$.
\end{corollary}

Appendix~\ref{app:common_field} proves the character and tree-decoding transfer and the reduction of projection to the prime field. It does not replace $\mathbb F_{p^r}$ by $\Z_{p^r}$ or introduce arbitrary nonzero matrix coefficients.

For the rest of this subsection, we return to Max-XORSAT, with $q=2$, $f(z)=z$, and $A=\{0\}$.

\paragraph{An initial local estimate.} The same product-state reduction applies when the decoder's initial estimate is selected using input-independent randomness. The measure used to approximate a symmetric state does not depend on the selected neighborhood, so the correctness probabilities can be averaged over the admissible seeds.

\begin{proposition}[Symmetric amplitude functions with an initial local estimate]\label{prop:first_local_symmetric}
Let $B$ be $(k,D)$-regular, assume that $\beta_d<\infty$, and let $\Phi$ be a normalized symmetric amplitude function chosen independently of the offset. Suppose that a decoder for the decoding problem with $\Phi$ has block success $1-\varepsilon$, with $\varepsilon<1$, and an initial estimate satisfying~\eqref{eq:first_estimate_law}. Let $g_{\mathrm{init}}>0$ be the probability of the admissible seeds, let $a_{\mathrm{rec}}$ be the probability in~\eqref{eq:first_estimate_repair}, and set $v=\max_{r\in\mathcal R_0}|N_{t_r}(i_r)|$. For every $\eta>0$, define $s_\eta=p_{\beta_d+\eta}(0)$ and $\kappa_\eta=1-P_\infty(p_{\beta_d+\eta})>0$. The violation pattern $x$ output by DQI satisfies
\begin{equation}\label{eq:first_local_symmetric}
  \E_{x_0}\E\Big[1-\frac{|x|}{m}\Big]\le s_\eta+\frac{1}{\kappa_\eta}\Big(\frac{\varepsilon+a_{\mathrm{rec}}}{g_{\mathrm{init}}}+\frac{4v}{m}\Big)+\frac{4v}{m}+2\sqrt\varepsilon,
\end{equation}
where the offset is uniformly random.
\end{proposition}

\begin{proof}
By Lemma~\ref{lem:definetti} and permutation invariance, there is a probability measure $\mu$ on single-qubit unit vectors such that the reduced state of $\ket\Phi$ on any $v$ qubits is within trace distance $4v/m$ of $\int(\ket\xi\!\bra\xi)^{\otimes v}\,d\mu(\xi)$. Tracing out qubits gives the same bound, with the same measure, on every neighborhood associated with an admissible seed. As in the proof of Theorem~\ref{thm:approx_sym}, absorbing the phases of $\xi$ into an input-independent diagonal unitary and applying the tree bound shows that
$$
  \Pr[Y_r=c_{i_r}\mid R=r]\le\int P_{t_r}(p_\xi)\,d\mu(\xi)+\frac{4v}{m},\qquad p_\xi=(|\xi_0|^2,|\xi_1|^2).
$$
Let $w=\mu\{\xi:|\xi_0|^2\ge s_\eta\}$. On this set, Lemma~\ref{lem:binary_spectrum} gives $P_{t_r}(p_\xi)\le1-\kappa_\eta$, so the right-hand side is at most $1-\kappa_\eta w+4v/m$. Splitting block recovery according to the seed and the correctness of the initial estimate, as in Proposition~\ref{prop:first_local_block}, gives
$$
  1-\varepsilon\le1-g_{\mathrm{init}}+g_{\mathrm{init}}\Big(1-\kappa_\eta w+\frac{4v}{m}\Big)+a_{\mathrm{rec}},
$$
and hence $w\le\kappa_\eta^{-1}\big((\varepsilon+a_{\mathrm{rec}})/g_{\mathrm{init}}+4v/m\big)$. The one-qubit marginal bound gives $\sum_x|\Phi(x)|^2(1-|x|/m)\le s_\eta+w+4v/m$. Part~2 of Theorem~\ref{thm:guarantee} adds at most $2\sqrt\varepsilon$ when passing to the output of DQI, which proves~\eqref{eq:first_local_symmetric}.
\end{proof}

\begin{corollary}[Approximation with an initial local estimate]\label{cor:first_local_symmetric}
Fix the degrees $k,D$ in the setting of Proposition~\ref{prop:first_local_symmetric}, and consider a sequence of instances with $m_n\to\infty$. If $v_n/m_n\to0$ and $(\varepsilon_n+a_{\mathrm{rec},n})/g_{\mathrm{init},n}\to0$, then the expected satisfied fraction that DQI reaches, averaged over a uniformly random offset, is at most $s_d+o(1)$.
\end{corollary}

\begin{proof}
Since $g_{\mathrm{init},n}\le1$, the hypotheses imply $\varepsilon_n\to0$. First let $n\to\infty$ in~\eqref{eq:first_local_symmetric}, and then let $\eta\to0$; continuity gives $s_\eta\to s_d$.
\end{proof}

For the fixed-coordinate and uniformly randomized sequential BPQM implementations of Section~\ref{sec:strategy}, the admissible-seed probability tends to one on typical instances, while $v_n/m_n\to0$ throughout $t_n\le c_0\log m_n$. Retaining the initial estimate on successful recovery, or using a fixed default output on failure, makes $a_{\mathrm{rec},n}\to0$. Thus reliable decoding gives the same bound $s_d$ for these implementations, even when joint locality under Definition~\ref{def:local_decoder} has not been established. This corollary concerns Max-XORSAT; it does not replace the stronger two-valued-cost statement of Theorem~\ref{thm:approx_sym}.

The quantum de Finetti theorem also shows that, for the optimal polynomial amplitude functions of Jordan et al., the state $\ket\Phi$ is locally close to a product state. For a symmetric amplitude function $\Phi$, write $\hat\Phi(e)=2^{-m/2}\sum_x\Phi(x)(-1)^{e\cdot x}$ for its Fourier transform. A polynomial amplitude function of degree $L$ in the number of satisfied constraints has $\hat\Phi$ supported on weights at most $L$, and the decoding problem then has errors of weight at most $L$. The value $\sum_x|\Phi(x)|^2(1-|x|/m)$ does not depend on the code. Jordan et al.\ show that its maximum over such amplitude functions tends to $\tfrac12+\sqrt{\rho(1-\rho)}$ when $L/m\to\rho\le\tfrac12$, which is the semicircle law, and that it equals the expected satisfied fraction that DQI reaches when the minimum distance of $C$ exceeds $2L+1$. The condition on the minimum distance is not needed for the value to be reached approximately: by part~2 of Theorem~\ref{thm:guarantee}, on average over the offset, the expected satisfied fraction that DQI reaches is within $2\sqrt\varepsilon$ of this value whenever the decoder has block success $1-\varepsilon>0$, and Jordan et al.\ prove a similar bound for imperfect decoders.

\begin{proposition}[Local structure of optimal polynomial amplitude functions]\label{prop:semicircle_local}
Let $q=2$, and let $\Phi_n$ be symmetric amplitude functions on $m_n\to\infty$ qubits whose Fourier transforms are supported on weights at most $L_n$, where $L_n/m_n\to\rho\in(0,\tfrac12)$. Then
$$
  \limsup_{n\to\infty}\sum_x|\Phi_n(x)|^2\Big(1-\frac{|x|}{m_n}\Big)\le\frac12+\sqrt{\rho(1-\rho)} .
$$
If $\sum_x|\Phi_n(x)|^2(1-|x|/m_n)$ converges to $\frac12+\sqrt{\rho(1-\rho)}$, then for every fixed $v$ the reduced state of $\ket{\Phi_n}$ on $v$ qubits converges in trace norm to $(\ket{\phi_\rho}\!\bra{\phi_\rho})^{\otimes v}$, where $\ket{\phi_\rho}=\sqrt{1-\rho^\perp}\ket0+\sqrt{\rho^\perp}\ket1$ is the single-qubit factor of the product state $\ket{\Phi_\beta}$ with $p_\beta(1)=\rho^\perp$.
\end{proposition}

\begin{proof}
For a unit vector $\xi\in\mathbb C^2$, let $a_\xi=|\xi_0|^2$, and let $\rho_\xi=\frac12|\xi_0-\xi_1|^2$ be the probability that measuring $\ket\xi$ in the Hadamard basis yields $\ket-$. Write $u=\xi_0\bar\xi_1$. Then $(a_\xi-\frac12)^2=\frac14-|u|^2$ and $\rho_\xi(1-\rho_\xi)=\frac14-(\operatorname{Re}u)^2$, so
\begin{equation}\label{eq:pointwise_semicircle}
  a_\xi\le\frac12+\sqrt{\rho_\xi(1-\rho_\xi)},
\end{equation}
with equality if and only if $u$ is real and $a_\xi\ge\frac12$. If in addition $\rho_\xi<\frac12$, equality forces $u>0$.

Fix $v$, and let $\mu_n$ be the measure of Lemma~\ref{lem:definetti} for the reduced state of $\ket{\Phi_n}$ on $v$ qubits. Its one-qubit marginal is within trace distance $4v/m_n$ of $\int\ket\xi\!\bra\xi\,d\mu_n$. Measuring one qubit of $\ket{\Phi_n}$ in the Hadamard basis yields $\ket-$ with probability equal to the expected weight of $e$ under $|\hat\Phi_n(e)|^2$, divided by $m_n$, which is at most $L_n/m_n$. Hence $\bar\rho_n=\int\rho_\xi\,d\mu_n\le L_n/m_n+4v/m_n$. Similarly, $A_n=\sum_x|\Phi_n(x)|^2(1-|x|/m_n)\le\int a_\xi\,d\mu_n+4v/m_n$. By~\eqref{eq:pointwise_semicircle} and Jensen's inequality for the concave function $y\mapsto\sqrt{y(1-y)}$,
$$
  A_n\le\frac12+\int\sqrt{\rho_\xi(1-\rho_\xi)}\,d\mu_n+\frac{4v}{m_n}\le\frac12+\sqrt{\bar\rho_n(1-\bar\rho_n)}+\frac{4v}{m_n}.
$$
The function $y\mapsto\sqrt{y(1-y)}$ is increasing on $[0,\frac12]$, and $\bar\rho_n\le L_n/m_n+4v/m_n<\frac12$ for large $n$, so the right-hand side tends to at most $\frac12+\sqrt{\rho(1-\rho)}$. This proves the first claim.

Now suppose that $A_n\to\frac12+\sqrt{\rho(1-\rho)}$. The unit sphere of $\mathbb C^2$ is compact, so every subsequence of $(\mu_n)$ has a further subsequence that converges weakly to a probability measure $\mu$. The functions $a_\xi$, $\rho_\xi$, and $\sqrt{\rho_\xi(1-\rho_\xi)}$ are continuous, so along this subsequence
$$
  \frac12+\sqrt{\rho(1-\rho)}\le\int a_\xi\,d\mu\le\frac12+\int\sqrt{\rho_\xi(1-\rho_\xi)}\,d\mu\le\frac12+\sqrt{\bar\rho(1-\bar\rho)}\le\frac12+\sqrt{\rho(1-\rho)},
$$
where $\bar\rho=\int\rho_\xi\,d\mu\le\rho$. All inequalities are therefore equalities. Since $y\mapsto\sqrt{y(1-y)}$ is strictly increasing on $[0,\frac12]$, $\bar\rho=\rho$. Since it is strictly concave, $\rho_\xi=\rho$ for $\mu$-almost every $\xi$. In the second inequality, the difference of the two sides is the integral of the nonnegative function $\frac12+\sqrt{\rho_\xi(1-\rho_\xi)}-a_\xi$, so equality holds in~\eqref{eq:pointwise_semicircle} for $\mu$-almost every $\xi$. For such $\xi$, $u>0$ and $a_\xi=\frac12+\sqrt{\rho(1-\rho)}=1-\rho^\perp$, so $\xi=e^{i\theta}\ket{\phi_\rho}$ for some phase $\theta$. Hence $\int(\ket\xi\!\bra\xi)^{\otimes v}d\mu=(\ket{\phi_\rho}\!\bra{\phi_\rho})^{\otimes v}$. This limit does not depend on the subsequence, and the integral is a weakly continuous function of the measure, so $\int(\ket\xi\!\bra\xi)^{\otimes v}d\mu_n\to(\ket{\phi_\rho}\!\bra{\phi_\rho})^{\otimes v}$ along the whole sequence. Lemma~\ref{lem:definetti} completes the proof.
\end{proof}

Proposition~\ref{prop:semicircle_local} explains why DQI reaches the same satisfied fraction with the optimal polynomial amplitude functions of Jordan et al.\ and with the product amplitude functions of Shutty et al. The satisfied fraction is determined by the one-qubit reduced state, and asymptotically the two states have the same reduced states on any fixed number of qubits. The difference between the two states matters for exact optimization, as Appendix~\ref{sec:exact_heavy} shows.

\subsubsection{Block amplitude functions}\label{sec:approx_block}

Theorem~\ref{thm:approx_sym} concerns amplitude functions that depend only on the total cost. For Max-XORSAT on Gallager's ensemble, Shutty et al.~\cite{shutty2026locallyquantumdecoders} use instead an amplitude function that is a product over blocks of constraints. In this subsection, we describe this block amplitude function and compare it with product amplitude functions. On four degree pairs, DQI reaches a smaller satisfied fraction with the block amplitude function and any decoder whose block success tends to one than with a product amplitude function and a local decoder (Proposition~\ref{prop:block_four}). For local decoders on the other degree pairs, we reduce the comparison to an inequality for an exact recursion on trees, which we have not proved, and we state the resulting bound as Conjecture~\ref{conj:block} in Section~\ref{sec:scope}. Throughout this subsection, $q=2$, $f(z)=z$, and $B$ is drawn from Gallager's ensemble.

In Gallager's ensemble, the checks of the first layer partition the $m$ code symbols into blocks $S_1,\dots,S_{m/D}$ of size $D$. In the primal problem, each block consists of the $D$ clauses that contain one variable of the first layer, which we call the variable of the block. The block amplitude function is the product $\Phi(x)=\prod_i\Phi^{(D)}(x_{S_i})$. Its factor $\Phi^{(D)}$ depends on a parameter $\alpha\in(0,1)$ and on a set of patterns that contains exactly one pattern from each pair $\{y,y+\mathbf 1\}$ and has the smallest possible total weight, so that the pattern chosen from the pair of $y$ has weight $\min(|y|,D-|y|)$. It assigns amplitude $\sqrt{1-\alpha}$ to the all-zero pattern, amplitude $\sqrt{\alpha/(2^{D-1}-1)}$ to each of the other $2^{D-1}-1$ patterns of this set, and amplitude zero to all other patterns. Flipping the variable of a block changes its pattern from $y$ to $y+\mathbf 1$, so this set consists of patterns on which the flip does not reduce the number of violated clauses of the block. Shutty et al.\ use this amplitude function together with a nonlocal decoder that measures each block with FGUM and recovers the remaining blocks by Gaussian elimination (Section~\ref{sec:fgum}). The statements below concern every decoder for the decoding problem induced by DQI with this amplitude function.

Since $\ket\Phi$ is a product over the blocks, the restriction $c_{S_i}$ of the codeword is sent through a channel on each block separately, as each symbol is for the product amplitude function. Each block is the support of a check, so $c_{S_i}$ has even weight. For every codeword, measuring a block in the computational basis yields the all-zero pattern with probability $1-\alpha$ and each other pattern of the set with probability $\alpha/(2^{D-1}-1)$. The pair of a uniformly random $y\in\Z_2^D$ is uniformly random among the $2^{D-1}$ pairs, and the all-zero pattern has weight zero, so the other patterns of the set have mean weight $2^{D-1}\mu_D/(2^{D-1}-1)$, where $\mu_D=2^{-D}\sum_{w=0}^D\binom Dw\min(w,D-w)$. The expected satisfied fraction in $\ket\Phi$ is therefore
\begin{equation}\label{eq:block_fraction}
  s(\alpha)=1-\frac{2^{D-1}\alpha}{2^{D-1}-1}\,\frac{\mu_D}D ,
\end{equation}
and by part~2 of Theorem~\ref{thm:guarantee}, DQI with a decoder of block success $1-\varepsilon$ reaches $s(\alpha)$ up to $2\sqrt\varepsilon$, on average over the offset.

Every decoder whose block success tends to one must satisfy an entropy condition, which is the analogue for blocks of the limit $\beta_{\mathrm H}$ of Section~\ref{sec:dqi}. The code $C$ has at least $2^{m-n}$ codewords, that is, at least $D-k$ bits for each of the $m/D$ blocks, whereas the pattern of a measured block has entropy
$$
  \mathcal H_D(\alpha)=H_2(\alpha)+\alpha\log_2\big(2^{D-1}-1\big)
$$
bits. By the Holevo bound and Fano's inequality, the block success can tend to one only if $\mathcal H_D(\alpha)\ge D-k-o(1)$ (Lemma~\ref{lem:block_entropy}). As $\alpha$ increases from $0$ to $1-2^{1-D}$, the entropy $\mathcal H_D(\alpha)$ increases, whereas $s(\alpha)$ decreases, so the entropy condition bounds $s(\alpha)$ from above. For four degree pairs, the resulting bound lies below a satisfied fraction that DQI reaches with a product amplitude function and a local decoder.

\begin{proposition}[Block amplitude functions on four degree pairs]\label{prop:block_four}
Let $(k,D)$ be one of the pairs in the following table, and let the offset be uniformly random.
\begin{center}
\begin{tabular}{@{}ccc@{}}
\toprule
$(k,D)$ & Block amplitude function, every decoder & Product amplitude function, local decoder\\
\midrule
$(3,6)$ & $0.8551$ & $0.8625$\\
$(3,7)$ & $0.8248$ & $0.8368$\\
$(3,8)$ & $0.7896$ & $0.8157$\\
$(4,8)$ & $0.8422$ & $0.8520$\\
\bottomrule
\end{tabular}
\end{center}
\begin{enumerate}
\item Let $(B_m)$ be a sequence of matrices from Gallager's ensemble with $m\to\infty$ rows. For each $m$, let a decoder for the decoding problem induced by DQI with the block amplitude function of some parameter $\alpha_m\in(0,1)$ have block success $1-\varepsilon_m$. If $\varepsilon_m\to0$, then the expected satisfied fraction that DQI reaches, averaged over the offset, is at most the value in the second column plus $o(1)$.
\item There is a finite $\beta$ at which the primal model has non-reconstruction such that, with probability $1-o(1)$ over random instances from Gallager's ensemble, DQI with the product amplitude function $\Phi_\beta$ and the two-stage decoder of Proposition~\ref{prop:sharp}, which is local, has block success at least $1-2m^{-3}$ and reaches an expected satisfied fraction, averaged over the offset, larger than the value in the third column.
\end{enumerate}
\end{proposition}

The proof is given in Appendix~\ref{app:block}. Part~1 follows from the entropy condition and four exact integer inequalities, and part~2 from Proposition~\ref{prop:sharp} and the certified lower bounds on $\beta_d$ of Appendix~\ref{app:certified}. Part~1 holds for every decoder, local or nonlocal, and for every matrix from Gallager's ensemble, because the entropy condition uses only the number of codewords of $C$. The inverse temperature $\beta$ in part~2 has no reconstruction, so the values in the third column are below $p_\beta(0)\le s_d$. Hence, on these four families, DQI with the block amplitude function reaches a satisfied fraction below $s_d$ with every decoder. For the other eleven benchmark pairs, the bound from the entropy condition lies above our estimates of $s_d$ in Table~\ref{tab:benchmark}, so the entropy condition does not decide the comparison there.

For local decoders, a bound at $s_d$ for the block amplitude function requires a comparison of the optimal local estimates for the block and product channels. The states of a block for any two distinct values of $c_{S_i}$ have the same overlap (Lemma~\ref{lem:block_channel}), and this makes the optimal local estimate computable exactly. A $t$-local estimate at a symbol $i$ is at most as accurate as the optimal measurement on the union of the blocks that meet $N_t(i)$, which lies in $N_{t+1}(i)$. If $\mathrm{LT}_{t+1}(i)$ and $\mathrm{TP}_{t+1}(i)$ hold, the restriction of a uniformly random codeword to this union is a uniformly random codeword of a tree code whose checks are the blocks and some of the other checks (Lemma~\ref{lem:block_completion}). On such trees, an exact recursion computes the largest probability of recovering the root symbol (Proposition~\ref{prop:block_tree}). At the checks outside the first layer, it applies the check-node operation of BPQM, and at each block, it combines the state of the block with the incoming messages by a convolution on the patterns of the block.

The optimal root success is nondecreasing in the depth and in $\alpha\le1-2^{1-D}$. Hence, if it stays bounded away from one as the depth tends to infinity at the parameter with $s(\alpha)=s_d$, then DQI with the block amplitude function and such a local decoder reaches at most $s_d+o(1)$. We have not proved this infinite-depth bound. The noise of the channel on blocks is not a product over symbols, so Theorem~\ref{thm:duality} does not apply to it. At the parameter corresponding to our numerical estimate of $s_d$, evaluations of the exact recursion with populations from $4000$ to $12000$ and depths from $60$ to $400$ gave root errors between $0.169$ and $0.316$ across all fifteen benchmark pairs. The largest variation for a fixed pair was below $0.01$ (Appendix~\ref{app:block}). These finite-population, finite-depth observations support the conjectured bound but do not establish that the infinite-depth error is positive.

Proposition~\ref{prop:block_four} and the recursion concern only the comparison with product amplitude functions. A bound at $s_d$ for local decoders would show that the block amplitude function does not improve on product amplitude functions with local decoders, and, as for symmetric amplitude functions, it would exclude an advantage over classical algorithms only under Conjecture~\ref{conj:classical} (Lemma~\ref{lem:gibbs_cost}). With FGUM followed by Gaussian elimination, Regev's reduction reaches with the block amplitude function exactly the value of Turbo Prange (Section~\ref{sec:fgum}), but this equality concerns one nonlocal decoder and bounds no other decoder. We discuss the evidence for Conjecture~\ref{conj:block} in Section~\ref{sec:scope}.

\subsubsection{Comparison with other decoders and with classical algorithms}\label{sec:approx_bench}

Table~\ref{tab:benchmark} compares $s_d$ with the values reported for the fifteen benchmark families of Shutty et al.~\cite{shutty2026locallyquantumdecoders}. These are $(k,D)$-regular Max-XORSAT instances, which they call $D$-regular Max-$k$-XORSAT (Remark~\ref{rem:units}). The first four value columns concern DQI with product amplitudes. The classical belief-propagation values are those reported in that work. Since a classical decoder uses the binary symmetric channel with crossover probability $\alpha^\perp$ and the code has rate at least $1-k/D$, its Shannon limit is $\tfrac12+\sqrt{\rho_{\mathrm{Sh}}(1-\rho_{\mathrm{Sh}})}$, where $H_2(\rho_{\mathrm{Sh}})=k/D$.

The local-decoding value $s_d$ is approached on both regular ensembles by Proposition~\ref{prop:sharp}. The tree BPQM values reported in~\cite{mandal2026blockerror} estimate this same quantity by Theorem~\ref{thm:duality}. The Holevo value $s_H$ bounds every decoder, and on the configuration model the tighter value is $s_c<s_H$ (Section~\ref{sec:condensation}). The remaining columns are classical optimization algorithms. Prange's algorithm has asymptotic value $\tfrac12+\tfrac k{2D}$, while the Turbo Prange values are the asymptotic predictions of Shutty et al., up to corrections exponentially small in $D$. Their Regev reduction with FGUM attains the same values as Turbo Prange (Section~\ref{sec:fgum}).

For simulated annealing, Table~\ref{tab:benchmark} retains the literature comparison at $2520$ variables and $10^6$ sweeps, with $\beta$ increasing linearly from $0$ to $4$. We report the mean largest satisfied fraction over six instances, with one run per instance; Shutty et al.\ instead take the best of four runs per instance. On fourteen pairs the reported values differ by at most $0.0023$. The larger and longer experiments are summarized separately in Table~\ref{tab:sim_summary}.

\begin{table}[ht]
\centering
\small
\begin{tabular}{@{}c cccc ccc@{}}
\toprule
& \multicolumn{4}{c}{DQI with a product amplitude function} & \multicolumn{3}{c}{Classical algorithms} \\
\cmidrule(lr){2-5}\cmidrule(l){6-8}
$(k,D)$ & BP decoder & Shannon & $s_d$ (local) & Holevo $s_H$ & SA & Prange & Turbo Prange \\
\midrule
$(3,4)$ & $0.8730$ & $0.910$ & $0.9345$ & $0.958$ & $\mathbf{0.9381}$ & $0.8750$ & $0.8930$ \\
$(3,5)$ & $0.8176$ & $0.853$ & $0.8946$ & $0.921$ & $\mathbf{0.9024}$ & $0.8000$ & $0.8379$ \\
$(3,6)$ & $0.7776$ & $0.813$ & $0.8628$ & $0.890$ & $\mathbf{0.8726}$ & $0.7500$ & $0.7857$ \\
$(3,7)$ & $0.7476$ & $0.783$ & $0.8372$ & $0.865$ & $\mathbf{0.8489}$ & $0.7143$ & $0.7621$ \\
$(3,8)$ & $0.7243$ & $0.759$ & $0.8161$ & $0.844$ & $\mathbf{0.8279}$ & $0.6875$ & $0.7312$ \\
$(4,5)$ & $0.8605$ & $0.929$ & $0.9275$ & $0.969$ & $\mathbf{0.9293}$ & $0.9000$ & $0.9216$ \\
$(4,6)$ & $0.8214$ & $0.879$ & $0.8983$ & $0.939$ & $\mathbf{0.9011}$ & $0.8333$ & $0.8616$ \\
$(4,7)$ & $0.7908$ & $0.842$ & $0.8734$ & $0.912$ & $\mathbf{0.8776}$ & $0.7857$ & $0.8267$ \\
$(4,8)$ & $0.7663$ & $0.813$ & $0.8523$ & $0.890$ & $\mathbf{0.8566}$ & $0.7500$ & $0.7905$ \\
$(5,6)$ & $0.8443$ & $0.941$ & $0.9162$ & $0.975$ & $0.9170$ & $0.9167$ & $\mathbf{0.9312}$ \\
$(5,7)$ & $0.8140$ & $0.897$ & $0.8924$ & $0.950$ & $\mathbf{0.8943}$ & $0.8571$ & $0.8853$ \\
$(5,8)$ & $0.7893$ & $0.863$ & $0.8718$ & $0.928$ & $\mathbf{0.8730}$ & $0.8125$ & $0.8441$ \\
$(6,7)$ & $0.8291$ & $0.950$ & $0.9046$ & $0.980$ & $0.9051$ & $0.9286$ & $\mathbf{0.9427}$ \\
$(6,8)$ & $0.8045$ & $0.910$ & $0.8845$ & $0.958$ & $0.8860$ & $0.8750$ & $\mathbf{0.8962}$ \\
$(7,8)$ & $0.8155$ & $0.956$ & $0.8936$ & $0.983$ & $0.8932$ & $0.9375$ & $\mathbf{0.9481}$ \\
\bottomrule
\end{tabular}
\caption{Expected satisfied fractions for $(k,D)$-regular Max-XORSAT, which Shutty et al.\ call $D$-regular Max-$k$-XORSAT. The values in the columns ``BP decoder'' and ``Prange'' are those that Shutty et al.~\cite{shutty2026locallyquantumdecoders} report for DQI with classical belief propagation and for the asymptotic value of Prange's algorithm. The column ``Turbo Prange'' gives their predicted values for Regev's reduction with FGUM, which they show to equal those of Turbo Prange. The column ``SA'' gives the largest satisfied fraction that simulated annealing reaches when it raises $\beta$ linearly from $0$ to $4$ over $10^6$ sweeps, on instances with $2520$ variables, averaged over six instances, with standard deviation below $0.0013$ (Appendix~\ref{app:dynamics}). ``Shannon'' is the limit for classical decoders, and ``Holevo $s_H$'' is the limit that Jordan et al.\ derived from the Holevo bound for every decoder (Section~\ref{sec:dqi}); Table~\ref{tab:condensation} gives the limit $s_c<s_H$ of every decoder on the configuration model. The column $s_d$ is computed by population dynamics for the spectral recursion with $2\times10^5$ samples and $1500$ levels; its entries are accurate to about $2\times10^{-4}$. Bold entries mark the best classical value in each row.}
\label{tab:benchmark}
\end{table}

The numerical comparison shows a genuine improvement in quantum decoding: $s_d$ exceeds the reported belief-propagation value on every benchmark pair and the Shannon limit on nine pairs. For optimization of the primal instance, however, the best classical value in every row exceeds $s_d$. It comes from Turbo Prange on four pairs and simulated annealing on the other eleven. Quantum improvements in the decoding problem therefore do not by themselves establish an optimization advantage.

\paragraph{Extended annealing comparison.} We tested $62$ binary degree pairs with $3\le k<D\le36$ and $105$ nonbinary tuples on instances with about $10080$ variables from Gallager's ensemble. With enough sweeps, the mean largest satisfied fraction exceeds the estimated $s_d$ on every family. Table~\ref{tab:sim_summary} gives both the common $10^6$-sweep comparison and the results at the largest budget used for each family. For the binary study, all $785$ survey and long-run trajectories completed in less than four hours of wall-clock time on a ten-core M1 Max MacBook Pro, using parallel processes ($19.7$ CPU-hours). Thus the finite-instance comparison with the local-decoding bound is achievable on a laptop, without assuming Conjecture~\ref{conj:classical}.

At $10^6$ sweeps, the mean over four instances exceeds $s_d$ on $51$ of the $62$ binary pairs, including all fifteen benchmarks. The other eleven fall short by at most $0.0013$. The selected $10^7$-sweep runs include all eleven and bring every family mean above $s_d$. Over $\Z_q$ with $q\in\{3,4,5,7,8\}$ and over $\mathbb F_4$ and $\mathbb F_8$, more sweeps are required, reaching $10^8$ on some families. At the largest budgets, the means are over one to four instances. The full degree-pair tables, fluctuations and selection protocol are given in Appendix~\ref{app:dynamics}.

\begin{table}[!htbp]
\centering
\small
\setlength{\tabcolsep}{5pt}
\begin{tabular}{@{}c c cc c c@{}}
\toprule
& & \multicolumn{2}{c}{Families with mean above $s_d$} & & \\
\cmidrule(lr){3-4}
Alphabet & Families & $10^6$ sweeps & Largest budget & \shortstack{Maximum\\sweeps} & \shortstack{Smallest margin\\at largest budget} \\
\midrule
$\Z_2$ & $62$ & $51$ & $62$ & $10^7$ & $2.3\times10^{-4}$ \\
$\Z_3$ & $15$ & $10$ & $15$ & $10^7$ & $4.6\times10^{-4}$ \\
$\Z_4$ & $15$ & $8$ & $15$ & $10^8$ & $5.5\times10^{-4}$ \\
$\Z_5$ & $15$ & $7$ & $15$ & $10^8$ & $2.6\times10^{-4}$ \\
$\Z_7$ & $15$ & $6$ & $15$ & $10^8$ & $5.7\times10^{-4}$ \\
$\Z_8$ & $15$ & $5$ & $15$ & $10^8$ & $6.4\times10^{-4}$ \\
$\mathbb F_4$ & $15$ & $8$ & $15$ & $10^7$ & $5.9\times10^{-4}$ \\
$\mathbb F_8$ & $15$ & $6$ & $15$ & $10^8$ & $5.2\times10^{-4}$ \\
\bottomrule
\end{tabular}
\caption{Simulated annealing versus the estimated local-decoding bound for unweighted Max-LINSAT, with about $10080$ variables and zero-one matrices from Gallager's ensemble. A family is a degree pair at the specified alphabet. The $10^6$-sweep column uses means over four instances; the largest-budget column uses the largest sweep count available for each family, with one to four instances at longer budgets. The maximum sweep count need not be used on every family. Margins are differences of the mean largest satisfied fraction and the estimated $s_d$, not confidence bounds. The binary schedule ends at $\beta=4$ and the others at $\beta=7$. Detailed results are in Tables~\ref{tab:sim_pairs} and~\ref{tab:sim_qsa}.}
\label{tab:sim_summary}
\end{table}

\paragraph{Relation to Gibbs sampling.} The energy during annealing follows its equilibrium value up to inverse temperatures that approach $\beta_d$ as the sweep count grows, and continues decreasing after annealing leaves equilibrium. This explains why the final satisfied fraction can exceed the Gibbs value at reconstruction. The behavior agrees with the cavity-method account of slow annealing~\cite{montanari2004cooling,krzakala2010following}, discussed in Appendix~\ref{app:sim_fraction}. Polynomial-time mixing remains unproved; the optimization comparison is directly observed at the simulated sizes and computational costs.

The optimal-decoder ceiling $s_c$ on the configuration model exceeds the best classical benchmark value in every row (Table~\ref{tab:condensation}). The classical values concern Gallager's ensemble, for which the same condensation ceiling is expected but not proved. Subject to that ensemble comparison, an efficient nonlocal decoder approaching $s_c$ could improve on these classical values.

DQI has also been compared numerically with classical Markov chains on other sparse instances. On the irregular sparse Max-XORSAT instances of Jordan et al., Gil-Fuster et al.~\cite{gilfuster2026approximate} run block Gibbs sampling, a Markov chain that resamples three variables at a time and whose stationary law is the target law of DQI with a polynomial amplitude function. Their simulations extend to more than $1000$ qubits, one for each variable. In these simulations, the number of steps after which this chain reaches the expected satisfied fraction that DQI reaches with belief propagation is well approximated by a power law of degree about five in the number of variables. For optimal polynomial intersection, another problem to which Jordan et al.\ apply DQI, the corresponding number of steps grows exponentially in the number of qubits, with base about $1.1$.

\paragraph{Classical post-processing.} Our bounds concern the output of DQI. Classical post-processing, such as greedy local improvement, can increase the satisfied fraction of any output. For product amplitude functions, part~1 of Theorem~\ref{thm:guarantee} shows that the output law is close to $\pi_\beta$ in expected total variation. Post-processing applied to DQI outputs therefore reaches, up to this distance, the same expected objective as when it is applied to samples from $\mu_\beta$. If Conjecture~\ref{conj:classical} holds, such samples can be generated classically at every $\beta$ without reconstruction, so post-processing does not create an advantage.

\subsection{Exact optimization}\label{sec:exact}

For exact optimization, quantum minimum finding uses repeated coherent preparations of a state and its inverse, with amplitude amplification over assignments of low cost. Its preparation count is governed by the probability of an optimum. We first compare quantum minimum finding on an ideal Gibbs state with classical sampling, conditionally throughout non-reconstruction and unconditionally in the Prange union-bound range (Sections~\ref{sec:conditional} and~\ref{sec:prange}). We then extend the comparison to all polynomial target states in a fixed subcritical-degree range (Section~\ref{sec:poly_exact}). Section~\ref{sec:prepared} explains when these target-state comparisons apply to the state actually prepared by DQI.

\subsubsection{Quantum and classical costs of finding an optimum}\label{sec:extraction}

Theorem~\ref{thm:only_nonrecon} limits the inverse temperatures that DQI with local decoders can reach, but a bound on the inverse temperature does not by itself bound the speedup. To compare DQI with classical algorithms, we need a task for which both use a Gibbs measure as a resource. We use the task of finding an exact minimizer of~\eqref{eq:primal_cost}. For this task, the quantum cost and the classical cost are both governed by one quantity, the probability that a sample from the Gibbs measure is a minimizer. Let $S^\star=\arg\min H$ and let $\lambda_\beta=\mu_\beta(S^\star)$. Since $H$ depends on $z$ only through $x=Bz+x_0$, $\lambda_\beta$ is also the probability under $\pi_\beta$ of the image of $S^\star$.

\begin{lemma}\label{lem:mass_monotone}
$\lambda_\beta$ is nondecreasing in $\beta$, and $\lambda_\beta\ge q^{-n}$.
\end{lemma}
\begin{proof}
Let $H^\star=\min H$. Differentiation gives
$$
  \frac{d}{d\beta}\ln\lambda_\beta=\E_{\mu_\beta}H-H^\star\ge0.
$$
For the second claim,
$$
  \lambda_\beta=\frac{|S^\star|e^{-\beta H^\star}}{\sum_ze^{-\beta H(z)}}
  \ge\frac{e^{-\beta H^\star}}{q^ne^{-\beta H^\star}}=q^{-n}.
$$
\end{proof}

Both algorithms below assume a lower bound $\lambda$ on $\lambda_\beta$ with $\lambda_\beta/\poly(m+n)\le\lambda\le\lambda_\beta$. Both algorithms use this bound to choose the number of repetitions.

\paragraph{Quantum minimum finding.} Suppose that a unitary $U$ prepares the state $\sum_x\sqrt{\pi_\beta(x)}\ket x$ (which we call the ideal Gibbs state), and that $H$ can be evaluated reversibly. Amplitude amplification with respect to the set of minimizers, combined with a standard search over cost thresholds~\cite{boyer1998tight,durr1996minimum}, returns an element of the image of $S^\star$ using $O(\lambda_\beta^{-1/2})$ calls to $U$ and $U^\dagger$, up to polynomial factors. A uniformly random preimage under $z\mapsto Bz+x_0$ is then a minimizer. The exponent is optimal in the worst-case oracle model: uniform preparation with an arbitrary marked set is unstructured search, which requires $\Omega(\lambda^{-1/2})$ queries~\cite{boyer1998tight,zalka1999grover}. This is not a pointwise lower bound for every preparation and objective. We take
\begin{equation}\label{eq:TQ}
  T_Q(\beta)\coloneqq\lambda_\beta^{-1/2}
\end{equation}
as the quantum cost of finding an optimum with the ideal Gibbs state at $\beta$.

\paragraph{Classical sampling.} The classical algorithm uses an approximate sampler as in Hypothesis~\ref{hyp:sampler}.

\begin{proposition}[Classical search by sampling]\label{prop:classical}
Assume Hypothesis~\ref{hyp:sampler} at some $\beta'\ge\beta$. Then a classical algorithm that is given a lower bound $\lambda$ with $\lambda_\beta/\poly(m+n)\le\lambda\le\lambda_\beta$ finds an element of $S^\star$ with probability at least $1-e^{-2}$ in time $\poly(m+n)\cdot\lambda_\beta^{-1}=\poly(m+n)\cdot T_Q(\beta)^2$.
\end{proposition}

The idea is that each sample from the Gibbs measure is optimal with probability $\lambda_\beta$, so about $\lambda_\beta^{-1}$ samples suffice, while amplitude amplification requires about $\lambda_\beta^{-1/2}$ preparations. By Lemma~\ref{lem:mass_monotone}, the sampler may also run at any larger inverse temperature.

\begin{proof}
Run the sampler at $\beta'$ with $\epsilon=q^{-n}/2$. By Lemma~\ref{lem:mass_monotone}, $\lambda_{\beta'}\ge\lambda_\beta\ge\lambda$ and $\epsilon\le\lambda_{\beta'}/2$, so each sample lies in $S^\star$ with probability at least $\lambda_{\beta'}-\epsilon\ge\lambda/2$. Draw $\lceil4/\lambda\rceil$ samples and output one of minimal cost. The algorithm fails with probability at most $(1-\lambda/2)^{4/\lambda}\le e^{-2}$. Each sample costs $\poly(m+n,n\ln q)$, which is polynomial in $m+n$ for fixed $q$, and $\lambda^{-1}\le\poly(m+n)\,\lambda_\beta^{-1}$.
\end{proof}

\subsubsection{A quadratic bound under efficient classical sampling}\label{sec:conditional}

Theorem~\ref{thm:only_nonrecon} shows that DQI with a local decoder whose block success tends to one prepares Gibbs states only in the non-reconstruction phase. If classical sampling is efficient there, as Conjecture~\ref{conj:classical} asserts, Proposition~\ref{prop:classical} bounds the classical cost of finding an optimum by the square of the quantum cost.

\begin{corollary}[Conditional quadratic bound]\label{cor:quadratic}
Assume Conjecture~\ref{conj:classical}. Let $q\ge2$, let $f$ be any cost, let $k\ge2$, $D\ge3$, and $k<D$, and let $c_0$ be as in Corollary~\ref{cor:typical_nogo}. Fix $\beta_Q\ge0$ and fixed symbols $i_j$. Suppose that, with probability $1-o(1)$ over random $(k,D)$-regular instances over $\Z_q$ from the configuration model or from Gallager's ensemble, a decoder for the decoding problem induced by DQI at $\beta_Q$ has block success $1-o(1)$ and makes a $t_j$-local estimate at $i_j$ with $t_j\le c_0\log m$. Then the primal model at $\beta_Q$ has non-reconstruction. Consequently, with probability $1-o(1)$ over the instance $(B,x_0)$ with uniformly random $x_0$, a classical algorithm that is given a lower bound $\lambda$ on $\lambda_{\beta_Q}$ as in Proposition~\ref{prop:classical} finds an exact minimizer in time $\poly(m+n)\cdot T_Q(\beta_Q)^2$.
\end{corollary}

\begin{proof}
By Corollary~\ref{cor:typical_nogo}, $\mathrm{LT}_{t_j}(i_j)$ and $\mathrm{TP}_{t_j}(i_j)$ hold with probability $1-o(1)$. Hence, for all large $j$, there are instances on which these conditions hold and the decoder has block success $1-o(1)$. Applying Theorem~\ref{thm:only_nonrecon} along these instances, we obtain non-reconstruction at $\beta_Q$. By Conjecture~\ref{conj:classical}, Hypothesis~\ref{hyp:sampler} then holds at $\beta_Q$ with probability $1-o(1)$, and Proposition~\ref{prop:classical} with $\beta'=\beta_Q$ finishes the proof.
\end{proof}

In words, quantum minimum finding on the ideal Gibbs state at $\beta_Q$ costs $T_Q(\beta_Q)$ preparations, while the classical algorithm costs at most $\poly(m+n)\,T_Q(\beta_Q)^2$. The speedup of quantum minimum finding over classical sampling is therefore at most quadratic up to polynomial factors. The corollary does not compare DQI with other quantum algorithms for optimization. Appendix~\ref{sec:exact_heavy} discusses the state that DQI actually prepares.

\subsubsection{An unconditional bound from Prange's algorithm}\label{sec:prange}

Proposition~\ref{prop:classical} uses the sampler of Hypothesis~\ref{hyp:sampler} with exponentially small error, because the probability $\lambda_\beta$ of an optimum is exponentially small. Its proof needs only a sampler that puts mass at least $\lambda_\beta/\poly(m+n)$ on $S^\star$. We show that the adaptation of Prange's algorithm to sampling by Zhou et al.~\cite{zhou2026gibbs} has this property whenever a union bound over the code $C$ converges. Throughout this subsection, $q=2$ and $f(z)=z$, and
$$
  \sigma_\beta=\tanh(\beta/2),\qquad \theta=1-e^{-\beta},\qquad U_C(\theta)=\sum_{c\in C\setminus\{0\}}\theta^{|c|},
$$
where $|c|$ is the Hamming weight. Note that $\sigma_\beta\le\theta$, because $\theta=2\sigma_\beta/(1+\sigma_\beta)$.

\emph{Prange's algorithm at $\beta$.} Draw $R\subseteq[m]$ by keeping each clause independently with probability $\sigma_\beta$. If the linear system $(Bz+x_0)_j=0$ for $j\in R$ has a solution, output a uniformly random solution $z\in\Z_2^n$. Otherwise output ``fail''. Gaussian elimination performs this in time $O((m+n)^3)$. Let $\nu(z)$ be the probability that the algorithm outputs $z$. This version differs from that of Zhou et al.\ only in that it does not reject rank-deficient systems.

The idea of the analysis, following Zhou et al., is that $e^{-\beta H}$ is a mixture over retained sets $R$ of uniform measures on the solutions of the system indexed by $R$. If every retained system had full row rank, Prange's algorithm would sample exactly from this mixture. The output law can differ from $\mu_\beta$ only through retained sets whose rows are linearly dependent. A dependency among the retained rows is exactly a nonzero codeword of $C$ whose support lies in $R$, and the weight enumerator $U_C$ bounds the probability of this event.

\begin{proposition}[Prange's algorithm]\label{prop:prange}
For every $B$, $x_0$, $\beta>0$ and $z\in\Z_2^n$,
$$
  \nu(z)\ge\big(1-U_C(\theta)\big)^2\,\mu_\beta(z)\qquad\text{whenever }U_C(\theta)<1 .
$$
\end{proposition}

\begin{proof}
Write $\mathrm{Sat}(z)$ for the set of clauses satisfied by $z$, and $B_R$ for the rows of $B$ indexed by $R$. A consistent system indexed by $R$ has $2^{n-\rank B_R}$ solutions. Hence, with $P(R)=\sigma_\beta^{|R|}(1-\sigma_\beta)^{m-|R|}$,
$$
  \nu(z)=\sum_{R\subseteq\mathrm{Sat}(z)}P(R)\,2^{\rank B_R-n}.
$$
Define $\tilde\pi(z)=\sum_{R\subseteq\mathrm{Sat}(z)}P(R)\,2^{|R|-n}=2^{-n}(1+\sigma_\beta)^{|\mathrm{Sat}(z)|}(1-\sigma_\beta)^{m-|\mathrm{Sat}(z)|}$. Since $(1-\sigma_\beta)/(1+\sigma_\beta)=e^{-\beta}$, we have $\tilde\pi(z)\propto e^{-\beta H(z)}$, so $\mu_\beta=\tilde\pi/\sum_w\tilde\pi(w)$.

\emph{Step 1: $\nu(z)\ge(1-U_C(\theta))\,\tilde\pi(z)$.} Let $Q_z$ be the law of $R$ that keeps each clause of $\mathrm{Sat}(z)$ independently with probability $2\sigma_\beta/(1+\sigma_\beta)=\theta$ and keeps no other clause. Then
$$
  Q_z(R)=\frac{P(R)2^{|R|}}{\sum_{R'\subseteq\mathrm{Sat}(z)}P(R')2^{|R'|}}\qquad\text{for }R\subseteq\mathrm{Sat}(z),
$$
so
$$
  \nu(z)=\tilde\pi(z)\,\E_{Q_z}\big[2^{-\mathrm{def}(R)}\big],\qquad \mathrm{def}(R)=|R|-\rank B_R .
$$
The deficiency $\mathrm{def}(R)$ is the dimension of the left kernel of $B_R$. A vector $y\in\Z_2^R$ with $y^\top B_R=0$, extended by zero, is exactly a codeword $c\in C$ with $\operatorname{supp}c\subseteq R$. So $\mathrm{def}(R)=0$ unless some nonzero codeword has support in $R$, and
$$
  \E_{Q_z}\big[2^{-\mathrm{def}(R)}\big]\ge\Pr_{Q_z}[\mathrm{def}(R)=0]\ge1-\sum_{c\in C\setminus\{0\}}\Pr_{Q_z}[\operatorname{supp}c\subseteq R]\ge1-U_C(\theta),
$$
because $\Pr_{Q_z}[\operatorname{supp}c\subseteq R]$ equals $\theta^{|c|}$ if $\operatorname{supp}c\subseteq\mathrm{Sat}(z)$ and $0$ otherwise.

\emph{Step 2: $\sum_w\tilde\pi(w)\ge1-U_C(\theta)$.} Let $g(0)=1+\sigma_\beta$ and $g(1)=1-\sigma_\beta$. Every $x\in x_0+K$ arises from exactly $|\ker B|$ vectors $w$, and $|\ker B|\,|K|=2^n$. By Poisson summation over $K$, with $\hat g(0)=2$ and $\hat g(1)=2\sigma_\beta$,
$$
  \sum_w\tilde\pi(w)=2^{-n}|\ker B|\sum_{x\in x_0+K}\prod_jg(x_j)=\sum_{c\in C}(-1)^{c\cdot x_0}\sigma_\beta^{|c|}\ge1-U_C(\sigma_\beta)\ge1-U_C(\theta).
$$

Combining the two steps, $\nu(z)\ge(1-U_C(\theta))\,\mu_\beta(z)\sum_w\tilde\pi(w)\ge(1-U_C(\theta))^2\mu_\beta(z)$.
\end{proof}

Prange's algorithm can therefore be used in place of the sampler of Hypothesis~\ref{hyp:sampler} in Proposition~\ref{prop:classical}: draw $\lceil4/((1-U_C(\theta))^2\lambda)\rceil$ samples and keep one of minimal cost. It remains to bound $U_C(\theta)$ for random codes. In the code $C$ of a $(k,D)$-regular instance, each symbol lies in $k$ checks and each check involves $D$ symbols. Define
$$
  G_{k,D}(u,x;\theta)=(1-k)\,h(u)+\frac k D\ln W_D(x)-k u\ln x+u\ln\theta ,
$$
where $h(u)=-u\ln u-(1-u)\ln(1-u)$ and $W_D(x)=\tfrac12\big((1+x)^D+(1-x)^D\big)$ is the generating function of the even-weight vectors in $\{0,1\}^D$, counted by weight.

\begin{theorem}[Union bound for random regular codes]\label{thm:union}
Let $k\ge3$, $D\ge2$, $\theta\in(0,1]$, $A=e^{1+2^{D-1}k/D}$ and $\epsilon=(2A)^{-1/(k/2-1)}$. Suppose there is $\eta>0$ such that for every $u\in[\epsilon,1]$ some $x>0$ satisfies $G_{k,D}(u,x;\theta)\le-\eta\,u$. Then for the configuration model of random $(k,D)$-regular instances, $U_C(\theta)\to0$ in probability.
\end{theorem}

\begin{proof}
Work in the configuration model. The transfer to the ensemble without multiple edges is as in Section~\ref{sec:typical}. Let $A_w$ be the number of codewords of weight $w$. A set of $w$ symbols is the support of a codeword exactly when every check meets its $k w$ half-edges an even number of times. Counting as in the proof of Theorem~\ref{thm:tp_typical}, for every $x>0$,
$$
  \E A_w\le\binom mw\frac{W_D(x)^{n}x^{-k w}}{\binom{k m}{k w}},\qquad n=\frac{k m}D .
$$
\emph{Small weights, $1\le w\le\epsilon m$.} Take $x=\sqrt{w/m}$. Then $W_D(x)\le1+2^{D-1}x^2$, so $W_D(x)^n\le\exp(2^{D-1}k w/D)$, and with $\binom mw\le(em/w)^w$ and $\binom{k m}{k w}\ge(m/w)^{k w}$ we get $\theta^w\E A_w\le\E A_w\le[A(w/m)^{k/2-1}]^w$. Since $k/2-1\ge1/2$, the argument in the proof of Theorem~\ref{thm:tp_typical} shows that the sum over $1\le w\le\epsilon m$ is $o(1)$.

\emph{Linear weights, $\epsilon m\le w\le m$.} With $u=w/m$, $\binom mw\le e^{mh(u)}$ and $\binom{k m}{k w}\ge e^{k mh(u)}/(k m+1)$, the hypothesis implies $\theta^w\E A_w\le(k m+1)e^{mG_{k,D}(u,x;\theta)}\le(k m+1)e^{-\eta\epsilon m}$. The sum over the at most $m$ values of $w$ is therefore $o(1)$.

Hence $\E U_C(\theta)=o(1)$, and Markov's inequality completes the proof.
\end{proof}

The hypothesis of Theorem~\ref{thm:union} becomes weaker as $\theta$ decreases, so it holds on an interval $\theta\le\theta_U$. We verify it by computer with $\eta=10^{-3}$. We partition $[\epsilon,1]$ into $2\times10^4$ cells and pick one $x$ per cell, namely the saddle point at the center of the cell. On a cell $[a,b]$ we bound $G_{k,D}(u,x;\theta)$ uniformly by $(1-k)\min\{h(a),h(b)\}+\frac k D\ln W_D(x)+u^\ast(\ln\theta-k\ln x)$, with $u^\ast=b$ if $\ln\theta\ge k\ln x$ and $u^\ast=a$ otherwise. This bound is valid because $h$ is concave and $1-k<0$. We require that the bound divided by $b$ is at most $-\eta$ on every cell. We locate the largest such $\beta$ by bisection in floating-point arithmetic and round it down to three decimals. At the rounded value we recheck every cell in interval arithmetic with outward rounding, with the grid starting at a lower bound for $\epsilon$, so that the inequality does not depend on rounding errors. The certified values $\beta_U=-\ln(1-\theta_U)$ are:
\begin{center}
\begin{tabular}{@{}lccccc@{}}
\toprule
$(k,D)$ & $(3,6)$ & $(6,12)$ & $(3,4)$ & $(4,6)$ & $(4,8)$\\
\midrule
$\beta_U$ & $0.525$ & $0.533$ & $1.136$ & $0.882$ & $0.532$\\
$\beta_d$ & $1.839$ & $1.549$ & & & \\
\bottomrule
\end{tabular}
\end{center}
For these degree pairs the constant $\epsilon$ is very small, for example about $10^{-223}$ for $(6,12)$, because the estimate for small weights is far from tight. Theorem~\ref{thm:union} still holds because it is an asymptotic statement, but the bound on $\E U_C(\theta)$ in its proof is small only for very large $m$.

\begin{corollary}[Unconditional quadratic bound]\label{cor:unconditional}
Let $q=2$ and $f(z)=z$, let $(k,D)$ be one of the degree pairs in the table above, and let $\beta_Q\le\beta_U(k,D)$. Then, with probability $1-o(1)$ over random $(k,D)$-regular instances from the configuration model and for every $x_0$, the following holds. Given a lower bound $\lambda$ on $\lambda_{\beta_Q}$ as in Proposition~\ref{prop:classical}, a classical algorithm finds an exact minimizer in time $\poly(m+n)\cdot T_Q(\beta_Q)^2$. Quantum minimum finding on the ideal Gibbs state at $\beta_Q$ uses $T_Q(\beta_Q)$ preparations of this state when the preparation is used as a black box. Hence, for finding exact optima, it is at most quadratically faster than this classical algorithm, up to polynomial factors, however the ideal Gibbs state is prepared.
\end{corollary}

\begin{proof}
If $\beta_Q=0$, then Prange's algorithm keeps no clause and samples exactly from $\mu_0$. Otherwise, by Theorem~\ref{thm:union}, $U_C(\theta)\le1/2$ with probability $1-o(1)$ at $\theta=1-e^{-\beta_Q}$. On this event, $\nu(S^\star)\ge\lambda_{\beta_Q}/4$ by Proposition~\ref{prop:prange}, and the argument of Proposition~\ref{prop:classical} applies with $\lceil16/\lambda\rceil$ samples.
\end{proof}

Corollary~\ref{cor:unconditional} requires no assumption on the decoder, because it bounds the classical cost directly. It covers about $29\%$ of the non-reconstruction interval $[0,\beta_d)$ for $(3,6)$ and about $34\%$ for $(6,12)$. For $\beta_U<\beta_Q<\beta_d$, the conclusion of Corollary~\ref{cor:quadratic} still depends on Conjecture~\ref{conj:classical}.

\subsubsection{Polynomial target states}\label{sec:poly_exact}

The polynomial amplitude functions of Jordan et al.~\cite{jordan2024optimization} do not have Gibbs target states. Nevertheless, their probability of an optimum admits the same comparison, up to a polynomial factor. The result holds for every complex polynomial within the degree bound, rather than only for the polynomial that maximizes the expected satisfied fraction.

Throughout this subsubsection, $q=2$ and $f(z)=z$. A polynomial amplitude function of degree at most $L$ is a normalized function $\Phi(x)=P(|x|)$ with $\deg P\le L$. Let $\lambda_\Phi(x_0)$ be the probability of an optimal pattern in its target state, and set $T_\Phi=\lambda_\Phi(x_0)^{-1/2}$ when $\lambda_\Phi(x_0)>0$. As in~\eqref{eq:TQ}, $T_\Phi$ counts target-state preparations in quantum minimum finding, not the cost of implementing a preparation. For $0\le\rho<\tfrac12$, write
$$
  \rho^\perp=\frac12-\sqrt{\rho(1-\rho)},\qquad
  \beta_\rho=\ln\frac{1-\rho^\perp}{\rho^\perp}.
$$
The product amplitude function at $\beta_\rho$ has independent errors of rate $\rho$ in the decoding problem, whereas a degree-$L$ polynomial has no errors of weight greater than $L$.

\begin{theorem}[Polynomial target states and Gibbs measures]\label{thm:poly_target}
Let $B\in\Z_2^{m\times n}$ have no zero row, let $0\le L\le\rho m$ with $\rho<\tfrac12$, and let $\Phi(x)=P(|x|)$ be a normalized complex polynomial amplitude function chosen independently of $x_0$. Suppose that some decoder has block success at least $1-\varepsilon$. For a uniformly random offset, outside a set of probability at most $2\sqrt\varepsilon+(m+1)^{-1}$, the target state is defined and
\begin{equation}\label{eq:poly_target}
  \lambda_\Phi(x_0)\le(m+1)^4\lambda_\beta(x_0)
  \qquad\text{for every }\beta\ge\beta_\rho.
\end{equation}
\end{theorem}

The proof is given in Appendix~\ref{app:poly_comparison}. A degree-$L$ polynomial has a Fourier expansion in Dicke states of weight at most $L$. Each Dicke state is a phase average of product states with only a polynomial normalization factor. This bounds the squared amplitude of every optimal pattern by a polynomial factor times its probability under a suitable product law. Reliable decoding controls the target state's coset normalization, and a union bound controls the product normalizations for most offsets. Neither a minimum-distance hypothesis nor a unique optimum is required.

For binary Max-XORSAT, put $\alpha_d=(1+e^{\beta_d})^{-1}$ and $\rho_d=\alpha_d^\perp$ when $\beta_d<\infty$. Thus $\beta_{\rho_d}=\beta_d$.

\begin{corollary}[Conditional quadratic comparison for polynomial target states]\label{cor:poly_quadratic}
Fix $3\le k<D$, assume $0<\beta_d<\infty$ and Conjecture~\ref{conj:classical}, and fix $0\le\rho<\rho_d$. For each $m$, let $\Phi_m$ be a normalized complex polynomial amplitude function of degree at most $\rho m$, possibly depending on $B$ but not on $x_0$. Suppose that, with probability $1-o(1)$ over $B$ from either regular ensemble, some decoder has block success $1-o(1)$. Then, with probability $1-o(1)$ over $(B,x_0)$ with uniformly random $x_0$, the following holds whenever $\lambda_{\Phi_m}(x_0)>0$. Given a bound $\underline\lambda$ with
$$
  \lambda_{\Phi_m}(x_0)/\poly(m+n)\le\underline\lambda\le\lambda_{\Phi_m}(x_0),
$$
a classical algorithm finds an exact optimum with probability at least $1-e^{-2}$ in time $\poly(m+n)T_{\Phi_m}^2$.
\end{corollary}

\begin{proof}
Since $\beta_\rho<\beta_d$, Corollary~\ref{cor:binary} and Conjecture~\ref{conj:classical} supply the sampler of Hypothesis~\ref{hyp:sampler} at $\beta_\rho$, with probability $1-o(1)$ over the instance. On the event in Theorem~\ref{thm:poly_target}, use sampling error $2^{-n-1}$. Lemma~\ref{lem:mass_monotone} gives $\lambda_{\beta_\rho}\ge2^{-n}$, so each independent sample is optimal with probability at least
$$
  \frac12\lambda_{\beta_\rho}\ge\frac{\lambda_{\Phi_m}}{2(m+1)^4}\ge\frac{\underline\lambda}{2(m+1)^4}.
$$
After $\lceil4(m+1)^4/\underline\lambda\rceil$ samples, the least-cost sample is optimal with probability at least $1-e^{-2}$. The logarithm of the inverse sampling error is $O(n)$, so each sample costs polynomial time. Averaging the exceptional-offset estimate over $B$ proves the stated high-probability conclusion.
\end{proof}

For the mean-optimal polynomials of Jordan et al., reliable local decoding forces a limiting degree ratio at most $\rho_d$ (Proposition~\ref{prop:poly_degree}). This is a consequence of their semicircle value and the local approximation bound, not a restriction on every polynomial that can be decoded reliably. At the endpoint, the classical cost is at most $\poly(m+n)e^{\eta m}T_\Phi^2$ for each fixed $\eta>0$, with an algorithm that may depend on $\eta$ (Corollary~\ref{cor:poly_endpoint}). A polynomial-overhead endpoint comparison remains open.

The Prange comparison also extends to polynomial target states in its union-bound range. Under exact decoding it holds for every offset, and under reliable decoding it holds for most offsets (Proposition~\ref{prop:poly_prange}). These statements do not assert that an exact decoder exists throughout that degree range.

\subsubsection{The state that DQI prepares}\label{sec:prepared}

With block success one, DQI prepares its target state exactly for every offset. The polynomial comparison then applies to its actual output; bounded-distance decoding with minimum distance greater than $2L$ is one sufficient case (Proposition~\ref{prop:bounded_distance}). The comparison still counts state preparations and does not assume that the exact decoder is efficient.

When the block success is only $1-\varepsilon$, Theorem~\ref{thm:guarantee} controls the output law up to additive error of order $\sqrt\varepsilon$, averaged over the offset. This error can be much larger than the exponentially small optimum probability, so reliable decoding alone does not extend Corollary~\ref{cor:poly_quadratic} to the prepared state. For a particular instance, the comparison does extend if its prepared law $\nu$ satisfies, for example,
$$
  \|\nu-\pi^\Phi\|_{\mathrm{TV}}\le\lambda_\Phi/2,
$$
where $\pi^\Phi$ is the target law: the optimum probability under $\nu$ then lies between $\lambda_\Phi/2$ and $3\lambda_\Phi/2$. More generally, an upper bound by a polynomial multiple of $\lambda_\Phi$ suffices for the classical comparison. The same qualification applies to Gibbs target states with $\lambda_\beta$ in place of $\lambda_\Phi$.

\section{Nonlocal decoders and the condensation threshold}\label{sec:outside}\label{sec:nonlocal}

Local decoders fail above the reconstruction threshold $\beta_d$ (Theorem~\ref{thm:only_nonrecon}), and no decoder succeeds above the limit $\beta_{\mathrm H}$ that Jordan et al.\ derived from the Holevo bound (Section~\ref{sec:dqi}). This limit depends on the code only through its rate, and for uniformly random linear codes the pretty-good measurement attains it (Section~\ref{sec:condensation}). In this section we determine the exact limit for the sparse codes induced by DQI. For binary codes from the configuration model with $3\le k<D$, the pretty-good measurement, which is an optimal decoder, has block success tending to one below the condensation threshold $\beta_c$ of the primal Gibbs measure, and every decoder has exponentially small block success above it (Theorem~\ref{thm:condensation}). Optimal decoding is thus characterized by a transition of the primal Gibbs measure, as local decoding is characterized by tree reconstruction. The resulting bound $s_c$ on the expected satisfied fraction that DQI reaches holds for every symmetric amplitude function (Corollary~\ref{cor:condensation_sym}). At every finite degree pair, $\beta_c<\beta_{\mathrm H}$, while at a fixed degree ratio the positive difference converges to zero exponentially in the degrees (Proposition~\ref{prop:condensation_holevo}). We also show that, below $\beta_c$, a coherent implementation of the pretty-good measurement is equivalent to a coherent preparation of the primal Gibbs states (Corollary~\ref{cor:pgm_equiv}), so an efficient coherent implementation of the optimal decoder there would itself be an efficient quantum algorithm for preparing these states.

Before these results, we recall in Section~\ref{sec:known_nonlocal} the efficient nonlocal decoders that have been analyzed for DQI and the nonlocal classical algorithms that reach the same values. In Section~\ref{sec:pgm} we compute the block success of the pretty-good measurement and prove the equivalence, and in Section~\ref{sec:condensation} we state the results on the condensation threshold, which are proved in Appendix~\ref{app:condensation}, and compare the three thresholds. Efficient decoders between $\beta_d$ and $\beta_c$, and the consequences that an efficient coherent implementation of the pretty-good measurement would have for planted instances, are discussed in Section~\ref{sec:window}. A conjecture about decoders that measure small sets of qudits is stated in Section~\ref{sec:small_sets}.

\subsection{Known nonlocal decoders and nonlocal classical algorithms}\label{sec:known_nonlocal}

Two efficient nonlocal decoders have been analyzed for DQI on random sparse instances: unambiguous discrimination followed by Gaussian elimination, and FGUM followed by Gaussian elimination. For each of them, a nonlocal classical algorithm that also uses Gaussian elimination reaches the values of DQI with that decoder. We recall these results and then compare them with the limit of local decoders at larger degrees.

\subsubsection{Unambiguous discrimination and Prange's algorithm}\label{sec:ud_prange}

Unambiguous discrimination applies to each qudit a measurement that either returns the codeword symbol with certainty or reports an erasure, and Gaussian elimination with all the checks then recovers the erased symbols~\cite{chailloux2024quantumdecoding,zhou2026gibbs}. Zhou et al.~\cite{zhou2026gibbs} show that, on random instances whose constraints form an Erd\H{o}s--R\'enyi hypergraph with large clause size, DQI with this decoder samples in the shattered phase, in which the Gibbs measure splits into exponentially many well-separated clusters and classical algorithms that are stable under small changes of the instance provably fail to sample from it. They also show that a classical sampler based on Prange's algorithm (Section~\ref{sec:prange}) reaches the same inverse temperatures. On regular instances, the two algorithms are limited by the same rank condition. For $q=2$, unambiguous discrimination erases each symbol independently with probability $\sigma_\beta$, the overlap of the two states of a symbol, so the set $R$ of erased symbols has the same law as the set of clauses that Prange's algorithm retains. Gaussian elimination recovers the codeword exactly when the rows of $B$ indexed by $R$ are linearly independent, and the same rows determine whether the linear system of Prange's algorithm has a solution. These rows can be linearly independent with high probability only if the expected number $\sigma_\beta m$ of retained clauses is at most the number $n=mk/D$ of variables, that is, for $\beta\le\beta_{\mathrm P}\coloneqq2\operatorname{atanh}(k/D)$. A stronger necessary condition follows from counting only the variables that occur in the retained clauses. Shutty et al.\ use the same count, for erasures of blocks of symbols, to estimate the erasure threshold of FGUM followed by Gaussian elimination~\cite{shutty2026locallyquantumdecoders}. In the following lemma, $\zeta(\sigma)m$ is the expected number of retained clauses minus the expected number of variables that occur in them, when each clause is retained with probability $\sigma$.

\begin{lemma}[Rank deficiency of random row subsets]\label{lem:rank}
Let $q=2$, $k\ge2$, and $D>k$, and let $\zeta(\sigma)=\sigma-\frac k D\big(1-(1-\sigma)^D\big)$ for $\sigma\in[0,1]$. Then $\zeta$ has a unique zero $\sigma^\ast=\sigma^\ast(k,D)$ in $(0,1)$, this zero lies in $(0,k/D)$, and $\zeta(\sigma)>0$ for every $\sigma\in(\sigma^\ast,1]$. Let $B$ be a $(k,D)$-regular matrix, let $\sigma\in(\sigma^\ast,1]$, and let $R\subseteq[m]$ contain each row of $B$ independently with probability $\sigma$. Let $B_R$ consist of the rows of $B$ indexed by $R$, and let $\mathrm{def}(R)=|R|-\rank B_R$, which is the dimension of the space of codewords of $C$ supported in $R$ (proof of Proposition~\ref{prop:prange}). Then
$$
  \Pr\Big[\mathrm{def}(R)<\tfrac12\zeta(\sigma)\,m\Big]\le\exp\Big(-\frac{\zeta(\sigma)^2m}{2k^2}\Big).
$$
In particular, the rows of $B_R$ are linearly dependent with probability at least $1-\exp\big(-\zeta(\sigma)^2m/(2k^2)\big)$.
\end{lemma}

For $(3,6)$ and $(6,12)$, $2\operatorname{atanh}(\sigma^\ast)$ is $1.07565$ and $1.09829$, while $\beta_{\mathrm P}=\ln3\approx1.09861$. Every decoder that uses only the outcomes of unambiguous discrimination recovers the codeword with probability at most $\E\,2^{-\mathrm{def}(R)}$, so by Lemma~\ref{lem:rank} the block success of this decoder tends to zero for $\beta>2\operatorname{atanh}(\sigma^\ast)$. On the configuration model, for the degree pairs of the table in Section~\ref{sec:prange}, its block success is $1-o(1)$ with probability $1-o(1)$ over the instance when $\sigma_\beta\le1-e^{-\beta_U}$, that is, for $\beta\le0.867$ for $(3,6)$ and $\beta\le0.878$ for $(6,12)$, by Theorem~\ref{thm:union}. The largest inverse temperature at which this decoder succeeds therefore lies in $[0.867,1.0757]$ for $(3,6)$ and in $[0.878,1.0983]$ for $(6,12)$, well below $\beta_d$ (Table~\ref{tab:landscape}). Wherever its block success tends to one, DQI with this decoder reaches the expected satisfied fraction $p_\beta(0)+o(1)=\tfrac12+\tfrac12\sigma_\beta+o(1)$ by part~2 of Theorem~\ref{thm:guarantee}, so by Lemma~\ref{lem:rank} it reaches at most $\tfrac12+\tfrac12\sigma^\ast+o(1)$. This bound is below the asymptotic value $\tfrac12+k/(2D)$ of the optimization version of Prange's algorithm, which solves a maximal set of linearly independent equations~\cite{shutty2026locallyquantumdecoders}; for $(3,6)$, the two values are $0.7457$ and $0.75$. The proofs of Lemma~\ref{lem:rank} and of these statements are given in Appendix~\ref{app:rank}.

\subsubsection{Fine-grained unambiguous measurements and Turbo Prange}\label{sec:fgum}

Shutty et al.~\cite{shutty2026locallyquantumdecoders} combine the block amplitude function of Section~\ref{sec:approx_block}, which is a product over the blocks of the first layer of Gallager's ensemble, with a decoder that measures blocks of symbols. The decoder applies to each block a fine-grained unambiguous measurement (FGUM)~\cite{buzet2025fine}, which either fails or returns $c_{S_i}$ with certainty, and then recovers the failed blocks by Gaussian elimination with all the checks, so it is nonlocal. With the block amplitude function and FGUM, Regev's reduction reaches a larger satisfied fraction than simulated annealing and Prange's algorithm for the four degree pairs $(5,6)$, $(6,7)$, $(6,8)$, and $(7,8)$. Shutty et al.\ also give a classical algorithm, Turbo Prange, that reaches exactly the same satisfied fraction for every $(k,D)$. Turbo Prange uses Gaussian elimination to satisfy all clauses in as many blocks as possible, and then, for each remaining block, flips the variable of the block whenever this increases the number of satisfied clauses in that block. The block amplitude function is supported only on the patterns that this final step can leave on a block, and Shutty et al.\ establish the equality of the two values by computing both. Neither hypothesis of Theorem~\ref{thm:approx_sym} holds for Regev's reduction with FGUM: the block amplitude function is not symmetric, and the decoder is nonlocal.

\subsubsection{Larger degrees}\label{sec:larger_degrees}

\begin{figure}[t]
\centering
\begin{tikzpicture}
\begin{axis}[
  width=0.85\textwidth, height=5.4cm,
  xmin=2, xmax=19, ymin=0.9, ymax=2.8,
  xtick={3,6,9,12,15,18},
  xticklabel style={font=\scriptsize},
  yticklabel style={font=\scriptsize, /pgf/number format/fixed, /pgf/number format/precision=1},
  xlabel={Clause size $k$}, xlabel style={font=\small},
  ylabel={Inverse temperature}, ylabel style={font=\small},
  legend columns=3,
  legend style={font=\scriptsize, draw=none, at={(0.5,1.02)}, anchor=south, /tikz/every even column/.append style={column sep=8pt}},
  every axis plot/.append style={line width=0.8pt, mark size=1.7pt},
]
\addplot[black, mark=*] coordinates {(3,1.8388) (4,1.7531) (5,1.6448) (6,1.5489) (7,1.4674) (8,1.3976) (9,1.3380) (10,1.2858) (12,1.1992) (14,1.1294) (16,1.0717) (18,1.0228)};
\addlegendentry{$\beta_d$, $k/D=1/2$}
\addplot[blue, mark=square*] coordinates {(4,2.1781) (6,1.8727) (8,1.6699) (10,1.5256) (12,1.4161)};
\addlegendentry{$\beta_d$, $k/D=2/3$}
\addplot[red!75!black, mark=diamond*] coordinates {(3,2.6574) (6,2.0358) (9,1.7155) (12,1.5196) (15,1.3838)};
\addlegendentry{$\beta_d$, $k/D=3/4$}
\addplot[black, dashed, domain=2:19, samples=2] {1.0986};
\addlegendentry{$\beta_{\mathrm P}$, $k/D=1/2$}
\addplot[blue, dashed, domain=2:19, samples=2] {1.6094};
\addlegendentry{$\beta_{\mathrm P}$, $k/D=2/3$}
\addplot[red!75!black, dashed, domain=2:19, samples=2] {1.9459};
\addlegendentry{$\beta_{\mathrm P}$, $k/D=3/4$}
\end{axis}
\end{tikzpicture}
\caption{Reconstruction threshold $\beta_d$ of binary $(k,D)$-regular instances against the clause size $k$ at the three ratios $k/D=\tfrac12,\tfrac23,\tfrac34$, with $\beta_{\mathrm P}=2\operatorname{atanh}(k/D)$ at each ratio as a dashed horizontal line. At each ratio, $\beta_d$ decreases as the degrees grow and falls below $\beta_{\mathrm P}$: between $k=6$ and $k=9$ for $k/D=\tfrac34$, between $k=8$ and $k=10$ for $k/D=\tfrac23$, and between $k=14$ and $k=16$ for $k/D=\tfrac12$. We computed $\beta_d$ as the threshold of the recursion of Definition~\ref{def:spectral} for $q=2$, which equals the reconstruction threshold by Theorem~\ref{thm:duality}, by bisection on $\tanh(\beta/2)$ with population dynamics ($2\times10^5$ samples, $1500$ levels); each value lies in a bisection interval of width at most $5\times10^{-4}$ (Table~\ref{tab:degrees}). Above $\beta_{\mathrm P}$, the retained system of Prange's algorithm has more equations than variables with high probability. At the clause sizes on both sides of each crossing, the bound $2\operatorname{atanh}(\sigma^\ast)$ of Lemma~\ref{lem:rank} differs from $\beta_{\mathrm P}$ by less than $10^{-4}$, so the crossings do not change when $\beta_{\mathrm P}$ is replaced by this bound (Section~\ref{sec:larger_degrees}).}
\label{fig:degrees}
\end{figure}
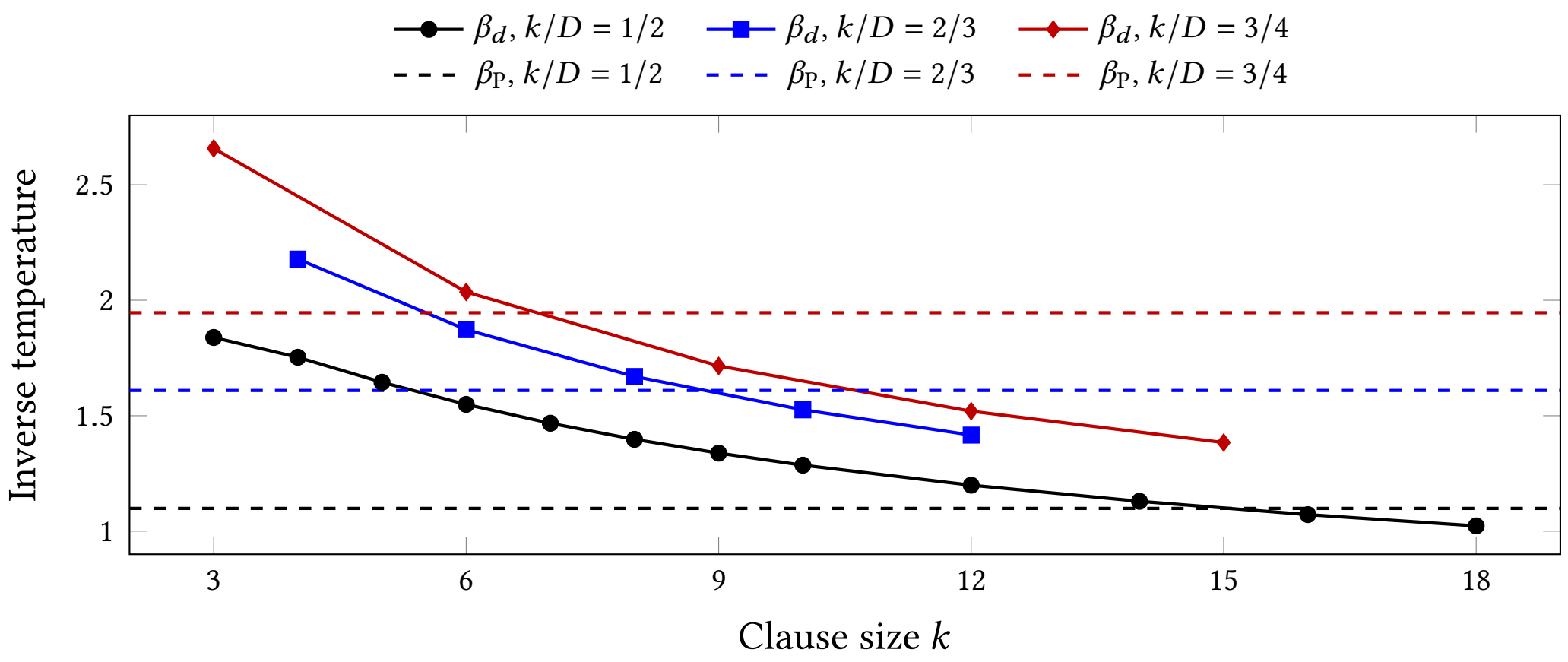

Nonlocal classical samplers, such as Prange's algorithm, are expected to succeed beyond $\beta_d$ at larger degrees. Zhou et al.~\cite{zhou2026gibbs} predict that $\beta_d$ tends to zero as the degrees grow at a fixed ratio $k/D$, whereas $\beta_{\mathrm P}$ depends only on $k/D$. Figure~\ref{fig:degrees} shows that, at three fixed ratios, $\beta_d$ decreases as the degrees grow, exceeds $\beta_{\mathrm P}$ for small degrees, and falls below it at a clause size between $6$ and $16$. At the clause sizes on both sides of each crossing, $2\operatorname{atanh}(\sigma^\ast)$ differs from $\beta_{\mathrm P}$ by less than $10^{-4}$, so $\beta_d$ also falls below $2\operatorname{atanh}(\sigma^\ast)$ between the same clause sizes. If the retained system of Prange's algorithm has full row rank with probability $1-\delta$, then the output law of the algorithm is within $3\delta$ of $\mu_\beta$ in total variation on average over the offset (Appendix~\ref{app:rank}). Classical sampling is predicted to be efficient up to $\beta_d$ (Conjecture~\ref{conj:classical}). At larger degrees, Prange's algorithm samples up to nearly $\beta_{\mathrm P}$ if its retained system has full row rank with high probability almost up to $\beta_{\mathrm P}$, as it does on Erd\H{o}s--R\'enyi hypergraphs~\cite{zhou2026gibbs}.

On the fifteen benchmark families of Shutty et al., the value of a known efficient nonlocal method exceeds $s_d$ exactly for the four pairs $(5,6)$, $(6,7)$, $(6,8)$, and $(7,8)$ (Table~\ref{tab:benchmark} and Figure~\ref{fig:fractions}). For all four, the value of Turbo Prange, which equals that of Regev's reduction with FGUM, exceeds $s_d$. The value of Prange's algorithm, an upper bound on the value of DQI with unambiguous discrimination, exceeds $s_d$ for $(5,6)$, $(6,7)$, and $(7,8)$, for $(5,6)$ only by about $0.0005$.

Table~\ref{tab:rates} restates the comparison of Table~\ref{tab:benchmark} in terms of error rates. To reach satisfied fraction $s$ with a product amplitude function, DQI requires a decoder that succeeds at error rate $(1-s)^\perp$. The rate needed to match the best classical value exceeds the reconstruction rate $\rho_d=(1-s_d)^\perp$, which is the limit of local decoders, in every row. In every row, it also lies below the condensation rate $\rho_c=\alpha_c^\perp$, which is the limit of every decoder on the configuration model. The certified lower bounds on $s_c$ in Appendix~\ref{app:certified} exceed the best classical values by at least $0.0025$, up to the difference between Gallager's ensemble, on which the classical values were obtained, and the configuration model (Section~\ref{sec:condensation}). By Proposition~\ref{prop:condensation_holevo}, $\rho_c<\rho_H=\alpha_H^\perp$.

\begin{table}[ht]
\centering
\small
\begin{tabular}{@{}c ccc@{\hspace{2em}}c ccc@{}}
\toprule
$(k,D)$ & $\rho_d$ & $\rho_{\mathrm{needed}}$ & $\rho_c$ & $(k,D)$ & $\rho_d$ & $\rho_{\mathrm{needed}}$ & $\rho_c$ \\
\midrule
$(3,4)$ & $0.2525$ & $0.2590$ & $0.2921$ & $(4,8)$ & $0.1452$ & $0.1495$ & $0.1839$ \\
$(3,5)$ & $0.1929$ & $0.2032$ & $0.2184$ & $(5,6)$ & $0.2228$ & $0.2469$ & $0.3448$ \\
$(3,6)$ & $0.1560$ & $0.1666$ & $0.1748$ & $(5,7)$ & $0.1901$ & $0.1925$ & $0.2819$ \\
$(3,7)$ & $0.1308$ & $0.1419$ & $0.1458$ & $(5,8)$ & $0.1657$ & $0.1670$ & $0.2401$ \\
$(3,8)$ & $0.1126$ & $0.1225$ & $0.1251$ & $(6,7)$ & $0.2063$ & $0.2676$ & $0.3591$ \\
$(4,5)$ & $0.2407$ & $0.2437$ & $0.3254$ & $(6,8)$ & $0.1804$ & $0.1950$ & $0.3000$ \\
$(4,6)$ & $0.1977$ & $0.2015$ & $0.2576$ & $(7,8)$ & $0.1917$ & $0.2782$ & $0.3702$ \\
$(4,7)$ & $0.1675$ & $0.1723$ & $0.2144$ & & & & \\
\bottomrule
\end{tabular}
\caption{Error rates of the decoding problem for $(k,D)$-regular Max-XORSAT with a product amplitude function. Local decoders can succeed only at error rates at most $\rho_d$, every decoder fails above $\rho_c$ (Theorem~\ref{thm:condensation}; midpoints of the certified intervals of Appendix~\ref{app:certified}), and DQI reaches the best classical value in Table~\ref{tab:benchmark} only with a decoder that succeeds at $\rho_{\mathrm{needed}}$.}
\label{tab:rates}
\end{table}

\subsection{The pretty-good measurement}\label{sec:pgm}

Optimal decoding is reliable exactly when the ambient distribution of violation patterns induces an asymptotically uniform law on the cosets of $K$. We first establish this criterion by computing the success of the pretty-good measurement (PGM), which is optimal for the decoding problem induced by DQI. For product amplitudes, we then identify its measurement vectors as Fourier superpositions of primal Gibbs states. This proves an equivalence between coherent implementation of the PGM and coherent preparation of these states whenever the success is at least inverse polynomial.

Fix $q\ge2$ and a matrix $B\in\{0,1\}^{m\times n}$ viewed over $\Z_q$. For an amplitude function $\Phi$ and a coset $X$ of $K$, let $W_\Phi(X)=\sum_{x\in X}|\Phi(x)|^2$, so that $W_\Phi$ is the law of the coset of a violation pattern drawn from $|\Phi|^2$. Let $U_B$ be the uniform law on the $|C|$ cosets of $K$, and let $L_\Phi=|C|\,W_\Phi$ be the density of $W_\Phi$ with respect to $U_B$.

\begin{lemma}[Block success of the optimal decoder]\label{lem:pgm_success}
For every amplitude function $\Phi$, the PGM is an optimal decoder for the decoding problem induced by DQI with amplitude function $\Phi$, and its block success is
\begin{equation}\label{eq:pgm_bhattacharyya}
  P_{\mathrm{PGM}}(\Phi)=\Big(\E_{X\sim U_B}\sqrt{L_\Phi(X)}\Big)^2=\frac1{|C|}\Big(\sum_X\sqrt{W_\Phi(X)}\Big)^2 .
\end{equation}
\end{lemma}

This formula is known. Rengaswamy and Pfister state it for binary linear codes on the pure-state channel as the square of the Bhattacharyya coefficient between the law of the coset and the uniform law~\cite[Theorem~15 of the full version]{rengaswamy2021duality}, and Chailloux and Tillich give it for every amplitude function over every prime field~\cite[Proposition~6]{chailloux2024quantumdecoding}. Blanvillain, Chailloux, and Tillich state it over every finite field~\cite[Proposition~5]{blanvillain2026tight}, and Chailloux and Hermouet compute the optimal success probability over $\Z_q$ for every modulus~\cite[Proposition~7]{chailloux2026equivalence}; see also~\cite{rengaswamy2021bpqm,piveteau2025efficient,renes2018duality}. We include a short proof in our notation, which holds for every $q\ge2$ and requires no assumption on the structure of $C$.

\begin{proof}
The states $\ket{\psi^\Phi_c}=Z^c\ket\Phi$ of~\eqref{eq:states_general} form a geometrically uniform family under the abelian group $\{Z^c:c\in C\}$. Their Gram matrix has entries
$$
  G_{c,c'}=\langle\psi^\Phi_c|\psi^\Phi_{c'}\rangle=\sum_x|\Phi(x)|^2\omega^{(c'-c)\cdot x},
$$
which depend only on $c'-c$. With equal prior probabilities, the PGM is optimal for this family~\cite{eldar2001square}.

\emph{The Gram spectrum.} The matrix $G$ is a convolution on $C$, so its eigenvectors are the characters $c\mapsto\omega^{-c\cdot x}$. Two vectors $x$ define the same character exactly when they lie in the same coset of $C^\perp=K$. For any representative $x_X\in X$, the corresponding eigenvalue is
$$
  \sum_{c\in C}G_{0,c}\,\omega^{-c\cdot x_X}=\sum_x|\Phi(x)|^2\sum_{c\in C}\omega^{c\cdot(x-x_X)}=|C|\,W_\Phi(X)
$$
by character orthogonality. Thus $\tr G^{1/2}=\sum_X\sqrt{|C|W_\Phi(X)}$.

\emph{The success probability.} Write $A=\sum_{c\in C}\ket{\psi^\Phi_c}\!\bra c$, so that $G=A^\dagger A$. The PGM vectors are $\ket{b_c}=(AA^\dagger)^{-1/2}A\ket c$, with the inverse taken on the range of $A$. The singular value decomposition gives
$$
  \langle b_c|\psi^\Phi_c\rangle=\bra cA^\dagger(AA^\dagger)^{-1/2}A\ket c=(G^{1/2})_{cc}.
$$
The square root $G^{1/2}$ is also a convolution, so each diagonal entry is $|C|^{-1}\tr G^{1/2}$. Averaging the squared overlaps over $c$ therefore yields $P_{\mathrm{PGM}}=(|C|^{-1}\tr G^{1/2})^2$, which is~\eqref{eq:pgm_bhattacharyya}.
\end{proof}

The quantity $\mathrm{BC}=\E_{X\sim U_B}\sqrt{L_\Phi(X)}$ is the Bhattacharyya coefficient of $W_\Phi$ and $U_B$, and it satisfies $1-\mathrm{BC}\le\|W_\Phi-U_B\|_{\mathrm{TV}}\le\sqrt{1-\mathrm{BC}^2}$. By Lemma~\ref{lem:pgm_success}, the block success of the optimal decoder is therefore close to one exactly when the coset law $W_\Phi$ is close to $U_B$ in total variation.

For the product amplitude function $\Phi_\beta$, the measurement vectors of the PGM are superpositions of normalized primal Gibbs states at the same inverse temperature, and this relation permits the exact preparation of the Gibbs state from a coherent implementation of the PGM. Fix a cost $f:\Z_q\to\mathbb R$ and a finite $\beta\ge0$. Consider the average state
$$
  \Psi=\frac1{|C|}\sum_{c\in C}\ket{\psi_c}\!\bra{\psi_c}.
$$
Averaging the phases $\omega^{c\cdot(x-x')}$ over $C$ leaves only pairs with $x-x'\in C^\perp=K$, so
$$
  \Psi=\sum_{X}\ket{\phi_X}\!\bra{\phi_X},
  \qquad
  \ket{\phi_X}=\sum_{x\in X}\sqrt{\textstyle\prod_jp(x_j)}\,\ket x ,
$$
where $X$ runs over the cosets of $K$ in $\Z_q^m$. Each $\ket{\phi_X}$ is the unnormalized Gibbs state on $X$, with squared norm $Z_X$. Positivity of $p$ implies $Z_X>0$ for every $X$, and normalization gives $\sum_XZ_X=1$. The states $\ket{\hat\phi_X}=\ket{\phi_X}/\sqrt{Z_X}$ are orthonormal because their supports are disjoint. Writing $\mathcal S$ for their span, we have $\Psi=\sum_XZ_X\ket{\hat\phi_X}\!\bra{\hat\phi_X}$.

For $x_X\in X$, the overlap $\langle\phi_X|\psi_c\rangle=\omega^{c\cdot x_X}Z_X$ gives $\Psi^{-1/2}\ket{\psi_c}=\sum_X\omega^{c\cdot x_X}\ket{\hat\phi_X}$, with the inverse taken on $\mathcal S$. The normalized PGM vectors $\ket{b_c}=|C|^{-1/2}\Psi^{-1/2}\ket{\psi_c}$ are therefore
\begin{equation}\label{eq:pgm_vectors}
  \ket{b_c}=\frac1{\sqrt{|C|}}\sum_X\omega^{c\cdot x_X}\ket{\hat\phi_X},\qquad c\in C.
\end{equation}
The phases do not depend on the representatives because $c\in K^\perp$, and character orthogonality shows that the $\ket{b_c}$ form an orthonormal basis of $\mathcal S$. Hence the PGM effects restricted to $\mathcal S$ are the projectors $\ket{b_c}\!\bra{b_c}$; their completion outside $\mathcal S$ will not matter. Since $W_{\Phi_\beta}(X)=Z_X$, Lemma~\ref{lem:pgm_success} shows that the block success of the PGM is $P_{\mathrm{PGM}}(B,\beta)\coloneqq P_{\mathrm{PGM}}(\Phi_\beta)=|C|^{-1}\big(\sum_X\sqrt{Z_X}\big)^2$, which is the quantity~\eqref{eq:pgm_success} in Proposition~\ref{prop:pgm_gibbs} below.

The coherent constructions below require only positive mass on every coset. They therefore also hold for any amplitude function $\Phi$ with $W_\Phi(X)>0$ for every $X$, with normalized target states in place of Gibbs states and with efficient preparation of $\ket\Phi$ assumed in the complexity statements.

A coherent implementation can map each $\ket{\hat\phi_X}$ to its coset label and apply a Fourier transform over the labels. Reversing this change of basis prepares a specified Gibbs state from a Fourier superposition of codeword labels. Proposition~\ref{prop:pgm_gibbs} applies more generally to any exact coherent PGM circuit that retains all work registers and has an available inverse. Regev's reduction compares the decoded and original codeword labels without measuring either one. On the successful branch, applying the inverse circuit prepares the Gibbs state of the specified offset. The success probability is $P_{\mathrm{PGM}}$, independently of the offset.

This construction has been analyzed for the zero offset~\cite[Section~6.3 of the full version]{chailloux2024quantumdecoding} and for arbitrary offsets over finite fields~\cite[Section~4.4, Algorithm~4.4]{blanvillain2026tight}. Related reductions solve bounded-distance instances on the dual code from an exact quantum decoder~\cite[Proposition~3]{garrido2026regev}, or from a coherent PGM and its inverse with an amplitude function chosen for the instance and a constant expected number of repetitions~\cite[Theorem~7]{garrido2026regev}. Here the amplitude function is fixed, its coset states are the primal Gibbs states, and the construction is stated over $\Z_q$ for every $q$.

Conversely, coherent preparation of all $\ket{\hat\phi_X}$ and its inverse implement the PGM by the change of basis~\eqref{eq:pgm_vectors} (Proposition~\ref{prop:gibbs_pgm}), as in~\cite{piveteau2025efficient,chailloux2026equivalence}. The two tasks are equivalent up to polynomial overhead whenever $P_{\mathrm{PGM}}$ is at least inverse polynomial (Corollary~\ref{cor:pgm_equiv}).

\begin{proposition}[Gibbs preparation from a coherent PGM]\label{prop:pgm_gibbs}
Suppose that a unitary $U$ implements the PGM coherently: for every unit vector $\ket v\in\mathcal S$, applying $U$ to $\ket v\ket0$ and measuring an outcome register $O$ yields $c\in C$ with probability $|\langle b_c|v\rangle|^2$. The ancillary input $\ket0$ includes all work registers and the outcome register, and all output registers are retained. Given access to $U$ and $U^\dagger$, for every offset $x_0\in\Z_q^m$ there is a heralded procedure whose successful output is exactly $\ket{\hat\phi_{x_0+K}}$, with all auxiliary registers returned to zero. Its success probability is
\begin{equation}\label{eq:pgm_success}
  P_{\mathrm{PGM}}\coloneqq\frac1{|C|}\left(\sum_X\sqrt{Z_X}\right)^2,
\end{equation}
which is also the block success of the PGM on the uniform codeword ensemble. Each attempt uses at most one application of $U$ and one of $U^\dagger$, together with polynomial overhead for preparing the DQI input and comparing and subtracting codeword labels, assuming efficient preparation of the single-symbol factors. For fixed $q$, an efficient implementation of $U$ with $P_{\mathrm{PGM}}\ge1/\poly(m+n)$ therefore yields polynomial expected-time preparation of the Gibbs state.
\end{proposition}

The proof of Proposition~\ref{prop:pgm_gibbs} is given in Appendix~\ref{app:pgm}.

To state the converse, we label the cosets of $K$ and the codewords by coordinates in which the pairing $(c,x)\mapsto\omega^{c\cdot x}$ is explicit. The group $C$ need not be isomorphic to $\Z_q^h$ for any $h$, so we use its invariant factors, which the Smith normal form computes. Let $UB^\top V=S$ be the Smith normal form of $B^\top$ over the integers, with $U\in GL_n(\mathbb Z)$, $V\in GL_m(\mathbb Z)$, and $S$ the $n\times m$ matrix with diagonal entries $e_1,\dots,e_n\ge0$ and zeros elsewhere, and set $e_j=0$ for $n<j\le m$. For $c\in\Z_q^m$ and $u=V^{-1}c$, we have $B^\top c=0$ in $\Z_q^n$ if and only if $e_ju_j=0$ in $\Z_q$ for every $j$, that is, if and only if $u_j$ is a multiple of $q/g_j$, where $g_j=\gcd(e_j,q)$ and $\gcd(0,q)=q$. Hence the map $y\mapsto c(y)=V\big((q/g_j)\,y_j\big)_j$ is an isomorphism from $\prod_j\Z_{g_j}$ onto $C$. For $x\in\Z_q^m$, let $\tau(x)\in\prod_j\Z_{g_j}$ have coordinates $\tau_j(x)=(V^\top x)_j\bmod g_j$. Then
\begin{equation}\label{eq:factor_pairing}
  \omega^{c(y)\cdot x}=\exp\Big(2\pi i\sum_j\frac{y_j\,\tau_j(x)}{g_j}\Big),
\end{equation}
so $\tau(x)=0$ exactly when $x$ is orthogonal to $C$, that is, when $x\in C^\perp=K$. The map $\tau$ is onto, because $|\Z_q^m/K|=q^m/|K|=|C|=\prod_jg_j$. The cosets of $K$ are therefore the sets $X_\tau=\{x\in\Z_q^m:\tau(x)=\tau\}$ with $\tau\in\prod_j\Z_{g_j}$. Factors with $g_j=1$ can be omitted. For fixed $q$, the Smith normal form, the maps $y\mapsto c(y)$ and $x\mapsto\tau(x)$, and the inverse of $y\mapsto c(y)$ on $C$ can be computed in polynomial time.

\begin{proposition}[Coherent PGM from coherent Gibbs states]\label{prop:gibbs_pgm}
Suppose that a unitary $V$ satisfies $V\ket\tau\ket0=\ket\tau\ket{\hat\phi_{X_\tau}}$ for every $\tau\in\prod_j\Z_{g_j}$, where the second register consists of $m$ qudits, and that $V^\dagger$ is available. Then the reversible computation of the coset label $\tau(x)$, one application of $V^\dagger$, the inverse quantum Fourier transform over $\prod_j\Z_{g_j}$, and the reversible computation of $c(y)$ from $y$, applied in this order, form a unitary $U$ with $U(\ket{b_c}\ket0)=\ket c_O\ket0$ for every $c\in C$, where $O$ is an outcome register. In particular, $U$ implements the PGM coherently in the sense of Proposition~\ref{prop:pgm_gibbs}. The inverse $U^\dagger$ is implemented in the same way with one application of $V$.
\end{proposition}

\begin{proof}
Let $c=c(y)\in C$. By~\eqref{eq:factor_pairing}, every $x\in X_\tau$ satisfies $\omega^{c\cdot x}=\chi_y(\tau)$, where $\chi_y(\tau)=\exp\big(2\pi i\sum_jy_j\tau_j/g_j\big)$, so by~\eqref{eq:pgm_vectors} and $|C|=\prod_jg_j$,
$$
  \ket{b_c}=|C|^{-1/2}\sum_{\tau}\chi_y(\tau)\ket{\hat\phi_{X_\tau}}.
$$
Every $x$ in the support of $\ket{\hat\phi_{X_\tau}}$ satisfies $\tau(x)=\tau$, so computing the coset label into a new register maps $\ket{\hat\phi_{X_\tau}}\ket0$ to $\ket{\hat\phi_{X_\tau}}\ket\tau$. Applying $V^\dagger$ to the label register and the $m$ qudits then maps $\ket{b_c}\ket0$ to the product of the state $\ket0$ of the qudits and the state $|C|^{-1/2}\sum_\tau\chi_y(\tau)\ket\tau$ of the label register. The inverse of the quantum Fourier transform $\ket y\mapsto|C|^{-1/2}\sum_\tau\chi_y(\tau)\ket\tau$ over $\prod_j\Z_{g_j}$ maps the label register to $\ket y$. Computing $c(y)$ into the outcome register $O$, and then uncomputing $y$ from $c(y)$ with the inverse of $y\mapsto c(y)$, leaves $\ket c_O$ with all other registers equal to zero. These steps are unitary, so their composition $U$ satisfies $U(\ket{b_c}\ket0)=\ket c_O\ket0$ for every $c\in C$. The vectors $\ket{b_c}$ form an orthonormal basis of $\mathcal S$, so by linearity, applying $U$ to $\ket v\ket0$ with a unit vector $\ket v\in\mathcal S$ and measuring $O$ yields $c$ with probability $|\langle b_c|v\rangle|^2$. Running the steps in reverse implements $U^\dagger$ with one application of $V$.
\end{proof}

Combining Propositions~\ref{prop:pgm_gibbs} and~\ref{prop:gibbs_pgm}, we obtain the following equivalence.

\begin{corollary}[Coherent PGM and coherent Gibbs states]\label{cor:pgm_equiv}
Let $q\ge2$ and let $B$, $f$, and $\beta$ be fixed, with $\tau(x)$ and $X_\tau$ as above. Up to polynomial overhead, a coherent implementation $U$ of the PGM as in Proposition~\ref{prop:pgm_gibbs}, with access to $U^\dagger$, and a coherent preparation $V$ of the states $\ket{\hat\phi_X}$ for all cosets $X$ as in Proposition~\ref{prop:gibbs_pgm}, with access to $V^\dagger$, are equivalent in the following sense.
\begin{enumerate}
\item One application of $V^\dagger$ implements $U$ exactly, and one application of $V$ implements $U^\dagger$ exactly.
\item One application each of $U$ and $U^\dagger$ implements a unitary $\tilde V$, which acts on a register that holds $\tau$, the $m$ qudits, auxiliary registers, and a flag qubit, with the following property. For every state $\sum_\tau u_\tau\ket\tau$, applying $\tilde V$ to $\sum_\tau u_\tau\ket\tau\ket0$ and measuring the flag indicates success with probability $P_{\mathrm{PGM}}$, which does not depend on the state, and after success the state is $\sum_\tau u_\tau\ket\tau\ket{\hat\phi_{X_\tau}}$, with all auxiliary registers returned to zero. The inverse $\tilde V^\dagger$ also uses one application each of $U$ and $U^\dagger$.
\end{enumerate}
\end{corollary}

\begin{proof}
Part~1 is Proposition~\ref{prop:gibbs_pgm}. For part~2, the procedure in Appendix~\ref{app:pgm} depends on $x_0$ only through the phases $\omega^{-c\cdot x_0}$. By~\eqref{eq:factor_pairing}, these phases are characters of $\tau(x_0)$, so they can be applied controlled on a coset-label register with polynomial overhead.

Let $\tilde V$ perform this controlled procedure, retaining rather than measuring the comparison flag, then subtract the decoded word from the original label and apply $U^\dagger$ without conditioning on the flag. The flag remains unchanged, so success and failure branches are orthogonal. For every $\tau$, the successful branch is $\sqrt{P_{\mathrm{PGM}}}\ket\tau\ket{\hat\phi_{X_\tau}}$ with all auxiliary registers zero. Its coefficient is independent of $\tau$ by~\eqref{eq:pgm_equal_overlap}. Thus the successful branch for an arbitrary superposition of labels is
$$
  \sqrt{P_{\mathrm{PGM}}}\sum_\tau u_\tau\ket\tau\ket{\hat\phi_{X_\tau}},
$$
which proves part~2 by linearity.
\end{proof}

Corollary~\ref{cor:pgm_equiv} holds for every instance. For $q=2$, $f(z)=z$, and $3\le k<D$, Theorem~\ref{thm:condensation} in Section~\ref{sec:condensation} shows that, at every $\beta<\beta_c$, $P_{\mathrm{PGM}}(B,\beta)=1-o(1)$ with probability $1-o(1)$ over random instances from the configuration model. At these inverse temperatures, an efficient coherent implementation of the optimal decoder, with access to its inverse, is therefore equivalent to an efficient coherent preparation of the primal Gibbs states at the same inverse temperature for all right-hand sides, up to a heralded failure probability $o(1)$. The coherent Gibbs state on assignments then follows as in Appendix~\ref{app:pgm}.

\subsection{The condensation threshold}\label{sec:condensation}

For binary Max-XORSAT on the configuration model, optimal decoding is reliable below $\beta_c$ and every decoder has exponentially small block success above it (Theorem~\ref{thm:condensation}). We locate the transition through the limiting relative entropy of the coset law. Concentration converts positive entropy per variable into failure above the threshold; added noise and linear minimum distance prove success below it. A mixture over weights then extends the resulting satisfied-fraction bound to every symmetric amplitude function.

Throughout, $q=2$, $f(z)=z$, and $B$ is drawn from the configuration model of Section~\ref{sec:typical} with $3\le k<D$, with or without conditioning on the absence of multiple edges. Without the conditioning, $B$ is the incidence matrix of the bipartite multigraph reduced modulo two. In both cases $n=mk/D$.

\paragraph{The relative entropy of the coset law.} For $\alpha\in(0,\tfrac12)$ we write $\sigma=1-2\alpha$, so that $\sigma=\sigma_\beta=\tanh(\beta/2)$ when $\alpha=p_\beta(1)$. Let $W_{B,\alpha}$ be the law of the coset $E+K$ when $E\in\Z_2^m$ has independent Bernoulli$(\alpha)$ entries. The law $|\Phi_\beta(x)|^2=\prod_jp_\beta(x_j)$ is the law of such a vector with $\alpha=p_\beta(1)$, so $W_{B,p_\beta(1)}=W_{\Phi_\beta}$, and by Lemma~\ref{lem:pgm_success}, $P_{\mathrm{PGM}}(B,\beta)=\big(\E_{X\sim U_B}\sqrt{|C|\,W_{B,\alpha}(X)}\big)^2$ with $\alpha=p_\beta(1)$. We measure the distance of $W_{B,\alpha}$ from uniform by the relative entropy
$$
  F_B(\alpha)=D_{\mathrm{KL}}\big(W_{B,\alpha}\,\big\|\,U_B\big)=\sum_XW_{B,\alpha}(X)\ln\big(|C|\,W_{B,\alpha}(X)\big).
$$
To describe its limit on random instances, let
$$
  \operatorname{cap}(t)=\tfrac12\big[(1+t)\ln(1+t)+(1-t)\ln(1-t)\big],\qquad t\in[-1,1],
$$
be the capacity in nats of the binary symmetric channel with crossover probability $(1-t)/2$, so that $\operatorname{cap}(\sigma)=(1-H_2(\alpha))\ln2$. For a law $\pi$ on $[-1,1]$ with mean zero, which we view as a law of magnetizations, let the ${\mathfrak m}_{i,j}$ be independent with law $\pi$, and let
$$
  a=\E\operatorname{cap}\big(\sigma {\mathfrak m}_{1,1}\cdots {\mathfrak m}_{1,k-1}\big),\qquad b=\E\operatorname{cap}\big(\sigma {\mathfrak m}_{1,1}\cdots {\mathfrak m}_{1,k}\big).
$$
Let $\mathcal I_D$ be the mutual information between a uniformly random sign $\xi$ and $D$ independent observations of $\xi$, where observation $i$ consists of $R_i=|\sigma {\mathfrak m}_{i,1}\cdots {\mathfrak m}_{i,k-1}|$ and of a sign that equals $\xi$ with probability $(1+R_i)/2$. We define
\begin{equation}\label{eq:functional}
  \mathcal B_\sigma(\pi)=D\,a-\mathcal I_D-\frac{D(k-1)}k\,b ,
\end{equation}
which vanishes at $\pi=\delta_0$. Up to the additive constant $\ln2$, $\mathcal B_\sigma$ is the functional that appears in the theorem of van den Brand and Jaafari on the mutual information of low-density generator-matrix codes~\cite{vandenbrand2017mutual}, specialized to the codes considered here (Appendix~\ref{app:condensation}).

\begin{proposition}[Relative entropy and mutual information]\label{prop:entropy}
Let $\alpha\in(0,\tfrac12)$.
\begin{enumerate}
\item For every matrix $B\in\{0,1\}^{m\times n}$ over $\Z_2$, let $Z$ be uniform on $\Z_2^n$, let $E$ be independent of $Z$ with independent Bernoulli$(\alpha)$ entries, and let $Y=BZ+E$. Then $F_B(\alpha)=m\operatorname{cap}(\sigma)-I(Z;Y\mid B)$, where the mutual information is taken for the fixed matrix $B$.
\item For every matrix $B$, $F_B(\alpha)$ is a nonincreasing function of $\alpha\in(0,\tfrac12)$.
\item For $B$ from the configuration model without the conditioning, the limit $\bar F(\alpha)=\lim_{n\to\infty}\frac1n\E_BF_B(\alpha)$ exists and equals $\sup_\pi\mathcal B_\sigma(\pi)$, where $\pi$ runs over the laws on $[-1,1]$ with mean zero.
\end{enumerate}
\end{proposition}

Part~1 says that $F_B(\alpha)$ is the amount by which the mutual information of the low-density generator-matrix (LDGM) code with generator matrix $B$, used on the binary symmetric channel with crossover probability $\alpha$, falls short of $m\operatorname{cap}(\sigma)$, the largest mutual information that $m$ uses of this channel can carry. Part~3 follows from a theorem of van den Brand and Jaafari on the mutual information of LDGM codes~\cite[Theorem~2.1]{vandenbrand2017mutual} (see also~\cite[Theorem~1.1 of the full version]{cojaoghlan2021inference}), combined with part~1. The proofs are given in Appendix~\ref{app:condensation}.

\paragraph{The condensation threshold.} By part~2, and since $F_B\ge0$, the limit $\bar F$ is nonnegative and nonincreasing. We define
$$
  \alpha_c=\sup\big\{\alpha\in(0,\tfrac12):\bar F(\alpha)>0\big\},\qquad\beta_c=\ln\frac{1-\alpha_c}{\alpha_c},\qquad s_c=1-\alpha_c,\qquad\rho_c=\alpha_c^\perp,
$$
so that $\alpha_c=p_{\beta_c}(1)$, $\bar F>0$ on $(0,\alpha_c)$, and $\bar F=0$ on $(\alpha_c,\tfrac12)$. The mutual information in part~1 is at most $H(Z)=n\ln2$, so $F_B(\alpha)\ge m\operatorname{cap}(\sigma)-n\ln2=m\big(1-\tfrac k D-H_2(\alpha)\big)\ln2$. This bound is positive for $\alpha<\alpha_H$, where $H_2(\alpha_H)=1-k/D$ as in Section~\ref{sec:dqi}. Hence $\alpha_c\ge\alpha_H$, and therefore $\beta_c\le\beta_{\mathrm H}$, $s_c\le s_H$, and $\rho_c\le\rho_H$; by Proposition~\ref{prop:condensation_holevo}, these inequalities are strict.

Given $B$, the vector $Y=BZ+E$ is uniformly distributed on its coset, because $BZ$ is uniform on $K$, so $F_B(\alpha)$ is also the relative entropy between the law of $Y$ and the uniform law on $\Z_2^m$. The vector $Y$ is the right-hand side of a planted instance, in which the uniformly random assignment $Z$ violates each equation independently with probability $\alpha$, and the uniform law is the law of a uniformly random right-hand side. Hence $\beta_c$ is the information-theoretic threshold of this planted model, which van den Brand and Jaafari study~\cite{vandenbrand2017mutual}: above $\beta_c$ the relative entropy between the planted and the uniformly random right-hand sides is linear in $m$ on average, and below $\beta_c$ it is $o(m)$. The cavity method identifies this threshold with the condensation threshold of the primal Gibbs measure with a uniformly random right-hand side~\cite{krzakala2007gibbs,krzakala2009hiding}. Coja-Oghlan, Krzakala, Perkins, and Zdeborov\'a~\cite{cojaoghlan2018information} proved this identification for random factor graphs of Erd\H{o}s--R\'enyi type, and Coja-Oghlan, Hahn-Klimroth, Loick, M\"uller, Panagiotou, and Pasch~\cite{cojaoghlan2021inference} proved it for random factor graphs with given degrees, which include the configuration model with the weights of Max-XORSAT (Theorems~17.1 and~17.2 and Section~16.1 of the full version). We therefore call $\beta_c$ the condensation threshold. In classical coding, the analogous picture is predicted for regular low-density parity-check codes: optimal decoding succeeds up to the static transition of the corresponding spin glass, and iterative decoding up to the dynamical transition~\cite{franz2002dynamic}.

For uniformly random linear codes, Chailloux and Tillich showed that, on average over the code, the PGM recovers the codeword with probability tending to one at every rate below the Holevo information of the single-symbol channel with uniform input (Section~\ref{sec:dqi}), and that every measurement fails above it~\cite[Theorem~5]{chailloux2024quantumdecoding}; Blanvillain, Chailloux, and Tillich extended both statements to every product amplitude function~\cite[Theorems~1 and~2]{blanvillain2026tight}. For uniformly random linear codes of rate $1-k/D$, the threshold of the PGM is therefore the limit $\beta_{\mathrm H}$ of Section~\ref{sec:dqi}, which Jordan et al.\ derived for DQI and which Zhou et al.\ and Piveteau, Quek, and Franco Garrido use as the limit of quantum decoders~\cite{zhou2026gibbs,piveteau2026gibbs}. For the sparse binary codes of random $(k,D)$-regular instances from the configuration model, the following theorem shows that the PGM succeeds below the condensation threshold $\beta_c<\beta_{\mathrm H}$, and that every decoder fails above it.

\begin{theorem}[The optimal decoder and the condensation threshold]\label{thm:condensation}
Let $q=2$, $f(z)=z$, $k\ge3$, and $k<D$, and let $B$ be drawn from the configuration model, with or without the conditioning on the absence of multiple edges.
\begin{enumerate}[label=(\alph*)]
\item If $0\le\beta<\beta_c$, then $\E_B\big[1-P_{\mathrm{PGM}}(B,\beta)\big]\to0$ as $m\to\infty$. In particular, the PGM has block success $1-o(1)$ with probability $1-o(1)$ over the instance.
\item If $\beta>\beta_c$, then there is $\epsilon>0$ such that, with probability $1-e^{-\Omega(m)}$ over the instance, every decoder for the decoding problem induced by DQI with $\Phi_\beta$ has block success at most $e^{-\epsilon m}$.
\end{enumerate}
\end{theorem}

By part~(a) and part~1 of Theorem~\ref{thm:guarantee}, at every $\beta<\beta_c$ and with probability $1-o(1)$ over the instance, DQI with the PGM outputs a violation pattern whose law is within $o(1)$ of $\pi_\beta$ in expected total variation over the offset. Above $\beta_c$, with probability $1-e^{-\Omega(m)}$ over the instance, Regev's reduction with any decoder requires $e^{\Omega(m)}$ attempts in expectation, because its check passes with probability equal to the block success of the decoder (Section~\ref{sec:dqi}). For the product amplitude functions, part~(b) also bounds the satisfied fraction: a decoder whose block success tends to one exists only for $\beta\le\beta_c$, and DQI with $\Phi_\beta$ and such a decoder then reaches at most $p_\beta(0)+o(1)\le s_c+o(1)$. The next corollary extends this bound to every symmetric amplitude function.

\begin{corollary}[Symmetric amplitude functions]\label{cor:condensation_sym}
Under the hypotheses of Theorem~\ref{thm:condensation}, write $s(\Phi)=\sum_x|\Phi(x)|^2(1-|x|/m)$ for the expected satisfied fraction in the state $\ket\Phi$ of an amplitude function $\Phi$. For every $\epsilon>0$ there are $\delta>0$, a sequence $\eta_m\to0$, and events of probability $1-e^{-\Omega(m)}$ over the instance on which the following holds: for every symmetric amplitude function $\Phi$ with $s(\Phi)\ge s_c+\epsilon$, every decoder for the decoding problem induced by DQI with $\Phi$ has block success at most $1-\delta+\eta_m$. Consequently, with probability $1-o(1)$ over the instance, the expected satisfied fraction that DQI reaches, averaged over the offset, with a symmetric amplitude function and a decoder whose block success tends to one is at most $s_c+o(1)$.
\end{corollary}

The second statement follows from the first because, by part~2 of Theorem~\ref{thm:guarantee}, $s(\Phi)$ differs by at most $2\sqrt\varepsilon$ from the expected satisfied fraction that DQI reaches with $\Phi$ and a decoder of block success $1-\varepsilon$, averaged over the offset. The bound $s_c$ is attained in the limit $\beta\uparrow\beta_c$. Indeed, by part~(a) of Theorem~\ref{thm:condensation} and part~2 of Theorem~\ref{thm:guarantee}, DQI with the product amplitude function $\Phi_\beta$, which is symmetric, and the PGM reaches the expected satisfied fraction $p_\beta(0)+o(1)$ at every $\beta<\beta_c$, and $p_\beta(0)\to1-\alpha_c=s_c$ as $\beta\uparrow\beta_c$.

\paragraph{Outline of the proofs.} Above $\beta_c$, the coset density $L=|C|W_{B,\alpha}$ separates two typical sets. Under the uniform law, $L$ rarely exceeds $e^{F_B(\alpha)/2}$; under the coset law, $\ln L$ is concentrated around its mean $F_B(\alpha)$. When $F_B(\alpha)$ is linear in $m$, the overlap of the two laws, and hence optimal decoding success, is exponentially small (Lemma~\ref{lem:entropy_bound}). The relative entropy also concentrates over the configuration-model instance, because one half-edge switching changes only two rows of $B$ (Lemma~\ref{lem:concentration}). Thus the positive limiting entropy locates the failure regime.

Below $\beta_c$, sublinear relative entropy alone does not imply coset uniformity. We start at a slightly smaller violation probability $\alpha_0$ where $\E_BF_B(\alpha_0)=o(m)$, and truncate its density at an exponential cutoff. The entropy bounds the removed mass by $o(1)$ and the cutoff controls the remaining second moment. Adding independent noise to reach $\alpha>\alpha_0$ multiplies the Fourier coefficient at a nonzero codeword $c$ by $\lambda^{|c|}$ for a fixed $\lambda<1$. Linear minimum distance makes every such coefficient decay exponentially; after truncation this makes the variance tend to zero. The coset law at $\alpha$ is therefore asymptotically uniform (Lemmas~\ref{lem:min_distance} and~\ref{lem:smoothing}).

For a symmetric amplitude function, its coset law is a mixture of the laws $W_w$ of uniformly random words of weight $w$. Each $W_w$ is bounded pointwise by $(m+1)W_{B,w/m}$. If the mean satisfied fraction exceeds $s_c$ by a fixed amount, a positive fraction of the mixture must lie at weights below $\alpha_cm$ by a fixed amount. The failure estimate is uniform over these weights and remains small after the polynomial factor. Their positive mass prevents block success from tending to one, proving Corollary~\ref{cor:condensation_sym}.

\begin{corollary}[Local and optimal decoders for Max-XORSAT]\label{cor:two_thresholds}
Let $q=2$, $f(z)=z$, and $3\le k<D$, let $B$ be drawn from the configuration model, and fix $\beta\ge0$. Consider the decoding problem induced by DQI with the product amplitude function $\Phi_\beta$.
\begin{enumerate}
\item If $\beta<\beta_d$, then with probability $1-o(1)$ over the instance, the two-stage BPQM decoder of Mandal et al.~\cite{mandal2026blockerror} in the form of Proposition~\ref{prop:sharp}, which is local in the sense of Definition~\ref{def:local_decoder}, has block success $1-o(1)$. If $\beta>\beta_d$, then for every instance, every decoder that makes a $t$-local estimate at a symbol $i$ at which $\mathrm{LT}_t(i)$ and $\mathrm{TP}_t(i)$ hold has block success at most $P_\infty<1$, where $P_\infty$ depends only on $k$, $D$, and $\beta$.
\item If $\beta<\beta_c$, then with probability $1-o(1)$ over the instance, the pretty-good measurement, which is an optimal decoder, has block success $1-o(1)$. If $\beta>\beta_c$, then there is $\epsilon>0$ such that, with probability $1-e^{-\Omega(m)}$ over the instance, every decoder has block success at most $e^{-\epsilon m}$.
\item The two thresholds satisfy $\beta_d\le\beta_c$.
\end{enumerate}
\end{corollary}

\begin{proof}
For part~1, the primal model has non-reconstruction at every $\beta<\beta_d$ by Corollary~\ref{cor:binary}, so Proposition~\ref{prop:sharp} applies. If $\beta>\beta_d$, the primal model has reconstruction by Definition~\ref{def:beta_d}, so Theorem~\ref{thm:locality} bounds the block success by $P_t\le P_\infty$, and $P_\infty<1$. Part~2 is Theorem~\ref{thm:condensation}, and the pretty-good measurement is an optimal decoder by Lemma~\ref{lem:pgm_success}. For part~3, suppose that $\beta_c<\beta_d$, and let $\beta_c<\beta<\beta_d$. By part~1, the two-stage BPQM decoder has block success $1-o(1)$ with probability $1-o(1)$ over the instance, and by part~2, every decoder has block success at most $e^{-\epsilon m}$ with probability $1-o(1)$. For large $m$, both events occur together with positive probability, and on their intersection $1-o(1)\le e^{-\epsilon m}$, which is impossible.
\end{proof}

\paragraph{Separation from the Holevo threshold.} The condensation threshold is strictly below the Holevo threshold at every finite degree pair, but converges to it exponentially at a fixed degree ratio. The comparison uses the exact variational identity of Proposition~\ref{prop:entropy}, without assuming that any particular trial law is an optimizer.

\begin{proposition}[Separation and convergence of the thresholds]\label{prop:condensation_holevo}
Consider binary Max-XORSAT in the configuration model with $3\le k<D$, and define $\beta_c$ by Proposition~\ref{prop:entropy}. Its operational interpretation for both the unconditioned model and the model conditioned on no multiple edges is given by Theorem~\ref{thm:condensation}.
\begin{enumerate}
\item At every fixed finite degree pair, $\beta_c<\beta_{\mathrm H}$.
\item Along admissible degree pairs with a fixed rational ratio $k/D\in(0,1)$, for all sufficiently large degrees,
\begin{equation}\label{eq:condensation_exp_gap}
  0<\beta_{\mathrm H}-\beta_c\le C_{k/D}e^{-k/48},
  \qquad C_{k/D}=\frac{8(k/D)}{\beta_{\mathrm H}(1-\sigma_H^2)},
  \qquad \sigma_H=\tanh(\beta_{\mathrm H}/2).
\end{equation}
\end{enumerate}
The instance-size limit defining $\beta_c$ is taken first at each fixed degree pair.
\end{proposition}

\begin{proof}[Proof of strict separation]
Take $\pi_*=(\delta_{-1}+\delta_1)/2$. In~\eqref{eq:functional}, $a=b=\operatorname{cap}(\sigma)$, and each observation has reliability $\sigma$. If $H_D(\alpha)$ is the conditional entropy of a uniform sign after $D$ independent binary symmetric channel observations with crossover probability $\alpha$, then
\begin{equation}\label{eq:condensation_trial}
  \mathcal B_\sigma(\pi_*)=\frac Dk\operatorname{cap}(\sigma)-\ln2+H_D(\alpha).
\end{equation}
At $\alpha=\alpha_H$, the first two terms cancel. Since $0<\alpha_H<1/2$ and $D$ is finite, every observation string has positive likelihood under both signs, so $H_D(\alpha_H)>0$. By continuity, the fixed-trial value remains positive at some $\beta<\beta_{\mathrm H}$. Proposition~\ref{prop:entropy} then implies $\beta_c<\beta_{\mathrm H}$. The exponential estimate is proved in Section~\ref{app:condensation_holevo}.
\end{proof}

Strict separation is therefore due to residual uncertainty about a variable after finitely many noisy observations. For the reverse comparison at large degrees, positivity at a single trial law is not enough: one must show that $\mathcal B_\sigma(\pi)\le0$ for every law $\pi$ below a suitable inverse temperature. With sufficiently reliable incoming observations, the conditional entropy of the hidden sign is exponentially small in $k$. With weak observations, we instead bound the redundancy among the observations by the negative clause term in~\eqref{eq:functional}. With these estimates and the intermediate cases in Section~\ref{app:condensation_holevo}, we establish exact nonpositivity when capacity is below its Holevo value by an exponentially small margin.

Combining Proposition~\ref{prop:condensation_holevo} with Corollary~\ref{cor:two_thresholds}, the ordering is $\beta_d\le\beta_c<\beta_{\mathrm H}$. We do not establish $\beta_d<\beta_c$ for every finite degree pair, but for each of the fifteen benchmark pairs of Table~\ref{tab:benchmark} we prove it with a computer-assisted computation, with a margin of at least $0.087$ in $\beta$ (Appendix~\ref{app:certified}). At a fixed degree ratio, however, $\beta_c$ converges to the positive constant $\beta_{\mathrm H}$, whereas $\beta_d$ is predicted to converge to zero (Section~\ref{sec:larger_degrees}). Under this prediction, $\beta_c-\beta_d\to\beta_{\mathrm H}$, so the first inequality is strict for all sufficiently large degrees.

For optimization, $s=p_\beta(0)$ has derivative at most $1/4$, so under the large-degree hypotheses of~\eqref{eq:condensation_exp_gap},
\begin{equation}\label{eq:condensation_exp_s}
  0<s_H-s_c\le\frac14 C_{k/D}e^{-k/48}.
\end{equation}
This comparison concerns the configuration model and does not establish the corresponding result for Gallager's ensemble.

\paragraph{Numerical comparison.} Table~\ref{tab:condensation} lists the three thresholds and the corresponding satisfied fractions for the fifteen benchmark pairs of Table~\ref{tab:benchmark}, together with the best classical value in each row. The values of $\beta_c$ and $s_c$ are obtained from certified intervals, which we compute in Appendix~\ref{app:certified} from the functional~\eqref{eq:functional}. They are accurate to within $0.011$ in $\beta_c$ and $6\times10^{-4}$ in $s_c$ for $k=3$, and to within $10^{-3}$ and $2\times10^{-4}$ for $k\ge4$. The certified intervals show that $\beta_d<\beta_c<\beta_{\mathrm H}$ for every pair (Table~\ref{tab:certified}). For $(3,6)$, $1.985\le\beta_c\le1.995$, which lies strictly between the certified upper bound $1.850$ on $\beta_d$ and $\beta_{\mathrm H}\approx2.090$; correspondingly, $s_c\approx0.880$ and $s_H\approx0.890$. The difference $s_H-s_c$ lies between $0.003$ and $0.014$ for $k=3$, is at most $0.003$ for $k=4$, and is below $10^{-3}$ for $k\ge5$. On every pair, the certified lower bound on $s_c$ exceeds the best classical value in Table~\ref{tab:benchmark} by at least $0.0025$.

The classical values were obtained on Gallager's ensemble, whereas Theorem~\ref{thm:condensation} and our bounds on $s_c$ concern the configuration model. Both ensembles are locally tree-like with the same degrees, and we expect them to have the same value of $s_c$, but we have not proved this. If they do, DQI with the optimal decoder exceeds the best classical value on every benchmark pair.

\begin{table}[ht]
\centering
\small
\begin{tabular}{@{}c ccc ccc c@{}}
\toprule
& \multicolumn{3}{c}{Inverse temperature} & \multicolumn{3}{c}{Satisfied fraction} & \\
\cmidrule(lr){2-4}\cmidrule(lr){5-7}
$(k,D)$ & $\beta_d$ & $\beta_c$ & $\beta_{\mathrm H}$ & $s_d$ & $s_c$ & $s_H$ & Best classical \\
\midrule
$(3,4)$ & $2.6574$ & $3.0488$ & $3.1348$ & $0.9345$ & $0.9547$ & $0.9583$ & $0.9381$ (SA) \\
$(3,5)$ & $2.1387$ & $2.3532$ & $2.4508$ & $0.8946$ & $0.9132$ & $0.9206$ & $0.9024$ (SA) \\
$(3,6)$ & $1.8388$ & $1.9902$ & $2.0905$ & $0.8628$ & $0.8798$ & $0.8900$ & $0.8726$ (SA) \\
$(3,7)$ & $1.6373$ & $1.7577$ & $1.8561$ & $0.8372$ & $0.8529$ & $0.8648$ & $0.8489$ (SA) \\
$(3,8)$ & $1.4901$ & $1.5915$ & $1.6872$ & $0.8161$ & $0.8308$ & $0.8439$ & $0.8279$ (SA) \\
$(4,5)$ & $2.5494$ & $3.4260$ & $3.4381$ & $0.9275$ & $0.9685$ & $0.9689$ & $0.9293$ (SA) \\
$(4,6)$ & $2.1781$ & $2.7047$ & $2.7254$ & $0.8983$ & $0.9373$ & $0.9385$ & $0.9011$ (SA) \\
$(4,7)$ & $1.9316$ & $2.3188$ & $2.3427$ & $0.8734$ & $0.9104$ & $0.9124$ & $0.8776$ (SA) \\
$(4,8)$ & $1.7531$ & $2.0645$ & $2.0905$ & $0.8523$ & $0.8874$ & $0.8900$ & $0.8566$ (SA) \\
$(5,6)$ & $2.3912$ & $3.6766$ & $3.6791$ & $0.9162$ & $0.9753$ & $0.9754$ & $0.9312$ (TP) \\
$(5,7)$ & $2.1156$ & $2.9433$ & $2.9478$ & $0.8924$ & $0.9499$ & $0.9502$ & $0.8943$ (SA) \\
$(5,8)$ & $1.9173$ & $2.5433$ & $2.5497$ & $0.8718$ & $0.9271$ & $0.9276$ & $0.8730$ (SA) \\
$(6,7)$ & $2.2496$ & $3.8785$ & $3.8790$ & $0.9046$ & $0.9797$ & $0.9797$ & $0.9427$ (TP) \\
$(6,8)$ & $2.0358$ & $3.1337$ & $3.1348$ & $0.8845$ & $0.9583$ & $0.9583$ & $0.8962$ (TP) \\
$(7,8)$ & $2.1281$ & $4.0495$ & $4.0497$ & $0.8936$ & $0.9829$ & $0.9829$ & $0.9481$ (TP) \\
\bottomrule
\end{tabular}
\caption{The reconstruction threshold $\beta_d$, which is the limit of local decoders, the condensation threshold $\beta_c$, which is the limit of every decoder on the configuration model (Theorem~\ref{thm:condensation}), and the Holevo threshold $\beta_{\mathrm H}$, for the fifteen benchmark pairs of Table~\ref{tab:benchmark}, with the corresponding expected satisfied fractions $s=p_\beta(0)$. The values of $\beta_d$ are computed in the same way as those of Table~\ref{tab:degrees}, and the values of $s_d$ are those of Table~\ref{tab:benchmark}. The values of $\beta_c$ and $s_c$ concern the configuration model and are the midpoints of the certified intervals of Table~\ref{tab:certified}, and those of $\beta_{\mathrm H}$ and $s_H$ solve $H_2(\alpha_H)=1-k/D$. The classical values were obtained on Gallager's ensemble. The last column gives the largest classical value in Table~\ref{tab:benchmark}, which comes from simulated annealing (SA) or Turbo Prange (TP). For $k\ge5$, $s_c$ and $s_H$ differ by less than $10^{-3}$, but $\beta_c<\beta_{\mathrm H}$ on every pair (Table~\ref{tab:certified}).}
\label{tab:condensation}
\end{table}

\subsection{Efficient decoders between the two thresholds}\label{sec:window}

For binary codes from the configuration model with $3\le k<D$, the PGM succeeds throughout $(\beta_d,\beta_c)$, whereas reliable local decoding is impossible there. With each efficient nonlocal decoder of Section~\ref{sec:known_nonlocal}, DQI or Regev's reduction reaches at most the values of a nonlocal classical algorithm that also uses Gaussian elimination. The existence of the PGM therefore identifies the information-theoretic range still available, rather than an efficient algorithm outperforming these classical methods.

DQI uses its decoder coherently and then in reverse (Section~\ref{sec:dqi}). An efficient coherent PGM with access to its inverse would consequently prepare the primal Gibbs states at the same inverse temperature for every right-hand side (Corollary~\ref{cor:pgm_equiv}). In particular, it would sample from the Gibbs measures in $(\beta_d,\beta_c)$, where the cavity method predicts that local Markov chains fail to equilibrate in polynomial time~\cite{montanari2006rigorous,krzakala2011melting}.

The existing numerical tests support these difficulties without proving a computational obstruction. For $(3,6)$ at $\beta=1.85$, BPQM decodes only about $7\%$ of the symbols with error below $10^{-3}$, whereas Gaussian elimination requires an information set containing half of the symbols (Appendix~\ref{sec:bpqm_elim}). In the fixed-temperature simulations at $1.03\beta_d$ to $1.1\beta_d$, Glauber dynamics did not relax within the simulated times (Appendix~\ref{app:dynamics}).

An efficient coherent implementation of the optimal decoder would also have consequences for planted instances. In a planted instance, the right-hand sides are generated from a uniformly random assignment $z^\star$ by violating each equation independently with some probability $\eta$. The output guarantee of Theorem~\ref{thm:guarantee} holds on average over a uniformly random offset, so it does not apply to planted right-hand sides when their law is far from uniform. The heralded preparation of Proposition~\ref{prop:pgm_gibbs}, which uses a coherent implementation of the PGM and its inverse, succeeds instead with probability $P_{\mathrm{PGM}}(B,\beta)$ for every specified offset (Section~\ref{sec:pgm}), so it prepares the Gibbs state at $\beta$ of a planted instance with the same probability as for any other right-hand sides.

At the violation probability $\eta=p_\beta(1)$, the Gibbs measure at $\beta$ is the posterior law of $z^\star$ given the instance, but below $\beta_c$, where $p_\beta(1)>\alpha_c$, a planted instance with this violation probability cannot be distinguished from an instance with uniformly random right-hand sides. Indeed, by Lemma~\ref{lem:pgm_success}, $P_{\mathrm{PGM}}(B,\beta)$ is the square of the Bhattacharyya coefficient between $W_{B,\alpha}$ and $U_B$ with $\alpha=p_\beta(1)$, which are the laws of the coset of the right-hand sides of a planted instance with violation probability $\alpha$ and of the coset of uniformly random right-hand sides. Given its coset, each of the two vectors of right-hand sides is uniformly distributed on it (Section~\ref{sec:condensation}), so $P_{\mathrm{PGM}}(B,\beta)$ is also the square of the Bhattacharyya coefficient between the laws of the two vectors, and it tends to one exactly when the total variation distance between these laws tends to zero. Theorem~\ref{thm:condensation}, applied at the inverse temperature $\ln((1-\eta)/\eta)$, at which $p(1)=\eta$, therefore shows that a planted instance with violation probability $\eta$ can be distinguished from an instance with uniformly random right-hand sides with probability tending to one if $\eta<\alpha_c$, and cannot be distinguished if $\eta>\alpha_c$.

We therefore consider a planted instance with $\eta<\alpha_c$, and run the PGM at a different inverse temperature $\beta$ slightly below $\beta_c$, where it succeeds. Thus $p_\beta(1)>\alpha_c>\eta$. A first-moment comparison motivates the prediction that this Gibbs measure is correlated with the planted assignment, even though its temperature is not the posterior temperature of that assignment.

The planted assignment violates about $\eta m$ equations, so its weight in the partition function is about $e^{-\beta\eta m}$. For comparison, with uniformly random right-hand sides, $Z=\sum_ze^{-\beta H(z)}$ has expectation $\E Z=2^n((1+e^{-\beta})/2)^m$. This identity holds because a fixed assignment violates the independent random right-hand sides with probability $1/2$ each. Since $m/n=D/k$, the planted assignment's weight is exponentially larger than this first moment when
$$
  \frac Dk\Big(\ln\frac2{1+e^{-\beta}}-\beta\eta\Big)>\ln2 .
$$
For $(k,D)=(3,6)$ and $\beta=1.9$, for which $p_\beta(1)\approx0.130$, this inequality holds for $\eta<0.109$, whereas $\alpha_c\approx0.120$. We expect, on the basis of the cavity method, that the Gibbs measure at $\beta$ then concentrates near the planted assignment (and, when $k$ is even, near its complement, which has the same violation pattern), and that, as $\beta$ increases to $\beta_c$, the same holds for $\eta$ up to close to $\alpha_c$. A proof would require control of the assignments at intermediate distance from the planted one, which the first-moment comparison does not provide.

If this prediction holds, an efficient coherent implementation of the PGM and its inverse at inverse temperatures close to $\beta_c$ would give an efficient quantum algorithm for weak recovery, that is, for finding an assignment that agrees with the planted one on a fraction of the variables bounded away from $\tfrac12$, in $(k,D)$-regular noisy XOR systems from the configuration model at violation probabilities up to close to $\alpha_c$, the threshold for distinguishing planted instances from instances with uniformly random right-hand sides. The algorithm would apply the procedure of Proposition~\ref{prop:pgm_gibbs}, which succeeds with probability $P_{\mathrm{PGM}}(B,\beta)=1-o(1)$ by Theorem~\ref{thm:condensation}, to the right-hand sides of the planted instance, and then measure the resulting state to obtain a sample from the Gibbs measure (Section~\ref{sec:primal}); by the prediction, this sample agrees with the planted assignment on such a fraction of the variables.

The known results on the hardness of this task concern uniformly random equations rather than regular instances. Consider, for $k\ge3$, linearly many uniformly random equations with $k$ variables each and a constant violation probability at which, given unbounded computation, a planted instance can be distinguished from a uniformly random one and its planted assignment can be weakly recovered. No polynomial-time classical or quantum algorithm is known for either task in this setting. Known polynomial-time algorithms require on the order of $n^{k/2}$ equations in $n$ variables, up to logarithmic factors~\cite{raghavendra2017strongly,schmidhuber2025quartic}, and lower bounds of the same order hold for sum-of-squares relaxations and for statistical-query algorithms~\cite{kothari2017sum,feldman2018complexity}. The quantum algorithm of Schmidhuber et al.\ for even $k$ is nearly quartically faster than the classical spectral algorithm based on Kikuchi matrices, but it requires the same number of equations~\cite{schmidhuber2025quartic}. The conjecture that planted noisy XOR systems are hard for polynomial-time algorithms is known as the sparse learning parity with noise (LPN) assumption~\cite{alekhnovich2011more,applebaum2010public}. The consequence of our prediction thus concerns the analogous problem for $(k,D)$-regular instances from the configuration model, for which no polynomial-time algorithm is known either.

\subsection{Decoders that measure small sets of qudits}\label{sec:small_sets}

For every efficient decoder that we have analyzed, a classical algorithm reaches the same inverse temperatures or satisfied fractions as DQI with that decoder (for BPQM, under Conjecture~\ref{conj:classical}). In each case the decoder produces a classical record whose law can be generated by a classical procedure on the primal problem: the labels of BPQM, the erasure pattern of unambiguous discrimination, and the erased blocks of FGUM. This suggests the following conjecture.

\begin{conjecture}[Measurements on small sets]\label{conj:small_sets}
Fix $(k,D,q,f)$ and $\beta\ge0$. Suppose that a decoder for the decoding problem with the product amplitude function $\Phi_\beta$ applies measurements that each act on $O(1)$ qudits and then processes all measurement outcomes classically, and that it has block success $1-o(1)$ on random $(k,D)$-regular instances from the configuration model or from Gallager's ensemble. Then, with probability $1-o(1)$ over the instance, a classical algorithm that runs in polynomial time outputs a sample from a law within $o(1)$ of $\mu_\beta$ in total variation.
\end{conjecture}

If, in addition, the estimate of the decoder at a typical symbol $i$ is the outcome of a measurement on the qudits of $N_t(i)$ with $t\le c_0\log m$, as in Definition~\ref{def:local}, then Conjecture~\ref{conj:small_sets} follows from Theorem~\ref{thm:only_nonrecon}, Corollary~\ref{cor:typical_nogo}, and Conjecture~\ref{conj:classical}. Local classical post-processing does not imply that the estimate at $i$ is such a measurement outcome, because the measurements made for other symbols may disturb the state on which the estimate at $i$ is based.

\section{Discussion}\label{sec:discussion}

Local and optimal decoding in DQI are characterized by two transitions of the primal Gibbs measure. For fixed cyclic alphabets and costs, reliable local decoding requires non-reconstruction; for $k\ge3$, two-stage BPQM attains every finite inverse temperature without reconstruction on both regular ensembles. The field counterpart is Corollary~\ref{cor:field_sharp}. For costs that are integer-valued up to an affine rescaling, optimizing a symmetric amplitude function does not improve the Gibbs expected-cost bound at $\beta_d$, and product amplitudes approach it.

For binary Max-XORSAT on the configuration model with $3\le k<D$, optimal decoding succeeds below $\beta_c$ and fails above it. This gives the tight symmetric-amplitude ceiling $s_c$ and the strict refinement $\beta_c<\beta_{\mathrm H}$ of the Holevo limit. Under Conjecture~\ref{conj:classical}, an advantage in sampling or approximate optimization with symmetric amplitudes therefore requires efficient nonlocal decoding between $\beta_d$ and $\beta_c$, together with an improvement over nonlocal classical algorithms. We compare with related work in Section~\ref{sec:related}, state the scope of these conclusions in Section~\ref{sec:scope}, and retain three central open problems in Section~\ref{sec:open}.

\subsection{Relation to other work}\label{sec:related}

M\'ezard and Montanari~\cite{mezard2006reconstruction} related tree reconstruction to the dynamical transition of the cavity method. By the binary case of Theorem~\ref{thm:duality}, optimal decoding on trees succeeds exactly when the primal Gibbs measure has no reconstruction. This case is closely related to Renes' duality between channels and codes~\cite{renes2018duality}. Renes shows that the dual of a binary symmetric channel is a pure-state channel and that the convolutions at symbol and check nodes are exchanged under this duality, and one proves the binary case by applying these facts node by node on a tree. Renes does not state the consequence for trees or thresholds, and our proof covers every $q$ and every cost. Rengaswamy and Pfister~\cite{rengaswamy2021duality} relate optimal decoding on the pure-state channel to the binary symmetric channel through this duality, and their formula for the optimal block success is the starting point of Section~\ref{sec:condensation}.

BPQM was introduced by Renes~\cite{renes2017bpqm} and developed further in~\cite{rengaswamy2021bpqm,piveteau2022messagepassing,brandsen2022bpqm}. For $q=2$, BPQM is optimal on trees~\cite{renes2017bpqm,rengaswamy2021bpqm,piveteau2022messagepassing}, and Piveteau and Renes~\cite{piveteau2022messagepassing} extended it to factor graphs with cycles by approximate cloning. Mandal and Pfister extended BPQM to prime alphabets and to finite abelian groups~\cite{mandal2026qary,mandal2026abelian}, with the node operations that we use in Section~\ref{sec:tree}, and Mandal et al.~\cite{mandal2026blockerror} show that it is optimal on trees for prime $q$ and analyze the two-stage decoder that we follow in Proposition~\ref{prop:sharp}. Piveteau, Quek, and Franco Garrido~\cite{piveteau2026gibbs} estimated the threshold of BPQM by density evolution and found that it exceeds the Shannon limit of classical decoders for certain dilute spin glasses. Their reported thresholds agree with our reconstruction thresholds (Table~\ref{tab:pqf}).

Zhou et al.~\cite{zhou2026gibbs} apply DQI to Ising Hamiltonians with random signs, which are equivalent to random Max-XORSAT. On random instances whose constraints form an Erd\H{o}s--R\'enyi hypergraph with large clause size, they show that DQI with unambiguous discrimination followed by Gaussian elimination samples in the shattered phase, as does a classical sampler based on Prange's algorithm (Section~\ref{sec:ud_prange}). Zhou et al., like Piveteau, Quek, and Franco Garrido, use the limit given by the Holevo bound, which Jordan et al.\ applied to DQI, as the information-theoretic limit of quantum decoders. On Erd\H{o}s--R\'enyi hypergraphs, they show that this limit lies above the replica-symmetry-breaking transition, which they define through the condensation threshold of random factor graphs, and they ask whether DQI can sample beyond this transition. For random regular instances from the configuration model and decoders whose block success tends to one, Theorem~\ref{thm:condensation} shows that DQI cannot sample beyond this transition: no decoder succeeds above the condensation threshold, which is strictly smaller than the limit given by the Holevo bound. Our results on local decoders concern a class of decoders complementary to the one that Zhou et al.\ analyze.

Shutty et al.~\cite{shutty2026locallyquantumdecoders} use FGUM in Regev's reduction, prove the error bounds that we use, and give the classical algorithm Turbo Prange (Section~\ref{sec:fgum}); we use their fifteen families of random regular Max-XORSAT instances as benchmarks. Chailloux and Tillich~\cite{chailloux2024quantumdecoding} formulated the quantum decoding problem and proposed unambiguous discrimination for it. They also showed that, for uniformly random linear codes, the pretty-good measurement attains the limit given by the Holevo bound~\cite[Theorem~5]{chailloux2024quantumdecoding}. Theorem~\ref{thm:condensation} is the analogue for the sparse codes induced by DQI, on which the pretty-good measurement succeeds exactly below the condensation threshold $\beta_c<\beta_{\mathrm H}$. In Section~6.3 of the full version of their paper, Chailloux and Tillich analyze Regev's reduction with the pretty-good measurement, which is the construction of Proposition~\ref{prop:pgm_gibbs} for the zero offset. Blanvillain, Chailloux, and Tillich~\cite{blanvillain2026tight} give tight achievability bounds for the quantum decoding problem over finite fields, which extend both directions of the result of Chailloux and Tillich to every product amplitude function~\cite[Theorems~1 and~2]{blanvillain2026tight}, together with the formula for the block success of the pretty-good measurement and its coherent use in Regev's reduction for a specified coset. Chailloux and Hermouet~\cite{chailloux2026equivalence} prove the reduction from quantum decoding to the preparation of coset states over $\Z_q$ for every modulus, and the converse construction of a decoder from a clean controlled preparation of these states. Piveteau and Renes~\cite{piveteau2025efficient} build optimal measurements from the inverse of a coherent preparation of coset states, as in Proposition~\ref{prop:gibbs_pgm}. Buzet and Chailloux~\cite{buzet2025fine} introduced fine-grained unambiguous measurements, and Franco Garrido and Chailloux~\cite{garrido2026regev} note that Regev's reduction can sometimes succeed when the decoding problem cannot be solved reliably.

In classical coding, optimal decoding of regular low-density parity-check codes is predicted to succeed up to the static transition of the corresponding spin glass, and iterative decoding up to the dynamical transition~\cite{franz2002dynamic}. For DQI, the analogous limits are the reconstruction threshold of the primal Gibbs measure for local decoders, which the cavity method identifies with the dynamical transition, and the condensation threshold for every decoder, which is the static transition.

Anschuetz, Gamarnik, and Lu~\cite{anschuetz2025decoded} bound the satisfied fraction that DQI reaches on random Max-XORSAT through the overlap gap property, for quantum algorithms that are stable under small changes of the right-hand sides. They prove that DQI with the polynomial amplitude functions of Jordan et al.\ is stable if the degree of the polynomial is at most $m/D$ and the decoder corrects every error of weight up to this degree, or if the decoder satisfies a Lipschitz condition, which holds for belief propagation with a fixed number of rounds. Under the first condition, on Gallager's ensemble and for large $k$ at fixed $D/k$, they obtain the bound $\tfrac12+\Theta(\sqrt{\ln k/D})$ on the satisfied fraction that DQI reaches with constant probability. Their results and ours are complementary, because the hypotheses restrict different parts of DQI. Stability bounds the dependence of the output on the instance, whereas locality bounds the dependence of one estimate on distant qudits; the overlap gap property concerns the near-optimal assignments, whereas reconstruction concerns the Gibbs measure. Their bounds hold with constant probability and are asymptotic in the degrees, whereas ours bound the expectation and are computed at fixed degrees. Prange's algorithm, which is neither stable nor local, reaches the asymptotic satisfied fraction $\tfrac12+k/(2D)$ and exceeds both bounds at large degrees.

In a setting with oracle access, Kramer et al.~\cite{kramer2026provable} prove a separation in query complexity between DQI with a nonlocal decoder and classical algorithms. They consider a folded version of optimal polynomial intersection, in which each constraint asks that a block of values of a polynomial over a finite field lie in a set accessed through a membership oracle. When these sets are independent and uniformly random among the subsets that contain half of the values of their blocks, DQI with Reed--Solomon unique decoding of the dual code reaches a larger expected satisfied fraction than Prange's algorithm at every fixed rational code rate in $(0,1)$, whereas every classical algorithm requires superpolynomially many queries to exceed the satisfied fraction of Prange's algorithm by a fixed amount with constant probability. Gil-Fuster et al.~\cite{gilfuster2026approximate} compare DQI numerically with block Gibbs sampling on the sparse instances of Jordan et al.\ (Section~\ref{sec:approx_bench}).

\subsection{Scope and limitations}\label{sec:scope}

\paragraph{Costs and alphabets.} The symmetric-amplitude approximation theorem fixes the alphabet and the cost, and uses zero-one matrices. It covers costs that are integer-valued up to an affine rescaling over cyclic alphabets and fields for $k\ge3$ and $v=o(\sqrt m)$. The cyclic two-valued theorem permits the stronger condition $v=o(m)$ without that new degree restriction. Arbitrary real costs, heterogeneous weights or costs, growing alphabets and other coefficient ensembles are not covered by Theorem~\ref{thm:common_cost_local} or its field corollary. Product amplitude functions retain their arbitrary-fixed-cost guarantees.

\paragraph{Locality on finite graphs.} Optimal decoding on a tree is characterized at every depth, including the infinite-depth limit. The logarithmic-radius restriction instead guarantees LT and TP around a typical symbol of a finite random instance. A deeper BPQM computation on a graph with cycles can involve correlations absent from that tree input, and its success above $\beta_d$ is not determined by the local theorem. A retained estimate from an LT/TP neighborhood still bounds block recovery regardless of later depth. The binary configuration-model condensation bounds remain independent of locality.

\paragraph{Exact optimization.} The conditional quadratic comparison holds for every complex polynomial target state in a fixed subcritical-degree range. Reliable local decoding bounds the degree only for mean-optimal polynomials; it does not exclude other reliably decoded polynomials of larger degree whose mean objective stays below $s_d$. At the endpoint, the result gives one algorithm for each fixed exponential slack, not uniform polynomial overhead or a single subexponential-overhead algorithm.

The comparison does not extend to an arbitrary imperfectly prepared state merely because its decoding failure tends to zero. An additive error can dominate an exponentially rare optimum. Exact decoding, or a suitable bound relative to the target optimum probability, suffices (Section~\ref{sec:prepared}). The block-amplitude and constant-failure settings below are separate from this prepared-state issue.

\paragraph{Block amplitude functions.} To our knowledge, the block amplitude function of Shutty et al.\ is the only amplitude function that is not symmetric for which an improvement over symmetric ones has been reported on random $(k,D)$-regular instances. By Proposition~\ref{prop:block_four}, on four degree pairs, DQI with the block amplitude function and any decoder whose block success tends to one reaches a smaller satisfied fraction than DQI with a product amplitude function and a local decoder. On the other degree pairs, the bound $s_d$ of Theorem~\ref{thm:approx_sym} is not known to hold for the block amplitude function with local decoders. We conjecture that it holds, under the hypotheses of the analysis on trees of blocks in Appendix~\ref{app:block}.

\begin{conjecture}[Block amplitude functions with local decoders]\label{conj:block}
Fix $3\le k<D$, and let $(B_m)$ be a sequence of matrices from Gallager's ensemble with $m\to\infty$ rows. For each $m$, consider DQI for Max-XORSAT with a uniformly random offset and the block amplitude function of Section~\ref{sec:approx_block} of some parameter $\alpha_m\in(0,1)$, together with a decoder whose block success tends to one. Suppose that the decoder makes a $t_m$-local estimate at a symbol $i_m$ at which $\mathrm{LT}_{t_m+1}(i_m)$ and $\mathrm{TP}_{t_m+1}(i_m)$ hold. Then the expected satisfied fraction that DQI reaches, averaged over the offset, is at most $s_d+o(1)$.
\end{conjecture}

For random instances from Gallager's ensemble, Theorem~\ref{thm:tp_typical} at depth $t_m+1$ supplies the hypotheses on $\mathrm{LT}$ and $\mathrm{TP}$ at a typical symbol when $t_m\le c_0\log m$, but not necessarily at a symbol that the decoder selects after inspecting the instance. If Conjectures~\ref{conj:block} and~\ref{conj:classical} hold, then DQI with the block amplitude function and a decoder satisfying the hypotheses of Conjecture~\ref{conj:block} gives no advantage in approximate optimization, because classical sampling from the Gibbs measure reaches satisfied fractions arbitrarily close to $s_d$ (Lemma~\ref{lem:gibbs_cost}).

Proposition~\ref{prop:block_four} establishes the conclusion of Conjecture~\ref{conj:block} for four degree pairs, even without restricting the decoder to be local. On all fifteen benchmark pairs, numerical evaluation of the optimal block-tree recursion is also consistent with the conjecture; the estimated root errors remain appreciably positive across the populations, depths, and independent random seeds tested in Appendix~\ref{app:block}. The remaining step is to show that the optimal root error stays positive in the infinite-depth limit at the block parameter corresponding to $s_d$. Theorem~\ref{thm:duality} does not establish this, because the channel noise is not a product over symbols. Regev's reduction with FGUM followed by Gaussian elimination provides further context: for that particular nonlocal decoder, the block amplitude function reaches exactly the values of Turbo Prange (Section~\ref{sec:fgum}). This equality is not an upper bound on BPQM or on other local quantum measurements.

Amplitude functions that are neither symmetric nor of this block form are not covered by our results. Some restriction on the amplitude function is necessary, however, because the decoding problem can be made trivial by an amplitude function whose state is as hard to prepare as the target states. By Lemma~\ref{lem:pgm_success}, the optimal decoder succeeds with probability one whenever all cosets of $K$ have the same mass under $|\Phi|^2$. For every finite $\beta$, this holds for $\Phi'(x)=\Phi_\beta(x)/\sqrt{|C|\,W_{\Phi_\beta}(x+K)}$, where $W_{\Phi_\beta}(X)=\sum_{x\in X}|\Phi_\beta(x)|^2$. Since $\Phi'$ is proportional to $\Phi_\beta$ on each coset, it has the same target states, so the optimal decoder for $\Phi'$ succeeds even above $\beta_c$, where no decoder succeeds for $\Phi_\beta$ (Theorem~\ref{thm:condensation}). However, $\ket{\Phi'}$ is the uniform superposition of the target states of all cosets of $K$, so a preparation of $\ket{\Phi'}$, followed by a measurement of the coset of $K$ that contains $x$, already yields the target state for a uniformly random offset, without any decoding. A bound that holds for every amplitude function must therefore account for the cost of preparing $\ket\Phi$.

\paragraph{Decoders with constant failure probability.} The reliable-decoding assumption is block success tending to one, not exact decoding at finite size. A decoder whose success stays bounded away from both zero and one can still be used with postselection, at $1/(1-\varepsilon)$ expected attempts in Regev's reduction. Such constant-failure decoders require a separate analysis.

For binary codes, a local estimate at just one symbol does not exclude this possibility. The construction in Section~\ref{sec:strategy} turns a codeword-valued decoder into one with a $0$-local estimate at one symbol, halving its success without changing the output law of Regev's reduction. When $D$ is even, the all-ones codeword also permits a variant with uniform estimates at all but one symbol. Definition~\ref{def:local_decoder} therefore requires the joint law of separate local measurements at $\omega(\log m)$ symbols. Applied independently at so many symbols, the artificial randomization reduces block success below every inverse polynomial.

To exclude efficient postselection for this stronger locality notion above $\beta_d$, one would like joint projection onto the product of the selected tree codes. Theorem~\ref{thm:locality} would then bound the joint correctness probability by a product of tree success probabilities, which is smaller than every inverse polynomial for $\omega(\log m)$ symbols. This joint TP statement is not proved and can fail when an outside check joins the balls; Theorem~\ref{thm:tp_typical} proves TP at a typical symbol one at a time. For binary product amplitudes on the configuration model with $3\le k<D$, exponentially small block success above $\beta_c$ already rules out efficient postselection there, independently of locality (Theorem~\ref{thm:condensation}). No analogous statement is known for the block amplitude function.

\subsection{Open problems}\label{sec:open}

We single out three problems that our results leave open.

\begin{enumerate}[leftmargin=*]
\item \emph{Nonlocal decoders between the two thresholds.} Is there an efficient nonlocal decoder that succeeds between $\beta_d$ and $\beta_c$ and with which DQI exceeds the best classical values? By Corollary~\ref{cor:pgm_equiv}, an efficient coherent implementation of the optimal decoder in this range is equivalent to an efficient coherent preparation of the primal Gibbs states in the phase in which local Markov chains are predicted to fail. Section~\ref{sec:window} discusses the consequences that such an implementation would have for planted noisy XOR systems if a prediction of the cavity method holds, and Table~\ref{tab:rates} lists the error rates at which a decoder must succeed to match the best classical values on the benchmark families.
\item \emph{Gallager's ensemble and the condensation threshold itself.} Do the results on the condensation threshold hold for Gallager's ensemble, and does the optimal decoder succeed at $\beta=\beta_c$? Theorem~\ref{thm:condensation} concerns the configuration model and $\beta\ne\beta_c$, and part~3 of Proposition~\ref{prop:entropy}, which determines $\beta_c$, rests on a theorem about the configuration model. The benchmark families are drawn from Gallager's ensemble, so a positive answer to the first question would place the bound $s_c$ on the same ensemble as the classical values with which we compare it.
\item \emph{Classical sampling up to the reconstruction threshold.} For random $(k,D)$-regular instances with $k\ge3$, is there a polynomial-time classical algorithm that samples from the Gibbs measure at every fixed inverse temperature below $\beta_d$? A positive answer would prove Conjecture~\ref{conj:classical} for $k\ge3$ and $\beta<\beta_d$, making the corresponding classical comparisons in Sections~\ref{sec:conditional} and~\ref{sec:poly_exact} unconditional below $\beta_d$. Rigorous mixing results cover only part of $[0,\beta_d)$ (Section~\ref{sec:status}). Sampling at $\beta_d$ itself is a separate question, which approximate optimization arbitrarily close to $s_d$ does not require.
\end{enumerate}

\section{AI methodology statement}
Large language models (LLMs) were used extensively to develop these results. The correspondence between the BPQM threshold and tree reconstruction was first pointed out by GPT-6 Sol. We then used GPT-6.1 Sol and Claude Opus 5.5 to flesh out the details of our research program, including the extension to local decoders, various applications of DQI, different amplitude functions and ensembles of instances, the characterization of optimal decoding by the condensation threshold, and the numerical simulations. They have also been used to draft and modify text and to provide an additional automated check on all cited references. We have tried to ensure that the presentation is readable and useful for experts in the field. We have checked all AI-generated proofs and simulations, and take full responsibility for the correctness of all results in the paper. All cited references have also been manually checked by the authors.

\section*{Disclaimer}
This paper was prepared for informational purposes by the Global Technology Applied Research center of JPMorgan Chase \& Co. 
This paper is not a product of the Research Department of JPMorgan Chase \& Co. or its affiliates. Neither JPMorgan Chase \& Co. nor any of its affiliates makes any explicit or implied representation or warranty and none of them accept any liability in connection with this paper, including, without limitation, with respect to the completeness, accuracy, or reliability of the information contained herein and the potential legal, compliance, tax, or accounting effects thereof. 
This document is not intended as investment research or investment advice, or as a recommendation, offer, or solicitation for the purchase or sale of any security, financial instrument, financial product or service, or to be used in any way for evaluating the merits of participating in any transaction.

\section*{Acknowledgments}
We thank Stephen Jordan for informing us of the concurrent result~\cite{JNZ26}. We thank Rob Otter for executive support of the project and invaluable feedback on this project and our colleagues at the Global Technology Applied Research center of JPMorganChase for their support.

\appendix
\addtocontents{toc}{\protect\setcounter{tocdepth}{1}}

\section{Numerical thresholds and simulations}\label{sec:numerics}

Our theorems identify the local-decoding threshold with reconstruction, while this appendix evaluates the thresholds and compares them with classical simulations. We first report values of $\beta_d$ for binary codes, including the fixed-ratio families of Figure~\ref{fig:degrees}, compare with the reported BPQM thresholds, and turn to larger alphabets. We also compare condensation estimates with the certified intervals and examine BPQM followed by elimination above $\beta_d$.

Except for the Holevo bound, the first two rows of Table~\ref{tab:landscape}, and the certified intervals in Appendix~\ref{app:certified}, the thresholds reported here are population-dynamics estimates. Appendix~\ref{app:dynamics} gives the extended classical study: fixed-temperature dynamics on $62$ binary degree pairs, annealing on the same pairs, and $105$ additional cyclic-alphabet and finite-field tuples. The benchmark relaxation table and the headline annealing summary are in the main text (Tables~\ref{tab:dynamics} and~\ref{tab:sim_summary}); detailed values, protocols, runtime and scope are given below.

\paragraph{Binary codes.} Table~\ref{tab:landscape} compares decoders for the degree pairs $(3,6)$ and $(6,12)$, both of rate $1/2$. By Theorem~\ref{thm:duality}, the tree threshold of BPQM equals the reconstruction threshold, and the two recursions agree numerically: with population $2\times10^5$ and $400$ levels, the spectral recursion gives $1.8384$ for $(3,6)$ and $1.5486$ for $(6,12)$, and the primal recursion gives $1.8384$ and $1.5491$. With $1500$ levels, the threshold of BPQM is $1.8388$ for $(3,6)$ and $1.5489$ for $(6,12)$ (Table~\ref{tab:degrees}), and Table~\ref{tab:landscape} and the main text use these values. The $(3,6)$ values match the cavity method: Rizzo~\cite{rizzo2022stochastic} reports the dynamical temperature $T_d=1.087815$ for the $3$-spin model with couplings $\pm1$ in which each spin lies in $6$ interactions. This model also uses $\pm1$ spins, so by Remark~\ref{rem:units} the value corresponds to $\beta_d=2/T_d=1.838548$ in our units, which lies between our two values for $(3,6)$.

The decoders in the first four rows apply a measurement to each qubit separately and then run classical belief propagation (BP) or Gaussian elimination. The error/erasure measurement has the outcomes $0$, $1$, and erasure, and it has the smallest error probability among such measurements with a given erasure probability~\cite{eldar2003inconclusive}. It interpolates between the Helstrom measurement, which never erases, and unambiguous discrimination, which never errs. Unambiguous discrimination erases each symbol with probability $\sigma_\beta$ and reveals the other symbols without error, so BP after it is described by the density evolution of erasures, $x\mapsto\sigma_\beta\big(1-(1-x)^{D-1}\big)^{k-1}$. We compute its threshold from this scalar recursion; the erasure probability at the threshold is $0.42944$ for $(3,6)$ and $0.30746$ for $(6,12)$. For the error/erasure measurement, with erasure probability $\sigma_\beta/2$, and for the Helstrom measurement, we estimated the threshold of BP by population dynamics of log-likelihood ratios with $2\times10^4$ messages and at most $200$ levels. A run succeeded when the fraction of edge messages with the wrong sign, with ties counted as one half, fell below $10^{-4}$, and we bisected on $\beta\in[0.05,2.5]$ to a width of $6\times10^{-4}$. The row of Gaussian elimination is explained in Section~\ref{sec:ud_prange} and derived in Appendix~\ref{app:rank}.
\begin{table}[H]
\centering
\begin{tabular}{@{}lcc@{}}
\toprule
Decoder & $(3,6)$ & $(6,12)$ \\
\midrule
Unambiguous discrimination + BP & 0.918 & 0.635 \\
Unambiguous discrimination + Gaussian elimination & $[0.867,1.0757]$ & $[0.878,1.0983]$ \\
Error/erasure measurement + BP & 1.203 & 0.945 \\
Helstrom measurement + BP & 1.251 & 1.042 \\
Tree BPQM $=\beta_d$ & 1.839 & 1.549 \\
Holevo bound $\beta_{\mathrm H}$ (upper bound for every decoder) & 2.090 & 2.090 \\
\bottomrule
\end{tabular}
\caption{Largest $\beta$ at which each decoder succeeds, for the two degree pairs of rate $1/2$. The first row comes from the density evolution of erasures, the third and fourth rows are estimates by population dynamics, and the fifth row is $\beta_d$. The second row gives an interval, with its ends rounded outward, that contains the largest $\beta$ at which this decoder succeeds: the lower end follows from Theorem~\ref{thm:union} for the configuration model, and the upper end from Lemma~\ref{lem:rank}. The last row is an upper bound for every decoder; for $(3,6)$, the largest $\beta$ at which some decoder succeeds on the configuration model is the condensation threshold $\beta_c\approx1.99$ (Table~\ref{tab:condensation}).}
\label{tab:landscape}
\end{table}

\paragraph{Larger degrees at fixed ratios.} Table~\ref{tab:degrees} lists the values of $\beta_d$ shown in Figure~\ref{fig:degrees}, for binary codes at the three ratios $k/D=\tfrac12,\tfrac23,\tfrac34$. Each value is the threshold of the spectral recursion of Definition~\ref{def:spectral} for $q=2$, which equals $\beta_d$ by Theorem~\ref{thm:duality}. We located it by bisection on $\tanh(\beta/2)\in[0,1]$ with $14$ steps, using population dynamics for the overlaps of the messages with population $2\times10^5$, and we counted a run as convergent when the mean overlap fell below $10^{-6}$ within $1500$ levels. The table reports the midpoint of the final bisection interval, whose width in $\beta$ is at most $5\times10^{-4}$. For $(3,6)$ and $(6,12)$, these values exceed the values $1.8384$ and $1.5486$ obtained above with $400$ levels by $4\times10^{-4}$ and $3\times10^{-4}$, and the value for $(3,6)$ exceeds the value $1.838548$ obtained from the dynamical temperature of Rizzo by about $2.5\times10^{-4}$. The values of $s_d$ in Table~\ref{tab:benchmark}, of $\beta_d$ in Table~\ref{tab:condensation}, and of $\beta_d/2$ in Table~\ref{tab:pqf} are computed in the same way.
\begin{table}[H]
\centering
\small
\setlength{\tabcolsep}{4pt}
\begin{tabular}{@{}l*{12}{c}@{}}
\toprule
 & \multicolumn{12}{c}{$k/D=\tfrac12$, $\beta_{\mathrm P}=1.0986$}\\
\cmidrule(l){2-13}
$k$ & $3$ & $4$ & $5$ & $6$ & $7$ & $8$ & $9$ & $10$ & $12$ & $14$ & $16$ & $18$\\
$\beta_d$ & 1.8388 & 1.7531 & 1.6448 & 1.5489 & 1.4674 & 1.3976 & 1.3380 & 1.2858 & 1.1992 & 1.1294 & 1.0717 & 1.0228\\
\midrule
 & \multicolumn{5}{c}{$k/D=\tfrac23$, $\beta_{\mathrm P}=1.6094$} & & \multicolumn{5}{c}{$k/D=\tfrac34$, $\beta_{\mathrm P}=1.9459$} & \\
\cmidrule(lr){2-6}\cmidrule(lr){8-12}
$k$ & $4$ & $6$ & $8$ & $10$ & $12$ & & $3$ & $6$ & $9$ & $12$ & $15$ & \\
$\beta_d$ & 2.1781 & 1.8727 & 1.6699 & 1.5256 & 1.4161 & & 2.6574 & 2.0358 & 1.7155 & 1.5196 & 1.3838 & \\
\bottomrule
\end{tabular}
\caption{Reconstruction thresholds $\beta_d$ of binary $(k,D)$-regular instances at three fixed ratios $k/D$, which Figure~\ref{fig:degrees} shows, with $\beta_{\mathrm P}=2\operatorname{atanh}(k/D)$ (Section~\ref{sec:larger_degrees}) at each ratio. Each entry is the midpoint of a bisection interval of width at most $5\times10^{-4}$, computed with population $2\times10^5$ and $1500$ levels.}
\label{tab:degrees}
\end{table}

\paragraph{Comparison with reported BPQM thresholds.} Table~\ref{tab:pqf} compares the BPQM thresholds that Piveteau, Quek, and Franco Garrido~\cite{piveteau2026gibbs} report for the ten degree pairs of their Table~1 with our reconstruction thresholds, computed with population $2\times10^5$ and $1500$ levels as for Table~\ref{tab:benchmark}. The two agree to within $0.002$ for every pair.
\begin{table}[H]
\centering
\begin{tabular}{@{}lcc@{}}
\toprule
$(k,D)$ & BPQM threshold reported in~\cite{piveteau2026gibbs} & our $\beta_d/2$\\
\midrule
$(3,4)$ & 1.327 & 1.329\\
$(3,5)$ & 1.068 & 1.069\\
$(3,6)$ & 0.920 & 0.919\\
$(3,7)$ & 0.818 & 0.819\\
$(4,5)$ & 1.274 & 1.275\\
$(4,6)$ & 1.088 & 1.089\\
$(4,7)$ & 0.965 & 0.966\\
$(5,6)$ & 1.196 & 1.196\\
$(5,7)$ & 1.058 & 1.058\\
$(6,7)$ & 1.125 & 1.125\\
\bottomrule
\end{tabular}
\caption{Thresholds in the units of~\cite{piveteau2026gibbs}, whose inverse temperature is half of ours (Remark~\ref{rem:units}).}
\label{tab:pqf}
\end{table}

\paragraph{Larger alphabets.} Table~\ref{tab:qary} reports thresholds over $\Z_q$ for the $(3,6)$ tree, for the Potts cost $f(z)=\mathbf 1[z\ne0]$ and the clock cost $f(z)=-\cos(2\pi z/q)$. For the Potts cost, the inverse temperatures without reconstruction form an interval (Section~\ref{sec:monotone}), so bisection locates $\beta_d$. For $q=3$ the clock cost is an affine function of the Potts cost, so the same holds. For the clock cost with $q\ge5$ we have not proved this, but for $q=5$ and $q=8$ the scans described in Section~\ref{sec:monotone} found a single change from success to failure. Above $\beta_d$, only nonlocal decoders can succeed. The window $(\beta_d,\beta_{\mathrm H})$ has width about $0.26$ for the Potts cost. For the clock cost, the window between the transition found by bisection and $\beta_{\mathrm H}$ widens as $q$ grows.
\begin{table}[H]
\centering
\begin{tabular}{@{}llrr@{}}
\toprule
$q$ & cost & $\beta_d$ & $\beta_{\mathrm H}$ \\
\midrule
2 & Potts & 1.834 & 2.090 \\
3 & Potts & 2.098 & 2.355 \\
5 & Potts & 2.452 & 2.712 \\
8 & Potts & 2.795 & 3.061 \\
3 & clock & 1.400 & 1.570 \\
5 & clock & 2.165 & 2.600 \\
8 & clock & 2.358 & 4.018 \\
\bottomrule
\end{tabular}
\caption{Thresholds for the $(3,6)$ tree over $\Z_q$, from the spectral recursion with population $10^4$, $120$ levels, bisection width $2\times10^{-3}$, and the success criterion of Appendix~\ref{app:numerics}. For $q=2$ the Potts cost coincides with $f(z)=z$. The value $1.834$ in the first row comes from this lower-precision run and differs from the value $1.8388$ obtained with population $2\times10^5$ and $1500$ levels by more than the bisection width, which we attribute to the smaller population and the smaller number of levels. For the clock cost with $q\ge5$, the reported $\beta_d$ is the point found by bisection, which equals the supremum of Definition~\ref{def:beta_d} if the inverse temperatures without reconstruction form an interval.}
\label{tab:qary}
\end{table}

\paragraph{Condensation thresholds.} The values of $\beta_c$ and $s_c$ in Table~\ref{tab:condensation} are the midpoints of the certified intervals of Appendix~\ref{app:certified}. As an independent check, we also estimated $\alpha_c$ by population dynamics, as the point at which the functional~\eqref{eq:functional}, evaluated at the fixed point of planted density evolution, changes sign. The limit $\bar F(\alpha)=\sup_\pi\mathcal B_\sigma(\pi)$ itself does not change sign: it is nonnegative, because $\mathcal B_\sigma(\delta_0)=0$, and by definition it is positive for $\alpha<\alpha_c$ and zero for $\alpha>\alpha_c$ (Section~\ref{sec:condensation}). For each pair and each value of $\alpha$, we evaluated $\mathcal B_\sigma(\pi)$ at the law $\pi$ of the magnetizations at the fixed point of planted density evolution. We started from a population of $2\times10^5$ magnetizations equal to one, which corresponds to complete information about the planted assignment, and applied $200$ sweeps of the update
$$
  {\mathfrak m}\leftarrow\tanh\Big(\sum_{i=1}^{D-1}\operatorname{atanh}\big(s_i\,\sigma\,{\mathfrak m}_{i,1}\cdots {\mathfrak m}_{i,k-1}\big)\Big),
$$
where the ${\mathfrak m}_{i,j}$ are drawn uniformly from the population, and $s_i=+1$ with probability $(1+\sigma)/2$ and $s_i=-1$ otherwise. We then gave the magnetizations independent uniformly random signs, so that $\pi$ has mean zero; the functional depends only on the law of $|{\mathfrak m}|$. We estimated $a$, $b$, and $\mathcal I_D$ from $4\times10^6$ samples each, using $\mathcal I_D=\E\ln(1+M)$, where $M=\tanh\big(\sum_{i=1}^DT_i\operatorname{atanh}R_i\big)$ is the posterior mean of $\xi$ given the observations when $\xi=+1$ and $T_i=+1$ with probability $(1+R_i)/2$. We located $\alpha_c$ by evaluating the functional on a grid of ten values of $\alpha$ in $[\alpha_H,\alpha_d]$ and on a finer grid of nine values around the change of sign. For three independent seeds, we then fitted a line to the values on a grid in a window of half-width $10^{-3}$ around the change of sign, clipped below at $\alpha_H$, and we report the mean of the three zeros of the fitted lines. The value at the planted fixed point is a lower bound on $\sup_\pi\mathcal B_\sigma(\pi)$, so, up to statistical error, $\bar F(\alpha)>0$ whenever this value is positive. Up to statistical error, the computed $\alpha_c$ is therefore a lower bound on the true value, and the computed $\beta_c$ and $s_c$ are upper bounds. The zeros obtained with the three seeds differ from their mean by at most $1.5\times10^{-4}$ in $\alpha$ for every pair, and by at most $1.2\times10^{-4}$ for the pairs with $k\ge4$. The resulting estimates of $\beta_c$ lie inside the certified intervals for nine pairs, including all pairs with $k=3$, and exceed their upper ends by at most $8\times10^{-4}$ for $(4,5)$, $(4,7)$, $(4,8)$, and $(5,8)$. For $(6,7)$ and $(7,8)$, where $\alpha_c$ lies within $2\times10^{-5}$ of $\alpha_H$, the change of sign could not be resolved by this method. For $(3,6)$, the value $\beta_c\approx1.9925$ agrees with an independent scan of the same functional with a population of $4\times10^5$ magnetizations, in which the sign changes between $\beta=1.990$ and $\beta=1.995$.

\paragraph{BPQM followed by elimination.}\label{sec:bpqm_elim} We tested numerically whether BPQM on the tree neighborhoods followed by elimination on the most reliable symbols can succeed above $\beta_d$. Above $\beta_d$, the error of BPQM on the tree neighborhood of a symbol, given its labels, is small only for a few percent of the symbols. For $(3,6)$ at $\beta=1.85$, we ran five independent runs of population dynamics for the binary spectral recursion, each with $10^5$ messages and $300$ levels, and computed the Helstrom error $\tfrac12\big(1-\sqrt{1-v^2}\big)$ for the overlap $v$ of the root message. In the five runs, the fraction of symbols with error below $10^{-3}$ lies between $0.066$ and $0.072$, the fraction with error below $10^{-6}$ between $0.003$ and $0.004$, and the mean error between $0.0708$ and $0.0722$. Gaussian elimination at rate $1/2$ requires an information set, which contains half of the symbols, so in this example the symbols that BPQM decodes with error below $10^{-3}$ are too few for elimination on them to recover the codeword.

\subsection{Numerical method}\label{app:numerics}

Population dynamics represents the law of a random message by a large population of independent samples. One level of a recursion replaces every member of the population by a new message, which is computed by the node rule from members drawn uniformly at random from the current population, with fresh randomness for the labels. After $T$ levels, the population approximates the law of the message at depth $T$.

We located thresholds by bisection on $\beta$, or on $\tanh(\beta/2)$ for $q=2$. For each value we iterated the relevant recursion with a population of independent messages and declared success when a statistic of the population fell below a tolerance. For the binary thresholds of BPQM (the row of BPQM in Table~\ref{tab:landscape}, and Tables~\ref{tab:degrees} and~\ref{tab:pqf}), the statistic is the population mean of the overlaps $|V_T|$ of the messages sent up from symbols (Section~\ref{sec:bpqm}), and the tolerance is $10^{-6}$. The BPQM value $1.8384$ was computed with $2\times10^5$ messages and $400$ levels, and Table~\ref{tab:degrees} with $2\times10^5$ messages, $1500$ levels, and $14$ bisection steps on $\tanh(\beta/2)$. For Table~\ref{tab:qary}, the statistic is the population mean of $1-S(V^{(T)})$ for the spectrum $V^{(T)}$ of Definition~\ref{def:spectral}, with $T=120$ levels and $10^4$ messages, the tolerance is $5\times10^{-3}$, and the bisection ends at a width of $2\times10^{-3}$. This statistic uses $V^{(T)}$ rather than the root spectrum $N^{(T)}$. The two have the same transition, because $N^{(T)}$ and $V^{(T)}$ are convolutions of $p$ with $k$ and $k-1$ independent copies of the same check spectrum, the Fourier coefficients of a convolution are the products of those of its factors, and a product of a fixed number of independent copies of a variable in $[0,1]$ tends to zero in probability if and only if one copy does. The primal values $1.8384$ for $(3,6)$ and $1.5491$ for $(6,12)$ come from population dynamics for the magnetizations of the messages that primal variables send to their parent clauses. These magnetizations follow the recursion of Zhou et al.~\cite[Eq.~(D.8)]{zhou2026gibbs} with the clause bias $\tanh(\beta/2)$, and we started from magnetization one at the leaves. We declared non-reconstruction at $\beta$ if the population mean of the magnetizations fell below $10^{-4}$ at some level or was at most $5\times10^{-3}$ after the last level, and we located the threshold by $16$ bisection steps on $\tanh(\beta/2)\in[0.05,0.99]$, with $2\times10^5$ messages and $400$ levels.

Two numerical artifacts appeared in the primal recursion for $q>2$. First, computing convolutions by the fast Fourier transform introduces round-off that can set the entry of a sharp message at the true value to zero, after which this entry stays zero. This produced a spurious reconstruction threshold above $\beta_{\mathrm H}$. Second, even with exact convolutions, finite populations for smooth costs drift collectively along $\Z_q$, which degrades statistics that depend on translation, such as $M(0)$, within a few levels. The spectral recursion of Definition~\ref{def:spectral} does not depend on translations and avoids both problems, so we use it to estimate $\beta_d$ for $q>2$.

\subsection{Certified bounds that separate the three thresholds}\label{app:certified}

We prove that the reconstruction, condensation, and Holevo thresholds are distinct on each of the fifteen benchmark pairs: for every pair, Table~\ref{tab:certified} gives an interval that contains $\beta_d$ and an interval that contains $\beta_c$, and these intervals satisfy $\beta_d<\beta_c<\beta_{\mathrm H}$ with the margins listed in the table. The proof is computer-assisted. Lemma~\ref{lem:cert_condensation} bounds $\beta_c$ from below by reducing the supremum in part~3 of Proposition~\ref{prop:entropy} to an inequality between explicit functions of one variable, and from above by evaluating the functional~\eqref{eq:functional} at an explicit law. Lemma~\ref{lem:cert_reconstruction} bounds $\beta_d$ from both sides by comparing the recursion of Definition~\ref{def:spectral} for $q=2$ with laws that have finitely many atoms. We then describe the computations, which use interval arithmetic and outward rounding. The values of $\beta_c$, $s_c$, and $\rho_c$ in Tables~\ref{tab:rates} and~\ref{tab:condensation} and in Figure~\ref{fig:fractions} are computed at the midpoint of the interval for $\beta_c$ in Table~\ref{tab:certified}.

\begin{table}[ht]
\centering
\small
\begin{tabular}{@{}c cc c cc c@{}}
\toprule
$(k,D)$ & $\beta_d$ & $\beta_c$ & $\beta_{\mathrm H}$ & $\beta_c-\beta_d\ge$ & $\beta_{\mathrm H}-\beta_c\ge$ & $s_c$ \\
\midrule
$(3,4)$ & $[2.6547,2.6974]$ & $[3.0380,3.0595]$ & $3.1348$ & $0.3405$ & $0.0753$ & $[0.9542,0.9552]$ \\
$(3,5)$ & $[2.1366,2.1587]$ & $[2.3474,2.3590]$ & $2.4508$ & $0.1886$ & $0.0917$ & $[0.9127,0.9137]$ \\
$(3,6)$ & $[1.8365,1.8499]$ & $[1.9854,1.9950]$ & $2.0905$ & $0.1354$ & $0.0955$ & $[0.8792,0.8803]$ \\
$(3,7)$ & $[1.6350,1.6574]$ & $[1.7542,1.7612]$ & $1.8561$ & $0.0968$ & $0.0949$ & $[0.8524,0.8534]$ \\
$(3,8)$ & $[1.4878,1.5011]$ & $[1.5887,1.5944]$ & $1.6872$ & $0.0875$ & $0.0928$ & $[0.8304,0.8313]$ \\
$(4,5)$ & $[2.5468,2.5695]$ & $[3.4255,3.4264]$ & $3.4381$ & $0.8559$ & $0.0118$ & $[0.9684,0.9686]$ \\
$(4,6)$ & $[2.1752,2.1892]$ & $[2.7038,2.7056]$ & $2.7254$ & $0.5146$ & $0.0198$ & $[0.9372,0.9374]$ \\
$(4,7)$ & $[1.9292,1.9517]$ & $[2.3182,2.3194]$ & $2.3427$ & $0.3665$ & $0.0233$ & $[0.9103,0.9105]$ \\
$(4,8)$ & $[1.7510,1.7731]$ & $[2.0638,2.0651]$ & $2.0905$ & $0.2907$ & $0.0253$ & $[0.8873,0.8875]$ \\
$(5,6)$ & $[2.3887,2.4022]$ & $[3.6763,3.6769]$ & $3.6791$ & $1.2740$ & $0.0022$ & $[0.9753,0.9754]$ \\
$(5,7)$ & $[2.1134,2.1357]$ & $[2.9431,2.9435]$ & $2.9478$ & $0.8073$ & $0.0043$ & $[0.9499,0.9500]$ \\
$(5,8)$ & $[1.9148,1.9284]$ & $[2.5431,2.5435]$ & $2.5497$ & $0.6147$ & $0.0062$ & $[0.9271,0.9272]$ \\
$(6,7)$ & $[2.2469,2.2607]$ & $[3.8783,3.8787]$ & $3.8790$ & $1.6176$ & $3.0\times10^{-4}$ & $[0.9797,0.9798]$ \\
$(6,8)$ & $[2.0331,2.0469]$ & $[3.1336,3.1339]$ & $3.1348$ & $1.0867$ & $1.0\times10^{-3}$ & $[0.9582,0.9583]$ \\
$(7,8)$ & $[2.1252,2.1391]$ & $[4.0493,4.0497]$ & $4.0497$ & $1.9101$ & $5.5\times10^{-5}$ & $[0.9828,0.9829]$ \\
\bottomrule
\end{tabular}
\caption{Certified bounds on the reconstruction threshold $\beta_d$ and the condensation threshold $\beta_c$ for the fifteen benchmark pairs of Table~\ref{tab:benchmark}, with the Holevo threshold $\beta_{\mathrm H}$ and the satisfied fraction $s_c=p_{\beta_c}(0)$. The intervals are proved by Lemmas~\ref{lem:cert_condensation} and~\ref{lem:cert_reconstruction} with the computations described in Appendix~\ref{app:certified}, and they are rounded outward. The fifth and sixth columns are lower bounds on the two differences, computed before rounding, so $\beta_d<\beta_c<\beta_{\mathrm H}$ on every pair even where the rounded upper end of the interval for $\beta_c$ reaches $\beta_{\mathrm H}$. The values of $\beta_c$ and $s_c$ concern the configuration model.}
\label{tab:certified}
\end{table}

\paragraph{The condensation threshold.} Both bounds on $\beta_c$ come from the functional~\eqref{eq:functional}: the lower bound compares it with an explicit function of one variable, and the upper bound evaluates it at an explicit law. Both use the following quantities. For $t\in[0,1]$, let $\mathcal J_D(t)$ be the mutual information between a uniformly random sign and $D$ independent observations of it through the binary symmetric channel with crossover probability $(1-t)/2$. Then $D\operatorname{cap}(t)-\mathcal J_D(t)$ is the relative entropy between the law of the $D$ observations and the uniform law on $\{\pm1\}^D$, which we use in the computation below. For $j\ge0$, let $H_j(\alpha)$ be the conditional entropy of a uniformly random sign given $j$ independent observations of it through the binary symmetric channel with crossover probability $\alpha$, which is given by~\eqref{eq:condensation_binomial} with $j$ in place of $D$.

\begin{lemma}[Bounds on the condensation threshold]\label{lem:cert_condensation}
Let $3\le k<D$.
\begin{enumerate}
\item Let $\sigma_0\in(0,1)$ and $\beta_0=2\operatorname{atanh}(\sigma_0)$. If
$$
  G(t)=D\operatorname{cap}(t)-\mathcal J_D(t)-\frac{D(k-1)}k\,\frac{\operatorname{cap}(t)^{k/(k-1)}}{\operatorname{cap}(\sigma_0)^{1/(k-1)}}\le0
  \qquad\text{for every } t\in[0,\sigma_0],
$$
then $\beta_c\ge\beta_0$.
\item For $w\in[0,1]$, let $\pi_w=(1-w)\delta_0+\tfrac w2(\delta_{-1}+\delta_1)$. Then
$$
  \mathcal B_\sigma(\pi_w)=\Big(Dw^{k-1}-\frac{D(k-1)}k\,w^k\Big)\operatorname{cap}(\sigma)-\sum_{j=0}^D\binom Dj w^{(k-1)j}\big(1-w^{k-1}\big)^{D-j}\big(\ln2-H_j(\alpha)\big).
$$
If $\mathcal B_\sigma(\pi_w)>0$ for some $w$, where $\sigma=\tanh(\beta/2)$, then $\beta_c\le\beta$.
\end{enumerate}
\end{lemma}

\begin{proof}
\emph{Part~1.} Let $\sigma=\sigma_0$, let $\pi$ be a law on $[-1,1]$ with mean zero, and let $a$, $b$, and $\mathcal I_D$ be as in~\eqref{eq:functional}. We first bound $b$ from below in terms of $a$. Write $v_j=\E{\mathfrak m}^{2j}$ for ${\mathfrak m}$ with law $\pi$, and let $\gamma_j$ be the coefficients of~\eqref{eq:condensation_capacity_series}. By independence, $a=\sum_j\gamma_j\sigma^{2j}v_j^{k-1}$ and $b=\sum_j\gamma_j\sigma^{2j}v_j^k$, as in Appendix~\ref{app:condensation_holevo}. The weights $\gamma_j\sigma^{2j}$ are positive and sum to $\operatorname{cap}(\sigma)$, so H\"older's inequality with exponents $k/(k-1)$ and $k$ gives $a\le b^{(k-1)/k}\operatorname{cap}(\sigma)^{1/k}$, that is,
$$
  b\ge a^{k/(k-1)}\operatorname{cap}(\sigma)^{-1/(k-1)}.
$$

Next we bound $\mathcal I_D$ from below. Since $\operatorname{cap}$ is increasing on $[0,1]$ and $a\le\operatorname{cap}(\sigma)$, there is $t\in[0,\sigma]$ with $\operatorname{cap}(t)=a$. We show that $\mathcal I_D\ge\mathcal J_D(t)$. For $z\in[0,1]$, the function $\phi_z(u)=\operatorname{cap}\big(z\operatorname{cap}^{-1}(u)\big)$ is concave on $[0,\ln2]$. Indeed, with $x=\operatorname{cap}^{-1}(u)$ and $h(x)=(1-x^2)\operatorname{atanh}(x)/x$, differentiation gives
$$
  \phi_z''(u)=\frac{z^2x\,\big[h(x)-h(zx)\big]}{(1-x^2)(1-z^2x^2)\operatorname{atanh}(x)^3},
$$
and $h(x)=1-\sum_{n\ge1}2x^{2n}/((2n-1)(2n+1))$ is decreasing on $(0,1)$, so $\phi_z''\le0$. Now let $O$ be any collection of observations of $\xi$ that is independent of the $i$-th observation given $\xi$, and let $\zeta=\E[\xi\mid O]$. Given $O$ and $R_i=r$, the sign in the $i$-th observation equals $+1$ with probability $(1+r\zeta)/2$. Since $R_i$ is independent of $(\xi,O)$ and $\operatorname{cap}$ is even, adding the $i$-th observation to $O$ increases the mutual information with $\xi$ by
$$
  \E\big[\operatorname{cap}(R_i)-\operatorname{cap}(R_i\zeta)\big]=a-\E\,\phi_{|\zeta|}\big(\operatorname{cap}(R_i)\big)\ge a-\E\,\phi_{|\zeta|}(a)=\operatorname{cap}(t)-\E\operatorname{cap}(t\zeta),
$$
by Jensen's inequality for the concave function $\phi_{|\zeta|}$, conditional on $\zeta$. The right-hand side is the increase obtained by adding instead an observation through the binary symmetric channel with crossover probability $(1-t)/2$. Replacing the $D$ observations by such observations one at a time therefore does not increase the mutual information, so $\mathcal I_D\ge\mathcal J_D(t)$.

With $D\,a=D\operatorname{cap}(t)$ and the two bounds, \eqref{eq:functional} gives $\mathcal B_{\sigma_0}(\pi)\le G(t)\le0$. By part~3 of Proposition~\ref{prop:entropy}, $\bar F(\alpha_0)=0$ for $\alpha_0=(1-\sigma_0)/2$. Since $\bar F$ is nonincreasing and positive on $(0,\alpha_c)$, this implies $\alpha_c\le\alpha_0$, that is, $\beta_c\ge\beta_0$.

\emph{Part~2.} Under $\pi_w$, each magnetization is $0$ with probability $1-w$ and $\pm1$ otherwise. Hence $|\sigma{\mathfrak m}_{i,1}\cdots{\mathfrak m}_{i,k-1}|$ equals $\sigma$ with probability $w^{k-1}$ and $0$ otherwise, so $a=w^{k-1}\operatorname{cap}(\sigma)$ and $b=w^k\operatorname{cap}(\sigma)$. An observation with $R_i=0$ carries no information about $\xi$, and an observation with $R_i=\sigma$ is an observation through the binary symmetric channel with crossover probability $\alpha$. Given that $j$ observations have $R_i=\sigma$, the mutual information is therefore $\ln2-H_j(\alpha)$, and averaging over the binomial law of $j$ proves the formula. If $\mathcal B_\sigma(\pi_w)>0$, then $\bar F(\alpha)>0$ by part~3 of Proposition~\ref{prop:entropy}, so $\alpha\le\alpha_c$ and $\beta_c\le\beta$.
\end{proof}

\paragraph{The reconstruction threshold.} For $q=2$ we describe the spectral recursion of Definition~\ref{def:spectral} by the absolute values of the overlaps, as in the paragraph on the binary case in Section~\ref{sec:spectral}. For $a,b\in[0,1]$, let $\widetilde\Gamma(a,b)$ be the random variable that equals $(a+b)/(1+ab)$ with probability $(1+ab)/2$ and $|a-b|/(1-ab)$ with probability $(1-ab)/2$. The law of the absolute value of the overlap of the output of the check-node operation $\Gamma$ depends only on the absolute values of the input overlaps, and it is the law of $\widetilde\Gamma$. For a law $P$ on $[0,1]$ and $s\in[0,1]$, let $v_1,\dots,v_{D-1}$ be independent with law $P$, let $c$ be the result of applying $\widetilde\Gamma$ to $v_1$ and $v_2$, then to the result and $v_3$, and so on up to $v_{D-1}$, with fresh randomness at each step, and let $\Psi_s(P)$ be the law of $s\,c_1\cdots c_{k-1}$, where $c_1,\dots,c_{k-1}$ are independent copies of $c$. At inverse temperature $\beta$, the overlap of $p_\beta$ is $s=\tanh(\beta/2)$, and $\Psi_s$ maps the law of the absolute overlap of $V^{(t)}$ to that of $V^{(t+1)}$.

For laws $P$ and $P'$ on $[0,1]$, we write $P\preceq P'$, and say that $P$ precedes $P'$ in the increasing convex order, if $\int g\,dP\le\int g\,dP'$ for every nondecreasing convex function $g$ on $[0,1]$. When $P$ and $P'$ have finitely many atoms, $P\preceq P'$ holds if $\int(v-\vartheta)^+dP\le\int(v-\vartheta)^+dP'$ for every $\vartheta\in[0,1]$. Indeed, the values of $g$ at the atoms agree with those of the piecewise linear interpolation of $g$ between consecutive atoms, extended by a constant to the left of the smallest atom, and this interpolation is a constant plus a nonnegative combination of functions $(v-\vartheta)^+$. Since both sides are piecewise linear in $\vartheta$ with breakpoints at the atoms, it suffices to check $\vartheta=0$ and the atoms of $P$ and $P'$.

\begin{lemma}[Bounds on the reconstruction threshold]\label{lem:cert_reconstruction}
Let $q=2$, $f(z)=z$, $k\ge3$, $D\ge2$, and $s=\tanh(\beta/2)$.
\begin{enumerate}
\item If $P\preceq P'$ and $s\le s'$, then $\Psi_s(P)\preceq\Psi_{s'}(P')$.
\item Suppose that there are laws $P_0,\dots,P_T$ on $[0,1]$ with $\delta_s\preceq P_0$ and $\Psi_s(P_{t-1})\preceq P_t$ for $1\le t\le T$, and that $\bar v=\int v\,dP_T$ satisfies $s\big((D-1)\bar v\big)^{k-1}<\bar v$. Then the primal model has non-reconstruction at $\beta$, so $\beta_d\ge\beta$.
\item Suppose that there is a law $Q\ne\delta_0$ on $[0,s]$ with $Q\preceq\Psi_s(Q)$. Then the primal model has reconstruction at $\beta$, so $\beta_d\le\beta$.
\end{enumerate}
\end{lemma}

\begin{proof}
\emph{Part~1.} Let $\xi$ be a uniformly random sign, and let $Y_1$ and $Y_2$ be signs that are independent given $\xi$, with $\Pr[Y_1=\xi]=(1+a)/2$ and $\Pr[Y_2=\xi]=(1+b)/2$. The posterior mean of $\xi$ is $(aY_1+bY_2)/(1+abY_1Y_2)$, so its absolute value has the law of $\widetilde\Gamma(a,b)$. Fix a nondecreasing convex function $g$ on $[0,1]$. If an observation $O'$ is obtained from an observation $O$ of $\xi$ and from randomness that is independent of $\xi$, $O$, and $Y_2$, then $\E[\xi\mid O',Y_2]=\E\big[\E[\xi\mid O,Y_2]\bigm| O',Y_2\big]$, and Jensen's inequality gives
$$
  \E g\big(|\E[\xi\mid O',Y_2]|\big)\le\E g\big(|\E[\xi\mid O,Y_2]|\big).
$$
Hence $\varphi(a)=\E g(\widetilde\Gamma(a,b))$ is nondecreasing in $a$: for $a'<a$, an observation that equals $\xi$ with probability $(1+a')/2$ is obtained by flipping $Y_1$ with probability $(1-a'/a)/2$. It is also convex: if $a=\lambda a_1+(1-\lambda)a_2$ with $\lambda\in[0,1]$, let $F$ equal $1$ with probability $\lambda$ and $2$ otherwise, independently of $\xi$, and let $\Pr[Y_1=\xi\mid F]=(1+a_F)/2$. The observation $(F,Y_1)$ gives the value $\lambda\varphi(a_1)+(1-\lambda)\varphi(a_2)$, and discarding $F$ leaves an observation that equals $\xi$ with probability $(1+a)/2$, which gives $\varphi(a)$. The law of $\widetilde\Gamma(a,b)$ is symmetric in $a$ and $b$, so the same holds in $b$. Consequently, if $A$ and $B$ are independent, $A'$ and $B'$ are independent, the law of $A$ precedes that of $A'$, and the law of $B$ precedes that of $B'$, then the law of $\widetilde\Gamma(A,B)$ precedes that of $\widetilde\Gamma(A',B')$: conditional on $B$, the function $x\mapsto\E g(\widetilde\Gamma(x,B))$ is nondecreasing and convex, and then the same argument applies to the second argument. Similarly, for independent nonnegative random variables $X$ and $Y$, the function $x\mapsto\E g(sxY)$ is nondecreasing and convex for every $s\ge0$, and $\E g(sXY)$ is nondecreasing in $s$. Applying these facts at each step in the definition of $\Psi_s$ proves part~1.

\emph{Part~2.} Let $\mathcal V_t$ be the law of the absolute overlap of $V^{(t)}$, so that $\mathcal V_0=\delta_s$ and $\mathcal V_t=\Psi_s(\mathcal V_{t-1})$. By part~1 and induction, $\mathcal V_t\preceq P_t$ for $t\le T$, and with $g(v)=v$ this gives $\int v\,d\mathcal V_T\le\bar v$. For fixed $a,b\in[0,1]$, $\E\,\widetilde\Gamma(a,b)=\tfrac12(a+b)+\tfrac12|a-b|=\max(a,b)\le a+b$. Hence, for every law $P$ on $[0,1]$, the mean of $c$ in the definition of $\Psi_s(P)$ is at most $(D-1)\int v\,dP$, and
$$
  \int v\,d\Psi_s(P)\le T_s\Big(\int v\,dP\Big),\qquad T_s(x)=s\big((D-1)x\big)^{k-1}.
$$
Since $k\ge3$, the ratio $T_s(x)/x$ is nondecreasing in $x$, and $T_s(\bar v)<\bar v$. The means of $\mathcal V_t$ for $t\ge T$ are therefore at most the iterates of $T_s$ started at $\bar v$, which decrease to zero. The overlap of the root spectrum $N^{(t)}$ is $s$ times a product of $k$ independent copies of $c$, so its mean also tends to zero. By the proof of part~2 of Theorem~\ref{thm:duality}, $P_t\to1$, and the primal model has non-reconstruction at $\beta$. Hence $\beta\in\mathcal N$ and $\beta_d\ge\beta$.

\emph{Part~3.} Since $Q$ lives on $[0,s]$ and $g$ is nondecreasing, $Q\preceq\delta_s=\mathcal V_0$. If $Q\preceq\mathcal V_t$, then by part~1, $Q\preceq\Psi_s(Q)\preceq\Psi_s(\mathcal V_t)=\mathcal V_{t+1}$. Hence the mean of $\mathcal V_t$ is at least the mean $\bar q>0$ of $Q$ for every $t$. Since $\E\,\widetilde\Gamma(a,b)\ge a$, the mean of $c$ is also at least $\bar q$, and the mean of the overlap of $N^{(t)}$ is at least $s\,\bar q^k>0$ for every $t$. Overlaps lie in $[0,1]$, so $N^{(t)}$ does not tend to $\unif$ in probability, $P_t\not\to1$, and the primal model has reconstruction at $\beta$ by Theorem~\ref{thm:duality}. By Proposition~\ref{prop:degrade} and the discussion that follows it, the set $\mathcal N$ of inverse temperatures without reconstruction is $[0,\beta_d)$ or $[0,\beta_d]$, so $\beta_d\le\beta$.
\end{proof}

\paragraph{Computation of the bounds on $\beta_c$.} For each pair, we found $\sigma_0$ by bisection in floating-point arithmetic, rounded it down to a multiple of $10^{-5}$, and verified the hypothesis of part~1 of Lemma~\ref{lem:cert_condensation} in interval arithmetic with $120$-bit precision and outward rounding. The function $\operatorname{cap}$ is nondecreasing on $[0,1]$, and so is $D\operatorname{cap}-\mathcal J_D$, because the relative entropy between the law of the observations and the uniform law does not increase when independent noise is added to each observation. Hence $G\le0$ on an interval $[t^-,t^+]$ follows from
$$
  D\operatorname{cap}(t^+)-\mathcal J_D(t^+)\le\frac{D(k-1)}k\,\frac{\operatorname{cap}(t^-)^{k/(k-1)}}{\operatorname{cap}(\sigma_0)^{1/(k-1)}}.
$$
Let $t_0$ be a power of two. We covered $[t_0,\sigma_0]$ by between $2655$ and $4462$ such intervals, obtained by repeated halving. On $[0,t_0]$, we bounded the relative entropy by the $\chi^2$-divergence $\tfrac12\big[(1+t^2)^D+(1-t^2)^D\big]-1$, a polynomial in $t^2$ with nonnegative coefficients and no terms of degree below $4$, which is therefore at most $(t/t_0)^4$ times its value at $t_0$. Together with $\operatorname{cap}(t)\ge t^2/2$, this reduces $G\le0$ on $[0,t_0]$ to an inequality at $t_0$. For the upper bound, we rounded an inverse temperature found by bisection up to a multiple of $10^{-5}$, chose $w$ by a search in floating-point arithmetic, and verified $\mathcal B_\sigma(\pi_w)>0$ in interval arithmetic. For $(6,7)$, $(6,8)$, and $(7,8)$, the chosen value is $w=1$, which is the trial law $\pi_*$ of~\eqref{eq:condensation_trial}.

\paragraph{Computation of the bounds on $\beta_d$.} We applied Lemma~\ref{lem:cert_reconstruction} to laws with finitely many atoms, stored as double-precision numbers, and we used the grid of double-precision numbers nearest to $x_i=\tfrac12\big(1-\cos(\pi i/600)\big)$, $i=0,\dots,600$, which is finer near $0$ and $1$. The overlap $s$ was enclosed in an interval with $80$-bit interval arithmetic, all other arithmetic was done in double precision with outward rounding after every operation, every sum of fewer than $2^{22}$ nonnegative terms was enlarged or reduced by the relative factor $2^{-30}$, and the means used in part~2 were computed exactly in rational arithmetic. For part~2, we computed $P_t$ from $P_{t-1}$ by the operations that define $\Psi_s$, one at a time, with $s$ rounded up. After each operation, we enclosed every output atom and its probability in intervals, kept the lower end of each probability, and replaced each atom by the two endpoints of the grid cell that contains the upper end of its interval, with weights whose mean is at least that upper end. The probability that was not kept was placed at $v=1$. Each of these changes can only increase $\int g$ for every nondecreasing convex $g$: moving mass upwards increases it, and replacing an atom by two points with a mean at least as large increases it by Jensen's inequality. By part~1, the resulting laws satisfy the hypothesis of part~2, and we stopped when the mean of $P_t$, computed in rational arithmetic, satisfied the condition of part~2. For part~3, we took for $Q$ an approximate fixed point of $\Psi_{s'}$ with $s'=\tanh((\beta-\delta)/2)$, computed in floating-point arithmetic at the slightly smaller inverse temperature $\beta-\delta$, with $\delta$ between $3.3\times10^{-3}$ and $1.3\times10^{-2}$; we removed atoms with probability below $10^{-20}$ and placed their mass at $v=0$, and we checked that the mean of $Q$ is positive. We then computed a law $R\preceq\Psi_s(Q)$ by the same operations as for part~2 with every rounding in the opposite direction: output atoms were rounded down, the atoms in each grid cell were merged into one atom at a lower bound on their mean, and the probability that was not kept was placed at $v=0$. Finally, we checked $\int(v-\vartheta)^+dQ\le\int(v-\vartheta)^+dR$, with the left-hand side rounded up and the right-hand side rounded down, at $\vartheta=0$ and at every atom of $Q$ and $R$, which gives $Q\preceq R\preceq\Psi_s(Q)$. The check requires a positive margin, which comes from $\delta>0$; close to $\beta_d$ this margin is smaller than the rounding errors, so our upper bounds lie between $0.010$ and $0.040$ above the estimates of $\beta_d$ in Table~\ref{tab:condensation}. Our lower bounds lie within $3\times10^{-3}$ of these estimates. For $(3,6)$, the interval contains the value $2/T_d=1.838548$ that follows from the dynamical temperature reported by Rizzo~\cite{rizzo2022stochastic}.

\paragraph{Results.} Table~\ref{tab:certified} lists the bounds. The difference $\beta_c-\beta_d$ is at least $0.087$ on every pair, with the smallest margin for $(3,8)$, and $\beta_{\mathrm H}-\beta_c$ is at least $5.5\times10^{-5}$, with the smallest margin for $(7,8)$. The certified lower bound on $s_c$ exceeds the best classical value in Table~\ref{tab:benchmark} by at least $0.0025$ on every pair.

\subsection{Local classical algorithms reach the reconstruction threshold in simulations}\label{app:dynamics}

We compare classical sampling and optimization with the reconstruction threshold on $62$ degree pairs of Max-XORSAT and on the fifteen benchmark pairs over seven larger alphabets. Conjecture~\ref{conj:classical} predicts an efficient classical Gibbs sampler at every fixed inverse temperature without reconstruction. The cavity method predicts more specifically that Glauber dynamics equilibrates below $\beta_d$. For $k\ge3$, efficient sampling arbitrarily close to $\beta_d$ would also reach satisfied fractions arbitrarily close to $s_d$, by Proposition~\ref{prop:sharp} and Lemma~\ref{lem:gibbs_cost}.

The fixed-temperature experiments support this prediction across all $62$ binary pairs. At $0.85\beta_d$ and $0.9\beta_d$, the number of sweeps required to decorrelate from an equilibrium assignment changes by at most $21\%$ when the number of variables is multiplied by ten. It grows rapidly near $\beta_d$, with fitted divergence points within a few percent of that threshold. From a uniformly random assignment, the dynamics also reaches energies close to their equilibrium values below $\beta_d$. Above the threshold, decorrelation and energy relaxation are much slower on the simulated timescales.

For optimization, simulated annealing attains a mean largest satisfied fraction above the estimated $s_d$ on all $62$ binary pairs and all $105$ larger-alphabet tuples. The binary comparison uses at most $10^7$ sweeps; the larger-alphabet comparison uses at most $10^8$. Its computational cost is practical: the $785$ trajectories underlying the binary survey and its long-run extensions required $19.7$ CPU-hours and completed in less than four hours of elapsed time on one MacBook Pro, using parallel runs. This finite-instance optimization result is distinct from the conjecture about mixing at fixed inverse temperature.

Section~\ref{app:sim_setup} describes the instances, updates and degree pairs. In Section~\ref{app:sim_glauber} we measure decorrelation from a planted equilibrium assignment, and in Section~\ref{app:sim_energy} we examine the energy reached from a uniformly random start. Section~\ref{app:sim_fraction} compares the satisfied fraction of simulated annealing with $s_d$ and the predictions of the cavity method. Section~\ref{app:sim_alphabets} extends the comparison to cyclic alphabets and finite fields. We state the scope of the evidence in Section~\ref{app:sim_summary}.

\subsubsection{Instances, updates, and degree pairs}\label{app:sim_setup}

All our simulations use instances from Gallager's ensemble (Section~\ref{sec:typical}), which Shutty et al.\ also use. Since $\beta_d$ is defined on the infinite tree, $\beta_d$ and $s_d$ are the same for both ensembles. A sweep consists of $n$ updates, each at a variable chosen uniformly at random. For Max-XORSAT, in a heat-bath update at inverse temperature $\beta$, the chosen variable is flipped with probability $1/(1+e^{\beta\Delta})$, where $\Delta$ is the change in the number of violated equations that the flip causes, and Glauber dynamics at $\beta$ consists of such updates. Simulated annealing uses Metropolis updates, which flip the chosen variable with probability $\min(1,e^{-\beta\Delta})$, and it raises $\beta$ linearly over its schedule.

We simulated $62$ degree pairs: the fifteen benchmark pairs of Table~\ref{tab:benchmark}, the other $30$ pairs with $3\le k<D\le12$, the five pairs $(k,k+1)$ with $12\le k\le16$, and the twelve pairs of Table~\ref{tab:degrees} with $D>12$, which lie on the lines $k/D=\tfrac12,\tfrac23,\tfrac34$ and reach $(18,36)$. On the benchmark pairs, simulated annealing comes closest to $s_d$ for the largest $k$ and for $D=k+1$ (Table~\ref{tab:benchmark}), so we added the pairs $(k,k+1)$ up to $k=16$; with the pairs of Table~\ref{tab:degrees}, we also test larger degrees at fixed ratios $k/D$. On every pair $(k,k+1)$ with $k\ge5$, the asymptotic value $\tfrac12+\tfrac k{2D}$ of Prange's algorithm, a nonlocal classical algorithm, already exceeds $s_d$. On these pairs, the simulations therefore concern only the comparison between DQI with local decoders and local classical algorithms, which is the subject of the last part of Question~\ref{q:main}.

\paragraph{Computational resources.} The binary annealing survey consists of four instances at each of $10^4$, $10^5$ and $10^6$ sweeps on all $62$ pairs, together with $41$ trajectories at $10^7$ sweeps on $29$ pairs, for $785$ trajectories in total. These runs used $19.7$ CPU-hours on a ten-core Apple M1 Max MacBook Pro. With parallel single-threaded processes, the campaign completed in less than four hours of wall-clock time. This timing concerns the binary annealing comparison; threshold computations, fixed-temperature tests and the larger-alphabet studies are separate computations.

For $30$ of these pairs, Tables~\ref{tab:condensation} and~\ref{tab:degrees} do not give $\beta_d$, and we computed it as for Table~\ref{tab:degrees}: by population dynamics for the spectral recursion of Definition~\ref{def:spectral}, with $2\times10^5$ messages, at most $1500$ levels, and $14$ bisection steps on $\sigma=\tanh(\beta/2)$ (Appendix~\ref{app:numerics}). Table~\ref{tab:sim_pairs} lists $\beta_d$ and $s_d$ for all $62$ pairs, computed from the midpoint of the final bisection interval in $\sigma$. This interval has width $2^{-14}$, which corresponds to at most $5\times10^{-4}$ in $\beta_d$ and to $2^{-15}\approx3.1\times10^{-5}$ in $s_d=(1+\sigma)/2$. This resolution is not a confidence interval, because the finite population and the finite number of levels add errors of their own; for $(3,6)$, the comparison with the value of Rizzo at the beginning of this appendix shows that these errors are comparable to the width of the bisection interval.

\begin{table}[p]
\centering
\footnotesize
\setlength{\tabcolsep}{3pt}
\begin{tabular}{@{}ccc rrrr @{\hspace{12pt}} ccc rrrr@{}}
\toprule
& & & \multicolumn{4}{c}{$10^3(\mathrm{SA}-s_d)$, sweeps} & & & & \multicolumn{4}{c}{$10^3(\mathrm{SA}-s_d)$, sweeps} \\
\cmidrule(lr){4-7}\cmidrule(l){11-14}
$(k,D)$ & $\beta_d$ & $s_d$ & $10^4$ & $10^5$ & $10^6$ & $10^7$ & $(k,D)$ & $\beta_d$ & $s_d$ & $10^4$ & $10^5$ & $10^6$ & $10^7$ \\
\midrule
$(3,4)$ & $2.657$ & $0.9345$ & $-1.1$ & $+2.1$ & $+4.0$ & -- & $(6,11)$ & $1.638$ & $0.8372$ & $-5.0$ & $-1.0$ & $+1.5$ & -- \\
$(3,5)$ & $2.139$ & $0.8946$ & $+0.9$ & $+5.7$ & $+8.1$ & -- & $(7,11)$ & $1.709$ & $0.8467$ & $-7.6$ & $-1.5$ & $+0.9$ & -- \\
$(4,5)$ & $2.549$ & $0.9275$ & $-6.0$ & $-1.2$ & $+1.7$ & -- & $(8,11)$ & $1.767$ & $0.8540$ & $-6.8$ & $-3.0$ & $+0.6$ & -- \\
$(3,6)$ & $1.839$ & $0.8628$ & $+4.4$ & $+8.1$ & $+9.8$ & $+11.1$ & $(9,11)$ & $1.815$ & $0.8600$ & $-7.4$ & $-3.0$ & $+0.0$ & $+0.7$ \\
$(4,6)$ & $2.178$ & $0.8983$ & $-2.4$ & $+0.5$ & $+2.9$ & -- & $(10,11)$ & $1.858$ & $0.8650$ & $-6.6$ & $-2.1$ & $-0.7$ & $+1.4$ \\
$(5,6)$ & $2.391$ & $0.9162$ & $-4.6$ & $-1.9$ & $+1.1$ & $+1.8$ & $(3,12)$ & $1.147$ & $0.7590$ & $+9.2$ & $+11.8$ & $+13.4$ & -- \\
$(3,7)$ & $1.637$ & $0.8372$ & $+6.5$ & $+10.0$ & $+12.1$ & -- & $(4,12)$ & $1.342$ & $0.7929$ & $+0.5$ & $+3.6$ & $+6.0$ & -- \\
$(4,7)$ & $1.932$ & $0.8734$ & $-2.7$ & $+1.9$ & $+3.9$ & -- & $(5,12)$ & $1.463$ & $0.8120$ & $-2.7$ & $+0.5$ & $+2.7$ & -- \\
$(5,7)$ & $2.116$ & $0.8924$ & $-5.8$ & $-0.7$ & $+1.6$ & -- & $(6,12)$ & $1.549$ & $0.8248$ & $-4.6$ & $-0.5$ & $+1.6$ & -- \\
$(6,7)$ & $2.250$ & $0.9046$ & $-7.4$ & $-2.4$ & $+0.7$ & $+1.9$ & $(7,12)$ & $1.616$ & $0.8342$ & $-5.2$ & $-1.3$ & $+0.9$ & -- \\
$(3,8)$ & $1.490$ & $0.8161$ & $+7.3$ & $+10.6$ & $+12.0$ & -- & $(8,12)$ & $1.670$ & $0.8416$ & $-5.5$ & $-1.8$ & $+0.4$ & $+2.0$ \\
$(4,8)$ & $1.753$ & $0.8523$ & $-0.8$ & $+2.9$ & $+4.5$ & -- & $(9,12)$ & $1.716$ & $0.8475$ & $-6.2$ & $-2.1$ & $-0.2$ & $+0.9$ \\
$(5,8)$ & $1.917$ & $0.8718$ & $-4.2$ & $-0.5$ & $+1.5$ & -- & $(10,12)$ & $1.755$ & $0.8525$ & $-5.6$ & $-2.7$ & $-0.3$ & $+1.5$ \\
$(6,8)$ & $2.036$ & $0.8845$ & $-5.0$ & $-1.3$ & $+0.6$ & $+2.1$ & $(11,12)$ & $1.789$ & $0.8569$ & $-6.3$ & $-2.5$ & $+0.7$ & $+1.6$ \\
$(7,8)$ & $2.128$ & $0.8936$ & $-6.3$ & $-2.4$ & $+0.6$ & $+2.2$ & $(12,13)$ & $1.729$ & $0.8492$ & $-7.7$ & $-3.3$ & $+0.0$ & $+1.2$ \\
$(3,9)$ & $1.377$ & $0.7984$ & $+7.7$ & $+11.2$ & $+13.3$ & -- & $(13,14)$ & $1.674$ & $0.8421$ & $-5.7$ & $-3.1$ & $-1.3$ & $+0.3$ \\
$(4,9)$ & $1.616$ & $0.8343$ & $-1.0$ & $+3.5$ & $+5.3$ & -- & $(14,15)$ & $1.625$ & $0.8355$ & $-7.5$ & $-3.2$ & $-0.3$ & $+1.7$ \\
$(5,9)$ & $1.765$ & $0.8539$ & $-4.3$ & $+0.1$ & $+2.0$ & -- & $(15,16)$ & $1.581$ & $0.8293$ & $-5.6$ & $-2.7$ & $-0.4$ & $+1.2$ \\
$(6,9)$ & $1.873$ & $0.8668$ & $-6.3$ & $-1.0$ & $+1.0$ & -- & $(16,17)$ & $1.540$ & $0.8234$ & $-8.1$ & $-3.1$ & $-0.4$ & $+1.2$ \\
$(7,9)$ & $1.957$ & $0.8762$ & $-7.1$ & $-2.1$ & $+0.3$ & $+2.2$ & $(7,14)$ & $1.467$ & $0.8127$ & $-4.7$ & $-1.0$ & $+0.7$ & -- \\
$(8,9)$ & $2.024$ & $0.8833$ & $-6.8$ & $-3.1$ & $-0.2$ & $+1.4$ & $(8,16)$ & $1.398$ & $0.8018$ & $-5.5$ & $-1.2$ & $+1.0$ & -- \\
$(3,10)$ & $1.286$ & $0.7834$ & $+8.7$ & $+11.8$ & $+13.1$ & -- & $(9,18)$ & $1.338$ & $0.7922$ & $-5.4$ & $-2.0$ & $+0.6$ & -- \\
$(4,10)$ & $1.507$ & $0.8187$ & $+0.1$ & $+3.4$ & $+5.4$ & -- & $(10,20)$ & $1.286$ & $0.7834$ & $-5.3$ & $-2.0$ & $+0.6$ & $+1.4$ \\
$(5,10)$ & $1.645$ & $0.8382$ & $-3.3$ & $+0.9$ & $+2.1$ & -- & $(12,24)$ & $1.199$ & $0.7684$ & $-5.5$ & $-2.1$ & $+0.6$ & $+0.6$ \\
$(6,10)$ & $1.744$ & $0.8512$ & $-4.7$ & $-0.5$ & $+1.0$ & -- & $(14,28)$ & $1.129$ & $0.7557$ & $-5.0$ & $-3.1$ & $-0.3$ & $+2.2$ \\
$(7,10)$ & $1.820$ & $0.8605$ & $-4.3$ & $-1.7$ & $+0.4$ & $+1.7$ & $(16,32)$ & $1.072$ & $0.7449$ & $-5.3$ & $-2.3$ & $+0.3$ & $+1.2$ \\
$(8,10)$ & $1.882$ & $0.8679$ & $-6.5$ & $-2.0$ & $+0.5$ & $+2.0$ & $(18,36)$ & $1.023$ & $0.7355$ & $-5.9$ & $-2.2$ & $-0.0$ & $+1.7$ \\
$(9,10)$ & $1.935$ & $0.8738$ & $-5.1$ & $-2.2$ & $+0.3$ & $+2.3$ & $(10,15)$ & $1.526$ & $0.8214$ & $-6.2$ & $-2.2$ & $+0.2$ & -- \\
$(3,11)$ & $1.211$ & $0.7704$ & $+9.0$ & $+11.9$ & $+13.5$ & -- & $(12,18)$ & $1.416$ & $0.8047$ & $-6.1$ & $-2.3$ & $+0.4$ & $+1.0$ \\
$(4,11)$ & $1.418$ & $0.8050$ & $+0.1$ & $+3.7$ & $+6.3$ & -- & $(12,16)$ & $1.520$ & $0.8205$ & $-6.4$ & $-1.7$ & $+0.3$ & $+1.6$ \\
$(5,11)$ & $1.546$ & $0.8243$ & $-2.9$ & $+1.1$ & $+3.0$ & -- & $(15,20)$ & $1.384$ & $0.7996$ & $-5.3$ & $-2.5$ & $-0.3$ & $+1.7$ \\
\bottomrule
\end{tabular}
\caption{Reconstruction thresholds and simulated annealing on the $62$ degree pairs of Max-XORSAT. The values of $\beta_d$ come from bisection intervals of width at most $5\times10^{-4}$, and $s_d=\tfrac12+\tfrac12\tanh(\beta_d/2)$. The other columns give $10^3$ times the difference between the largest satisfied fraction of simulated annealing (SA) and $s_d$, for the number of sweeps in the column header, with $\beta$ raised linearly from $0$ to $4$ on instances with about $10080$ variables. The values are means over four instances, and over one or two instances with $10^7$ sweeps; a dash means that we did not run $10^7$ sweeps.}
\label{tab:sim_pairs}
\end{table}

\subsubsection{Relaxation of Glauber dynamics at fixed inverse temperatures}\label{app:sim_glauber}

Glauber dynamics is the local Markov chain to which the prediction of the cavity method refers, so its relaxation at fixed inverse temperatures tests the prediction behind Conjecture~\ref{conj:classical} most directly. We simulated it on all $62$ pairs. On the benchmark pairs, the simulations use the value of $\beta_d$ in Table~\ref{tab:dynamics}, which was computed as $2\operatorname{atanh}(2s_d-1)$ from an earlier estimate of $s_d$ and differs from the value in Table~\ref{tab:condensation} by less than $1.5\times10^{-3}$. On the other pairs, they use the value in Table~\ref{tab:sim_pairs}.

\begin{table}[t]
\centering
\footnotesize
\setlength{\tabcolsep}{4pt}
\begin{tabular}{@{}c rr cc @{\hspace{14pt}} c rr cc@{}}
\toprule
& \multicolumn{2}{c}{$t_{\mathrm{rel}}$ at $0.9\beta_d$} & & & & \multicolumn{2}{c}{$t_{\mathrm{rel}}$ at $0.9\beta_d$} & & \\
\cmidrule(lr){2-3}\cmidrule(lr){7-8}
$(k,D)$ & $n$ & $10n$ & $\beta_{\mathrm{div}}/\beta_d$ & $Q$ at $1.1\beta_d$ & $(k,D)$ & $n$ & $10n$ & $\beta_{\mathrm{div}}/\beta_d$ & $Q$ at $1.1\beta_d$ \\
\midrule
$(3,9)$ & $1049$ & $988$ & $0.996\text{--}1.037$ & $0.91$ & $(6,12)$ & $1660$ & $1751$ & $0.996\text{--}1.014$ & $0.99$ \\
$(4,9)$ & $1474$ & $1627$ & $0.999\text{--}1.032$ & $0.97$ & $(7,12)$ & $2002$ & $1949$ & $1.000\text{--}1.028$ & $0.99$ \\
$(5,9)$ & $1932$ & $2027$ & $1.011\text{--}1.056$ & $0.98$ & $(8,12)$ & $2242$ & $2231$ & $0.992\text{--}1.015$ & $0.99$ \\
$(6,9)$ & $2392$ & $2500$ & $1.007\text{--}1.039$ & $0.99$ & $(9,12)$ & $2333$ & $2353$ & $1.003\text{--}1.012$ & $1.00$ \\
$(7,9)$ & $2849$ & $2979$ & $1.005\text{--}1.023$ & $0.99$ & $(10,12)$ & $2594$ & $2543$ & $1.006\text{--}1.030$ & $1.00$ \\
$(8,9)$ & $3631$ & $3408$ & $1.013\text{--}1.031$ & $0.99$ & $(11,12)$ & $2891$ & $2839$ & $0.998\text{--}1.021$ & $1.00$ \\
$(3,10)$ & $978$ & $957$ & $0.993\text{--}1.011$ & $0.91$ & $(12,13)$ & $2767$ & $2916$ & $1.016\text{--}1.042$ & $1.00$ \\
$(4,10)$ & $1403$ & $1372$ & $0.997\text{--}1.004$ & $0.97$ & $(13,14)$ & $2484$ & $2981$ & $1.006\text{--}1.013$ & $1.00$ \\
$(5,10)$ & $1830$ & $1818$ & $1.002\text{--}1.020$ & $0.98$ & $(14,15)$ & $2605$ & $2663$ & $1.019\text{--}1.038$ & $1.00$ \\
$(6,10)$ & $2170$ & $2012$ & $0.994\text{--}1.017$ & $0.99$ & $(15,16)$ & $2671$ & $2748$ & $0.996\text{--}1.006$ & $1.00$ \\
$(7,10)$ & $2468$ & $2599$ & $1.011\text{--}1.022$ & $0.99$ & $(16,17)$ & $2892$ & $2733$ & $1.009\text{--}1.020$ & $1.00$ \\
$(8,10)$ & $3039$ & $2828$ & $0.995\text{--}1.011$ & $0.99$ & $(7,14)$ & $1601$ & $1758$ & $1.012\text{--}1.015$ & $0.99$ \\
$(9,10)$ & $2885$ & $3003$ & $1.000\text{--}1.022$ & $1.00$ & $(8,16)$ & $1572$ & $1619$ & $1.029\text{--}1.042$ & $0.99$ \\
$(3,11)$ & $894$ & $816$ & $1.013\text{--}1.017$ & $0.91$ & $(9,18)$ & $1654$ & $1748$ & $1.011\text{--}1.033$ & $0.99$ \\
$(4,11)$ & $1220$ & $1332$ & $0.997\text{--}1.001$ & $0.96$ & $(10,20)$ & $1942$ & $1850$ & $1.002\text{--}1.016$ & $0.99$ \\
$(5,11)$ & $1799$ & $1697$ & $1.005\text{--}1.014$ & $0.98$ & $(12,24)$ & $1896$ & $1786$ & $0.998\text{--}1.014$ & $1.00$ \\
$(6,11)$ & $1983$ & $1869$ & $1.002\text{--}1.018$ & $0.99$ & $(14,28)$ & $1712$ & $1819$ & $0.997\text{--}1.055$ & $1.00$ \\
$(7,11)$ & $2245$ & $2333$ & $1.006\text{--}1.019$ & $0.99$ & $(16,32)$ & $1696$ & $1946$ & $0.991\text{--}1.037$ & $1.00$ \\
$(8,11)$ & $2719$ & $2586$ & $1.008\text{--}1.016$ & $0.99$ & $(18,36)$ & $1817$ & $1980$ & $1.009\text{--}1.025$ & $1.00$ \\
$(9,11)$ & $2964$ & $2671$ & $1.009\text{--}1.047$ & $1.00$ & $(10,15)$ & $2079$ & $2169$ & $1.005\text{--}1.019$ & $1.00$ \\
$(10,11)$ & $2968$ & $3025$ & $1.012\text{--}1.047$ & $1.00$ & $(12,18)$ & $2164$ & $2116$ & $1.005\text{--}1.015$ & $1.00$ \\
$(3,12)$ & $924$ & $797$ & $0.999\text{--}1.034$ & $0.90$ & $(12,16)$ & $2046$ & $2376$ & $0.998\text{--}1.057$ & $1.00$ \\
$(4,12)$ & $1235$ & $1207$ & $1.007\text{--}1.019$ & $0.96$ & $(15,20)$ & $2360$ & $2407$ & $0.987\text{--}1.031$ & $1.00$ \\
$(5,12)$ & $1503$ & $1510$ & $1.008\text{--}1.026$ & $0.98$ & &&&& \\
\bottomrule
\end{tabular}
\caption{Glauber dynamics on planted instances of the $47$ degree pairs that are not benchmark pairs, with the quantities of Table~\ref{tab:dynamics}. The relaxation time at $0.9\beta_d$ is the geometric mean over three instances with $n\approx30240$ variables, and it is measured on one instance with $10n$ variables.}
\label{tab:dynamics_more}
\end{table}

\paragraph{Planted instances.} To start Glauber dynamics in equilibrium, we use planted instances. We draw $z^\star\in\{0,1\}^n$ uniformly at random and choose the right-hand sides so that each equation is violated by $z^\star$ independently with probability $p_\beta(1)$. The joint probability of the right-hand sides and $z^\star$ is then proportional to $e^{-\beta H(z^\star)}$, so conditionally on the instance, $z^\star$ is a sample from $\mu_\beta$, and Glauber dynamics started from $z^\star$ stays in equilibrium. We record the overlap $Q(t)=1-2|Z(t)-z^\star|/n$ between the assignment $Z(t)$ after $t$ sweeps and $z^\star$, where $|Z(t)-z^\star|$ is the number of variables on which they differ, and we call the first $t$ with $Q(t)\le0.1$ the relaxation time $t_{\mathrm{rel}}$. It estimates the correlation time of Glauber dynamics in equilibrium, which the cavity method predicts not to grow with the number of variables below $\beta_d$. For planted instances, Krzakala and Zdeborov\'a show that the dynamics started from the planted configuration can be mapped exactly to the equilibrium dynamics, and that the dynamical transition is the point above which the planted configuration stops ``melting'', that is, above which the dynamics started from it no longer decorrelates from it~\cite{krzakala2011melting}. Their result concerns $p$-spin models on the Nishimori line, which include planted instances of Max-XORSAT at the inverse temperature used to generate them.

\paragraph{Planted and uniformly random right-hand sides.} Relative to instances with uniformly random right-hand sides, planted instances are reweighted by the partition function, so the two ensembles are expected to be indistinguishable as long as the partition function is concentrated around its expectation. The cavity method predicts that this holds up to the condensation threshold, which lies above $\beta_d$~\cite{krzakala2009hiding,krzakala2011melting}. For the configuration model, Lemma~\ref{lem:pgm_success} and Theorem~\ref{thm:condensation} show that the planted and the uniformly random right-hand sides are asymptotically indistinguishable below $\beta_c$ (Section~\ref{sec:window}). For $(3,6)$, $(3,7)$, and $(3,8)$, the inverse temperature $1.1\beta_d$ used below exceeds the certified upper bound on $\beta_c$ for the configuration model in Table~\ref{tab:certified}, and for $(3,5)$ it lies inside the certified interval. For these four pairs, the runs on planted instances at $1.1\beta_d$ therefore need not reflect instances with uniformly random right-hand sides, whereas the inverse temperature $1.03\beta_d$ lies below the certified lower bound on $\beta_c$ for every benchmark pair. We did not compute $\beta_c$ for the other $47$ pairs, so for them the same reservation applies to every run above $\beta_d$.

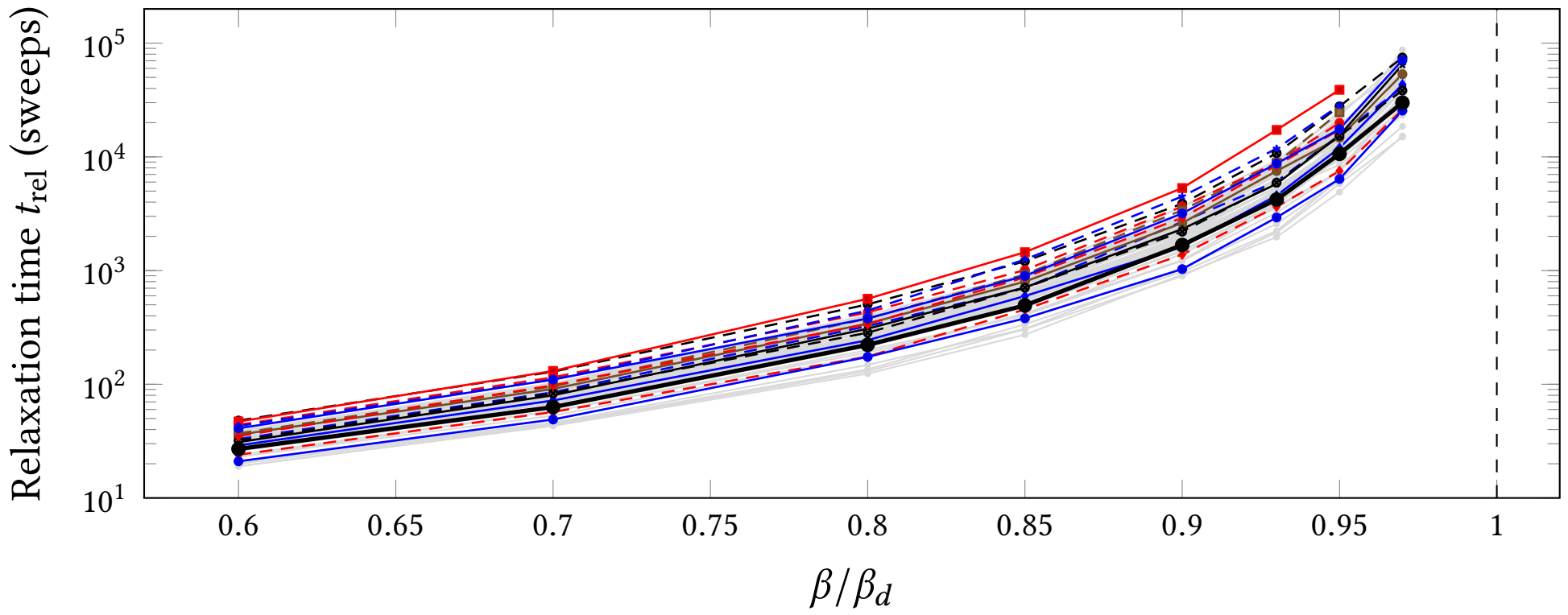
\begin{figure}[t]
\centering
\begin{tikzpicture}
\begin{semilogyaxis}[
  width=0.8\textwidth, height=5.6cm,
  xmin=0.57, xmax=1.02, ymin=10, ymax=2e5,
  xlabel={$\beta/\beta_d$}, ylabel={Relaxation time $t_{\mathrm{rel}}$ (sweeps)},
  xlabel style={font=\small}, ylabel style={font=\small},
  tick label style={font=\scriptsize},
  every axis plot/.append style={gray!65, line width=0.5pt, mark=*, mark size=0.9pt},
]
\addplot[gray!30, line width=0.4pt, mark=*, mark size=0.6pt] coordinates {(0.60,21) (0.70,46) (0.80,148) (0.85,307) (0.90,1049) (0.93,2573) (0.95,6188) (0.97,18573)};
\addplot[gray!30, line width=0.4pt, mark=*, mark size=0.6pt] coordinates {(0.60,26) (0.70,62) (0.80,217) (0.85,483) (0.90,1474) (0.93,3789) (0.95,10007) (0.97,28307)};
\addplot[gray!30, line width=0.4pt, mark=*, mark size=0.6pt] coordinates {(0.60,29) (0.70,72) (0.80,262) (0.85,650) (0.90,1932) (0.93,6386) (0.95,14373)};
\addplot[gray!30, line width=0.4pt, mark=*, mark size=0.6pt] coordinates {(0.60,33) (0.70,87) (0.80,306) (0.85,755) (0.90,2392) (0.93,6883) (0.95,16876) (0.97,53780)};
\addplot[gray!30, line width=0.4pt, mark=*, mark size=0.6pt] coordinates {(0.60,36) (0.70,92) (0.80,362) (0.85,908) (0.90,2849) (0.93,6838) (0.95,18434) (0.97,65657)};
\addplot[gray!30, line width=0.4pt, mark=*, mark size=0.6pt] coordinates {(0.60,39) (0.70,101) (0.80,405) (0.85,922) (0.90,3631) (0.93,9324) (0.95,23068) (0.97,73361)};
\addplot[gray!30, line width=0.4pt, mark=*, mark size=0.6pt] coordinates {(0.60,20) (0.70,45) (0.80,135) (0.85,337) (0.90,978) (0.93,2234) (0.95,6346) (0.97,25594)};
\addplot[gray!30, line width=0.4pt, mark=*, mark size=0.6pt] coordinates {(0.60,24) (0.70,58) (0.80,200) (0.85,405) (0.90,1403) (0.93,3577) (0.95,7500) (0.97,36247)};
\addplot[gray!30, line width=0.4pt, mark=*, mark size=0.6pt] coordinates {(0.60,27) (0.70,68) (0.80,223) (0.85,537) (0.90,1830) (0.93,4568) (0.95,14196) (0.97,42139)};
\addplot[gray!30, line width=0.4pt, mark=*, mark size=0.6pt] coordinates {(0.60,30) (0.70,78) (0.80,267) (0.85,646) (0.90,2170) (0.93,6254) (0.95,17090)};
\addplot[gray!30, line width=0.4pt, mark=*, mark size=0.6pt] coordinates {(0.60,33) (0.70,91) (0.80,302) (0.85,780) (0.90,2468) (0.93,7030) (0.95,20680)};
\addplot[gray!30, line width=0.4pt, mark=*, mark size=0.6pt] coordinates {(0.60,37) (0.70,95) (0.80,329) (0.85,835) (0.90,3039) (0.93,7319) (0.95,20686)};
\addplot[gray!30, line width=0.4pt, mark=*, mark size=0.6pt] coordinates {(0.60,38) (0.70,102) (0.80,391) (0.85,837) (0.90,2885) (0.93,7543) (0.95,21903)};
\addplot[gray!30, line width=0.4pt, mark=*, mark size=0.6pt] coordinates {(0.60,19) (0.70,44) (0.80,130) (0.85,305) (0.90,894) (0.93,2160) (0.95,5803) (0.97,14941)};
\addplot[gray!30, line width=0.4pt, mark=*, mark size=0.6pt] coordinates {(0.60,23) (0.70,58) (0.80,190) (0.85,407) (0.90,1220) (0.93,3235) (0.95,7415) (0.97,33530)};
\addplot[gray!30, line width=0.4pt, mark=*, mark size=0.6pt] coordinates {(0.60,27) (0.70,63) (0.80,231) (0.85,492) (0.90,1799) (0.93,4641) (0.95,12189) (0.97,43711)};
\addplot[gray!30, line width=0.4pt, mark=*, mark size=0.6pt] coordinates {(0.60,29) (0.70,74) (0.80,253) (0.85,634) (0.90,1983) (0.93,4836) (0.95,12337) (0.97,44652)};
\addplot[gray!30, line width=0.4pt, mark=*, mark size=0.6pt] coordinates {(0.60,32) (0.70,82) (0.80,283) (0.85,694) (0.90,2245) (0.93,6303) (0.95,18092) (0.97,57049)};
\addplot[gray!30, line width=0.4pt, mark=*, mark size=0.6pt] coordinates {(0.60,34) (0.70,86) (0.80,324) (0.85,797) (0.90,2719) (0.93,6900) (0.95,18279) (0.97,67376)};
\addplot[gray!30, line width=0.4pt, mark=*, mark size=0.6pt] coordinates {(0.60,36) (0.70,97) (0.80,350) (0.85,864) (0.90,2964) (0.93,8014) (0.95,21919)};
\addplot[gray!30, line width=0.4pt, mark=*, mark size=0.6pt] coordinates {(0.60,37) (0.70,102) (0.80,349) (0.85,886) (0.90,2968) (0.93,8400) (0.95,25039) (0.97,59634)};
\addplot[gray!30, line width=0.4pt, mark=*, mark size=0.6pt] coordinates {(0.60,19) (0.70,43) (0.80,124) (0.85,273) (0.90,924) (0.93,1970) (0.95,4903) (0.97,15240)};
\addplot[gray!30, line width=0.4pt, mark=*, mark size=0.6pt] coordinates {(0.60,23) (0.70,53) (0.80,179) (0.85,421) (0.90,1235) (0.93,3693) (0.95,7437) (0.97,27732)};
\addplot[gray!30, line width=0.4pt, mark=*, mark size=0.6pt] coordinates {(0.60,26) (0.70,61) (0.80,214) (0.85,478) (0.90,1503) (0.93,3908) (0.95,9865) (0.97,31173)};
\addplot[gray!30, line width=0.4pt, mark=*, mark size=0.6pt] coordinates {(0.60,28) (0.70,69) (0.80,241) (0.85,554) (0.90,1660) (0.93,5051) (0.95,11625) (0.97,52132)};
\addplot[gray!30, line width=0.4pt, mark=*, mark size=0.6pt] coordinates {(0.60,30) (0.70,77) (0.80,258) (0.85,606) (0.90,2002) (0.93,5391) (0.95,12837) (0.97,48224)};
\addplot[gray!30, line width=0.4pt, mark=*, mark size=0.6pt] coordinates {(0.60,31) (0.70,82) (0.80,284) (0.85,635) (0.90,2242) (0.93,5996) (0.95,13993)};
\addplot[gray!30, line width=0.4pt, mark=*, mark size=0.6pt] coordinates {(0.60,34) (0.70,88) (0.80,323) (0.85,757) (0.90,2333) (0.93,6375) (0.95,16105) (0.97,64452)};
\addplot[gray!30, line width=0.4pt, mark=*, mark size=0.6pt] coordinates {(0.60,36) (0.70,92) (0.80,333) (0.85,786) (0.90,2594) (0.93,6940) (0.95,18586) (0.97,56460)};
\addplot[gray!30, line width=0.4pt, mark=*, mark size=0.6pt] coordinates {(0.60,37) (0.70,98) (0.80,365) (0.85,855) (0.90,2891) (0.93,8428) (0.95,21098)};
\addplot[gray!30, line width=0.4pt, mark=*, mark size=0.6pt] coordinates {(0.60,37) (0.70,99) (0.80,351) (0.85,823) (0.90,2767) (0.93,9004) (0.95,18760)};
\addplot[gray!30, line width=0.4pt, mark=*, mark size=0.6pt] coordinates {(0.60,37) (0.70,98) (0.80,363) (0.85,823) (0.90,2484) (0.93,7553) (0.95,16643)};
\addplot[gray!30, line width=0.4pt, mark=*, mark size=0.6pt] coordinates {(0.60,38) (0.70,99) (0.80,360) (0.85,904) (0.90,2605) (0.93,8115) (0.95,19495) (0.97,48293)};
\addplot[gray!30, line width=0.4pt, mark=*, mark size=0.6pt] coordinates {(0.60,37) (0.70,99) (0.80,380) (0.85,856) (0.90,2671) (0.93,7686) (0.95,15782) (0.97,87266)};
\addplot[gray!30, line width=0.4pt, mark=*, mark size=0.6pt] coordinates {(0.60,37) (0.70,95) (0.80,341) (0.85,808) (0.90,2892) (0.93,7155) (0.95,12968) (0.97,57940)};
\addplot[gray!30, line width=0.4pt, mark=*, mark size=0.6pt] coordinates {(0.60,29) (0.70,68) (0.80,237) (0.85,545) (0.90,1601) (0.93,5215) (0.95,12043) (0.97,38310)};
\addplot[gray!30, line width=0.4pt, mark=*, mark size=0.6pt] coordinates {(0.60,29) (0.70,71) (0.80,233) (0.85,553) (0.90,1572) (0.93,5039) (0.95,8477) (0.97,23316)};
\addplot[gray!30, line width=0.4pt, mark=*, mark size=0.6pt] coordinates {(0.60,29) (0.70,72) (0.80,247) (0.85,561) (0.90,1654) (0.93,4754) (0.95,11222)};
\addplot[gray!30, line width=0.4pt, mark=*, mark size=0.6pt] coordinates {(0.60,30) (0.70,72) (0.80,246) (0.85,587) (0.90,1942) (0.93,4287) (0.95,11123) (0.97,42026)};
\addplot[gray!30, line width=0.4pt, mark=*, mark size=0.6pt] coordinates {(0.60,30) (0.70,75) (0.80,266) (0.85,568) (0.90,1896) (0.93,5024) (0.95,9840) (0.97,46790)};
\addplot[gray!30, line width=0.4pt, mark=*, mark size=0.6pt] coordinates {(0.60,30) (0.70,77) (0.80,261) (0.85,620) (0.90,1712) (0.93,4716) (0.95,9363) (0.97,26147)};
\addplot[gray!30, line width=0.4pt, mark=*, mark size=0.6pt] coordinates {(0.60,30) (0.70,79) (0.80,265) (0.85,629) (0.90,1696) (0.93,4945) (0.95,10267) (0.97,37987)};
\addplot[gray!30, line width=0.4pt, mark=*, mark size=0.6pt] coordinates {(0.60,31) (0.70,81) (0.80,289) (0.85,660) (0.90,1817) (0.93,4890) (0.95,9855) (0.97,34162)};
\addplot[gray!30, line width=0.4pt, mark=*, mark size=0.6pt] coordinates {(0.60,32) (0.70,80) (0.80,291) (0.85,686) (0.90,2079) (0.93,5415) (0.95,11981)};
\addplot[gray!30, line width=0.4pt, mark=*, mark size=0.6pt] coordinates {(0.60,33) (0.70,82) (0.80,270) (0.85,686) (0.90,2164) (0.93,5362) (0.95,12555)};
\addplot[gray!30, line width=0.4pt, mark=*, mark size=0.6pt] coordinates {(0.60,34) (0.70,86) (0.80,311) (0.85,742) (0.90,2046) (0.93,6467) (0.95,15474) (0.97,46101)};
\addplot[gray!30, line width=0.4pt, mark=*, mark size=0.6pt] coordinates {(0.60,34) (0.70,90) (0.80,311) (0.85,734) (0.90,2360) (0.93,6386) (0.95,19210)};
\addplot coordinates {(0.60,48) (0.70,129) (0.80,504) (0.85,1206) (0.90,3877) (0.93,10762) (0.95,27858) (0.97,74776)};
\addplot coordinates {(0.60,33) (0.70,85) (0.80,325) (0.85,705) (0.90,2687) (0.93,6077) (0.95,15530) (0.97,40972)};
\addplot coordinates {(0.60,24) (0.70,57) (0.80,175) (0.85,455) (0.90,1381) (0.93,3646) (0.95,7509) (0.97,26087)};
\addplot coordinates {(0.60,21) (0.70,49) (0.80,174) (0.85,380) (0.90,1030) (0.93,2931) (0.95,6388) (0.97,25514)};
\addplot coordinates {(0.60,47) (0.70,131) (0.80,565) (0.85,1451) (0.90,5317) (0.93,17257) (0.95,38884)};
\addplot coordinates {(0.60,36) (0.70,91) (0.80,342) (0.85,800) (0.90,2620) (0.93,7508) (0.95,14728) (0.97,53355)};
\addplot coordinates {(0.60,31) (0.70,80) (0.80,308) (0.85,710) (0.90,2318) (0.93,5779) (0.95,15456) (0.97,64560)};
\addplot coordinates {(0.60,29) (0.70,72) (0.80,244) (0.85,597) (0.90,1621) (0.93,4575) (0.95,11968) (0.97,43260)};
\addplot coordinates {(0.60,44) (0.70,115) (0.80,429) (0.85,1011) (0.90,3662) (0.93,8885) (0.95,19908)};
\addplot coordinates {(0.60,37) (0.70,96) (0.80,379) (0.85,924) (0.90,3397) (0.93,8852) (0.95,24688)};
\addplot coordinates {(0.60,32) (0.70,83) (0.80,284) (0.85,707) (0.90,2199) (0.93,5942) (0.95,15336) (0.97,38325)};
\addplot coordinates {(0.60,43) (0.70,110) (0.80,445) (0.85,1250) (0.90,4490) (0.93,11823) (0.95,28407)};
\addplot coordinates {(0.60,35) (0.70,98) (0.80,338) (0.85,873) (0.90,2915) (0.93,8434) (0.95,17298)};
\addplot coordinates {(0.60,41) (0.70,110) (0.80,379) (0.85,898) (0.90,3183) (0.93,8736) (0.95,17555) (0.97,70934)};
\addplot[black, line width=1pt, mark=*, mark size=1.2pt] coordinates {(0.60,27) (0.70,63) (0.80,222) (0.85,494) (0.90,1677) (0.93,4209) (0.95,10627) (0.97,29954)};
\draw[dashed] (axis cs:1,10) -- (axis cs:1,2e5);
\end{semilogyaxis}
\end{tikzpicture}
\caption{Relaxation time of Glauber dynamics on planted instances with about $30240$ variables, against $\beta/\beta_d$, for the $62$ degree pairs: the fifteen benchmark pairs in darker gray, with $(3,6)$ in black, and the other $47$ pairs in lighter gray. Each point is the geometric mean over three instances, and points at which some run did not relax within $10^5$ sweeps are omitted. The dashed line marks $\beta=\beta_d$. On every pair, the relaxation time grows rapidly as $\beta$ approaches $\beta_d$.}
\label{fig:relaxation}
\end{figure}

\paragraph{Below the reconstruction threshold.} For each pair, we ran Glauber dynamics on three planted instances with about $30240$ variables at $\beta=x\beta_d$ for $x\in\{0.6,0.7,0.8,0.85,0.9,0.93,0.95,0.97\}$, and stopped each run when $Q(t)\le0.1$, or after $2\times10^5$ sweeps ($10^5$ sweeps for $x=0.97$). Every run with $x\le0.95$ reached $Q(t)\le0.1$, after at most $56293$ sweeps on the benchmark pairs and $33060$ sweeps on the other pairs, and $161$ of the $186$ runs with $x=0.97$ did. Table~\ref{tab:dynamics} lists the relaxation times on the benchmark pairs, Table~\ref{tab:dynamics_more} lists those at $0.9\beta_d$ on the other pairs, and Figure~\ref{fig:relaxation} shows them against $\beta/\beta_d$ for all $62$ pairs. To check that $t_{\mathrm{rel}}$ does not grow with the size, we repeated the runs at $x=0.85$ and $x=0.9$ on one instance with ten times as many variables for each pair. These relaxation times differ from the geometric means over the smaller instances by at most $15\%$ on the benchmark pairs and by at most $21\%$ on the other pairs. We also fitted $\log t_{\mathrm{rel}}=a-\gamma\log(\beta_{\mathrm{div}}-\beta)$ by least squares to the relaxation times with $x\ge0.7$, both for each instance and for the geometric means over the three instances. The fits to the geometric means place $\beta_{\mathrm{div}}$ between $1.004\beta_d$ and $1.039\beta_d$ on every benchmark pair and between $0.998\beta_d$ and $1.034\beta_d$ on the other pairs, with $\gamma$ between $2.7$ and $4.2$, and the fits to single instances place it between $0.98\beta_d$ and $1.10\beta_d$. Such fits depend on the range of $\beta$ used, so they locate the divergence only to within a few percent.

\paragraph{Above the reconstruction threshold.} At $x=1.03$ and $x=1.1$, we ran the dynamics on the same planted instances for $2\times10^4$ sweeps, more than three times the relaxation time at $x=0.9$ for every pair and comparable to that at $x=0.95$. At the end, $Q(t)$ was above $0.84$ at $x=1.03$ and above $0.91$ at $x=1.1$ in every run on the benchmark pairs, and above $0.82$ and $0.90$ on the other pairs, with the smallest values for $k=3$ and $D\ge9$. The dynamics thus stayed close to the planted assignment.

These observations agree with the prediction that $\beta_d$ is the threshold of Glauber dynamics on these instances: on all $62$ pairs, the fits locate the divergence of the relaxation time within a few percent of $\beta_d$, and above $\beta_d$ the dynamics stays close to the planted assignment.

\subsubsection{Equilibration from a uniformly random assignment}\label{app:sim_energy}

The relaxation times of Section~\ref{app:sim_glauber} are measured from an equilibrium sample of a planted instance, whereas a sampler must start from an assignment that it can generate, such as a uniformly random one. We therefore ran Glauber dynamics at fixed inverse temperatures and simulated annealing from uniformly random assignments, on instances with uniformly random right-hand sides, and compared the fraction of violated equations with its equilibrium value. Below $\beta_d$, the expected fraction of violated equations under $\mu_\beta$, averaged over uniformly random right-hand sides, is $p_\beta(1)+o(1)$. Indeed, since $k\ge3$ on all our pairs, Proposition~\ref{prop:sharp} shows that, for every $\beta<\beta_d$ and with probability $1-o(1)$ over the instance, some decoder recovers the codeword with probability $1-o(1)$, and the claim follows from Lemma~\ref{lem:gibbs_cost}.

\paragraph{Glauber dynamics.} For each pair, we ran Glauber dynamics for $5\times10^4$ sweeps from a uniformly random assignment, on one instance with about $30240$ variables, at $\beta=x\beta_d$ for $x\in\{0.9,0.95,1.05,1.1\}$, and averaged the fraction of violated equations over the last fifth of each run. At $x=0.9$ and $x=0.95$, this average differs from $p_\beta(1)$ by at most $0.0024$ on the benchmark pairs and by at most $0.0040$ on the other pairs, whose largest deviations occur at $0.95\beta_d$, where the energy has a long correlation time and its average over one run fluctuates more. At $x=1.05$ and $x=1.1$, the average exceeds $p_\beta(1)$ by between $0.010$ and $0.026$. Started from a uniformly random assignment, Glauber dynamics thus reaches the equilibrium fraction $p_\beta(1)$ of violated equations at $0.9\beta_d$ and $0.95\beta_d$, whereas at $1.05\beta_d$ and $1.1\beta_d$ its fraction of violated equations stays well above $p_\beta(1)$.

\paragraph{Simulated annealing.} We ran simulated annealing on instances with $n=k\cdot\mathrm{round}(10080/k)$ variables. In sweep $t$ of $S$, the inverse temperature is $\beta=4(t-\tfrac12)/S$, so it increases linearly from $0$ to $4$, as for Table~\ref{tab:benchmark}. For every pair, we ran $S=10^4$, $10^5$, and $10^6$ sweeps on the same four instances, and for $29$ pairs also $S=10^7$ sweeps on one or two of these instances (Section~\ref{app:sim_fraction}). We recorded the mean fraction $e_{\mathrm{SA}}(\beta)$ of violated equations at the ends of the sweeps in each of $400$ bins of width $0.01$ in $\beta$.

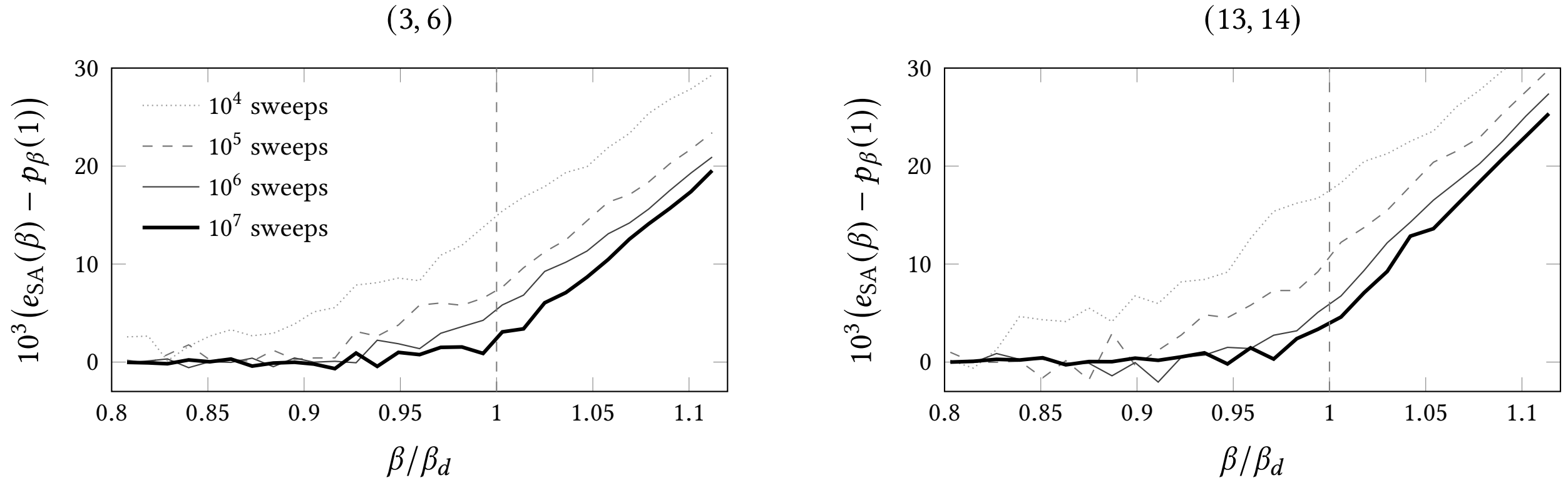
\begin{figure}[t]
\centering
\begin{tikzpicture}
\begin{axis}[
  width=0.49\textwidth, height=5cm,
  xmin=0.8, xmax=1.12, ymin=-3, ymax=30,
  xlabel={$\beta/\beta_d$}, ylabel={$10^3\big(e_{\mathrm{SA}}(\beta)-p_\beta(1)\big)$},
  xlabel style={font=\small}, ylabel style={font=\small}, title style={font=\small},
  tick label style={font=\scriptsize}, every axis plot/.append style={no marks},
  title={$(3,6)$},
  legend style={font=\scriptsize, at={(0.03,0.97)}, anchor=north west, draw=none, fill=none},
]
\addplot[black!40, densely dotted] coordinates {(0.808,2.57) (0.819,2.65) (0.829,0.14) (0.840,1.61) (0.851,2.65) (0.862,3.30) (0.873,2.68) (0.884,2.94) (0.895,3.88) (0.905,5.12) (0.916,5.56) (0.927,7.86) (0.938,8.10) (0.949,8.57) (0.960,8.30) (0.971,10.91) (0.982,11.92) (0.993,13.76) (1.003,15.41) (1.014,16.84) (1.025,17.91) (1.036,19.34) (1.047,19.94) (1.058,21.89) (1.069,23.33) (1.079,25.37) (1.090,26.80) (1.101,27.87) (1.112,29.30)};
\addplot[black!55, dashed] coordinates {(0.808,0.20) (0.819,-0.15) (0.829,0.76) (0.840,1.76) (0.851,0.22) (0.862,0.28) (0.873,0.07) (0.884,1.20) (0.895,0.28) (0.905,0.42) (0.916,0.43) (0.927,3.12) (0.938,2.64) (0.949,3.78) (0.960,5.81) (0.971,6.03) (0.982,5.79) (0.993,6.53) (1.003,7.65) (1.014,9.62) (1.025,11.24) (1.036,12.47) (1.047,14.42) (1.058,16.32) (1.069,17.08) (1.079,18.37) (1.090,20.24) (1.101,21.72) (1.112,23.40)};
\addplot[black!75] coordinates {(0.808,-0.11) (0.819,0.12) (0.829,0.32) (0.840,-0.57) (0.851,0.04) (0.862,-0.03) (0.873,0.40) (0.884,-0.47) (0.895,0.42) (0.905,-0.01) (0.916,0.08) (0.927,-0.07) (0.938,2.23) (0.949,1.87) (0.960,1.37) (0.971,2.94) (0.982,3.61) (0.993,4.26) (1.003,5.83) (1.014,6.83) (1.025,9.25) (1.036,10.19) (1.047,11.33) (1.058,13.09) (1.069,14.19) (1.079,15.61) (1.090,17.51) (1.101,19.29) (1.112,20.92)};
\addplot[black, line width=1.1pt] coordinates {(0.808,-0.03) (0.819,-0.08) (0.829,-0.17) (0.840,0.22) (0.851,0.03) (0.862,0.31) (0.873,-0.41) (0.884,-0.10) (0.895,-0.02) (0.905,-0.20) (0.916,-0.67) (0.927,0.92) (0.938,-0.44) (0.949,0.99) (0.960,0.76) (0.971,1.50) (0.982,1.54) (0.993,0.88) (1.003,3.08) (1.014,3.39) (1.025,6.05) (1.036,7.09) (1.047,8.67) (1.058,10.47) (1.069,12.54) (1.079,14.10) (1.090,15.68) (1.101,17.39) (1.112,19.54)};
\legend{$10^4$ sweeps, $10^5$ sweeps, $10^6$ sweeps, $10^7$ sweeps}
\draw[dashed, black!50] (axis cs:1,-3) -- (axis cs:1,30);
\end{axis}
\end{tikzpicture}\hfill
\begin{tikzpicture}
\begin{axis}[
  width=0.49\textwidth, height=5cm,
  xmin=0.8, xmax=1.12, ymin=-3, ymax=30,
  xlabel={$\beta/\beta_d$}, ylabel={$10^3\big(e_{\mathrm{SA}}(\beta)-p_\beta(1)\big)$},
  xlabel style={font=\small}, ylabel style={font=\small}, title style={font=\small},
  tick label style={font=\scriptsize}, every axis plot/.append style={no marks},
  title={$(13,14)$},
]
\addplot[black!40, densely dotted] coordinates {(0.803,0.18) (0.815,-0.62) (0.827,1.26) (0.839,4.65) (0.851,4.32) (0.863,4.14) (0.875,5.51) (0.887,4.13) (0.899,6.75) (0.911,5.97) (0.923,8.19) (0.935,8.43) (0.947,9.18) (0.959,12.63) (0.971,15.38) (0.983,16.21) (0.994,16.72) (1.006,18.29) (1.018,20.47) (1.030,21.26) (1.042,22.53) (1.054,23.57) (1.066,26.05) (1.078,27.74) (1.090,29.74) (1.102,31.32) (1.114,33.63)};
\addplot[black!55, dashed] coordinates {(0.803,1.00) (0.815,-0.02) (0.827,-0.02) (0.839,0.10) (0.851,-1.60) (0.863,0.14) (0.875,-1.81) (0.887,2.92) (0.899,-0.32) (0.911,1.20) (0.923,2.76) (0.935,4.83) (0.947,4.53) (0.959,5.80) (0.971,7.30) (0.983,7.30) (0.994,9.22) (1.006,12.22) (1.018,13.73) (1.030,15.48) (1.042,17.95) (1.054,20.39) (1.066,21.52) (1.078,22.91) (1.090,25.36) (1.102,27.59) (1.114,29.85)};
\addplot[black!75] coordinates {(0.803,-0.02) (0.815,-0.04) (0.827,0.87) (0.839,0.28) (0.851,0.28) (0.863,-0.11) (0.875,-0.12) (0.887,-1.40) (0.899,-0.05) (0.911,-2.04) (0.923,0.49) (0.935,0.71) (0.947,1.50) (0.959,1.38) (0.971,2.74) (0.983,3.19) (0.994,5.08) (1.006,6.75) (1.018,9.35) (1.030,12.17) (1.042,14.27) (1.054,16.52) (1.066,18.37) (1.078,20.24) (1.090,22.55) (1.102,25.09) (1.114,27.39)};
\addplot[black, line width=1.1pt] coordinates {(0.803,0.00) (0.815,0.08) (0.827,0.28) (0.839,0.21) (0.851,0.44) (0.863,-0.28) (0.875,0.05) (0.887,0.04) (0.899,0.40) (0.911,0.18) (0.923,0.53) (0.935,0.94) (0.947,-0.19) (0.959,1.45) (0.971,0.31) (0.983,2.39) (0.994,3.35) (1.006,4.60) (1.018,7.08) (1.030,9.25) (1.042,12.85) (1.054,13.61) (1.066,16.01) (1.078,18.39) (1.090,20.75) (1.102,23.04) (1.114,25.35)};
\draw[dashed, black!50] (axis cs:1,-3) -- (axis cs:1,30);
\end{axis}
\end{tikzpicture}
\caption{Excess of the fraction of violated equations during simulated annealing over its equilibrium value $p_\beta(1)$, against $\beta/\beta_d$, for $(3,6)$ and $(13,14)$ with about $10080$ variables. The curves are means over four instances, and over one instance for $(3,6)$ and two for $(13,14)$ with $10^7$ sweeps. The dashed lines mark $\beta=\beta_d$. With more sweeps, simulated annealing follows the equilibrium value up to inverse temperatures closer to $\beta_d$.}
\label{fig:sim_energy}
\end{figure}

Figure~\ref{fig:sim_energy} shows the excess $e_{\mathrm{SA}}(\beta)-p_\beta(1)$ for two pairs, and the other pairs behave in the same way. Up to some inverse temperature, the excess fluctuates around zero, so simulated annealing follows the equilibrium value. The excess then grows, and above $\beta_d$ it grows almost linearly, because $p_\beta(1)$ keeps decreasing while the energy of simulated annealing decreases more slowly. As $S$ grows, the inverse temperature at which the excess starts to grow approaches $\beta_d$.

To measure this inverse temperature, let $\beta_S$ be the infimum of the inverse temperatures $\beta\ge0.6\beta_d$ at which the excess, averaged over the bins with centers in $[\beta,\beta+0.03\beta_d]$, exceeds $0.002$. The threshold $0.002$ and the window $0.03\beta_d$ are only one possible choice. The median of $\beta_S/\beta_d$ over the $62$ pairs is $0.79$, $0.89$, and $0.94$ for $S=10^4$, $10^5$, and $10^6$, with ranges $0.71$--$0.85$, $0.84$--$0.93$, and $0.92$--$0.97$. Over the $29$ pairs with $S=10^7$, the median increases from $0.94$ with $S=10^6$ to $0.97$ with $S=10^7$, and all values with $S=10^7$ lie between $0.93$ and $0.99$. For each of the thresholds $0.001$, $0.003$, and $0.005$, the medians also increase towards one as $S$ grows.

The medians approach one roughly as a power of $S$, at a rate consistent with the relaxation times of Section~\ref{app:sim_glauber}. Heuristically, if the relaxation time diverges as $(\beta_{\mathrm{div}}-\beta)^{-\gamma}$, a linear schedule leaves equilibrium where the relaxation time is comparable to the number of sweeps in which $\beta$ changes by $\beta_{\mathrm{div}}-\beta$, which is proportional to $S(\beta_{\mathrm{div}}-\beta)$. The two are comparable when $\beta_{\mathrm{div}}-\beta$ is of order $S^{-1/(\gamma+1)}$, so the fitted exponents $\gamma$ between $2.7$ and $4.2$ predict that $\beta_{\mathrm{div}}-\beta_S$ decreases by a factor between about $1.6$ and $1.9$ for each factor of ten in $S$. With $\beta_{\mathrm{div}}$ the divergence point of the fit to the geometric means for each pair, the median of $(\beta_{\mathrm{div}}-\beta_S)/\beta_d$ decreases by factors of $1.86$, $1.66$, and $1.79$ for the steps from $S=10^4$ to $10^5$, from $10^5$ to $10^6$, and from $10^6$ to $10^7$, where the last factor is computed over the $29$ pairs with $S=10^7$. This agreement is only qualitative, because the observed rate depends on the threshold in the definition of $\beta_S$. We see no sign that $\beta_S$ stops short of $\beta_d$ as $S$ grows.

Simulated annealing therefore reaches the equilibrium energy from a uniformly random start up to inverse temperatures that approach $\beta_d$ as the number of sweeps grows, as a local Markov chain that equilibrates below $\beta_d$ would. With $10^6$ sweeps, its fraction of violated equations differs from the equilibrium value by at most $3.2\times10^{-4}$ on average over $[0.8\beta_d,0.9\beta_d]$ on every pair.

\subsubsection{Simulated annealing and the bound for local decoders}\label{app:sim_fraction}

With enough sweeps, the largest satisfied fraction of simulated annealing, averaged over instances, exceeds $s_d$ on all $62$ pairs, so in our simulations a local classical algorithm reaches the bound that Theorem~\ref{thm:approx_sym} places on DQI with symmetric amplitude functions and local decoders. As for Table~\ref{tab:benchmark}, we record the largest satisfied fraction over the ends of all sweeps of a run; it is close to the final satisfied fraction, as we show below.

\begin{figure}[t]
\centering
\begin{tikzpicture}
\begin{axis}[
  width=0.82\textwidth, height=6.2cm,
  xmode=log, xmin=6e3, xmax=1.6e7, ymin=-9, ymax=15,
  xtick={1e4,1e5,1e6,1e7},
  xlabel={Number of sweeps}, ylabel={$10^3(\mathrm{SA}-s_d)$},
  xlabel style={font=\small}, ylabel style={font=\small},
  tick label style={font=\scriptsize},
  legend style={font=\scriptsize, at={(0.02,0.98)}, anchor=north west, draw=none, fill=none},
  every axis plot/.append style={line width=0.5pt},
]
\addlegendimage{red!70!black, mark=*, mark size=0.8pt}
\addlegendentry{$k=3$}
\addlegendimage{blue!70!black, mark=*, mark size=0.8pt}
\addlegendentry{$k=4$}
\addlegendimage{black!35, mark=*, mark size=0.8pt}
\addlegendentry{$k\ge5$}
\addplot[red!70!black, mark=*, mark size=0.8pt] coordinates {(10000,-1.06) (100000,2.07) (1000000,3.97)};
\addplot[red!70!black, mark=*, mark size=0.8pt] coordinates {(10000,0.90) (100000,5.73) (1000000,8.10)};
\addplot[blue!70!black, mark=*, mark size=0.8pt] coordinates {(10000,-6.03) (100000,-1.21) (1000000,1.73)};
\addplot[red!70!black, mark=*, mark size=0.8pt] coordinates {(10000,4.40) (100000,8.10) (1000000,9.84) (10000000,11.05)};
\addplot[blue!70!black, mark=*, mark size=0.8pt] coordinates {(10000,-2.35) (100000,0.49) (1000000,2.94)};
\addplot[black!35, mark=*, mark size=0.8pt] coordinates {(10000,-4.61) (100000,-1.88) (1000000,1.13) (10000000,1.75)};
\addplot[red!70!black, mark=*, mark size=0.8pt] coordinates {(10000,6.53) (100000,10.04) (1000000,12.13)};
\addplot[blue!70!black, mark=*, mark size=0.8pt] coordinates {(10000,-2.72) (100000,1.88) (1000000,3.87)};
\addplot[black!35, mark=*, mark size=0.8pt] coordinates {(10000,-5.77) (100000,-0.69) (1000000,1.56)};
\addplot[black!35, mark=*, mark size=0.8pt] coordinates {(10000,-7.40) (100000,-2.36) (1000000,0.65) (10000000,1.93)};
\addplot[red!70!black, mark=*, mark size=0.8pt] coordinates {(10000,7.31) (100000,10.60) (1000000,11.96)};
\addplot[blue!70!black, mark=*, mark size=0.8pt] coordinates {(10000,-0.78) (100000,2.92) (1000000,4.50)};
\addplot[black!35, mark=*, mark size=0.8pt] coordinates {(10000,-4.19) (100000,-0.47) (1000000,1.55)};
\addplot[black!35, mark=*, mark size=0.8pt] coordinates {(10000,-5.02) (100000,-1.34) (1000000,0.63) (10000000,2.10)};
\addplot[black!35, mark=*, mark size=0.8pt] coordinates {(10000,-6.30) (100000,-2.39) (1000000,0.65) (10000000,2.19)};
\addplot[red!70!black, mark=*, mark size=0.8pt] coordinates {(10000,7.70) (100000,11.16) (1000000,13.31)};
\addplot[blue!70!black, mark=*, mark size=0.8pt] coordinates {(10000,-1.04) (100000,3.46) (1000000,5.32)};
\addplot[black!35, mark=*, mark size=0.8pt] coordinates {(10000,-4.35) (100000,0.08) (1000000,2.02)};
\addplot[black!35, mark=*, mark size=0.8pt] coordinates {(10000,-6.33) (100000,-1.00) (1000000,1.05)};
\addplot[black!35, mark=*, mark size=0.8pt] coordinates {(10000,-7.12) (100000,-2.10) (1000000,0.25) (10000000,2.22)};
\addplot[black!35, mark=*, mark size=0.8pt] coordinates {(10000,-6.84) (100000,-3.11) (1000000,-0.20) (10000000,1.43)};
\addplot[red!70!black, mark=*, mark size=0.8pt] coordinates {(10000,8.70) (100000,11.81) (1000000,13.07)};
\addplot[blue!70!black, mark=*, mark size=0.8pt] coordinates {(10000,0.06) (100000,3.37) (1000000,5.36)};
\addplot[black!35, mark=*, mark size=0.8pt] coordinates {(10000,-3.31) (100000,0.86) (1000000,2.15)};
\addplot[black!35, mark=*, mark size=0.8pt] coordinates {(10000,-4.72) (100000,-0.50) (1000000,1.02)};
\addplot[black!35, mark=*, mark size=0.8pt] coordinates {(10000,-4.27) (100000,-1.66) (1000000,0.39) (10000000,1.74)};
\addplot[black!35, mark=*, mark size=0.8pt] coordinates {(10000,-6.49) (100000,-2.04) (1000000,0.54) (10000000,2.05)};
\addplot[black!35, mark=*, mark size=0.8pt] coordinates {(10000,-5.11) (100000,-2.21) (1000000,0.29) (10000000,2.32)};
\addplot[red!70!black, mark=*, mark size=0.8pt] coordinates {(10000,9.02) (100000,11.94) (1000000,13.47)};
\addplot[blue!70!black, mark=*, mark size=0.8pt] coordinates {(10000,0.06) (100000,3.73) (1000000,6.25)};
\addplot[black!35, mark=*, mark size=0.8pt] coordinates {(10000,-2.87) (100000,1.10) (1000000,3.01)};
\addplot[black!35, mark=*, mark size=0.8pt] coordinates {(10000,-5.01) (100000,-0.98) (1000000,1.46)};
\addplot[black!35, mark=*, mark size=0.8pt] coordinates {(10000,-7.55) (100000,-1.54) (1000000,0.94)};
\addplot[black!35, mark=*, mark size=0.8pt] coordinates {(10000,-6.77) (100000,-3.00) (1000000,0.57)};
\addplot[black!35, mark=*, mark size=0.8pt] coordinates {(10000,-7.44) (100000,-3.00) (1000000,0.00) (10000000,0.71)};
\addplot[black!35, mark=*, mark size=0.8pt] coordinates {(10000,-6.60) (100000,-2.07) (1000000,-0.74) (10000000,1.43)};
\addplot[red!70!black, mark=*, mark size=0.8pt] coordinates {(10000,9.17) (100000,11.75) (1000000,13.44)};
\addplot[blue!70!black, mark=*, mark size=0.8pt] coordinates {(10000,0.54) (100000,3.62) (1000000,5.98)};
\addplot[black!35, mark=*, mark size=0.8pt] coordinates {(10000,-2.74) (100000,0.50) (1000000,2.67)};
\addplot[black!35, mark=*, mark size=0.8pt] coordinates {(10000,-4.60) (100000,-0.53) (1000000,1.65)};
\addplot[black!35, mark=*, mark size=0.8pt] coordinates {(10000,-5.16) (100000,-1.28) (1000000,0.86)};
\addplot[black!35, mark=*, mark size=0.8pt] coordinates {(10000,-5.46) (100000,-1.75) (1000000,0.43) (10000000,1.95)};
\addplot[black!35, mark=*, mark size=0.8pt] coordinates {(10000,-6.20) (100000,-2.11) (1000000,-0.25) (10000000,0.89)};
\addplot[black!35, mark=*, mark size=0.8pt] coordinates {(10000,-5.57) (100000,-2.72) (1000000,-0.28) (10000000,1.48)};
\addplot[black!35, mark=*, mark size=0.8pt] coordinates {(10000,-6.26) (100000,-2.53) (1000000,0.74) (10000000,1.63)};
\addplot[black!35, mark=*, mark size=0.8pt] coordinates {(10000,-7.72) (100000,-3.28) (1000000,0.02) (10000000,1.23)};
\addplot[black!35, mark=*, mark size=0.8pt] coordinates {(10000,-5.67) (100000,-3.14) (1000000,-1.30) (10000000,0.29)};
\addplot[black!35, mark=*, mark size=0.8pt] coordinates {(10000,-7.47) (100000,-3.23) (1000000,-0.32) (10000000,1.74)};
\addplot[black!35, mark=*, mark size=0.8pt] coordinates {(10000,-5.64) (100000,-2.66) (1000000,-0.43) (10000000,1.15)};
\addplot[black!35, mark=*, mark size=0.8pt] coordinates {(10000,-8.06) (100000,-3.11) (1000000,-0.35) (10000000,1.16)};
\addplot[black!35, mark=*, mark size=0.8pt] coordinates {(10000,-4.68) (100000,-0.96) (1000000,0.75)};
\addplot[black!35, mark=*, mark size=0.8pt] coordinates {(10000,-5.50) (100000,-1.18) (1000000,1.00)};
\addplot[black!35, mark=*, mark size=0.8pt] coordinates {(10000,-5.38) (100000,-1.96) (1000000,0.60)};
\addplot[black!35, mark=*, mark size=0.8pt] coordinates {(10000,-5.26) (100000,-2.01) (1000000,0.57) (10000000,1.39)};
\addplot[black!35, mark=*, mark size=0.8pt] coordinates {(10000,-5.49) (100000,-2.14) (1000000,0.56) (10000000,0.61)};
\addplot[black!35, mark=*, mark size=0.8pt] coordinates {(10000,-5.00) (100000,-3.12) (1000000,-0.29) (10000000,2.21)};
\addplot[black!35, mark=*, mark size=0.8pt] coordinates {(10000,-5.29) (100000,-2.26) (1000000,0.32) (10000000,1.16)};
\addplot[black!35, mark=*, mark size=0.8pt] coordinates {(10000,-5.93) (100000,-2.16) (1000000,-0.03) (10000000,1.73)};
\addplot[black!35, mark=*, mark size=0.8pt] coordinates {(10000,-6.17) (100000,-2.17) (1000000,0.23)};
\addplot[black!35, mark=*, mark size=0.8pt] coordinates {(10000,-6.06) (100000,-2.25) (1000000,0.36) (10000000,1.02)};
\addplot[black!35, mark=*, mark size=0.8pt] coordinates {(10000,-6.42) (100000,-1.66) (1000000,0.32) (10000000,1.62)};
\addplot[black!35, mark=*, mark size=0.8pt] coordinates {(10000,-5.26) (100000,-2.51) (1000000,-0.28) (10000000,1.66)};
\draw[black] (axis cs:6e3,0) -- (axis cs:1.6e7,0);
\end{axis}
\end{tikzpicture}
\caption{Difference between the mean largest satisfied fraction of simulated annealing and $s_d$ against the number of sweeps, for the $62$ degree pairs of Table~\ref{tab:sim_pairs}, with about $10080$ variables. With $10^7$ sweeps, the means are over one or two instances. On every pair, the difference is positive with $10^6$ or $10^7$ sweeps.}
\label{fig:sim_sweeps}
\end{figure}
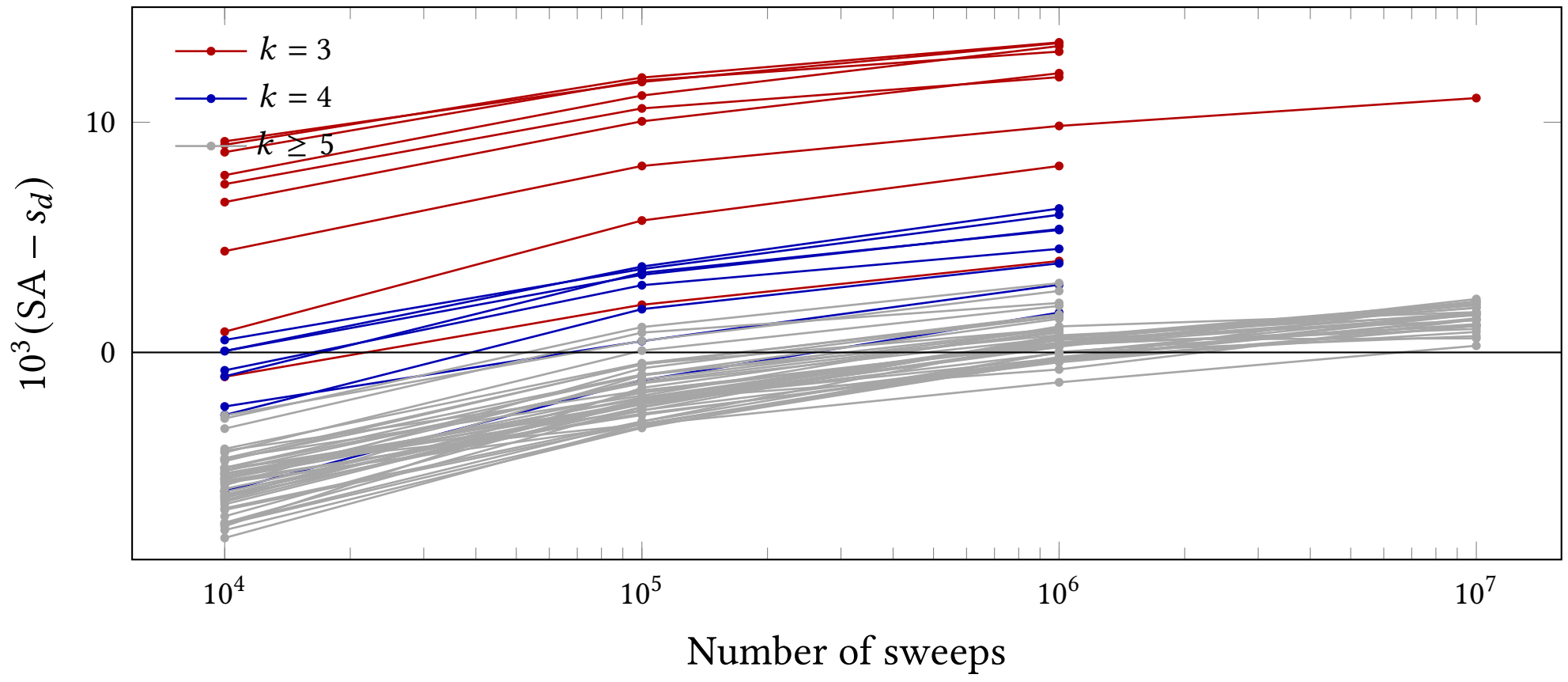

Table~\ref{tab:sim_pairs} and Figure~\ref{fig:sim_sweeps} give the results with about $10080$ variables and the schedules of Section~\ref{app:sim_energy}. On every pair, the mean over the four instances increases from $10^4$ to $10^5$ and from $10^5$ to $10^6$ sweeps, and the gain for each factor of ten decreases slowly: for $k\ge5$, the mean gain over the pairs is $0.0039$ from $10^4$ to $10^5$ sweeps, $0.0024$ from $10^5$ to $10^6$ sweeps, and $0.0014$ from $10^6$ to $10^7$ sweeps on the same instances. With $10^6$ sweeps, the mean over the four instances exceeds $s_d$ on $51$ of the $62$ pairs, including all fifteen benchmark pairs, and the standard deviation over instances is below $0.0014$. On the other eleven pairs, all of which have $k\ge8$, the mean falls short of $s_d$ by at most $0.0013$. We chose the $29$ pairs for $10^7$ sweeps from preliminary results so that they include these eleven, and with $10^7$ sweeps the mean exceeds $s_d$ on all $29$, by $0.0003$ to $0.0023$ for $k\ge5$. The largest satisfied fraction exceeds $s_d$ on $40$ of these $41$ instances; on the remaining one, an instance of $(13,14)$, it falls short by $0.0006$, while on the other instance of this pair it exceeds $s_d$ by $0.0012$. From $10^6$ to $10^7$ sweeps, the largest satisfied fraction increased on $38$ of the $41$ instances; it stayed the same on one instance of $(8,10)$ and decreased by $0.0002$ on one instance of $(9,11)$ and by $0.0005$ on one instance of $(12,24)$. Hence, on every one of the $62$ pairs, the mean exceeds $s_d$ with $10^6$ or $10^7$ sweeps.

The results depend little on the protocol. The column ``SA'' of Table~\ref{tab:benchmark} uses the instance size and the number of sweeps of Shutty et al., $2520$ variables and $10^6$ sweeps, with $\beta$ raised linearly from $0$ to $4$. It gives the mean over six instances with uniformly random right-hand sides, with one run for each instance, and the standard deviation over instances is below $0.0013$. Shutty et al.\ instead record the best of four runs for each instance. For fourteen pairs, our means differ from the values that they report by at most $0.0023$. For $(7,8)$, they list $0.8155$, which is also their value for DQI with belief propagation in that row, whereas we obtain $0.8932$. On the benchmark pairs, the means with about $10080$ variables and $10^6$ sweeps differ from the values in Table~\ref{tab:benchmark} by at most $0.0011$. On the ten pairs $(3,6)$, $(5,6)$, $(7,8)$, $(8,10)$, $(10,11)$, $(13,14)$, $(16,17)$, $(12,16)$, $(12,24)$, and $(18,36)$, the means with about $2520$, $40320$, or $161280$ variables, on four, two, and one instance, differ from those with about $10080$ variables and the same number of sweeps by at most $0.0017$. On six of these pairs, ending the schedule at $\beta=3$ or $\beta=5$ instead of $4$ changes the mean with $10^6$ sweeps by at most $0.0013$, with no systematic direction. Finally, the final satisfied fraction is close to the largest one: in the runs of Table~\ref{tab:sim_pairs} with at least $10^6$ sweeps, their difference has median $1.7\times10^{-4}$ and is at most $0.0018$.

For $k=3$, simulated annealing exceeds $s_d$ by much more: with $10^6$ sweeps, by $0.0040$ at $(3,4)$ and by up to $0.0135$, at $(3,11)$; for $k=4$, it exceeds $s_d$ by $0.0017$ to $0.0063$. The cavity method predicts such an excess, although the energy of simulated annealing leaves its equilibrium value near $\beta_d$.

\paragraph{Comparison with the cavity method.} Krzakala and Zdeborov\'a~\cite{krzakala2010following} describe the limit of a slow annealing as follows: in the limit of many variables, an annealing that is slow enough equilibrates down to the dynamical transition and then stays in one of the clusters into which the Gibbs measure splits there (Section~\ref{sec:status}), so the energy of this cluster, followed to lower temperatures, is the limiting energy of simulated annealing. This energy keeps decreasing as the temperature is lowered, so the limiting satisfied fraction exceeds $s_d$.

Montanari and Ricci-Tersenghi~\cite{montanari2004cooling} studied this behavior numerically for the $3$-spin model with couplings $\pm1$ on random regular hypergraphs in which each variable lies in $4$, $5$, or $8$ interactions, which is Max-XORSAT with uniformly random right-hand sides and $(k,D)=(3,4)$, $(3,5)$, and $(3,8)$ in the units with $\pm1$ spins of Remark~\ref{rem:units}. They cooled instances with $10^5-1$ variables from random configurations. Their dynamical temperatures $T_d=0.757$, $0.936$, and $1.345$, each with uncertainty $0.005$, correspond to $\beta_d=2/T_d=2.642$, $2.137$, and $1.487$, which agree with our values $2.657$, $2.139$, and $1.490$ within this uncertainty. Extrapolated to infinitely slow cooling, the energies in their simulations agree, within their errors, with the iso-complexity energy, which they computed by the 1RSB cavity method: the energy of the metastable clusters whose complexity, the logarithm of their number divided by the number of variables, equals that of the clusters at the dynamical transition. Both works argue that this energy is a lower bound on the energy that a slow annealing reaches~\cite{montanari2004cooling,krzakala2010following}. At zero temperature, it is $-1.17565$, $-1.35122$, and $-1.76036$ per variable. With $nD/3$ equations of energy $\pm1$, an energy $e$ per variable corresponds to the satisfied fraction $1-(3e+D)/(2D)$, so these energies correspond to the satisfied fractions $0.9409$, $0.9054$, and $0.8301$. These upper estimates of the limit of slow annealing exceed $s_d=0.9345$, $0.8946$, and $0.8161$ by $0.006$, $0.011$, and $0.014$, and our means with $10^6$ sweeps, $0.9384$, $0.9027$, and $0.8280$, lie between $s_d$ and these values.

The data fit this picture on every pair that we simulated. Simulated annealing follows the equilibrium energy up to an inverse temperature that approaches $\beta_d$ as the number of sweeps grows, and its energy keeps decreasing after it leaves equilibrium, which with enough sweeps brings its satisfied fraction above $s_d$. The excess over $s_d$ for small $k$ therefore does not contradict Conjecture~\ref{conj:classical}, because simulated annealing does not sample from $\mu_\beta$ at a fixed inverse temperature. For large $k$, simulated annealing gains little after it leaves equilibrium, so reaching $s_d$ requires it to stay in equilibrium nearly up to $\beta_d$. Its close agreement with $s_d$ for large $k$ is therefore further evidence for the conjecture, in agreement with the energies of Section~\ref{app:sim_energy}.

\subsubsection{Larger alphabets}\label{app:sim_alphabets}

For $3\le k<D$, Theorem~\ref{thm:common_cost_local} and Corollary~\ref{cor:common_cost_field} bound the expected cost per constraint for costs that are integer-valued up to an affine rescaling, over cyclic alphabets and finite fields. For two-valued costs over $\Z_q$, Theorem~\ref{thm:approx_sym} gives the stronger neighborhood range. We repeated the numerical comparison of Section~\ref{app:sim_fraction} for unweighted Max-LINSAT, whose cost $f(z)=\mathbf 1[z\ne0]$, the Potts cost, counts the violated equations. For this cost, the upper bound on the expected satisfied fraction is
$$
  s_d=p_{\beta_d}(0)=\frac1{1+(q-1)e^{-\beta_d}},
$$
where $\beta_d$ is the reconstruction threshold for this cost over the alphabet; for $q=2$, this is the value $s_d$ of Max-XORSAT. We ran simulated annealing for this problem over $\Z_q$ with $q\in\{3,4,5,7,8\}$, on the fifteen benchmark pairs and with all nonzero entries of $B$ equal to one, for $75$ tuples $(k,D,q)$ in total. We also simulated the same problem over the fields $\mathbb F_4$ and $\mathbb F_8$, for $30$ further tuples. Since all nonzero entries of $B$ equal one, the equations use only the addition of the field, which is that of $\Z_2^2$ and $\Z_2^3$, so with these alphabets we test whether the results depend on the group.

\begin{table}[t]
\centering
\small
\setlength{\tabcolsep}{4.5pt}
\begin{tabular}{@{}c ccccc cc@{}}
\toprule
& \multicolumn{5}{c}{$\Z_q$} & \multicolumn{2}{c}{Fields} \\
\cmidrule(lr){2-6}\cmidrule(l){7-8}
$(k,D)$ & $q=3$ & $q=4$ & $q=5$ & $q=7$ & $q=8$ & $\mathbb F_4$ & $\mathbb F_8$ \\
\midrule
$(3,4)$ & $2.929$ & $3.131$ & $3.291$ & $3.541$ & $3.643$ & $3.131$ & $3.642$ \\
$(3,5)$ & $2.409$ & $2.609$ & $2.768$ & $3.016$ & $3.116$ & $2.609$ & $3.116$ \\
$(3,6)$ & $2.102$ & $2.298$ & $2.454$ & $2.697$ & $2.797$ & $2.298$ & $2.797$ \\
$(3,7)$ & $1.893$ & $2.084$ & $2.237$ & $2.476$ & $2.574$ & $2.084$ & $2.574$ \\
$(3,8)$ & $1.739$ & $1.925$ & $2.075$ & $2.309$ & $2.405$ & $1.925$ & $2.405$ \\
$(4,5)$ & $2.801$ & $2.989$ & $3.141$ & $3.379$ & $3.476$ & $2.989$ & $3.476$ \\
$(4,6)$ & $2.429$ & $2.617$ & $2.767$ & $3.004$ & $3.101$ & $2.617$ & $3.101$ \\
$(4,7)$ & $2.179$ & $2.363$ & $2.514$ & $2.748$ & $2.844$ & $2.364$ & $2.844$ \\
$(4,8)$ & $1.996$ & $2.178$ & $2.325$ & $2.556$ & $2.651$ & $2.177$ & $2.651$ \\
$(5,6)$ & $2.633$ & $2.815$ & $2.963$ & $3.194$ & $3.289$ & $2.815$ & $3.289$ \\
$(5,7)$ & $2.356$ & $2.537$ & $2.684$ & $2.914$ & $3.009$ & $2.537$ & $3.009$ \\
$(5,8)$ & $2.154$ & $2.334$ & $2.478$ & $2.708$ & $2.801$ & $2.333$ & $2.801$ \\
$(6,7)$ & $2.484$ & $2.661$ & $2.806$ & $3.034$ & $3.128$ & $2.661$ & $3.128$ \\
$(6,8)$ & $2.269$ & $2.445$ & $2.588$ & $2.815$ & $2.908$ & $2.444$ & $2.908$ \\
$(7,8)$ & $2.357$ & $2.532$ & $2.674$ & $2.898$ & $2.991$ & $2.532$ & $2.991$ \\
\bottomrule
\end{tabular}
\caption{Reconstruction thresholds $\beta_d$ for unweighted Max-LINSAT on the fifteen benchmark pairs, over $\Z_q$ and over the fields $\mathbb F_4$ and $\mathbb F_8$, with all nonzero entries of $B$ equal to one. The values are midpoints of final bisection intervals of width $3.7\times10^{-4}$, and $s_d=1/\big(1+(q-1)e^{-\beta_d}\big)$, with $q=4$ and $q=8$ for the fields.}
\label{tab:sim_qbeta}
\end{table}

\begin{table}[t]
\centering
\footnotesize
\setlength{\tabcolsep}{2.6pt}
\begin{tabular}{@{}c rr rr rr rr rr rr rr@{}}
\toprule
& \multicolumn{2}{c}{$\Z_3$} & \multicolumn{2}{c}{$\Z_4$} & \multicolumn{2}{c}{$\Z_5$} & \multicolumn{2}{c}{$\Z_7$} & \multicolumn{2}{c}{$\Z_8$} & \multicolumn{2}{c}{$\mathbb F_4$} & \multicolumn{2}{c}{$\mathbb F_8$} \\
\cmidrule(lr){2-3}\cmidrule(lr){4-5}\cmidrule(lr){6-7}\cmidrule(lr){8-9}\cmidrule(lr){10-11}\cmidrule(lr){12-13}\cmidrule(l){14-15}
$(k,D)$ & $10^6$ & longer & $10^6$ & longer & $10^6$ & longer & $10^6$ & longer & $10^6$ & longer & $10^6$ & longer & $10^6$ & longer \\
\midrule
$(3,4)$ & $+4.4$ & -- & $+4.8$ & -- & $+3.6$ & -- & $+1.2$ & $+4.7$ & $+0.4$ & $+5.5$ & $+4.1$ & -- & $+0.8$ & $+6.0$ \\
$(3,5)$ & $+8.6$ & -- & $+8.1$ & -- & $+7.7$ & -- & $+6.5$ & -- & $+5.7$ & -- & $+7.6$ & -- & $+5.8$ & -- \\
$(3,6)$ & $+11.4$ & -- & $+11.0$ & -- & $+10.4$ & -- & $+9.2$ & -- & $+8.9$ & -- & $+11.0$ & -- & $+8.7$ & -- \\
$(3,7)$ & $+12.5$ & -- & $+12.5$ & -- & $+11.9$ & -- & $+10.8$ & -- & $+10.0$ & -- & $+12.3$ & -- & $+10.7$ & -- \\
$(3,8)$ & $+13.6$ & -- & $+13.4$ & -- & $+12.8$ & -- & $+11.3$ & -- & $+10.9$ & -- & $+13.8$ & -- & $+11.0$ & -- \\
$(4,5)$ & $-0.2$ & $+2.1$ & $-0.9$ & $+2.5$ & $-1.4$ & $+1.7$ & $-3.4$ & $+0.6$ & $-4.7$ & $+1.7$ & $-1.3$ & $+1.1$ & $-4.6$ & $+0.5$ \\
$(4,6)$ & $+2.4$ & -- & $+0.8$ & $+4.2$ & $-0.3$ & $+4.5$ & $-0.3$ & $+3.1$ & $-2.2$ & $+3.6$ & $+1.4$ & $+4.6$ & $-2.5$ & $+3.6$ \\
$(4,7)$ & $+3.4$ & -- & $+2.3$ & -- & $+2.4$ & -- & $-0.7$ & $+5.0$ & $-0.6$ & $+4.4$ & $+1.8$ & $+5.0$ & $-0.9$ & $+4.2$ \\
$(4,8)$ & $+4.4$ & -- & $+3.0$ & -- & $+2.4$ & -- & $+0.8$ & $+5.0$ & $-0.0$ & $+4.6$ & $+3.4$ & -- & $+0.0$ & $+3.5$ \\
$(5,6)$ & $-0.2$ & $+1.2$ & $-2.9$ & $+2.0$ & $-3.0$ & $+2.2$ & $-4.7$ & $+2.2^{\ast}$ & $-5.3$ & $+1.9^{\ast}$ & $-2.6$ & $+0.6$ & $-6.0$ & $+3.1^{\ast}$ \\
$(5,7)$ & $+0.6$ & $+2.1$ & $-0.8$ & $+1.9$ & $-1.9$ & $+0.7$ & $-4.3$ & $+1.5$ & $-4.0$ & $+3.7^{\ast}$ & $-1.0$ & $+3.1$ & $-5.2$ & $+4.1^{\ast}$ \\
$(5,8)$ & $+1.3$ & $+3.3$ & $-0.4$ & $+2.4$ & $-1.5$ & $+2.8$ & $-2.7$ & $+0.7$ & $-3.4$ & $+1.7$ & $-0.1$ & $+2.9$ & $-4.2$ & $+1.4$ \\
$(6,7)$ & $-2.3$ & $+0.5$ & $-2.8$ & $+3.5^{\ast}$ & $-4.3$ & $+0.3$ & $-5.2$ & $+1.0^{\ast}$ & $-6.8$ & $+1.7^{\ast}$ & $-2.1$ & $+1.4$ & $-6.3$ & $+1.6^{\ast}$ \\
$(6,8)$ & $-0.5$ & $+1.9$ & $-1.3$ & $+0.5$ & $-3.0$ & $+1.9$ & $-5.8$ & $+0.8$ & $-6.2$ & $+0.6$ & $-1.9$ & $+1.7$ & $-5.4$ & $+0.6$ \\
$(7,8)$ & $-1.8$ & $+1.6$ & $-3.2$ & $+0.7$ & $-3.7$ & $+1.9^{\ast}$ & $-6.8$ & $+1.9^{\ast}$ & $-6.6$ & $+3.4^{\ast}$ & $-3.1$ & $+1.1$ & $-6.9$ & $+2.2^{\ast}$ \\
\bottomrule
\end{tabular}
\caption{Simulated annealing for unweighted Max-LINSAT over larger alphabets: $10^3$ times the difference between the mean largest satisfied fraction and $s_d$, with $\beta$ raised linearly from $0$ to $7$ on instances with about $10080$ variables. For each alphabet, the first column is the mean over four instances with $10^6$ sweeps, and the second gives the longest runs: $10^7$ sweeps on one instance, or $10^8$ sweeps on one to four instances where marked with an asterisk (Section~\ref{app:sim_alphabets}). A dash means that we did not run more than $10^6$ sweeps, because the mean with $10^6$ sweeps exceeds $s_d$ by at least $0.002$. With the longest runs, the mean exceeds $s_d$ on every tuple.}
\label{tab:sim_qsa}
\end{table}

By Theorem~\ref{thm:duality}, $\beta_d$ is the threshold of the spectral recursion of Definition~\ref{def:spectral}. For the Potts cost, the inverse temperatures without reconstruction form an interval (Section~\ref{sec:monotone}), so bisection locates the transition. Corollary~\ref{cor:field_sharp} and its proof in Appendix~\ref{app:common_field} establish the same tree correspondence and monotonicity over finite fields. We iterated the recursion by population dynamics with $10^5$ messages, counted a run as convergent when the statistic used for Table~\ref{tab:qary} (Appendix~\ref{app:numerics}) fell below $10^{-6}$ within $1000$ levels, and used $14$ bisection steps on $\beta\in[0.5,6.5]$, which leave final intervals of width $3.7\times10^{-4}$. Over $\mathbb F_4$ and $\mathbb F_8$, we used the corresponding recursion on the additive groups $\Z_2^2$ and $\Z_2^3$. Table~\ref{tab:sim_qbeta} lists the midpoints of the final intervals. With these parameters and $q=2$, the bisection for $(3,6)$ ends in $[1.8381,1.8385]$, slightly below the value $1.8388$ of Table~\ref{tab:degrees}, which uses more messages and levels. For $(3,6)$ with $q=3$, $5$, and $8$, the midpoints $2.102$, $2.454$, and $2.797$ lie slightly above the values of Table~\ref{tab:qary}, which come from $10^4$ messages and $120$ levels. On every pair, $\beta_d$ increases with $q$, by $0.63$ to $0.71$ from $q=3$ to $q=8$, and $s_d$ decreases; at $(3,6)$, for example, it decreases from $0.804$ for $q=3$ to $0.701$ for $q=8$. The thresholds over $\mathbb F_4$ and $\mathbb F_8$ differ from those over $\Z_4$ and $\Z_8$ by at most $7.4\times10^{-4}$, two bisection widths, and the final intervals coincide on eight of the fifteen pairs for each field.

The simulations of simulated annealing follow Section~\ref{app:sim_energy}, with two changes. A Metropolis update chooses a variable uniformly at random, proposes to add to it a uniformly random nonzero element of the alphabet, and accepts the change with probability $\min(1,e^{-\beta\Delta})$, where $\Delta$ is the change in the number of violated equations. Because $\beta_d$ is larger than for $q=2$, reaching $3.643$ for $(3,4)$ with $q=8$, the inverse temperature increases linearly from $0$ to $7$ instead of $4$. The instances have about $10080$ variables. For each of the $105$ tuples, we ran $10^4$, $10^5$, and $10^6$ sweeps on four instances. We then ran $10^7$ sweeps on one of these instances for each tuple on which the mean with $10^6$ sweeps is below $s_d+0.002$ ($45$ tuples over $\Z_q$ and $20$ over the fields), and $10^8$ sweeps on the same instance for each tuple on which this instance did not exceed $s_d$ with $10^7$ sweeps. We added a second instance with $10^8$ sweeps for the two tuples over $\Z_q$ with the smallest positive margins at $10^8$ sweeps and for the four tuples over $\mathbb F_8$ that we ran with $10^8$ sweeps, and three further instances for $(7,8)$ with $q=5$, whose first instance remained below $s_d$.

Table~\ref{tab:sim_qsa} gives the results. With $10^6$ sweeps, simulated annealing exceeds $s_d$ on every pair with $k=3$, over every alphabet, but for $k\ge4$ it falls short of $s_d$ on more pairs as $q$ grows. Over $\Z_q$, its mean exceeds $s_d$ on $10$, $8$, $7$, $6$, and $5$ of the fifteen pairs for $q=3$, $4$, $5$, $7$, and $8$, and its largest shortfall grows from $0.0023$ for $q=3$ to $0.0068$ for $q=7$ and $q=8$, whereas for $q=2$ the mean exceeds $s_d$ on all fifteen pairs. More sweeps also help more for larger $q$: on the benchmark pairs with $k\ge5$, the mean gain from $10^5$ to $10^6$ sweeps is $0.0046$ for $q=3$ and $0.0090$ for $q=8$, against $0.0026$ for $q=2$. On the $45$ tuples over $\Z_q$ that we ran with $10^7$ sweeps, simulated annealing gained from $0.0003$ to $0.0068$ between $10^6$ and $10^7$ sweeps on the same instance, and with $10^7$ sweeps it exceeds $s_d$ on $36$ of them. On the other nine, all with $k\ge5$ and $q\ge4$, it exceeds $s_d$ with $10^8$ sweeps on the first instance, except on $(7,8)$ with $q=5$, where the first instance reached the same largest satisfied fraction with $10^7$ and $10^8$ sweeps, $0.0005$ below $s_d$. On the three further instances of this tuple, simulated annealing exceeds $s_d$ by $0.0014$, $0.0030$, and $0.0037$ with $10^8$ sweeps, so the mean over the four instances exceeds $s_d$ by $0.0019$.

Over $\mathbb F_4$ and $\mathbb F_8$, the means with $10^6$ sweeps differ from those over $\Z_4$ and $\Z_8$ by at most $0.0012$. With $10^7$ sweeps, simulated annealing exceeds $s_d$ on all nine tuples over $\mathbb F_4$ and on seven of the eleven over $\mathbb F_8$ that we ran with this number of sweeps, and on the other four it exceeds $s_d$ with $10^8$ sweeps on both instances. Within the accuracy of these computations, neither $\beta_d$ nor the satisfied fraction of simulated annealing depends on whether the alphabet is the cyclic group or the additive group of the field of the same size. Hence, on each of the $105$ tuples, the mean over the instances with the largest number of sweeps that we ran exceeds $s_d$.

These margins are small. On twelve tuples, nine over $\Z_q$ and three over the fields, the margin at the largest number of sweeps is below $0.001$ and comes from one instance with $10^7$ sweeps, so it is comparable to the standard deviation over instances, which is below $0.0017$ with $10^6$ sweeps. In every comparison on the same instance from $10^6$ sweeps on, however, the margin grew with the number of sweeps, except on the first instance of $(7,8)$ with $q=5$, where it stayed the same from $10^7$ to $10^8$ sweeps.

Simulated annealing also leaves equilibrium farther below $\beta_d$ for larger $q$, in agreement with the larger number of sweeps that it requires. With the excess now taken over the equilibrium fraction $1-p_\beta(0)$ of violated equations, the median of $\beta_S/\beta_d$ over the benchmark pairs with $10^6$ sweeps decreases from $0.914$ for $q=3$ to $0.884$ for $q=8$, against $0.942$ for $q=2$. With more sweeps, $\beta_S/\beta_d$ approaches one as for $q=2$: over the $45$ tuples over $\Z_q$ with $10^7$ sweeps, its median increases from $0.893$ with $10^6$ sweeps to $0.933$ with $10^7$ sweeps, and over the nine tuples with $10^8$ sweeps, from $0.939$ with $10^7$ sweeps to $0.956$ with $10^8$ sweeps.

Part of the larger number of sweeps is due to the Metropolis update, which proposes a uniformly random value among the $q-1$ others, but not all of it. A heat-bath update, which draws the new value of the chosen variable from its conditional law under $\mu_\beta$, reaches more in $10^6$ sweeps for $q=8$: on $(4,5)$, $(5,6)$, $(6,7)$, and $(7,8)$, its mean over two instances falls short of $s_d$ by $0.0002$ to $0.0029$, against $0.0047$ to $0.0068$ for Metropolis updates on four instances. For $q=8$, however, a heat-bath sweep costs between $4.2$ and $5.3$ times as much computing time as a Metropolis sweep, and Metropolis updates with $10^7$ sweeps reach more on all four pairs, at about twice the computing time. Finally, on $(3,6)$, $(5,6)$, and $(7,8)$ with $q\in\{3,8\}$, ending the schedule at $\beta=5$ or $\beta=9$ instead of $7$ changes the mean with $10^6$ sweeps by between $-0.0023$ and $0.0015$.

\subsubsection{Scope of the numerical evidence}\label{app:sim_summary}

The annealing experiments establish a practical finite-instance comparison: the mean largest satisfied fraction exceeds the estimated local-decoding bound on every degree pair and alphabet tested. More sweeps improve the mean from $10^4$ to $10^6$ on every family, and in $121$ of the $125$ comparisons with a longer run on the same instance, the largest satisfied fraction increases. The longer runs were selected from preliminary results and include every family whose shorter runs fell below $s_d$. Instance counts at the largest budgets range from one to four, and the smallest margins are comparable to fluctuations between instances, as reported above.

The fixed-temperature results support the predicted relation between reconstruction and classical relaxation, but do not prove polynomial-time mixing. We measure decorrelation from equilibrium and the energy from a uniformly random start on sizes up to about $3\times10^5$ variables. These observables are not total-variation mixing times. The optimization comparison requires no inference about mixing: annealing continues to lower the energy after leaving equilibrium, so attaining a satisfied fraction above $s_d$ is compatible with slow fixed-temperature dynamics above $\beta_d$.

All simulated instances come from Gallager's ensemble. The larger-alphabet experiments use simulated annealing with the Potts cost; they do not test Gibbs sampling over those alphabets or costs with more than two values. The theorem for costs that are integer-valued up to an affine rescaling has broader scope than these experiments.

\clearpage

\section{Proof of Proposition~\ref{prop:sharp}}\label{app:sharp}

Proposition~\ref{prop:sharp} says that, for $3\le k<D$, over every $\Z_q$, for every cost, and for both ensembles, the two-stage decoder of Mandal et al.\ with the node operations of Lemma~\ref{lem:nodes} recovers the codeword with probability tending to one at every finite inverse temperature without reconstruction, and that it is local. The proof combines three properties. First, the optimal error for the root of the tree code tends to zero doubly exponentially in the depth, so that depth $t_m=O(\log\log m)$ already gives error at most $m^{-8}$ (Appendix~\ref{app:decay}, Lemmas~\ref{lem:checkbound} and~\ref{lem:decay}). Second, by Theorem~\ref{thm:tp_typical}, the local conditions $\mathrm{LT}_{t_m}$ and $\mathrm{TP}_{t_m}$ fail only at a vanishing fraction of the symbols. Third, the minimum distance of $C$ is linear in $m$ (Appendix~\ref{app:distance}, Lemma~\ref{lem:distance}), so the values on the symbols where the local conditions hold determine the codeword. In Appendix~\ref{app:sequential} we measure the symbols one after another, bound the probability that some measurement errs by the sum of the square roots of their errors, complete the codeword by solving linear congruences, and bound the effect of finite precision. Appendix~\ref{app:random_order_local} then proves the joint locality of uniformly random sequential schedules in Corollary~\ref{cor:random_order_local}.

In the proof of Proposition~\ref{prop:sharp}, $\beta$ is finite, so $p=p_\beta$ is strictly positive. For a probability vector $\nu$ on $\Z_q$, recall that $\hat\nu(u)=\sum_z\nu(z)\omega^{uz}$, and let
$$
  d_1(\nu)=\|\nu-\unif\|_1,\qquad d_F(\nu)=\sum_{u\ne0}|\hat\nu(u)| .
$$
For $u\ne0$, $|\hat\nu(u)|=|\sum_z(\nu(z)-1/q)\omega^{uz}|\le d_1(\nu)$, and Fourier inversion gives $|\nu(z)-1/q|\le d_F(\nu)/q$, so $d_1(\nu)\le d_F(\nu)\le(q-1)d_1(\nu)$. Both quantities are invariant under translation and under the reflection $R$.

\subsection{Doubly exponential convergence of the tree error}\label{app:decay}

The tree error converges doubly exponentially because, at a symbol node, the nontrivial Fourier coefficients of the incoming spectra multiply, so they decay at a rate of degree $k-1\ge2$ once they are small, while a check node increases their expected sum by at most a constant factor. Mandal et al.\ prove doubly exponential convergence for prime $q$ throughout the region in which BPQM succeeds on the tree~\cite[Theorem~26]{mandal2026blockerror}, and the following two lemmas give it for every $q$ under non-reconstruction.

\begin{lemma}[Expected distance after a check node]\label{lem:checkbound}
For probability vectors $\lambda,\mu$ on $\Z_q$,
$$
  \E\,d_1\big(\Gamma(\lambda,\mu)\big)\le d_1(\lambda)+d_1(\mu)+\min\{d_1(\lambda),d_1(\mu)\}\le2\big(d_1(\lambda)+d_1(\mu)\big).
$$
For the spectra of Definition~\ref{def:spectral}, let $v_t=\E\,d_F(V^{(t)})$ and $c_{q,D}=(q-1)(D-1)2^{D-2}$. Then $\E\,d_F(C^{(t)})\le c_{q,D}v_t$.
\end{lemma}

\begin{proof}
Let $J(\delta,l)=\lambda(l+\delta)\mu(l)$ and $P=P_{\lambda,\mu}$, so that $P(\delta)=\sum_lJ(\delta,l)$ and $\Gamma(\lambda,\mu)=\nu_\Delta$ with $\Delta\sim P$ and $\nu_\delta=J(\delta,\cdot)/P(\delta)$ (Lemma~\ref{lem:nodes}). Hence
$$
  \E\,d_1\big(\Gamma(\lambda,\mu)\big)=\sum_\delta P(\delta)\sum_l\Big|\frac{J(\delta,l)}{P(\delta)}-\frac1q\Big|=\sum_{\delta,l}\Big|J(\delta,l)-\frac{P(\delta)}q\Big| .
$$
By the triangle inequality, this is at most $\sum_{\delta,l}|J(\delta,l)-q^{-2}|+\sum_{\delta,l}|q^{-2}-P(\delta)/q|$. The map $(\delta,l)\mapsto(l+\delta,l)$ is a bijection of $\Z_q^2$, so the first sum is $\|\lambda\otimes\mu-\unif\otimes\unif\|_1\le d_1(\lambda)+d_1(\mu)$. The second sum is $d_1(P)$. Since $P=\lambda*R\mu$ and convolution with a probability vector does not increase the distance from $\unif$, $d_1(P)\le\min\{d_1(\lambda),d_1(\mu)\}$. This proves the first claim. The check-node operation with $D-1$ arguments applies $\Gamma$ to the output of the previous step and a new argument, with fresh randomness. Applying the weaker bound conditionally at each step and taking expectations shows that the expected $d_1$ of the output is at most $2^{D-2}$ times the sum of the $d_1$ of the arguments. With $d_F\le(q-1)d_1$, $d_1\le d_F$, and the invariance under $R$, this gives $\E\,d_F(C^{(t)})\le(q-1)2^{D-2}(D-1)v_t$.
\end{proof}

\begin{lemma}[Doubly exponential convergence]\label{lem:decay}
Let $q\ge2$ and $3\le k<D$, and let $p$ be a strictly positive probability vector on $\Z_q$ at which the primal model has non-reconstruction. There are constants $A_0,c_\ast>0$ such that $1-P_t\le A_0\exp\big(-c_\ast(k-1)^t\big)$ for all sufficiently large $t$. In particular, for $t_m=2\lceil\log_{k-1}(1+\ln m)\rceil$ and all sufficiently large $m$, $1-P_{t_m}\le m^{-8}$.
\end{lemma}

\begin{proof}
Write $h=k-1\ge2$ and $c=c_{q,D}$. By Definition~\ref{def:spectral}, the spectrum $V^{(t+1)}$ is the convolution of $p$ with $h$ independent copies of $C^{(t)}$, and Fourier coefficients multiply under convolution. Since $|\hat p(u)|\le1$ and $\sum_ux_u^h\le(\sum_ux_u)^h$ for nonnegative $x_u$, Lemma~\ref{lem:checkbound} gives
\begin{equation}\label{eq:decay_recursion}
  v_{t+1}=\sum_{u\ne0}|\hat p(u)|\Big(\E\big|\hat C^{(t)}(u)\big|\Big)^h\le\Big(\E\,d_F\big(C^{(t)}\big)\Big)^h\le(c\,v_t)^h .
\end{equation}
By part~1 of Theorem~\ref{thm:duality}, $d_F(C^{(t)})$ has the same law as $d_F(M^{(t+1)})$, which tends to zero in probability under non-reconstruction. Since $d_F\le q-1$, $\E\,d_F(C^{(t)})\to0$, and the middle term of~\eqref{eq:decay_recursion} shows that $v_t\to0$. Let $y_t=c^{h/(h-1)}v_t$. Then~\eqref{eq:decay_recursion} gives $y_{t+1}\le c^{h/(h-1)+h}v_t^h=y_t^h$. Choose $t_0$ with $y_{t_0}\le\tfrac12$. Then $y_{t_0+s}\le2^{-h^s}$ for every $s\ge0$, so $v_t\le c^{-h/(h-1)}2^{-h^{t-t_0}}$ for $t\ge t_0$.

The root spectrum $N^{(t)}$ is the convolution of $p$ with $k$ independent copies of $C^{(t-1)}$, and $\E|\hat C^{(t-1)}(u)|\le1$, so $\E\,d_F(N^{(t)})\le v_t$ for $t\ge1$. The function $S(\nu)$ of Definition~\ref{def:spectral} is the square of the Bhattacharyya coefficient $\mathrm{BC}$ of $\nu$ and $\unif$, so $1-S(\nu)\le2(1-\mathrm{BC})\le2\|\nu-\unif\|_{\mathrm{TV}}=d_1(\nu)\le d_F(\nu)$. Hence $1-P_t=\E\big(1-S(N^{(t)})\big)\le v_t$, which proves the first claim with $A_0=c^{-h/(h-1)}$ and $c_\ast=(\ln2)\,h^{-t_0}$. For the second claim, $h^{t_m}\ge(1+\ln m)^2$, so $A_0\exp(-c_\ast h^{t_m})\le m^{-8}$ for all sufficiently large $m$. The radius $t_m$ does not depend on $t_0$.
\end{proof}

\subsection{Linear minimum distance}\label{app:distance}

Let $m_{\min}(C)$ be the smallest weight of a nonzero codeword of $C$, with $m_{\min}(C)=\infty$ if $C=\{0\}$. The following lemma uses only that a check cannot meet the support of a codeword in exactly one half-edge, so it holds for every $q$. Let $E_D(x)=(1+x)^D-Dx$, as $E_k$ in Section~\ref{sec:typical} with $D$ in place of $k$, and $J_D=2^D-D-1$. For $0<x\le1$, every term of $E_D(x)$ other than the constant term has degree at least two, so $E_D(x)\le1+J_Dx^2\le\exp(J_Dx^2)$.

\begin{lemma}[Linear minimum distance]\label{lem:distance}
For every $q\ge2$ and $3\le k<D$, there is $\gamma_0>0$ such that, for random $(k,D)$-regular instances over $\Z_q$ from the configuration model or from Gallager's ensemble, $\Pr[m_{\min}(C)<\gamma_0m]=o(1)$.
\end{lemma}

\begin{proof}
Let $N_w$ be the number of codewords of weight $w$, and write $\varrho=k/2-1\ge\tfrac12$.

\emph{Configuration model.} Work in the unconditioned model, in which $B$ is the incidence matrix of the multigraph, with multiplicities, reduced modulo $q$. If a word $c$ with support $Y$ of size $w$ lies in $C$, then no check meets the $k w$ half-edges of $Y$ in exactly one half-edge, since at such a check the sum of the values of $c$ would be a single nonzero value. These half-edges are matched to a uniformly random set of $k w$ of the $mk$ half-edges of the checks, and the number of such sets that meet no check in exactly one half-edge is the coefficient of $x^{k w}$ in $E_D(x)^n$, with $n=mk/D$. There are at most $(q-1)^w$ words with a given support, so, for every $x>0$,
$$
  \E N_w\le(q-1)^w\binom mw\binom{mk}{k w}^{-1}E_D(x)^nx^{-k w}.
$$
With $x=\sqrt{w/m}$, $E_D(x)^n\le\exp(J_Dk w/D)$, $\binom mw\le(em/w)^w$, and $\binom{mk}{k w}\ge(m/w)^{k w}$, this gives $\E N_w\le\big[A_{\mathrm{conf}}(w/m)^{\varrho}\big]^w$ with $A_{\mathrm{conf}}=(q-1)\exp(1+J_Dk/D)$.

\emph{Gallager's ensemble.} Let $A_w$ be the number of sets of $w$ symbols that meet no check of the first layer in exactly one symbol, which is the coefficient of $x^w$ in $E_D(x)^{m/D}$. The support of a codeword meets no check of any layer in exactly one symbol. Each other layer is the image of the first under an independent uniformly random permutation, so a fixed set of size $w$ meets no check of that layer in exactly one symbol with probability $A_w/\binom mw$. Hence
$$
  \E N_w\le(q-1)^w\frac{A_w^k}{\binom mw^{k-1}}\le\big[A_{\mathrm{Gal}}(w/m)^{\varrho}\big]^w,\qquad A_{\mathrm{Gal}}=(q-1)\exp(J_Dk/D),
$$
where the second inequality uses $A_w\le E_D(x)^{m/D}x^{-w}$ with $x=\sqrt{w/m}$ and $\binom mw\ge(m/w)^w$.

\emph{Both ensembles.} Let $A=\max\{A_{\mathrm{conf}},A_{\mathrm{Gal}}\}$, and choose $\gamma_0\in(0,1)$ with $A\gamma_0^{\varrho}\le\tfrac12$. For $w\le\ln^2m$, the base $A(w/m)^\varrho$ is at most $b_m=A(\ln^2m/m)^\varrho$, which tends to zero, so these terms sum to at most $b_m/(1-b_m)$. For $\ln^2m<w<\gamma_0m$, the base is at most $\tfrac12$, so these terms sum to at most $2^{1-\ln^2m}$. By Markov's inequality, $\Pr[m_{\min}(C)<\gamma_0m]=o(1)$. Conditioning the configuration model on the absence of multiple edges, an event of probability bounded below by a positive constant, multiplies this probability by at most a constant.
\end{proof}

For $q=2$ and the configuration model, Lemma~\ref{lem:min_distance} gives an explicit $\gamma_0$ and the rate $O(m^{-(k-2)/2})$.

\subsection{Sequential measurements and completion}\label{app:sequential}

\begin{proof}[Proof of Proposition~\ref{prop:sharp}]
Let $\mathcal G=\{i:\mathrm{LT}_{t_m}(i)\text{ and }\mathrm{TP}_{t_m}(i)\}$. Since $t_m=O(\log\log m)$, it satisfies the hypothesis of Theorem~\ref{thm:tp_typical} for all sufficiently large $m$, so $\E|[m]\setminus\mathcal G|=o(m)$, and by Markov's inequality $|[m]\setminus\mathcal G|<\gamma_0m$ with probability $1-o(1)$, for the $\gamma_0$ of Lemma~\ref{lem:distance}. By Lemma~\ref{lem:distance}, $m_{\min}(C)\ge\gamma_0m$ with probability $1-o(1)$. We work on an instance with both properties, on which $\mathcal G$ is nonempty.

\emph{The measurements.} At every $i\in\mathcal G$, the decoder applies the optimal measurement for the root of the tree code of depth $t_m$ to the qudits of $N_{t_m}(i)$, that is, the node operations of Lemma~\ref{lem:nodes} followed by the square-root measurement (Proposition~\ref{prop:tree_optimal}). Under $\mathrm{TP}_{t_m}(i)$ and $\mathrm{LT}_{t_m}(i)$, the restriction of a uniformly random codeword to $N_{t_m}(i)$ is a uniformly random codeword of the tree code (Section~\ref{sec:strategy}), so the error of this measurement, applied to the original state and averaged over the codeword, is $e_m=1-P_{t_m}\le m^{-8}$ by Lemma~\ref{lem:decay}. The decoder applies these measurements one after another, in any order fixed before processing the received input. For each, it uses fresh ancillas for a Naimark dilation, copies the outcome into a record register, and undoes the dilation. The balls of different symbols may overlap.

\emph{The probability that some measurement errs.} Fix a codeword $c$, let $\Psi_c$ be the original state with all ancillary registers, and let $e_{c,i}$ be the error of the measurement at $i$ applied to this state alone. For its dilation $V_i$, define $\Pi_i=V_i^\dagger(\ket{c_i}\!\bra{c_i}\otimes I)V_i$, with the projector acting on the outcome register.

Applying $V_i$, copying the outcome $o$, and undoing $V_i$ maps $\psi$ to $\sum_oV_i^\dagger(\ket o\!\bra o\otimes I)V_i\psi\otimes\ket o$. The branch with all records correct is therefore $\Pi_g\cdots\Pi_1\Psi_c$ with the correct records, where $g=|\mathcal G|$ and the symbols are numbered in their processing order. Its probability is the squared norm of this vector. Since projections do not increase norms, telescoping gives
$$
  \|\Psi_c-\Pi_g\cdots\Pi_1\Psi_c\|\le\sum_{i\in\mathcal G}\|(I-\Pi_i)\Psi_c\|=\sum_{i\in\mathcal G}\sqrt{e_{c,i}} .
$$
If the right-hand side is at most one, the reverse triangle inequality and $1-y^2\le2(1-y)$ for $y\in[0,1]$ bound the failure probability by twice that quantity. If it exceeds one, the same bound holds trivially. Averaging over codewords and applying Jensen's inequality gives
$$
  \Pr[\text{some recorded value is incorrect}]\le2\sum_{i\in\mathcal G}\E_c\sqrt{e_{c,i}}\le2m\sqrt{e_m}\le2m^{-3}.
$$

\emph{Completion and locality.} The decoder then solves the linear congruences $B^\top\hat c=0$ for the coordinates outside $\mathcal G$, given the recorded values on $\mathcal G$. If all recorded values are correct, $c$ is a solution, and it is the only one, because two solutions differ by a codeword supported on $[m]\setminus\mathcal G$, whose weight is less than $\gamma_0m\le m_{\min}(C)$. For fixed $q$, the Smith normal form solves the congruences in polynomial time. If the system has no solution, the decoder outputs zeros outside $\mathcal G$, and if it has several, it outputs a fixed one; in every case it keeps the recorded values on $\mathcal G$. Hence the block success is at least $1-2m^{-3}$, uniformly over all input-independent orders, because the preceding failure bound is a sum over the original-input errors and does not depend on their order.

If the order starts with a maximal set $S\subseteq\mathcal G$ of symbols whose balls are pairwise disjoint, their recorded values are the outcomes of separate measurements on the original input, each with its own ancillas. The decoder outputs them on every branch, so Lemma~\ref{lem:disjoint_measurements} gives $t_m$-local estimates on $S$. Since $S$ is maximal, every symbol of $\mathcal G$ lies within distance $4t_m$ of $S$, so $|S|\ge|\mathcal G|/\max_i|N_{2t_m}(i)|\ge(1-\gamma_0)m/\max_i|N_{2t_m}(i)|$. The denominator is polylogarithmic in $m$, so $|S|=m^{1-o(1)}$. For a uniformly random order of $\mathcal G$, Corollary~\ref{cor:random_order_local} instead gives joint locality with $s_m=\lceil(\log m)^2\rceil$ and exceptional probability at most $4^{-s_m}$. The reliability bound is uniform over the order, and completion retains every record, so both schedules satisfy the proposition.

\emph{Polynomial time.} Each ball contains $((k-1)(D-1))^{O(t_m)}$ symbols, which is polylogarithmic in $m$. For fixed $q$, Lemma~\ref{lem:nodes} implements each node on a fixed number of qudits, with parameters computed reversibly from the earlier labels, followed by a Fourier measurement at the root.

Let $p_{\min}>0$ be the smallest entry of $p$. A symbol spectrum contains $p$ as a convolution factor, so all its entries are at least $p_{\min}$. At a check, the unnormalized products of at most $D-1$ such inputs are at least $p_{\min}^{D-1}$, and their normalizing sums are at least $qp_{\min}^{D-1}$. The conditional probabilities used by the isometry in Lemma~\ref{lem:nodes}(1) are at least $p_{\min}^D$. Thus a constant $\eta_0>0$, depending only on $p,q,D$, bounds every probability and denominator needed by the implementation away from zero.

On the compact sets where these quantities are at least $\eta_0/2$, the normalized products, convolutions, and state-preparation parameters are Lipschitz, with a constant $L_0\ge1$. A tree level uses a bounded number of these operations, so errors of size $2^{-b}$ are amplified by at most $L_0^{c't_m}$, where $c'$ depends only on the degrees. Choose $b(m)=\lceil\log_2^2(m+1)\rceil$, without requiring the values of $p_{\min}$, $L_0$, or $t_0$. For all sufficiently large $m$, propagated errors remain in the stated compact sets. If $M_m$ is a polynomial bound on the number of arithmetic and circuit operations, including elementary-gate approximation to $b(m)$ bits, the total error is at most a constant times $M_mL_0^{c't_m}2^{-b(m)}=o(m^{-3})$.

When $p_\beta$ is computable to $b$ bits in time polynomial in $b$, the state preparations and reversible parameter computations take polynomial time. Ball exploration, the projection test comparing $C|_{N_{t_m}(i)}$ with $\mathcal C_{t_m}(i)$, and completion use polynomial-time computations with linear congruences. The telescoping estimate for approximate operations adds only $o(1)$ to the failure probability. Every approximated call still acts on the same received ball and fresh ancillas and preserves its record, so the locality arguments for both supported schedules remain valid.

\emph{Consequences.} Since $p_\beta$ is positive, every coset has positive mass, and part~1 of Theorem~\ref{thm:guarantee} shows that the output law is within $\sqrt{2m^{-3}}$ of $\pi_\beta$ in expected total variation over $x_0$, and within $o(1)$ for the efficient implementation. Apply part~2 of Theorem~\ref{thm:guarantee} to $h(x)=\frac1m\sum_j(f(x_j)-\min f)/(\max f-\min f)$ when $f$ is not constant. Since $\sum_x|\Phi_\beta(x)|^2h(x)=(\langle f\rangle_{p_\beta}-\min f)/(\max f-\min f)$, the expected cost per constraint differs from $\langle f\rangle_{p_\beta}$ by $o(1)$. For Max-XORSAT, $\langle f\rangle_{p_\beta}=p_\beta(1)$, so the expected satisfied fraction is $p_\beta(0)-o(1)$.
\end{proof}

\subsection{Locality of uniformly random sequential schedules}\label{app:random_order_local}

We prove Corollary~\ref{cor:random_order_local} by finding sufficiently many early calls whose neighborhoods are untouched by earlier calls. Their joint measurement law follows from Lemma~\ref{lem:disjoint_measurements}; it remains to bound the probability that too few such calls exist. This argument applies to every amplitude function and does not require a product received state.

\begin{proof}[Proof of Corollary~\ref{cor:random_order_local}]
Let $\mathcal G_m$ be the symbols satisfying $\mathrm{LT}_{t_m}$ and $\mathrm{TP}_{t_m}$. By Corollary~\ref{cor:typical_nogo} and Markov's inequality, there is a deterministic sequence $\eta_m\to0$ such that, with probability $1-o(1)$ over the instance, $|[m]\setminus\mathcal G_m|\le\eta_m m$. Intersect this event with $|J_m|\ge\lambda m$, and fix an instance in the intersection. Put $\delta=\tfrac12-c_0\log((k-1)(D-1))>0$.

Breadth-first exploration bounds the number of symbols in a ball, for every nonnegative integer $u$, by
$$
  |N_u(i)|\le1+k(D-1)\sum_{h=0}^{u-1}((k-1)(D-1))^h\le5((k-1)(D-1))^u.
$$
If $N_{t_m}(i)$ and $N_{t_m}(j)$ intersect, then $j\in N_{2t_m}(i)$. Thus each ball meets at most $5m^{1-2\delta}$ balls with other centers.

Expose the first $2s_m$ symbols of the uniformly random order, where $s_m=\lceil(\log m)^2\rceil$. Keep a symbol if it belongs to $\mathcal G_m$ and its ball meets no ball of a previously processed symbol, including previously rejected symbols. The kept balls are pairwise disjoint, and no earlier call touches their received qudits or fresh ancillary registers. For each fixed order, Lemma~\ref{lem:disjoint_measurements} therefore gives the joint local measurement law at all kept symbols.

Conditioned on any prefix of length $r-1<2s_m$, the next symbol is uniform among the remaining symbols of $J_m$. At most $\eta_m m$ possible choices are outside $\mathcal G_m$, and at most $5(r-1)m^{1-2\delta}$ have balls meeting an earlier ball. Consequently, for all sufficiently large $m$, the conditional rejection probability is at most
$$
  \frac{\eta_m m+10s_m m^{1-2\delta}}{\lambda m-2s_m}=o(1)\le\frac1{16}.
$$
For any specified $s_m$ positions among the first $2s_m$, repeated conditioning bounds the probability that all of them are rejected by $16^{-s_m}$. If fewer than $s_m$ symbols are kept, at least $s_m$ positions are rejected, so a union bound gives
$$
  \Pr[\text{fewer than }s_m\text{ symbols are kept}\mid B]\le\binom{2s_m}{s_m}16^{-s_m}\le4^{-s_m}\le2^{-s_m}.
$$
The kept set depends only on the instance and the order, not on the received input or on measurement outcomes. Since every recorded estimate is output on every branch, every nonexceptional order gives $t_m$-local estimates on at least $s_m$ symbols. The order is the input-independent random choice in Definition~\ref{def:local_decoder}, and $s_m/\log m\to\infty$, which proves the corollary.
\end{proof}

\section{Coherent Gibbs preparation from optimal decoding}\label{app:pgm}

A coherent PGM and its inverse prepare the primal Gibbs state of any specified offset, with heralded success equal to the PGM's block success (Proposition~\ref{prop:pgm_gibbs}). We first show that the correct branches have the same overlap for every codeword. Comparing the decoded and original labels without measuring either label preserves their superposition; uncomputation then selects the desired coset. Finally, a computable section of $B$ converts the state on violation patterns into the coherent Gibbs state on assignments.

\begin{proof}
\emph{Equal correct-branch overlap.} The received states have the expansion
$$
  \ket{\psi_c}=\sum_X\sqrt{Z_X}\,\omega^{c\cdot x_X}\ket{\hat\phi_X}.
$$
Their overlaps with the corresponding measurement vectors are therefore
\begin{equation}\label{eq:pgm_equal_overlap}
  \langle b_c|\psi_c\rangle=\frac1{\sqrt{|C|}}\sum_X\sqrt{Z_X}\coloneqq a>0,
\end{equation}
independently of $c$. In particular, the block success is $a^2=P_{\mathrm{PGM}}$.

\emph{A coherent implementation.} Let $\Pi_c$ project onto outcome $c$ in register $O$. The implementation assumption implies that there is a normalized state $\ket{\eta_c}$ of the remaining output registers such that
$$
  U\big(\ket{b_c}\ket0\big)=\ket c_O\ket{\eta_c}.
$$
Indeed, this input has outcome $c$ with probability one, while every $\ket{b_{c'}}$ with $c'\ne c$ has outcome $c$ with probability zero. Expanding any $\ket v\in\mathcal S$ in this basis shows that
\begin{equation}\label{eq:pgm_branch}
  \Pi_cU\big(\ket v\ket0\big)=\langle b_c|v\rangle\,\ket c_O\ket{\eta_c}.
\end{equation}
The states $\ket{\eta_c}$ may depend arbitrarily on $c$ and include all residual work systems and any phases of the implementation.

\emph{Preserving the codeword superposition.} Prepare the DQI input with a separate codeword register $A$,
$$
  \frac1{\sqrt{|C|}}\sum_{c\in C}\omega^{-c\cdot x_0}\ket c_A\ket{\psi_c}\ket0,
$$
apply $U$ to the received and ancillary registers, and compute a single flag that tests whether $A=O$. Measure only this flag, preserving the superposition of codeword labels. By~\eqref{eq:pgm_equal_overlap} and~\eqref{eq:pgm_branch}, the unnormalized successful state is
$$
  \frac a{\sqrt{|C|}}\sum_{c\in C}\omega^{-c\cdot x_0}\ket c_A\ket c_O\ket{\eta_c}.
$$
Its squared norm is $a^2$, independently of $x_0$.

\emph{Uncomputation.} Uncompute the comparison flag, subtract $O$ from $A$, and apply $U^\dagger$ to all registers on which $U$ acted. On the successful branch, these operations return all auxiliary registers to zero and leave the normalized received state
$$
  \frac1{\sqrt{|C|}}\sum_{c\in C}\omega^{-c\cdot x_0}\ket{b_c}
  =\sum_X\left(\frac1{|C|}\sum_{c\in C}\omega^{c\cdot(x_X-x_0)}\right)\ket{\hat\phi_X}
  =\ket{\hat\phi_{x_0+K}}.
$$
The last equality follows from character orthogonality. Repeating independent attempts until success takes $P_{\mathrm{PGM}}^{-1}$ attempts in expectation. The DQI input preparation is efficient under the stated assumption, as explained after~\eqref{eq:dqi_identity}, so the complexity claim follows.
\end{proof}

The proposition prepares the Gibbs state on violation patterns for each offset individually, without averaging over $x_0$. To obtain the coherent state on assignments, choose a computable section $s_B:K\to\Z_q^n$ satisfying $Bs_B(x)=x$. For fixed $q$, the Smith normal form computes this section in polynomial time and expresses $\ker B$ as a product of cyclic groups, whose uniform superpositions can be prepared efficiently. A linear right inverse need not exist for composite $q$, and is not required.

Prepare the uniform superposition over $\ker B$ in a new register, add $s_B(x-x_0)$, and uncompute $x$ as $Bz+x_0$ from the resulting assignment $z$. This implements the isometry
$$
  \ket x\ket0\longmapsto\ket0\,\frac1{\sqrt{|\ker B|}}\sum_{u\in\ker B}\ket{s_B(x-x_0)+u},\qquad x\in x_0+K.
$$
Applied to $\ket{\hat\phi_{x_0+K}}$, it prepares $\sum_z\sqrt{\mu_\beta(z)}\ket z$, since every violation pattern has exactly $|\ker B|$ preimages of equal energy.

Proposition~\ref{prop:pgm_gibbs} assumes a reversible circuit for the exact PGM, including all measurement records and work registers, and access to its inverse. A destructive measurement that returns only a classical outcome does not by itself provide this access. The converse construction is given in Proposition~\ref{prop:gibbs_pgm}, so a coherent implementation of the PGM and a coherent preparation of the Gibbs states of all cosets are equivalent up to polynomial overhead whenever $P_{\mathrm{PGM}}$ is at least inverse polynomial (Corollary~\ref{cor:pgm_equiv}).

\section{Polynomial amplitude functions and exact search}\label{sec:exact_heavy}

The target state of a polynomial amplitude function assigns an optimum at most polynomially more probability than a suitable Gibbs measure does, for most offsets under reliable decoding (Theorem~\ref{thm:poly_target}). We prove the pointwise bound first, then compare the coset normalizations. The bounded-distance formula and its Dicke-tail consequence describe exact decoding as a special case. We conclude with the degree restriction for mean-optimal polynomials, the endpoint comparison, and the unconditional consequence of the Prange union bound.

Let $q=2$, let $H^\star=\min_{x\in x_0+K}|x|$ be the optimal number of violated constraints, and let $S^\star_x$ be the set of patterns of weight $H^\star$ in this coset. Recall that $\lambda_\Phi$ is their probability in the normalized target state. By Fourier inversion, $\Phi(x)=2^{-m/2}\sum_e\hat\Phi(e)(-1)^{e\cdot x}$, and $|\hat\Phi(e)|^2$ is the probability of the error $e$ in the decoding problem. Polynomial degree restricts these Fourier weights; reliable decoding will control the coset normalization.

\subsection{A pointwise product bound}\label{app:poly_envelope}

A bound on polynomial degree becomes a bound on Fourier weight. Write $\mathsf H$ for the single-qubit Hadamard operator and $\ket{D_u}=\binom mu^{-1/2}\sum_{|e|=u}\ket e$ for the Dicke state of weight $u$. A normalized function $\Phi(x)=P(|x|)$ with $\deg P\le L$ has the expansion
\begin{equation}\label{eq:poly_dicke}
  \mathsf H^{\otimes m}\ket\Phi=\sum_{u=0}^La_u\ket{D_u},\qquad\sum_{u=0}^L|a_u|^2=1.
\end{equation}
Indeed, $|x|=\tfrac12(m-\sum_j(-1)^{x_j})$, so a degree-$L$ polynomial is a linear combination of characters of weight at most $L$. Permutation invariance makes its Fourier coefficients constant at each weight. Conversely, $\sum_{|e|=u}(-1)^{e\cdot x}=\sum_j(-1)^j\binom{|x|}j\binom{m-|x|}{u-j}$ is a polynomial of degree at most $u$ in $|x|$, so every expansion~\eqref{eq:poly_dicke} defines such a polynomial.

For $0\le u\le m$, put $\tau=u/m$ and
$$
  b_{m,u}=\binom mu\tau^u(1-\tau)^{m-u},\qquad
  A_{m,L}=\sum_{u=0}^Lb_{m,u}^{-1},
$$
with $0^0=1$. The quantity $b_{m,u}$ is the probability that a binomial random variable with mean $u$ equals $u$; this is its largest point probability, so $b_{m,u}\ge(m+1)^{-1}$ and $A_{m,L}\le(L+1)(m+1)$.

\begin{lemma}[Dicke phase average]\label{lem:poly_phase}
Let $0\le u\le m$, $\tau=u/m$, and $\ket{\xi_\varphi}=\sqrt{1-\tau}\ket0+e^{i\varphi}\sqrt\tau\ket1$. Then
\begin{equation}\label{eq:poly_phase}
  \ket{D_u}=\frac{b_{m,u}^{-1/2}}{2\pi}\int_0^{2\pi}e^{-iu\varphi}\ket{\xi_\varphi}^{\otimes m}\,d\varphi.
\end{equation}
Measuring $\mathsf H^{\otimes m}\ket{\xi_\varphi}^{\otimes m}$ gives independent bits equal to one with probability $\gamma(\varphi)=\tfrac12-\sqrt{\tau(1-\tau)}\cos\varphi$.
\end{lemma}

\begin{proof}
In the tensor expansion of $\ket{\xi_\varphi}^{\otimes m}$, the coefficient of a word $e$ has phase $e^{i|e|\varphi}$. Integration against $e^{-iu\varphi}$ removes every weight except $u$ and leaves $b_{m,u}^{1/2}\ket{D_u}$. Also, $\langle1|\mathsf H|\xi_\varphi\rangle=(\sqrt{1-\tau}-e^{i\varphi}\sqrt\tau)/\sqrt2$, whose squared modulus is $\gamma(\varphi)$. The product structure gives independence.
\end{proof}

\begin{lemma}[Pointwise bound for every polynomial]\label{lem:poly_envelope}
Let $0\le L\le\rho m$ with $\rho<\tfrac12$, and let $\Phi$ be a normalized complex polynomial amplitude function of degree at most $L$. For $0\le w\le m/2$, set $\gamma_w=\max\{\rho^\perp,w/m\}$ and $g_w=\gamma_w^w(1-\gamma_w)^{m-w}$. Every $x$ of weight $w$ satisfies
\begin{equation}\label{eq:poly_envelope}
  |\Phi(x)|^2\le A_{m,L}g_w.
\end{equation}
Moreover, $g_w$ is the largest probability of $x$ under an independent-bit law whose one-bit probability lies in $[\rho^\perp,1-\rho^\perp]$.
\end{lemma}

\begin{proof}
The function $\gamma^w(1-\gamma)^{m-w}$ increases up to $\gamma=w/m$ and decreases afterwards. Its maximum on the stated interval is therefore attained at $\gamma_w$. For each $u\le L$, the phase-average factors in Lemma~\ref{lem:poly_phase} have $u/m\le\rho<\tfrac12$, so their one-bit probabilities belong to this interval. The triangle inequality in~\eqref{eq:poly_phase} gives
$$
  |\langle x|\mathsf H^{\otimes m}|D_u\rangle|\le b_{m,u}^{-1/2}g_w^{1/2}.
$$
Substituting~\eqref{eq:poly_dicke} and applying Cauchy--Schwarz,
$$
  |\Phi(x)|\le g_w^{1/2}\sum_{u=0}^L|a_u|b_{m,u}^{-1/2}
  \le g_w^{1/2}\Big(\sum_{u=0}^L|a_u|^2\Big)^{1/2}A_{m,L}^{1/2}.
$$
The coefficient vector is normalized, which proves~\eqref{eq:poly_envelope}.
\end{proof}

\subsection{Coset normalization and optimum probability}\label{app:poly_comparison}

The numerator of the optimum probability is controlled by Lemma~\ref{lem:poly_envelope}. To compare the normalized target states, we retain their coset masses. For a coset $X$ of $K$, write
$$
  W_\Phi(X)=\sum_{x\in X}|\Phi(x)|^2,\qquad
  W_\beta(X)=\sum_{x\in X}\prod_jp_\beta(x_j),
$$
and let $U$ be the uniform law on the $|C|$ cosets. These are ambient masses, not probabilities under the distributions already conditioned on $X$.

\begin{lemma}[Coset mass under reliable decoding]\label{lem:poly_cosets}
If some decoder for $\Phi$ has block success at least $1-\varepsilon$, then
$$
  \|W_\Phi-U\|_{\mathrm{TV}}\le\sqrt\varepsilon,\qquad
  \Pr_{x_0}\big[|C|W_\Phi(x_0+K)<\tfrac12\big]\le2\sqrt\varepsilon.
$$
If the block success is one, $|C|W_\Phi(X)=1$ for every coset.
\end{lemma}

\begin{proof}
By Lemma~\ref{lem:pgm_success}, the optimal block success is the square of the Bhattacharyya coefficient
$$
  \operatorname{BC}(W_\Phi,U)=\sum_X\sqrt{W_\Phi(X)U(X)}.
$$
Cauchy--Schwarz gives
$$
\begin{aligned}
  \|W_\Phi-U\|_{\mathrm{TV}}
  &\le\frac12\sqrt{\sum_X(\sqrt{W_\Phi(X)}-\sqrt{U(X)})^2\sum_X(\sqrt{W_\Phi(X)}+\sqrt{U(X)})^2}\\
  &=\sqrt{1-\operatorname{BC}(W_\Phi,U)^2}\le\sqrt\varepsilon.
\end{aligned}
$$
Each coset of mass below $1/(2|C|)$ contributes more than $1/(2|C|)$ to the total deficit from $U$. A uniformly random offset chooses a uniformly random coset, so the probability of these cosets is at most twice the total variation distance. Exact success forces that distance to vanish.
\end{proof}

\begin{proof}[Proof of Theorem~\ref{thm:poly_target}]
\emph{The optimal weight.} Since every row of $B$ is nonzero, a uniformly random assignment violates each constraint with probability $1/2$. Some assignment therefore has at most $m/2$ violations, so $H^\star\le m/2$.

For each integer $0\le w\le m/2$, set $\beta^{(w)}=\ln((1-\gamma_w)/\gamma_w)\in[0,\beta_\rho]$. All optimal patterns have weight $H^\star$, so Lemma~\ref{lem:poly_envelope}, summed over these patterns, gives for $X=x_0+K$ with $W_\Phi(X)>0$
\begin{equation}\label{eq:poly_fixed_offset}
  \lambda_\Phi(x_0)\le A_{m,L}\frac{W_{\beta^{(H^\star)}}(X)}{W_\Phi(X)}\lambda_{\beta^{(H^\star)}}(x_0).
\end{equation}
The same inverse temperature serves every optimal pattern because their weights agree.

\emph{The normalizations.} For each $w$, averaging over a uniformly random coset gives $\E_{x_0}|C|W_{\beta^{(w)}}(x_0+K)=1$. Markov's inequality and a union bound over the at most $m+1$ possible weights show that
$$
  |C|W_{\beta^{(w)}}(x_0+K)\le(m+1)^2\quad\text{for every }w
$$
except on a set of offsets of probability at most $(m+1)^{-1}$. This event is independent of $\Phi$. By Lemma~\ref{lem:poly_cosets}, $|C|W_\Phi(X)\ge1/2$ outside another exceptional set of probability at most $2\sqrt\varepsilon$.

On the intersection,~\eqref{eq:poly_fixed_offset} and Lemma~\ref{lem:mass_monotone} imply
$$
  \lambda_\Phi\le2(m+1)^2A_{m,L}\lambda_{\beta^{(H^\star)}}
  \le(m+1)^4\lambda_\beta\qquad(\beta\ge\beta_\rho),
$$
because $A_{m,L}\le(L+1)(m+1)$ and $2(L+1)\le m+1$. The lower bound on $W_\Phi(X)$ also makes the target state well-defined.
\end{proof}

\subsection{Exact decoding and Dicke tails}\label{app:poly_exact_decoding}

When every Fourier error has weight at most $L$ and the code has minimum distance greater than $2L$, decoding is exact for every offset. The following formula then identifies the best possible optimum probability under that cutoff.

For $0\le w\le m$, let $\mathcal D_w$ be the law of the weight of the outcome when the Dicke state $\binom mw^{-1/2}\sum_{|x|=w}\ket x$ is measured in the Hadamard basis. Let $\bar N(w)=|K|\binom mw/2^m$ be the expected number of $x\in x_0+K$ with $|x|=w$ for a uniformly random offset $x_0$.

\begin{proposition}[Probability of an optimum under bounded-distance decoding]\label{prop:bounded_distance}
Let $q=2$, and let $\Phi$ be a symmetric amplitude function whose Fourier transform is supported on weights at most $L$, where $2L$ is smaller than the minimum distance of $C$. Then the states $\ket{\psi^\Phi_c}$, $c\in C$, are pairwise orthogonal, a decoder that corrects every error of weight at most $L$ has block success one, and DQI with such a decoder prepares the target state exactly. Moreover,
$$
  \lambda_\Phi\le\frac{|S^\star_x|}{\bar N(H^\star)}\,\Pr_{u\sim\mathcal D_{H^\star}}[u\le L],
$$
with equality when $\hat\Phi$ is proportional to the projection of the Hadamard transform of the Dicke state of weight $H^\star$ onto weights at most $L$.
\end{proposition}

\begin{proof}
For $c,c'\in\Z_2^m$, by Fourier inversion,
$$
  \langle\psi^\Phi_{c'}|\psi^\Phi_c\rangle=\sum_x|\Phi(x)|^2(-1)^{(c+c')\cdot x}=\sum_e\hat\Phi(e)\,\overline{\hat\Phi(e+c+c')} .
$$
If $c\ne c'$ are codewords, then $|c+c'|>2L$, so $e$ and $e+c+c'$ cannot both have weight at most $L$, and the inner product vanishes. In the Hadamard basis, $\ket{\psi^\Phi_c}$ is a superposition of the words $c+e$ with $|e|\le L$, so a decoder that corrects every error of weight at most $L$ recovers $c$ with certainty, and DQI prepares the target state exactly. The identity for $\mathbf 1[x\in x_0+K]$ in Section~\ref{sec:dqi} shows
$$
  \sum_{x\in x_0+K}|\Phi(x)|^2=\frac1{|C|}\sum_{c\in C}(-1)^{c\cdot x_0}\sum_x|\Phi(x)|^2(-1)^{c\cdot x}=\frac1{|C|},
$$
because the inner sum is $\langle\psi^\Phi_0|\psi^\Phi_c\rangle$, which vanishes for $c\ne0$ and equals one for $c=0$. Hence $\lambda_\Phi=|C|\sum_{x\in S^\star_x}|\Phi(x)|^2$.

Let $\hat\phi(u)$ be the value of $\hat\Phi$ on weight $u$. For $|x|=w$, $\sum_{|e|=u}(-1)^{e\cdot x}=K_u(w)$, where $K_u(w)=\sum_j(-1)^j\binom wj\binom{m-w}{u-j}$ is the Krawtchouk polynomial. Hence, for $x\in S^\star_x$,
$$
  \Phi(x)=2^{-m/2}\sum_{u\le L}\hat\phi(u)\,K_u(H^\star).
$$
Since $\sum_u\binom mu|\hat\phi(u)|^2=1$, the Cauchy--Schwarz inequality implies
$$
  |\Phi(x)|^2\le2^{-m}\sum_{u\le L}\frac{K_u(H^\star)^2}{\binom mu},
$$
with equality when $\hat\phi(u)$ is proportional to $K_u(H^\star)/\binom mu$ for $u\le L$. The Hadamard transform of the Dicke state of weight $w$ has amplitude $2^{-m/2}\binom mw^{-1/2}K_w(|e|)$ at $e$, so $\Pr_{\mathcal D_w}[u]=\binom mu K_w(u)^2/(2^m\binom mw)$. Using the identity $\binom mwK_u(w)=\binom muK_w(u)$, we obtain
$$
  \sum_{u\le L}\frac{K_u(H^\star)^2}{\binom mu}=\frac{2^m}{\binom m{H^\star}}\Pr_{u\sim\mathcal D_{H^\star}}[u\le L].
$$
The same identity shows that the equality case is the stated projection. Combining the bounds gives
$$
  \lambda_\Phi\le\frac{|C|\,|S^\star_x|}{\binom m{H^\star}}\Pr_{\mathcal D_{H^\star}}[u\le L].
$$
Since $|C|=2^m/|K|$, this is the stated bound in terms of $\bar N(H^\star)$.
\end{proof}

\begin{corollary}[Dicke lower tail]\label{cor:poly_tail}
Let $L\le\rho m$ with $0\le\rho<\tfrac12$, and let $0\le w\le m/2$. With $g_w$ as in Lemma~\ref{lem:poly_envelope},
$$
  \Pr_{u\sim\mathcal D_w}[u\le L]\le A_{m,L}\binom mwg_w.
$$
If $w/m\le\rho^\perp$, this is at most $(m+1)^2\exp(-mD_{\mathrm{KL}}(w/m\,\|\,\rho^\perp))$, where $D_{\mathrm{KL}}(h\|a)=h\ln(h/a)+(1-h)\ln((1-h)/(1-a))$.
\end{corollary}

\begin{proof}
For a fixed $x$ of weight $w$, the Krawtchouk identity used in Proposition~\ref{prop:bounded_distance} gives
$$
  \Pr_{u\sim\mathcal D_w}[u\le L]=\binom mw\sum_{u=0}^L|\langle x|\mathsf H^{\otimes m}|D_u\rangle|^2.
$$
The proof of Lemma~\ref{lem:poly_envelope} bounds each summand by $b_{m,u}^{-1}g_w$, which proves the first inequality. If $h=w/m\le\rho^\perp$, then $g_w=(\rho^\perp)^w(1-\rho^\perp)^{m-w}$. The elementary bound $\binom mwh^w(1-h)^{m-w}\le1$ implies $\binom mwg_w\le e^{-mD_{\mathrm{KL}}(h\|\rho^\perp)}$, and $A_{m,L}\le(m+1)^2$ finishes the proof.
\end{proof}

In particular, for a fixed violated fraction $h<\tfrac12$, the lower tail is exponentially small whenever $L/m$ stays below $h^\perp$ by a fixed positive amount. To see this, choose a fixed $\rho<h^\perp$ above $L/m$. The decreasing involution $a\mapsto a^\perp$ gives $h<\rho^\perp$, so the relative-entropy exponent is strictly positive. No replacement of a coset mass by its expectation is involved.

\subsection{Mean-optimal degrees and the reconstruction endpoint}\label{app:poly_degree}

The comparison in Theorem~\ref{thm:poly_target} holds for every polynomial within its degree bound. A restriction on the degree follows from local decoding only when the polynomial also attains the optimal mean value.

\begin{proposition}[Degree ratios of mean-optimal polynomials]\label{prop:poly_degree}
Fix $3\le k<D$ with $0<\beta_d<\infty$. Let $L_m/m\to\rho\in[0,\tfrac12]$, and let $\Phi_m$ maximize $\bar s(\Phi)=\sum_x|\Phi(x)|^2(1-|x|/m)$ among normalized polynomial amplitude functions of degree at most $L_m$. Suppose that a decoder has block success $1-\varepsilon_m$ and makes a local estimate at a symbol satisfying LT and TP on a neighborhood of size $v_m$. If $\varepsilon_m\to0$ and $v_m=o(m)$, then $\rho\le\rho_d$.
\end{proposition}

\begin{proof}
The maximum of $\bar s$ converges to $\tfrac12+\sqrt{\rho(1-\rho)}$, the semicircle value~\cite{jordan2024optimization}. This follows directly from~\eqref{eq:poly_dicke}: counting single-bit flips between adjacent Dicke weights gives
$$
  \bar s(\Phi)=\frac12+\frac1m\sum_{u=0}^{L_m-1}\operatorname{Re}(\bar a_ua_{u+1})\sqrt{(u+1)(m-u)}.
$$
For $L_m\le m/2$, Cauchy--Schwarz bounds the sum by $\sqrt{L_m(m-L_m+1)}$. If $\rho>0$, choose equal coefficients on the last $\lceil\sqrt{L_m}\rceil$ weights below $L_m$ and zero elsewhere. Their adjacent contributions attain the same limiting value. The case $\rho=0$ follows from the upper bound and the constant polynomial, while at $\rho=1/2$ the trivial upper bound $\bar s\le1$ suffices.

By Theorem~\ref{thm:guarantee}, the expected satisfied fraction of the output differs from $\bar s(\Phi_m)$ by $o(1)$. The local approximation theorem, Theorem~\ref{thm:approx_sym}, bounds that expectation by $s_d+o(1)$. Consequently $\tfrac12+\sqrt{\rho(1-\rho)}\le s_d=\tfrac12+\sqrt{\rho_d(1-\rho_d)}$. The function on the left is strictly increasing on $[0,\tfrac12]$, so $\rho\le\rho_d$.
\end{proof}

On either random regular ensemble, the local hypotheses hold with high probability for a fixed estimated symbol and the logarithmic radii in Corollary~\ref{cor:typical_nogo}. Thus the proposition applies to reliably locally decoded mean-optimal polynomial amplitude functions on these ensembles. It does not bound the degree of a nonoptimal polynomial whose expected satisfied fraction remains below $s_d$.

\begin{corollary}[Comparison with fixed exponential slack]\label{cor:poly_endpoint}
Fix $3\le k<D$ with $0<\beta_d<\infty$, and assume Conjecture~\ref{conj:classical} at every fixed $\beta<\beta_d$. Let $\Phi_m$ be polynomial amplitude functions of degrees $L_m$ with $\limsup_m L_m/m\le\rho_d$, chosen independently of $x_0$ and possibly depending on $B$. Suppose that, with probability $1-o(1)$ over $B$ from either regular ensemble, some decoder has block success $1-o(1)$. For every fixed $\eta>0$, there is a classical algorithm which, with probability $1-o(1)$ over $(B,x_0)$, has the following guarantee whenever $\lambda_{\Phi_m}(x_0)>0$. Given a polynomial-factor lower bound on $\lambda_{\Phi_m}(x_0)$ as in Corollary~\ref{cor:poly_quadratic}, it finds an exact optimum with probability at least $1-e^{-2}$ in time
$$
  \poly(m+n)e^{\eta m}T_{\Phi_m}^2.
$$
The algorithm and the polynomial may depend on $\eta$.
\end{corollary}

\begin{proof}
Choose fixed $0\le\beta_-<\beta_d<\beta_+$ with $\beta_+-\beta_-<\eta$, and put $\rho_+=\alpha_{\beta_+}^\perp$. Then $\rho_+>\rho_d$ and $L_m\le\rho_+m$ for all large $m$. Theorem~\ref{thm:poly_target} gives $\lambda_{\Phi_m}\le(m+1)^4\lambda_{\beta_+}$ outside an event of probability $o(1)$.

The derivative in Lemma~\ref{lem:mass_monotone} is at most $m$, since the objective lies in $[0,m]$. Hence
$$
  \lambda_{\beta_+}\le e^{(\beta_+-\beta_-)m}\lambda_{\beta_-}\le e^{\eta m}\lambda_{\beta_-}.
$$
At the fixed inverse temperature $\beta_-$, Corollary~\ref{cor:binary} and Conjecture~\ref{conj:classical} supply an efficient sampler with probability $1-o(1)$. The proof of Corollary~\ref{cor:poly_quadratic}, with the factor $(m+1)^4e^{\eta m}$ in place of $(m+1)^4$, proves the result.
\end{proof}

Corollary~\ref{cor:poly_endpoint} supplies one algorithm for each fixed slack. It asserts neither uniform polynomial overhead at the threshold nor a single algorithm with an $e^{o(m)}$ factor.

\subsection{The Prange comparison with true normalizations}\label{app:poly_prange}

The same codeword union bound used in Section~\ref{sec:prange} controls the Gibbs coset masses for every offset. This yields an all-offset polynomial comparison when decoding is exact.

\begin{proposition}[Polynomial target states and Prange's algorithm]\label{prop:poly_prange}
Let $B$ have no zero row, let $\Phi$ be a normalized complex polynomial amplitude function of degree at most $L\le\rho m$ with $\rho<\tfrac12$, and let $\beta'\ge\beta_\rho$. Put $U=U_C(1-e^{-\beta'})<1$. For every offset with $\kappa=|C|W_\Phi(x_0+K)>0$, the output law $\nu_{\beta'}$ of Prange's algorithm satisfies
$$
  \nu_{\beta'}(S^\star)\ge\frac{\kappa(1-U)^2}{(1+U)A_{m,L}}\lambda_\Phi(x_0).
$$
If decoding is exact, $\kappa=1$ for every offset. With block success $1-\varepsilon$, $\kappa\ge1/2$ outside a set of offsets of probability at most $2\sqrt\varepsilon$.
\end{proposition}

\begin{proof}
Character orthogonality gives, for every $\beta\ge0$,
$$
  |C|W_\beta(x_0+K)=\sum_{c\in C}(-1)^{c\cdot x_0}\tanh(\beta/2)^{|c|}.
$$
For $\beta\le\beta'$, its value is at most $1+U$, since $\tanh(\beta/2)\le1-e^{-\beta'}$ and $U_C$ is nondecreasing. Apply this bound in~\eqref{eq:poly_fixed_offset} at $\beta=\beta^{(H^\star)}\le\beta_\rho\le\beta'$. Monotonicity of the optimum probability gives
$$
  \lambda_\Phi\le\frac{A_{m,L}(1+U)}\kappa\lambda_{\beta'}.
$$
By Proposition~\ref{prop:prange}, $\nu_{\beta'}(S^\star)\ge(1-U)^2\lambda_{\beta'}$, which proves the inequality. The statements about $\kappa$ follow from Lemma~\ref{lem:poly_cosets}.
\end{proof}

When decoding is exact and $U\le1/2$, $\lceil12A_{m,L}/\underline\lambda\rceil$ independent Prange runs suffice for success at least $1-e^{-2}$, given $0<\underline\lambda\le\lambda_\Phi$. Since $A_{m,L}\le(m+1)^2$, a polynomial-factor lower bound gives running time $\poly(m+n)T_\Phi^2$. The conclusion holds for every offset simultaneously. Theorem~\ref{thm:union} supplies $U\le1/2$ with probability $1-o(1)$ for the configuration-model pairs and inverse temperatures certified in Section~\ref{sec:prange}, provided $\beta_\rho\le\beta'\le\beta_U$. It does not supply an exact decoder; that is a separate hypothesis. Under merely reliable decoding, the comparison remains a statement about target states on most offsets.

\section{Proofs for the condensation threshold}\label{app:condensation}

On binary configuration-model instances, the PGM succeeds below $\beta_c$ and every decoder fails above it. We first derive the limiting entropy of the coset law from the mutual information of the primal channel (Proposition~\ref{prop:entropy}). For the failure direction, a linear entropy and concentration imply exponentially small decoding success (Lemmas~\ref{lem:entropy_bound} and~\ref{lem:concentration}). For the success direction, linear minimum distance and independent noise convert sublinear entropy into coset uniformity (Lemmas~\ref{lem:min_distance} and~\ref{lem:smoothing}).

These ingredients prove Theorem~\ref{thm:condensation}. A mixture over Hamming weights then proves the symmetric-amplitude ceiling in Corollary~\ref{cor:condensation_sym}. Finally, in Section~\ref{app:condensation_holevo}, a positive trial value proves strict separation from the Holevo threshold, while uniform nonpositivity over all trial laws proves convergence at large degrees.

Throughout, $q=2$, $f(z)=z$, $\alpha\in(0,\tfrac12)$, $\sigma=1-2\alpha$, and $\beta=\ln((1-\alpha)/\alpha)$, so that $\alpha=p_\beta(1)$. We write $L_{B,\alpha}=|C|\,W_{B,\alpha}$, so that $P_{\mathrm{PGM}}(B,\beta)=(\E_{U_B}\sqrt{L_{B,\alpha}})^2$ by Lemma~\ref{lem:pgm_success}. Expectations of functions of cosets are taken under $U_B$ unless stated otherwise, and all logarithms are natural.

\paragraph{The relative entropy and the mutual information.} Parts~1 and~2 of Proposition~\ref{prop:entropy} hold for every matrix, and part~3 specializes a theorem of van den Brand and Jaafari to the configuration model.

\begin{proof}[Proof of Proposition~\ref{prop:entropy}]
\emph{Part~1: the entropy identity.} Fix $B$. Since $Z$ is uniform on $\Z_2^n$, the vector $BZ$ is uniform on $K$, because every element of $K$ has $|\ker B|$ preimages. Hence, given its coset, $Y=BZ+E$ is uniformly distributed on the coset, and the coset $E+K$ has law $W_{B,\alpha}$. The entropy of $Y$ is therefore $H(Y)=H(W_{B,\alpha})+\ln|K|$, where $H(W_{B,\alpha})$ is the entropy of the coset. Since $|C|\,|K|=2^m$,
$$
  F_B(\alpha)=\ln|C|-H(W_{B,\alpha})=m\ln2-H(Y).
$$
Given $Z$, the vector $Y$ is a translate of $E$, so $H(Y\mid Z)=H(E)=m\big(\ln2-\operatorname{cap}(\sigma)\big)$, because a Bernoulli$(\alpha)$ variable has entropy $\ln2-\operatorname{cap}(1-2\alpha)$. Hence $I(Z;Y\mid B)=H(Y)-H(Y\mid Z)=m\operatorname{cap}(\sigma)-F_B(\alpha)$. No assumption on the rank of $B$ is needed.

\emph{Part~2: degradation.} Let $\alpha\le\alpha'<\tfrac12$, and let $\alpha''=(\alpha'-\alpha)/(1-2\alpha)\in[0,\tfrac12)$, so that $1-2\alpha'=(1-2\alpha)(1-2\alpha'')$. If $E$ and $E''$ are independent with independent Bernoulli$(\alpha)$ and Bernoulli$(\alpha'')$ entries, then $E+E''$ has independent Bernoulli$(\alpha')$ entries, because $\E(-1)^{E_j+E''_j}=(1-2\alpha)(1-2\alpha'')$ for every $j$. Hence $W_{B,\alpha'}$ is the image of $W_{B,\alpha}$ under the Markov kernel that maps a coset $X$ to the coset $X+E''$. This kernel maps $U_B$ to itself, because translation by a fixed vector permutes the cosets. By the data-processing inequality for relative entropy, $F_B(\alpha')\le F_B(\alpha)$.

\emph{Part~3.} Van den Brand and Jaafari~\cite{vandenbrand2017mutual} consider the following planted model with spins in $\{\pm1\}$. A planted assignment $x^\ast\in\{\pm1\}^n$ is drawn uniformly. Independently of $x^\ast$, a factor graph with $n$ variable nodes of degree $D$ and factor nodes of arity $k$ is drawn by arranging the $Dn$ half-edges of the variable nodes in a uniformly random order and grouping consecutive $k$-tuples into factor nodes~\cite[Definitions~3.1 and~3.2]{vandenbrand2017mutual}. Each factor node $j$ then receives a weight function $\psi_j:\{\pm1\}^k\to(0,2)$ from a finite set, chosen with probability proportional to $p(\psi)\,\psi(x^\ast_{\partial j})$ for a prior $p$, where $x^\ast_{\partial j}$ is the vector of spins of the neighbors of $j$. Their Theorem~2.1, which allows degrees drawn from a general law, determines $\lim_{n\to\infty}\frac1nI(x^\ast;G^\ast)$, where $G^\ast$ is the factor graph together with its weight functions, under two conditions SYM and POS on the weight functions.

We take the uniform prior on the two weight functions $\psi_s(x)=1+s\sigma x_1\cdots x_k$ with $s\in\{\pm1\}$, and $x^\ast_a=(-1)^{Z_a}$. If $\psi_{s_j}$ is the weight function of factor node $j$, then $s_j=\prod_{a\in\partial j}x^\ast_a$ with probability $(1+\sigma)/2=1-\alpha$, independently for different factor nodes. Hence $(-1)^{Y_j}=s_j$ defines $Y=BZ+E$, where the factor nodes are the rows of $B$ and the variable nodes are its columns. A factor node that contains a variable node twice contributes the square of its spin, which is one, so $B$ is the incidence matrix reduced modulo two, and the factor graph has the law of the configuration model without the conditioning. The factor graph is independent of $x^\ast$, so $I(x^\ast;G^\ast)=\E_BI(Z;Y\mid B)$.

\emph{Checking the hypotheses SYM and POS.} The condition SYM requires that the mean of $\psi(x)$ under the prior does not depend on $x$; here it equals one for every $x$. To state POS, describe a probability distribution $\mu$ on $\{\pm1\}$ by its magnetization $\mu(+1)-\mu(-1)$, and note that $\sum_{x\in\{\pm1\}^k}\psi_s(x)\prod_i\mu_i(x_i)=1+s\sigma\prod_i{\mathfrak m}_i$ when $\mu_i$ has magnetization ${\mathfrak m}_i$. For two laws $\pi$ and $\pi'$ on $[-1,1]$ with mean zero, let ${\mathfrak m}_1,\dots,{\mathfrak m}_k$ have law $\pi$, let ${\mathfrak m}'_1,\dots,{\mathfrak m}'_k$ have law $\pi'$, and let $s$ be a uniformly random sign, all independent. The condition POS then reads
$$
\begin{aligned}
  &\E\Big[\Lambda\big(1+s\sigma {\mathfrak m}_1\cdots {\mathfrak m}_k\big)+(k-1)\Lambda\big(1+s\sigma {\mathfrak m}'_1\cdots {\mathfrak m}'_k\big)\\
  &\qquad-k\Lambda\big(1+s\sigma {\mathfrak m}_1{\mathfrak m}'_2\cdots {\mathfrak m}'_k\big)\Big]\ge0,
\end{aligned}
$$
where $\Lambda(u)=u\ln u$. For $|u|\le1$, $\Lambda(1+u)=u+\sum_{p\ge2}(-1)^pu^p/(p(p-1))$, with an absolutely convergent series. Since $s$ is uniform and independent of the magnetizations, the terms with odd $p$ have mean zero, and the term with even $p$ has mean $\sigma^p/(p(p-1))$ times $M_p^k+(k-1)(M'_p)^k-k M_p(M'_p)^{k-1}$, where $M_p=\E {\mathfrak m}_1^p\ge0$ and $M'_p=\E({\mathfrak m}'_1)^p\ge0$. This is nonnegative by the inequality of arithmetic and geometric means, applied to $M_p^k$ and $k-1$ copies of $(M'_p)^k$. The weight functions take the values $1\pm\sigma\in(0,2)$, as the theorem requires.

For degree $D$, arity $k$, and these weight functions, the functional of~\cite[Theorem~2.1]{vandenbrand2017mutual} at a law $\pi$ of magnetizations with mean zero equals $\tfrac12\E[\mathcal A\ln\mathcal A]-\frac{D(k-1)}k\,b$, where
$$
  \mathcal A=\sum_{x=\pm1}\prod_{i=1}^D\big(1+xs_i\sigma {\mathfrak m}_{i,1}\cdots {\mathfrak m}_{i,k-1}\big)
$$
with independent ${\mathfrak m}_{i,j}$ of law $\pi$ and independent uniformly random signs $s_i$, and the additive term of the theorem equals $\frac Dk\operatorname{cap}(\sigma)$. Indeed, for distributions $\mu_j$ with magnetizations ${\mathfrak m}_j$, the sum of $\psi_s(x)\prod_{j\ne h}\mu_j(x_j)$ over the $x$ with $x_h=\pm1$ equals $1\pm s\sigma\prod_{j\ne h}{\mathfrak m}_j$, the sum of $\psi_s(x)\prod_j\mu_j(x_j)$ over all $x$ equals $1+s\sigma\prod_j{\mathfrak m}_j$, and $\E_s\Lambda(1+st)=\operatorname{cap}(t)$ for a uniformly random sign $s$. By the theorem, therefore,
$$
  \lim_{n\to\infty}\frac1n\E_BI(Z;Y\mid B)=-\sup_\pi\Big(\tfrac12\E[\mathcal A\ln\mathcal A]-\frac{D(k-1)}k\,b\Big)+\ln2+\frac Dk\operatorname{cap}(\sigma).
$$
By part~1 and $m/n=D/k$, the limit $\bar F(\alpha)$ exists and equals the supremum over $\pi$ of $\tfrac12\E[\mathcal A\ln\mathcal A]-\ln2-\frac{D(k-1)}k\,b$.

\emph{The mutual-information term.} It remains to identify $\tfrac12\E[\mathcal A\ln\mathcal A]$ with $\ln2+Da-\mathcal I_D$. Let $\xi$ be a uniformly random sign, let $R_1,\dots,R_D$ be independent copies of $|\sigma {\mathfrak m}_{1,1}\cdots {\mathfrak m}_{1,k-1}|$, independent of $\xi$, and given $\xi$ and $R=(R_i)_{i\le D}$, let $T_i$ be independent signs satisfying $\Pr[T_i=\xi\mid\xi,R]=(1+R_i)/2$. Put $\mathcal A_R(t)=\sum_{x=\pm1}\prod_i(1+xt_iR_i)$. Given $R$, the vector $T$ has probability $2^{-D-1}\mathcal A_R(t)$ at $t$, and hence
$$
  H(T\mid R)=(D+1)\ln2-\E\ln\mathcal A_R(T).
$$
Its density relative to uniform signs is $\mathcal A_R/2$. Consequently, for uniform $T'$ independent of $R$,
$$
  \E\ln\mathcal A_R(T)=\tfrac12\E[\mathcal A_R(T')\ln\mathcal A_R(T')]
  =\tfrac12\E[\mathcal A\ln\mathcal A].
$$
The last equality holds because $T'_iR_i$ and $s_i\sigma{\mathfrak m}_{i,1}\cdots{\mathfrak m}_{i,k-1}$ have the same joint law: each is an independent uniform sign times the same reliability.

Conditional independence gives $H(T\mid\xi,R)=\sum_i\E[\ln2-\operatorname{cap}(R_i)]=D\ln2-Da$. Since $R$ is independent of $\xi$, the mutual information between $\xi$ and the observations is therefore
$$
  \mathcal I_D=H(T\mid R)-H(T\mid\xi,R)=\ln2+D\,a-\tfrac12\E[\mathcal A\ln\mathcal A],
$$
which proves part~3.
\end{proof}

\paragraph{Failure above the threshold.} Above $\beta_c$, the relative entropy $F_B(\alpha)$ is linear in $m$ on average. The next lemma shows that, for every matrix $B$, the block success is then exponentially small.

\begin{lemma}[Failure under a linear relative entropy]\label{lem:entropy_bound}
For every matrix $B$ over $\Z_2$ and every $\alpha\in(0,\tfrac12)$, with $F=F_B(\alpha)$ and $\beta=\ln((1-\alpha)/\alpha)$,
$$
  \sqrt{P_{\mathrm{PGM}}(B,\beta)}\le e^{-F/4}+e^{-F^2/(4m\beta^2)} .
$$
\end{lemma}

\begin{proof}
\emph{Separating the two laws.} Write $W=W_{B,\alpha}$ and $L=|C|\,W$, which is positive on every coset, and let $Q=\{X:\ln L(X)\ge F/2\}$. Since $\E[L\mathbf 1_S]=W(S)\le1$ for every set $S$ of cosets, the Cauchy--Schwarz inequality on $Q$ and on its complement $Q^c$ yields
$$
  \E\sqrt L\le\sqrt{U_B(Q)}\sqrt{W(Q)}+\sqrt{U_B(Q^c)}\sqrt{W(Q^c)}\le\sqrt{U_B(Q)}+\sqrt{W(Q^c)}.
$$
\emph{Uniform-law bound.} Since $\E L=1$, Markov's inequality gives $U_B(Q)\le e^{-F/2}$.

\emph{Coset-law bound.} Under $W$, the coset is $X=E+K$, where $E$ has independent Bernoulli$(\alpha)$ entries. Thus $\ln L(E+K)$ is a function of $m$ independent bits with mean $F$. Changing bit $j$ replaces $X$ by $X+e_j$. The corresponding terms in
$$
  W(X+e_j)=\sum_{x\in X}\alpha^{|x+e_j|}(1-\alpha)^{m-|x+e_j|}
$$
differ from those of $W(X)$ by factors between $e^{-\beta}$ and $e^\beta$, because $\alpha/(1-\alpha)=e^{-\beta}$. Hence the change in $\ln L$ is at most $\beta$. McDiarmid's bounded-differences inequality~\cite{mcdiarmid1998concentration} gives
$$
  W(Q^c)=\Pr\big[\ln L(E+K)-F<-F/2\big]\le\exp\Big(-\frac{2(F/2)^2}{m\beta^2}\Big)=e^{-F^2/(2m\beta^2)} .
$$
The three bounds and Lemma~\ref{lem:pgm_success} prove the lemma.
\end{proof}

\begin{lemma}[Concentration over the instance]\label{lem:concentration}
Let $B$ be drawn from the configuration model without the conditioning, and fix $\alpha\in(0,\tfrac12)$. There is a constant $\kappa_k>0$ that depends only on $k$ such that, for every $t>0$,
$$
  \Pr_B\big[|F_B(\alpha)-\E_BF_B(\alpha)|\ge t\big]\le2e^{-\kappa_k t^2/m} .
$$
For the configuration model conditioned on the absence of multiple edges, the same bound holds after its right-hand side is divided by the probability of this event, which is bounded below by a positive constant, with $\E_B$ still the expectation without the conditioning.
\end{lemma}

\begin{proof}
Label the $mk$ half-edges of the symbols and the $nD=mk$ half-edges of the checks. The configuration model without the conditioning is then determined by a uniformly random bijection $\varpi$ between the two sets of half-edges, that is, by a uniformly random permutation of $mk$ elements. Composing $\varpi$ with the transposition of two half-edges of the symbols exchanges the checks to which these two half-edges are attached. This changes only the edges at the at most two symbols that own these half-edges, so the matrix $B$, reduced modulo two, changes in at most two rows. Suppose that $B$ and $B'$ differ only in row $j$, and write $Y_{-j}$ for the entries of $Y$ other than $Y_j$. By the chain rule,
$$
  I(Z;Y\mid B)=I(Z;Y_{-j}\mid B)+I(Z;Y_j\mid Y_{-j},B).
$$
The first term depends only on the rows of $B$ other than $j$, so it is the same for $B'$, and the second term lies in $[0,\ln2]$, because $Y_j$ takes two values. Hence $|I(Z;Y\mid B)-I(Z;Y\mid B')|\le\ln2$, and by part~1 of Proposition~\ref{prop:entropy}, the transposition changes $F_B(\alpha)$ by at most $2\ln2$. By the bounded-differences inequality for uniformly random permutations~\cite[Section~3.2]{mcdiarmid1998concentration}, there is an absolute constant $c'>0$ such that every function $g$ of a uniformly random permutation $\varpi$ of $N$ elements with $|g(\varpi)-g(\varpi\circ\tau)|\le\theta$ for every transposition $\tau$ satisfies $\Pr[|g(\varpi)-\E g(\varpi)|\ge t]\le2e^{-c't^2/(N\theta^2)}$ for every $t>0$. An inequality of the same form holds for random pairings, with a proof by switchings~\cite[Theorem~2.19]{wormald1999models}. With $N=mk$ and $\theta=2\ln2$, we obtain the bound with $\kappa_k=c'/(4k\ln^22)$. For the conditioned model, let $S$ be the event that there are no multiple edges. Then $\Pr[A\mid S]\le\Pr[A]/\Pr[S]$ for every event $A$, and $\Pr[S]$ is bounded below by a positive constant~\cite{richardson2008modern}.
\end{proof}

\paragraph{Success below the threshold.} Below $\beta_c$, the relative entropy is $o(m)$ on average, which by itself does not show that $W_{B,\alpha}$ is close to uniform. The next two lemmas show this at a slightly larger violation probability, using the minimum distance of $C$.

\begin{lemma}[Minimum distance]\label{lem:min_distance}
Let $3\le k<D$, let $a_D=2^{D-1}-1$, $\Gamma=e^{1+k/2}(2a_D/D)^{k/2}$, and $\gamma=(2\Gamma)^{-2/(k-2)}$, and let $m_{\min}(C)$ be the smallest weight of a nonzero codeword of $C$. For random $(k,D)$-regular instances from the configuration model, with or without the conditioning on the absence of multiple edges, $\Pr[m_{\min}(C)<\gamma m]=O\big(m^{-(k-2)/2}\big)$.
\end{lemma}

\begin{proof}
Work in the unconditioned model, with $B$ reduced modulo two. A vector $c\in\Z_2^m$ of weight $w$ lies in $C$ if and only if every check receives an even number of the $k w$ half-edges of the symbols in the support of $c$. These half-edges are matched to a uniformly random set of $k w$ of the $mk$ half-edges of the checks, and the number of such sets that meet every check in an even number of half-edges is the coefficient of $z^{k w}$ in $W_D(z)^n$, where $W_D(z)=\tfrac12\big((1+z)^D+(1-z)^D\big)$ is the generating function of Section~\ref{sec:prange}. Hence the expected number $N_w$ of codewords of weight $w$ satisfies
$$
  \E N_w=\binom mw\binom{mk}{k w}^{-1}[z^{k w}]W_D(z)^n\le\Big(\frac{em}w\Big)^w\Big(\frac wm\Big)^{k w}\frac{W_D(x)^n}{x^{k w}}
$$
for every $x\in(0,1]$, because $W_D$ has nonnegative coefficients, $\binom mw\le(em/w)^w$, and $\binom{mk}{k w}\ge(m/w)^{k w}$. For $x\le1$, $W_D(x)=1+\sum_{j\ge2\text{ even}}\binom Djx^j\le1+a_Dx^2\le e^{a_Dx^2}$. Let $1\le w\le\gamma m$, and choose $x^2=wD/(2a_Dm)$, which is at most one because $\Gamma>1$, so that $\gamma<1\le2a_D/D$. With $n=mk/D$ and this choice, $W_D(x)^nx^{-k w}\le e^{k w/2}\big(2a_Dm/(Dw)\big)^{k w/2}$, and therefore
$$
  \E N_w\le\Big(\Gamma\Big(\frac wm\Big)^{(k-2)/2}\Big)^w .
$$
For $w\le\gamma m$, the base is at most $\Gamma\gamma^{(k-2)/2}=\tfrac12$. For $w\le\log^2m$, the base is at most $\varrho_m=\Gamma(\log^2m/m)^{(k-2)/2}$, which tends to zero, so
$$
  \sum_{1\le w\le\log^2m}\E N_w\le\Gamma m^{-(k-2)/2}\sum_{w\ge1}w^{(k-2)/2}\varrho_m^{\,w-1}=O\big(m^{-(k-2)/2}\big),
$$
and $\sum_{\log^2m<w\le\gamma m}\E N_w\le\sum_{w>\log^2m}2^{-w}\le2^{1-\log^2m}$. Markov's inequality proves the lemma for the unconditioned model, and conditioning on an event whose probability is bounded below by a positive constant multiplies the probability by at most a constant. For $(k,D)=(3,6)$, $a_6=31$, $\Gamma=e^{5/2}(31/3)^{3/2}$, and $\gamma=1/(4\Gamma^2)\approx1.5\times10^{-6}$. The proof of Theorem~\ref{thm:union} uses the same computation with $x=\sqrt{w/m}$.
\end{proof}

\begin{lemma}[Smoothing]\label{lem:smoothing}
Let $B$ be a matrix over $\Z_2$, let $0<\alpha_0<\alpha<\tfrac12$, let $\lambda=(1-2\alpha)/(1-2\alpha_0)$, and let $u>0$ and $\Delta\ge1$. If every nonzero codeword of $C$ has weight at least $\Delta$, then
$$
  1-P_{\mathrm{PGM}}(B,\beta)\le\frac{2\big(F_B(\alpha_0)+e^{-1}\big)}u+2e^u\lambda^{2\Delta},
$$
where $\beta=\ln((1-\alpha)/\alpha)$.
\end{lemma}

\begin{proof}
\emph{Fourier contraction.} We use Fourier analysis on the group of cosets of $K$. Its characters are $\chi_c(X)=(-1)^{c\cdot x}$, $x\in X$, indexed by $c\in C$. For a function $g$ on cosets, write $\hat g(c)=\E[g\chi_c]$, so Parseval's identity gives $\E g^2=\sum_{c\in C}\hat g(c)^2$. For $L_{B,\alpha}$,
$$
  \hat L_{B,\alpha}(c)=\sum_XW_{B,\alpha}(X)\chi_c(X)=\E(-1)^{c\cdot E}=\sigma^{|c|},
$$
where $E$ has independent Bernoulli$(\alpha)$ entries. Write $L_0=L_{B,\alpha_0}$ and $L=L_{B,\alpha}$. As in the proof of part~2 of Proposition~\ref{prop:entropy}, Bernoulli$(\alpha)$ noise is the sum of Bernoulli$(\alpha_0)$ noise and an independent vector $E'$ with independent Bernoulli$(\alpha')$ entries, where $1-2\alpha'=\lambda$. Hence $L=TL_0$, where $(Tg)(X)=\E\,g(X+E')$. The operator $T$ maps nonnegative functions to nonnegative functions, preserves the mean under $U_B$, and multiplies the Fourier coefficient at $c$ by $\lambda^{|c|}$.

\emph{Density truncation.} Let $\tilde L=L_0\mathbf 1[L_0\le e^u]$ and $L_1=T\tilde L$. Since $0\le\tilde L\le L_0$, we have $0\le L_1\le L$, so $\E\sqrt L\ge\E\sqrt{L_1}$. On the event $L_0>e^u$ we have $L_0\le L_0\ln L_0/u$, and $L_0\ln L_0\ge-e^{-1}$ everywhere. The removed mass therefore satisfies
$$
  \varepsilon_1\coloneqq1-\E L_1=\E\big[L_0\mathbf 1[L_0>e^u]\big]\le\frac{\E[L_0\ln L_0]+e^{-1}}u=\frac{F_B(\alpha_0)+e^{-1}}u .
$$
\emph{Variance after added noise.} Every nonzero $c\in C$ has $|c|\ge\Delta$, so Parseval's identity bounds the variance of $L_1$ under $U_B$ by
$$
  \mathrm{Var}(L_1)=\sum_{c\ne0}\lambda^{2|c|}\hat{\tilde L}(c)^2\le\lambda^{2\Delta}\E\tilde L^2\le\lambda^{2\Delta}e^u\E\tilde L\le e^u\lambda^{2\Delta} .
$$
\emph{From variance to success.} Let $\mu_1=\E L_1=1-\varepsilon_1$. H\"older's inequality with exponents $3/2$ and $3$ gives $\mu_1=\E[L_1^{1/3}L_1^{2/3}]\le(\E\sqrt{L_1})^{2/3}(\E L_1^2)^{1/3}$, so
$$
  \big(\E\sqrt{L_1}\big)^2\ge\frac{\mu_1^3}{\mu_1^2+\mathrm{Var}(L_1)}\ge\mu_1-\frac{\mathrm{Var}(L_1)}{\mu_1} .
$$
If $\mu_1\ge\tfrac12$, then $1-P_{\mathrm{PGM}}(B,\beta)=1-(\E\sqrt L)^2\le1-(\E\sqrt{L_1})^2\le\varepsilon_1+2\,\mathrm{Var}(L_1)$, which implies the claim. If $\mu_1<\tfrac12$, then $2\varepsilon_1>1$, and the claim holds because $1-P_{\mathrm{PGM}}\le1$.
\end{proof}

\paragraph{Proofs of the theorem and the corollary.} The proofs of Theorem~\ref{thm:condensation} and Corollary~\ref{cor:condensation_sym} combine the four lemmas.

\begin{proof}[Proof of Theorem~\ref{thm:condensation}]
\emph{Success below $\beta_c$.} Work first in the unconditioned configuration model. If $\beta=0$, then $W_{B,1/2}=U_B$ and $P_{\mathrm{PGM}}(B,0)=1$. For $0<\beta<\beta_c$, choose $\alpha_0\in(\alpha_c,\alpha)$ with $\alpha=p_\beta(1)$. Then $\E_BF_B(\alpha_0)=o(n)$. Put $\lambda=(1-2\alpha)/(1-2\alpha_0)<1$ and let $\gamma$ be the constant in Lemma~\ref{lem:min_distance}. Apply Lemma~\ref{lem:smoothing} with $\Delta=\lceil\gamma m\rceil$ and $u=\Delta\ln(1/\lambda)$, so $e^u\lambda^{2\Delta}=\lambda^\Delta$. Since minimum distance is integer-valued, it is at least $\Delta$ unless it is smaller than $\gamma m$. Therefore
$$
  \E_B\big[1-P_{\mathrm{PGM}}(B,\beta)\big]\le\Pr\big[m_{\min}(C)<\gamma m\big]+\frac{2\big(\E_BF_B(\alpha_0)+e^{-1}\big)}{\Delta\ln(1/\lambda)}+2\lambda^{\Delta} .
$$
The first term tends to zero by Lemma~\ref{lem:min_distance}, the second because $\E_BF_B(\alpha_0)=o(n)$ and $\Delta\ge\gamma m=\gamma nD/k$, and the third because $\lambda<1$ and $\Delta\to\infty$. By Markov's inequality, the probability over the instance that $1-P_{\mathrm{PGM}}(B,\beta)$ exceeds $\sqrt{\E_B[1-P_{\mathrm{PGM}}(B,\beta)]}$ is at most this square root, which proves the second statement of part~(a).

\emph{Failure above $\beta_c$.} Let $\beta>\beta_c$, so $\alpha=p_\beta(1)<\alpha_c$. By the definition of $\alpha_c$ and monotonicity of $\bar F$, we have $\bar F(\alpha)>0$. Thus $\E_BF_B(\alpha)\ge\tfrac12\bar F(\alpha)n$ for all large $n$. Lemma~\ref{lem:concentration}, with $t=\tfrac14\bar F(\alpha)n$, gives $F_B(\alpha)\ge c_2m$ outside an event of probability $e^{-\Omega(m)}$, where $c_2=\bar F(\alpha)k/(4D)>0$. On this event, Lemma~\ref{lem:entropy_bound} implies
$$
  \sqrt{P_{\mathrm{PGM}}(B,\beta)}\le2e^{-\epsilon m},\qquad
  \epsilon=\min\{c_2/4,c_2^2/(4\beta^2)\}>0.
$$
Hence $P_{\mathrm{PGM}}(B,\beta)\le4e^{-2\epsilon m}\le e^{-\epsilon m}$ for all large $m$. Optimality of the PGM bounds the success of every decoder by this value.

\emph{Conditioning on no multiple edges.} Let $S$ be this event, whose probability is bounded below by a positive constant~\cite{richardson2008modern}. Since $1-P_{\mathrm{PGM}}\ge0$, its conditional expectation is at most its unconditional expectation divided by $\Pr[S]$. The same bound applies to the exceptional-event probability in part~(b). Both conclusions therefore transfer to the conditioned model.
\end{proof}

\begin{proof}[Proof of Corollary~\ref{cor:condensation_sym}]
For a nonnegative function $W$ on the cosets of $K$, let
$$
  \mathrm{BC}(W)=\E_{X\sim U_B}\sqrt{|C|\,W(X)},
$$
which for a law $W$ is its Bhattacharyya coefficient with $U_B$. Then $\mathrm{BC}(tW)=\sqrt t\,\mathrm{BC}(W)$ for $t\ge0$, $\mathrm{BC}$ is nondecreasing in $W$, $\mathrm{BC}(W+W')\le\mathrm{BC}(W)+\mathrm{BC}(W')$ because $\sqrt{v+v'}\le\sqrt v+\sqrt{v'}$, and $\mathrm{BC}(W)\le1$ for every law $W$ by the Cauchy--Schwarz inequality. By Lemma~\ref{lem:pgm_success}, $P_{\mathrm{PGM}}(\Phi)=\mathrm{BC}(W_\Phi)^2$, and $P_{\mathrm{PGM}}(B,\beta)=\mathrm{BC}(W_{B,\alpha})^2$ for $\alpha=p_\beta(1)$.

\emph{Weight mixtures and product domination.} Let $\Phi$ be symmetric, and let $P_\Phi(w)=\sum_{|x|=w}|\Phi(x)|^2$ be the probability of the weight $w$ under $|\Phi|^2$. Since $|\Phi(x)|^2=P_\Phi(|x|)/\binom m{|x|}$,
$$
  W_\Phi=\sum_{w=0}^mP_\Phi(w)\,W_w,
$$
where $W_w$ is the coset law of a uniformly random word of weight $w$. A vector with independent Bernoulli$(w/m)$ entries is, given its weight $w'$, uniformly distributed on the words of weight $w'$, so $W_{B,w/m}=\sum_{w'}\Pr[\mathrm{Bin}(m,w/m)=w']\,W_{w'}$. For $1\le w\le m-1$, the most likely value of the binomial law $\mathrm{Bin}(m,w/m)$ is its mean $w$, which therefore has probability at least $1/(m+1)$. Hence $W_w\le(m+1)\,W_{B,w/m}$, and
$$
  \mathrm{BC}(W_w)\le\sqrt{m+1}\,\mathrm{BC}(W_{B,w/m})=\sqrt{m+1}\,\sqrt{P_{\mathrm{PGM}}(B,\beta_w)},\qquad\beta_w=\ln\frac{m-w}w .
$$
For $w=0$, the law $W_0$ is the point mass at $K$, so $\mathrm{BC}(W_0)=|C|^{-1/2}\le2^{-(m-n)/2}$, because $C$ has dimension at least $m-n$.

Fix $\epsilon>0$. If $\epsilon>\alpha_c$, no amplitude function satisfies $s(\Phi)\ge s_c+\epsilon=1+(\epsilon-\alpha_c)$, so we may assume $\epsilon\le\alpha_c$. Let $\alpha'=\alpha_c-\epsilon/2\in(0,\alpha_c)$. As in the proof of part~(b) of Theorem~\ref{thm:condensation}, $\bar F(\alpha')>0$, and by Lemma~\ref{lem:concentration}, the event $\mathcal E_m$ that $F_B(\alpha')\ge c_1n$, with $c_1=\tfrac14\bar F(\alpha')$, has probability $1-e^{-\Omega(m)}$, with or without the conditioning. On $\mathcal E_m$, part~2 of Proposition~\ref{prop:entropy} yields $F_B(w/m)\ge c_1n$ for every $w$ with $1\le w\le\alpha'm$, and since $\beta_w\le\ln m$, Lemma~\ref{lem:entropy_bound} yields
$$
  \mathrm{BC}(W_w)\le\sqrt{m+1}\,\Big(e^{-c_1n/4}+e^{-c_1^2n^2/(4m\ln^2m)}\Big)
$$
for these $w$. Let $\vartheta_m$ be the larger of this bound and $2^{-(m-n)/2}$, so that $\mathrm{BC}(W_w)\le\vartheta_m$ for every $w\le\alpha'm$ on $\mathcal E_m$. Since $n=mk/D$ and $k<D$, $\sqrt{m+1}\,\vartheta_m\to0$.

\emph{Positive mass below the threshold.} Let $\Phi$ be symmetric with $s(\Phi)\ge s_c+\epsilon$. The expected violated fraction is $1-s(\Phi)\le\alpha_c-\epsilon$. Markov's inequality therefore gives $\delta'=\sum_{w\le\alpha'm}P_\Phi(w)\ge\delta=\epsilon/(2\alpha_c-\epsilon)>0$. By the properties of $\mathrm{BC}$,
$$
  \mathrm{BC}(W_\Phi)\le\sum_{w\le\alpha'm}\sqrt{P_\Phi(w)}\,\mathrm{BC}(W_w)+\mathrm{BC}\Big(\sum_{w>\alpha'm}P_\Phi(w)\,W_w\Big)\le\sqrt{\delta'}\,\sqrt{m+1}\,\vartheta_m+\sqrt{1-\delta'},
$$
where the second step uses the Cauchy--Schwarz inequality over the at most $m+1$ weights $w\le\alpha'm$, and the fact that $\sum_{w>\alpha'm}P_\Phi(w)W_w$ is $1-\delta'$ times a law. On $\mathcal E_m$, therefore, $\mathrm{BC}(W_\Phi)\le\sqrt{1-\delta}+\sqrt{m+1}\,\vartheta_m$, and once $\sqrt{m+1}\,\vartheta_m\le1$,
$$
  P_{\mathrm{PGM}}(\Phi)\le1-\delta+3\sqrt{m+1}\,\vartheta_m .
$$
By Lemma~\ref{lem:pgm_success}, every decoder has block success at most $P_{\mathrm{PGM}}(\Phi)$. This proves the first statement with $\eta_m=3\sqrt{m+1}\,\vartheta_m$ for all large $m$, and with $\eta_m=\delta$ for the other $m$.

\emph{Transfer to the output.} Suppose that symmetric amplitude functions $\Phi_m$ have decoders with block success at least $1-\varepsilon_m$, where $\varepsilon_m\to0$. On $\mathcal E_m$, once $\varepsilon_m+\eta_m<\delta$, the first statement implies $s(\Phi_m)<s_c+\epsilon$. Theorem~\ref{thm:guarantee} bounds the offset-averaged expected satisfied fraction by $s(\Phi_m)+2\sqrt{\varepsilon_m}<s_c+\epsilon+2\sqrt{\varepsilon_m}$. Since $\epsilon>0$ is arbitrary and $\Pr[\mathcal E_m]\to1$, the second statement follows.
\end{proof}

\subsection{Separation from the Holevo threshold and convergence at large degrees}\label{app:condensation_holevo}

The positive difference $\beta_{\mathrm H}-\beta_c$ has an explicit finite-degree lower bound and an exponentially small upper bound at a fixed degree ratio (Proposition~\ref{prop:condensation_holevo}). For the lower bound, the trial law in~\eqref{eq:condensation_trial} leaves positive conditional entropy after finitely many noisy observations. For the upper bound, we show that~\eqref{eq:functional} is nonpositive for every mean-zero law once the capacity is below its Holevo value by a suitable exponentially small amount. The configuration-model identity is part~3 of Proposition~\ref{prop:entropy}, with the instance-size limit taken before the degree limit.

\paragraph{An explicit lower bound.} Write $h(p)=-p\ln p-(1-p)\ln(1-p)$ for the binary entropy in nats. For the trial law $\pi_*=(\delta_{-1}+\delta_1)/2$, the conditional entropy in~\eqref{eq:condensation_trial} has the finite formula
\begin{equation}\label{eq:condensation_binomial}
  H_D(\alpha)=\sum_{j=0}^D\binom Dj\alpha^j(1-\alpha)^{D-j}
  \ln\left[1+\left(\frac{\alpha}{1-\alpha}\right)^{D-2j}\right].
\end{equation}
Indeed, given the true sign, the number $j$ of flipped observations has law $\mathrm{Bin}(D,\alpha)$, and the likelihood ratio of the opposite sign to the true sign is $(\alpha/(1-\alpha))^{D-2j}$. Averaging the negative logarithm of the posterior probability of the true sign proves~\eqref{eq:condensation_binomial}.

Retaining only the central observation counts in the posterior entropy yields
\begin{equation}\label{eq:condensation_central}
  0<L_D(\alpha)\le H_D(\alpha)\le\ln2,\qquad
  L_D(\alpha)=
  \begin{cases}
    \displaystyle\binom D{D/2}[\alpha(1-\alpha)]^{D/2}\ln2,&D\text{ even},\\[5pt]
    \displaystyle\binom D{(D-1)/2}[\alpha(1-\alpha)]^{(D-1)/2}h(\alpha),&D\text{ odd}.
  \end{cases}
\end{equation}
For even $D$, the strings with equally many signs of each kind have posterior entropy $\ln2$ and total probability $\binom D{D/2}[\alpha(1-\alpha)]^{D/2}$. For odd $D$, the two central counts have posterior entropy $h(\alpha)$ and combined probability $\binom D{(D-1)/2}[\alpha(1-\alpha)]^{(D-1)/2}$. This proves~\eqref{eq:condensation_central}.

Set $g(\beta)=\operatorname{cap}(\tanh(\beta/2))$, so that $g(\beta_{\mathrm H})=(k/D)\ln2$. Differentiation gives
\begin{equation}\label{eq:condensation_capacity_derivative}
  g'(\beta)=\frac{\beta}{4\cosh^2(\beta/2)}\le\frac14
  \qquad(\beta\ge0),
\end{equation}
because $\cosh^2(\beta/2)\ge1+\beta^2/4\ge\beta$. Let $L_*=L_D(\alpha_H)$ and put $\beta_*=\beta_{\mathrm H}-(2k/D)L_*$. This temperature is positive, since $g(0)=0$ and~\eqref{eq:condensation_capacity_derivative} imply
$$
  0<\frac{2k}{D}L_*\le\frac{2k}{D}\ln2
  =2g(\beta_{\mathrm H})\le\frac{\beta_{\mathrm H}}2.
$$
For $0\le\beta\le\beta_{\mathrm H}$, the binary symmetric channel at $\alpha(\beta)=(1+e^\beta)^{-1}$ is a degradation of the channel at $\alpha_H$. The same holds for $D$ independent observations, so data processing gives $H_D(\alpha(\beta))\ge H_D(\alpha_H)\ge L_*$. Integrating~\eqref{eq:condensation_capacity_derivative} in~\eqref{eq:condensation_trial}, we obtain
$$
  \mathcal B_{\tanh(\beta/2)}(\pi_*)\ge
  L_*-\frac{D}{4k}(\beta_{\mathrm H}-\beta).
$$
At $\beta=\beta_*$, the right-hand side equals $L_*/2>0$. By continuity it remains positive at some smaller positive inverse temperature. The variational identity and the definition of $\beta_c$ therefore imply
\begin{equation}\label{eq:condensation_gap_lower}
  \beta_{\mathrm H}-\beta_c>\frac{2k}{D}L_D(\alpha_H).
\end{equation}

\paragraph{Sufficient finite-degree conditions for the upper bound.} The upper estimate in~\eqref{eq:condensation_exp_gap} holds whenever
\begin{equation}\label{eq:condensation_degree_conditions}
  k\ge8,\qquad
  3(\ln2)^2D(7/8)^{k-2}\le1,\qquad
  C_{k/D}e^{-k/48}\le\frac{\beta_{\mathrm H}}2.
\end{equation}
We first show that the first two conditions imply $\mathcal B_\sigma(\pi)\le0$ for every mean-zero law $\pi$ whenever $\sigma\in(0,1)$ satisfies the capacity margin
\begin{equation}\label{eq:condensation_capacity_margin}
  \operatorname{cap}(\sigma)\le\frac{k}{D}\big(\ln2-e^{-k/48}\big).
\end{equation}
We then convert this margin to a temperature bound using the third condition.

\paragraph{Capacity and moment estimates.} The capacity has the convergent expansion
\begin{equation}\label{eq:condensation_capacity_series}
  \operatorname{cap}(t)=\sum_{j\ge1}\gamma_jt^{2j},
  \qquad \gamma_j=\frac1{2j(2j-1)},\qquad
  \frac12t^2\le\operatorname{cap}(t)\le(\ln2)t^2
  \quad (|t|\le1).
\end{equation}
The series follows by integrating the expansion of $\operatorname{cap}'(t)=\operatorname{atanh}(t)$ for $|t|<1$. The coefficients are positive and sum to $\ln2$, by telescoping $\gamma_j=1/(2j-1)-1/(2j)$, which also proves convergence at the endpoints and both inequalities.

Fix a mean-zero law $\pi$, and let $\mathfrak m$ have law $\pi$. Introduce its even moments $v_j=\E\mathfrak m^{2j}$, let $v=v_1$, and set
$$
  y=v^{k-1},\qquad s_0=\sigma^2y=\E R_i^2,\qquad x=Ds_0,
  \qquad A=Da-\frac{D(k-1)}k b.
$$
Thus $x$ is the total expected squared reliability of the observations, and $\mathcal B_\sigma(\pi)=A-\mathcal I_D$. If $v=0$, then $\pi=\delta_0$ and the functional is zero. Assume henceforth that $v>0$, so $x>0$. By independence and~\eqref{eq:condensation_capacity_series},
$$
  a=\sum_{j\ge1}\gamma_j\sigma^{2j}v_j^{k-1},\qquad
  b=\sum_{j\ge1}\gamma_j\sigma^{2j}v_j^k,\qquad
  A=\frac Dk\sum_{j\ge1}\gamma_j\sigma^{2j}G_k(v_j),
$$
where $G_k(u)=ku^{k-1}-(k-1)u^k$. On $[0,1]$, $G_k'(u)=k(k-1)u^{k-2}(1-u)\ge0$ and $G_k(1)=1$. Since $0\le v_j\le v$, we have
\begin{equation}\label{eq:condensation_A_bounds}
  \begin{aligned}
    A&\le\frac Dk\operatorname{cap}(\sigma)G_k(v)
       \le\frac Dk\operatorname{cap}(\sigma),\\
    A&\le(\ln2)x\left(1-v+\frac vk\right),\\
    A&\le\frac Dk\operatorname{cap}(\sigma)y(1-\ln y).
  \end{aligned}
\end{equation}
For the second inequality, use $G_k(v)=y[k-(k-1)v]$ and $\operatorname{cap}(\sigma)\le(\ln2)\sigma^2$. For the third, use $(k-1)(1-v)\le-(k-1)\ln v=-\ln y$.

\paragraph{Information and conditional entropy.} Let $O=((R_i,T_i))_{i\le D}$ be the observations in~\eqref{eq:functional}, where $\Pr[T_i=\xi\mid\xi,R_i]=(1+R_i)/2$. Let $M=\E[\xi\mid O]$ and $S=\sum_iR_iT_i$. Conditional independence gives
$$
  \E[\xi S]=x,\qquad
  \E S^2=x+D(D-1)s_0^2=x+(1-1/D)x^2.
$$
Since $\E[\xi S]=\E[MS]$, the Cauchy--Schwarz inequality implies $\E M^2\ge x/[1+(1-1/D)x]\ge x/(1+x)$. The posterior probability of $\xi=+1$ is $(1+M)/2$, so $\mathcal I_D=\E\operatorname{cap}(M)$ and
\begin{equation}\label{eq:condensation_information}
  \mathcal I_D\ge\frac{x}{2(1+x)}.
\end{equation}

For reliable observations we instead use the conditional-entropy estimate
\begin{equation}\label{eq:condensation_residual}
  \ln2-\mathcal I_D=H(\xi\mid O)\le e^{-x/2}.
\end{equation}
To prove it, concavity of the logarithm gives, for $p\in[0,1]$,
$$
  h(p)\le2\ln\big(\sqrt p+\sqrt{1-p}\big)
  =\ln\big(1+2\sqrt{p(1-p)}\big)\le2\sqrt{p(1-p)}.
$$
If $P_+$ and $P_-$ are the conditional laws of $O$ given the two signs, averaging this inequality over the uniform-prior posterior yields
$$
  H(\xi\mid O)\le\int\sqrt{dP_+\,dP_-}
  =\big(\E\sqrt{1-R_1^2}\big)^D
  \le(1-s_0/2)^D\le e^{-x/2}.
$$
The equality follows because the observations are conditionally independent and the Bhattacharyya coefficient of a single observation, conditional on its revealed reliability $R_i=r$, is $\sqrt{1-r^2}$. This proves~\eqref{eq:condensation_residual}.

\paragraph{Weak observations.} When $v\le7/8$, we compare the redundancy $Da-\mathcal I_D$ directly with the negative term $D(k-1)b/k$. Let $I_i$ be the mutual information between $\xi$ and the first $i$ observations, and $M_i$ its posterior mean, with $I_0=M_0=0$. Given the first $i-1$ observations and $R_i$, the sign $T_i$ has mean $R_iM_{i-1}$. Since $R_i$ is independent of the past and of $\xi$, the entropy formula for this binary channel gives
$$
  I_i-I_{i-1}=a-\E\operatorname{cap}(R_iM_{i-1})\le a.
$$
Thus $I_i\le ia$. By~\eqref{eq:condensation_capacity_series}, $\E M_{i-1}^2\le2I_{i-1}$, and $R_i$ is independent of $M_{i-1}$. Summing the increments, we obtain the quadratic redundancy bound
\begin{equation}\label{eq:condensation_redundancy}
  \begin{aligned}
    Da-\mathcal I_D
    &=\sum_{i=1}^D\E\operatorname{cap}(R_iM_{i-1})\\
    &\le(\ln2)s_0\sum_{i=1}^D\E M_{i-1}^2
    \le2(\ln2)s_0\sum_{i=1}^DI_{i-1}\\
    &\le(\ln2)s_0aD(D-1)
    \le(\ln2)^2D(D-1)s_0^2.
  \end{aligned}
\end{equation}
The first capacity term also gives $b\ge\sigma^2v^k/2>0$. Consequently,
$$
  \frac{Da-\mathcal I_D}{D(k-1)b/k}
  \le\frac{2k}{k-1}(\ln2)^2(D-1)\sigma^2v^{k-2}
  \le3(\ln2)^2D(7/8)^{k-2}\le1.
$$
Here $2k/(k-1)\le3$ for $k\ge3$, and the final inequality is a hypothesis in~\eqref{eq:condensation_degree_conditions}. Hence $\mathcal B_\sigma(\pi)\le0$ for every law with $v\le7/8$, without using the capacity margin.

\paragraph{Strong observations and the intermediate cases.} Suppose $v>7/8$. The following four cases cover every $x>0$, $y\in(0,1]$, and $\sigma\in(0,1)$. We use $\ln2<7/10$, which follows by retaining the first four terms of the series for $e^{7/10}$.
\begin{enumerate}[label=(\alph*)]
\item If $x\le2$, then $k\ge8$ implies
$$
  1-v+\frac vk<\frac18+\frac7{8k}\le\frac{15}{64}.
$$
By~\eqref{eq:condensation_A_bounds}, $A<(21/128)x<x/6$, whereas~\eqref{eq:condensation_information} implies $\mathcal I_D\ge x/6$. Thus $\mathcal B_\sigma(\pi)<0$.

\item If $x\ge2$ and $y\le1/8$, then the capacity margin implies $(D/k)\operatorname{cap}(\sigma)\le\ln2$. Since $y(1-\ln y)$ is nondecreasing on $(0,1]$, \eqref{eq:condensation_A_bounds} gives
$$
  A\le\frac{(\ln2)(1+\ln8)}8<\frac{217}{800}<\frac13.
$$
On the other hand, \eqref{eq:condensation_information} implies $\mathcal I_D\ge1/3$, so the functional is negative.

\item If $x\ge2$, $y\ge1/8$, and $\sigma^2\le k/(3D)$, then
$$
  A\le\frac Dk\operatorname{cap}(\sigma)
  \le\frac Dk(\ln2)\sigma^2\le\frac{\ln2}3
  <\frac13\le\mathcal I_D.
$$
Again the functional is negative.

\item In the remaining case, $x\ge2$, $y\ge1/8$, and $\sigma^2\ge k/(3D)$. Thus $x=D\sigma^2y\ge k/24$. Combining~\eqref{eq:condensation_A_bounds}, \eqref{eq:condensation_residual}, and~\eqref{eq:condensation_capacity_margin}, we have
$$
  \mathcal B_\sigma(\pi)
  \le\frac Dk\operatorname{cap}(\sigma)-\ln2+e^{-x/2}
  \le-e^{-k/48}+e^{-k/48}=0.
$$
\end{enumerate}
Together with the weak-observation case and $\pi=\delta_0$, this proves exact nonpositivity for every mean-zero law, not merely an exponentially small upper bound. Since $\mathcal B_\sigma(\delta_0)=0$, part~3 of Proposition~\ref{prop:entropy} implies $\bar F((1-\sigma)/2)=0$ under~\eqref{eq:condensation_capacity_margin}.

\paragraph{Conversion to inverse temperature.} Assume all three conditions in~\eqref{eq:condensation_degree_conditions} and let $\beta_L=\beta_{\mathrm H}-C_{k/D}e^{-k/48}\ge\beta_{\mathrm H}/2>0$. On $[\beta_{\mathrm H}/2,\beta_{\mathrm H}]$, \eqref{eq:condensation_capacity_derivative} implies the lower bound
$$
  g'(\beta)\ge\frac{\beta_{\mathrm H}(1-\sigma_H^2)}8.
$$
Integrating over $[\beta_L,\beta_{\mathrm H}]$ and using the definition of $C_{k/D}$ in~\eqref{eq:condensation_exp_gap}, we obtain
$$
  g(\beta_{\mathrm H})-g(\beta_L)
  \ge\frac{\beta_{\mathrm H}(1-\sigma_H^2)}8 C_{k/D}e^{-k/48}
  =\frac kD e^{-k/48}.
$$
Since $g(\beta_{\mathrm H})=(k/D)\ln2$, the capacity margin holds at $\sigma=\tanh(\beta_L/2)$. It also holds at every smaller positive inverse temperature because $g$ is increasing. Uniform nonpositivity therefore implies $\beta_L\le\beta_c$, proving the upper bound in~\eqref{eq:condensation_exp_gap}. At any fixed rational ratio $k/D\in(0,1)$, $\beta_{\mathrm H}$ and $C_{k/D}$ are fixed, while $k$ tends to infinity along admissible pairs. Thus $D(7/8)^{k-2}\to0$ and $C_{k/D}e^{-k/48}\to0$, so all conditions in~\eqref{eq:condensation_degree_conditions} eventually hold. This completes Proposition~\ref{prop:condensation_holevo}.

\paragraph{Two-sided exponential estimates.} At a fixed ratio, $\alpha_H$ is fixed in $(0,1/2)$. Stirling's formula in~\eqref{eq:condensation_central}, applied to either parity of $D$, gives
$$
  L_D(\alpha_H)=\Theta_{k/D}\!\left(D^{-1/2}(1-\sigma_H^2)^{D/2}\right),
$$
because $2\sqrt{\alpha_H(1-\alpha_H)}=\sqrt{1-\sigma_H^2}$. Together, \eqref{eq:condensation_gap_lower} and~\eqref{eq:condensation_exp_gap} bound the positive gap below and above by exponentially decaying quantities, with a polynomial prefactor in the lower bound. These estimates do not identify a limiting exponential rate.

\section{Rank deficiency and unambiguous discrimination}\label{app:rank}

This appendix proves Lemma~\ref{lem:rank} and the statements of Section~\ref{sec:known_nonlocal} that rest on it. Lemma~\ref{lem:rank} says that, when each row of a $(k,D)$-regular matrix is kept independently with probability $\sigma>\sigma^\ast$, the kept rows are linearly dependent with high probability. We first prove the lemma and describe the limit of $2\operatorname{atanh}(\sigma^\ast)$ at larger degrees. We then show that the output law of Prange's algorithm is close to $\mu_\beta$ when its retained system has full row rank with high probability. Finally, we analyze the block success of unambiguous discrimination followed by Gaussian elimination, and the satisfied fraction that DQI reaches with this decoder.

\begin{proof}[Proof of Lemma~\ref{lem:rank}]
For $\sigma<1$, $\zeta''(\sigma)=k(D-1)(1-\sigma)^{D-2}>0$, so $\zeta$ is strictly convex on $[0,1]$. Moreover $\zeta(0)=0$, $\zeta'(0)=1-k<0$, and $\zeta(k/D)=\frac k D\big(1-\frac k D\big)^D>0$. A strictly convex function with these properties is negative on an interval $(0,\sigma^\ast)$, vanishes at a unique point $\sigma^\ast\in(0,k/D)$, and is positive on $(\sigma^\ast,1]$.

Let $n_R$ be the number of variables that occur in at least one clause of $R$. Every column of $B_R$ indexed by another variable is zero, so $\rank B_R\le n_R$ and $\mathrm{def}(R)\ge F$, where $F=|R|-n_R$. We have $\E|R|=\sigma m$. Each variable lies in $D$ clauses, so it occurs in no clause of $R$ with probability $(1-\sigma)^D$, and $\E n_R=n\big(1-(1-\sigma)^D\big)$. Since $n=mk/D$, it follows that $\E F=\zeta(\sigma)m$. The random variable $F$ is a function of the $m$ independent indicators of the events $j\in R$. Adding a clause to $R$ increases $|R|$ by one and $n_R$ by at most $k$, so it changes $F$ by an amount in $[1-k,1]$, whose absolute value is at most $k$. By McDiarmid's bounded-differences inequality, $\Pr[F\le\E F-t]\le\exp\big(-2t^2/(mk^2)\big)$ for every $t>0$. With $t=\frac12\zeta(\sigma)m$ and $\mathrm{def}(R)\ge F$, this proves the bound. On the complementary event, $\mathrm{def}(R)\ge\frac12\zeta(\sigma)m>0$, so the rows of $B_R$ are linearly dependent.
\end{proof}

\paragraph{The rank threshold at larger degrees.} Lemma~\ref{lem:rank} holds for every $(k,D)$-regular matrix, and the probability is only over $R$. For Prange's algorithm, $R$ keeps each clause with probability $\sigma_\beta$, so above $2\operatorname{atanh}(\sigma^\ast)$ its retained system is rank deficient with probability $1-e^{-\Omega(m)}$, and an argument that requires full row rank with high probability, such as that of Section~\ref{sec:prange}, applies only below this value. As the degrees grow at a fixed ratio $k/D$, $\zeta(k/D-\eta)=-\eta+\frac k D\big(1-\frac k D+\eta\big)^D$ tends to $-\eta$ for every fixed $\eta\in(0,k/D)$, so $k/D-\eta<\sigma^\ast<k/D$ for large degrees. Hence $\sigma^\ast\to k/D$ and $2\operatorname{atanh}(\sigma^\ast)\to\beta_{\mathrm P}$. Whether Prange's algorithm samples up to $2\operatorname{atanh}(\sigma^\ast)$ on regular instances is not known. For Erd\H{o}s--R\'enyi hypergraphs, the exact threshold for full row rank is known from the analysis of random XORSAT, and Zhou et al.\ use it to prove that Prange's algorithm samples below this threshold~\cite[Theorem~3.2 and Proposition~B.2]{zhou2026gibbs}. This threshold is at most the value obtained by counting rows and columns, and it approaches this value as the clause size grows~\cite[Lemma~3.3]{zhou2026gibbs}.

\paragraph{The output law of Prange's algorithm.} Prange's algorithm of Section~\ref{sec:prange} keeps each clause independently with probability $\sigma_\beta$ and outputs a uniformly random solution of the linear system of the kept clauses. Its output law differs from $\mu_\beta$ only through retained systems without full row rank, and on average over the offset it is close to $\mu_\beta$ if the retained system has full row rank with high probability. Indeed, let $\delta$ be the probability that the retained system does not have full row rank, and let $\mathrm{def}(R)=|R|-\rank B_R$ as in Lemma~\ref{lem:rank}. For a uniformly random offset, the system indexed by $R$ is consistent with probability $2^{-\mathrm{def}(R)}$, so the algorithm outputs ``fail'' with probability at most $\delta$ on average over the offset. By the formula for $\nu(z)$ in the proof of Proposition~\ref{prop:prange}, $\nu\le A\mu_\beta$, where $A=\E_R\big[2^{\mathrm{def}(R)}\mathbf 1\{\text{the system indexed by $R$ is consistent}\}\big]$ has mean one over the offset and differs from one by at most $2\delta$ on average. Counting ``fail'' as an outcome, the output law is therefore within $3\delta$ of $\mu_\beta$ in total variation on average over the offset.

\paragraph{Unambiguous discrimination followed by Gaussian elimination.} For $q=2$, unambiguous discrimination erases each symbol independently with probability $\sigma_\beta$ and reveals the other symbols without error, so the set $R$ of erased symbols has the same law as the set of retained clauses in Prange's algorithm. Given the outcomes, a uniformly random codeword $c$ is uniformly distributed on a coset of the space of codewords supported in $R$, because $R$ does not depend on $c$, and this space has dimension $\mathrm{def}(R)$. Every decoder that uses only these outcomes therefore recovers $c$ with probability at most $\E\,2^{-\mathrm{def}(R)}$, and Gaussian elimination recovers $c$ with certainty exactly when $\mathrm{def}(R)=0$. By Lemma~\ref{lem:rank}, if $\sigma_\beta>\sigma^\ast$, then $\E\,2^{-\mathrm{def}(R)}\le\exp\big(-\zeta(\sigma_\beta)^2m/(2k^2)\big)+2^{-\zeta(\sigma_\beta)m/2}$, so the block success tends to zero for $\beta>2\operatorname{atanh}(\sigma^\ast)$, which is $1.07565$ for $(3,6)$ and $1.09829$ for $(6,12)$. The same rank condition determines whether Prange's algorithm returns an assignment. For a uniformly random offset, the retained system is consistent with probability $2^{-\mathrm{def}(R)}$, so the algorithm of Section~\ref{sec:prange} returns an assignment with probability $\E\,2^{-\mathrm{def}(R)}$ on average over the offset, which tends to one exactly when $\Pr[\mathrm{def}(R)>0]$ tends to zero. The block success of this decoder lies between $\Pr[\mathrm{def}(R)=0]$ and $\E\,2^{-\mathrm{def}(R)}$, so it tends to one under the same condition.

Conversely, $\Pr[\mathrm{def}(R)>0]\le U_C(\sigma_\beta)$, where $U_C(\theta)=\sum_{c\in C\setminus\{0\}}\theta^{|c|}$ is the weight enumerator of Section~\ref{sec:prange}, and $U_C$ is increasing. By Theorem~\ref{thm:union}, on the configuration model, the block success is therefore $1-o(1)$ with probability $1-o(1)$ over the instance when $\sigma_\beta\le1-e^{-\beta_U}$, that is, for $\beta\le0.867$ for $(3,6)$ and $\beta\le0.878$ for $(6,12)$. At rate $1/2$, the largest inverse temperature at which this decoder succeeds therefore lies in $[0.867,1.0757]$ for $(3,6)$ and in $[0.878,1.0983]$ for $(6,12)$, where the endpoints are rounded outward (Table~\ref{tab:landscape}). These intervals lie well below the values $1.839$ and $1.549$ of $\beta_d$.

\paragraph{The satisfied fraction.} For Max-XORSAT, part~2 of Theorem~\ref{thm:guarantee} shows that, at every $\beta$ at which the block success of this decoder tends to one, the expected satisfied fraction that DQI reaches with it, averaged over the offset, is $p_\beta(0)+o(1)=\tfrac12+\tfrac12\sigma_\beta+o(1)$. By Lemma~\ref{lem:rank}, this fraction is at most $\tfrac12+\tfrac12\sigma^\ast(k,D)+o(1)$. The optimization version of Prange's algorithm differs from the sampler of Section~\ref{sec:prange}. It solves by Gaussian elimination a maximal set of linearly independent equations, which has $\rank B$ elements. Each other equation is then satisfied with probability $\tfrac12$ on average over the offset, so the algorithm reaches $\tfrac12+\rank B/(2m)$~\cite{shutty2026locallyquantumdecoders}. This value tends to $\tfrac12+k/(2D)$ when $B$ has rank $n-o(m)$, which Shutty et al.\ expect for random instances. The bound for DQI with unambiguous discrimination is strictly smaller, for example $0.7457$ against $0.75$ for $(3,6)$, because unambiguous discrimination erases each symbol independently instead of choosing the equations. At larger degrees, $\sigma^\ast$ approaches $k/D$, and the order of $\beta_d$ and $\beta_{\mathrm P}$ is reversed (Figure~\ref{fig:degrees}). On the benchmark families, both the value of Prange's algorithm and the bound $\tfrac12+\tfrac12\sigma^\ast$ exceed $s_d$ exactly for $(5,6)$, $(6,7)$, and $(7,8)$ (Table~\ref{tab:benchmark}). DQI with this nonlocal decoder can therefore exceed the limit of local decoders on these families only for these three pairs. We expect it to do so. Lemma~\ref{lem:rank} does not exclude it, because $k/D-\sigma^\ast$ is below $2\times10^{-5}$ for these pairs. A proof requires that the block success tend to one at an inverse temperature with $\tfrac12+\tfrac12\sigma_\beta>s_d$, which for $(5,6)$ means $\sigma_\beta$ within about $10^{-3}$ of $\sigma^\ast$.

\section{Block amplitude functions and local decoding}\label{app:block}

This appendix proves Proposition~\ref{prop:block_four} and the statements of Section~\ref{sec:approx_block} about local decoders. Proposition~\ref{prop:block_four} says that, for four degree pairs, DQI with the block amplitude function and any decoder whose block success tends to one reaches a smaller satisfied fraction than DQI with a product amplitude function and a local decoder. We first describe the channel that the block amplitude function induces on each block (Lemma~\ref{lem:block_channel}). We then derive an entropy condition for every decoder from the Holevo bound (Lemma~\ref{lem:block_entropy}) and deduce Proposition~\ref{prop:block_four} from it and from the certified bounds on $\beta_d$ of Appendix~\ref{app:certified}. For local decoders, we compute the optimal estimate of one symbol from a set of blocks (Lemma~\ref{lem:block_helstrom}), complete the ball around a symbol to whole blocks (Lemma~\ref{lem:block_completion}), and evaluate the optimal estimate on trees of blocks by a lossless recursion (Proposition~\ref{prop:block_tree}). We end with a numerical evaluation of this recursion. Throughout, $q=2$, $f(z)=z$, and $B$ is a matrix from Gallager's ensemble with blocks $S_1,\dots,S_{m/D}$.

\paragraph{The channel on a block.} Let $\mathcal P=\{c\in\Z_2^D:|c|\text{ is even}\}$, and let $\mathcal Q=\Z_2^D/\{0,\mathbf 1\}$ be the group of pairs $\{y,y+\mathbf 1\}$. Both groups have $2^{D-1}$ elements. For $c\in\mathcal P$ and $y\in\mathcal Q$, the product $c\cdot y$ does not depend on the representative of $y$, because $c\cdot\mathbf 1=|c|$ is even, and it identifies each of the two groups with the group of characters of the other. The restriction $c_S$ of a codeword to a block $S$ lies in $\mathcal P$, because $S$ is the support of a check. Each pair contains exactly one pattern of the set of Section~\ref{sec:approx_block}, so we index by $\mathcal Q$ the computational basis states of a block on which $\Phi^{(D)}$ is supported. Let
\begin{equation}\label{eq:block_overlap}
  \theta=1-\frac{2^{D-1}\alpha}{2^{D-1}-1},\qquad \nu=\theta\,\delta_0+(1-\theta)\,\unif_{\mathcal Q},
\end{equation}
where $\unif_{\mathcal Q}$ is the uniform law on $\mathcal Q$, so that $\nu(0)=1-\alpha$ and $\nu(y)=\alpha/(2^{D-1}-1)$ for $y\ne0$. Then $\theta\in[0,1]$ exactly when $\alpha\le1-2^{1-D}$.

\begin{lemma}[The channel on a block]\label{lem:block_channel}
For $c\in\mathcal P$, the state of a block on which the codeword has restriction $c$ is $\ket{\psi_c}=Z^c\ket{\Phi^{(D)}}=\sum_{y\in\mathcal Q}\sqrt{\nu(y)}\,(-1)^{c\cdot y}\ket y$, and $\langle\psi_c|\psi_{c'}\rangle=\theta$ for $c\ne c'$. The expected number of violated clauses of a block in the state $\ket{\Phi^{(D)}}$ is $(1-\theta)\mu_D$, so the expected satisfied fraction in $\ket\Phi$ is $s(\alpha)=1-(1-\theta)\mu_D/D$, as in~\eqref{eq:block_fraction}. These statements do not depend on which pattern is chosen from a pair of weight $D/2$.
\end{lemma}

\begin{proof}
The phase of $Z^c$ on a pattern $y$ is $(-1)^{c\cdot y}$, which depends only on the pair of $y$. For $c\ne c'$ in $\mathcal P$, the character $y\mapsto(-1)^{(c+c')\cdot y}$ of $\mathcal Q$ is not trivial, so its sum over $\mathcal Q$ vanishes, and $\langle\psi_c|\psi_{c'}\rangle=\sum_y\nu(y)(-1)^{(c+c')\cdot y}=\theta$. The pattern chosen from the pair of $y\in\Z_2^D$ has weight $\min(|y|,D-|y|)$, and the pair of a uniformly random $y$ is uniformly random on $\mathcal Q$, so under $\unif_{\mathcal Q}$ the chosen pattern has mean weight $\mu_D$. The pattern of the zero pair has weight zero, so under $\nu$ the mean weight is $(1-\theta)\mu_D$. Dividing by $D$ and using $1-\theta=2^{D-1}\alpha/(2^{D-1}-1)$ gives~\eqref{eq:block_fraction}. Only the weights of the chosen patterns enter, and the two patterns of a pair of weight $D/2$ have the same weight.
\end{proof}

By Lemma~\ref{lem:block_channel}, the Gram matrix of the states of a block is $(1-\theta)I+\theta J$, where $J$ is the all-ones matrix. For $\alpha\ge1-2^{1-D}$, that is, $\theta\le0$, the satisfied fraction $s(\alpha)$ is at most $1-\mu_D/D$, and a classical algorithm reaches $1-\mu_D/D$ on every matrix from Gallager's ensemble, in expectation over a uniformly random offset. Starting from the zero assignment, this algorithm flips the variable of each block if the flip reduces the number of violated clauses of the block. Flipping the variable of a block changes only the clauses of this block, and at the zero assignment the pattern of each block is uniformly random, so at the end each block has $\mu_D$ violated clauses in expectation. We therefore consider $\theta\in[0,1]$ below.

The value $1-\mu_D/D$ is strictly below $s_d$ for every $k\ge3$. For independent uniform signs $X_1,\dots,X_D$, Cauchy--Schwarz gives $1-2\mu_D/D=\E|\sum_jX_j|/D\le1/\sqrt D$. For the binary check-node operation, the conditional second moment of the outgoing absolute overlap is at most the sum of the squared incoming overlaps, as follows by substituting its two outcomes and their probabilities. On a $(2,D)$ tree with single-symbol overlap $\sigma$, the mean squared symbol-message overlap therefore contracts whenever $(D-1)\sigma^2<1$, and the optimal root error tends to zero. A tree with only its displayed parity constraints and no boundary constraints, which we call a free tree, has the same number of extensions of each assignment on any pruned subtree. Pruning additional child checks from a free $(k,D)$ tree gives a free $(2,D)$ tree, and the restriction of the uniform codeword remains uniform. Discarding the other qubits shows that the optimal root success on the $(k,D)$ tree is at least that on the $(2,D)$ tree. By the product-channel duality and the definition of $s_d$, it follows that $s_d\ge\tfrac12+1/(2\sqrt{D-1})>1-\mu_D/D$. In particular, the block parameter with $s(\alpha)=s_d$ lies in $[0,1-2^{1-D}]$.

\paragraph{The entropy condition.} Part~1 of Proposition~\ref{prop:block_four} rests on the following condition, which holds for every decoder.

\begin{lemma}[Entropy condition]\label{lem:block_entropy}
Let $B$ have $m$ rows, and let a decoder for the decoding problem induced by DQI with the block amplitude function of parameter $\alpha$ have block success $1-\varepsilon$. Then
$$
  (1-\varepsilon)(D-k)-\frac DmH_2(\varepsilon)\le\mathcal H_D(\alpha)=H_2(\alpha)+\alpha\log_2\big(2^{D-1}-1\big).
$$
\end{lemma}

\begin{proof}
Measure entropies in bits, and let $\bar\rho$ be the average of $\ket{\psi^\Phi_c}\!\bra{\psi^\Phi_c}$ over a uniformly random $c\in C$. Since $|C|\ge2^{m-n}$ and $n=mk/D$, we have $\log_2|C|\ge(m/D)(D-k)$. The states are pure, so the Holevo information of this ensemble is the von Neumann entropy $S(\bar\rho)$. By subadditivity, $S(\bar\rho)$ is at most the sum of the entropies of the reduced states of $\bar\rho$ on the $m/D$ blocks. For every codeword, the diagonal of the state of a block in the computational basis is $|\Phi^{(D)}|^2$, and replacing a state by its diagonal does not decrease its entropy, so each reduced state has entropy at most the entropy of $\nu$, which is $\mathcal H_D(\alpha)$. By the Holevo bound, the mutual information between $c$ and the output $\hat c$ of the decoder is at most $S(\bar\rho)\le(m/D)\,\mathcal H_D(\alpha)$, and by Fano's inequality it is at least $(1-\varepsilon)\log_2|C|-H_2(\varepsilon)$. Dividing by $m/D$ proves the claim.
\end{proof}

\begin{proof}[Proof of Proposition~\ref{prop:block_four}]
\emph{Part~1.} The derivative of $\mathcal H_D$ is $\log_2\big((2^{D-1}-1)(1-\alpha)/\alpha\big)$, which is positive for $\alpha<1-2^{1-D}$, so $\mathcal H_D$ is increasing on $[0,1-2^{1-D}]$, and $s(\alpha)$ is decreasing on $[0,1]$. For each of the four pairs, let $\bar s$ be the value in the second column of the table, and let $\bar\alpha=a/b$, in lowest terms, be the parameter with $s(\bar\alpha)=\bar s$. With $\mu_6=33/16$, $\mu_7=77/32$, and $\mu_8=93/32$, the parameter $\bar\alpha$ is $44919/110000$ for $(3,6)$, $13797/27500$ for $(3,7)$, $33401/58125$ for $(3,8)$, and $33401/77500$ for $(4,8)$, and each is below $1-2^{1-D}$. Multiplying the inequality $\mathcal H_D(a/b)<D-k$ by $b$ and exponentiating shows that it is equivalent to
$$
  b^b\,\big(2^{D-1}-1\big)^a<2^{b(D-k)}\,(b-a)^{b-a}\,a^a ,
$$
an inequality between integers, which we verified for the four pairs in exact integer arithmetic. Now let $\varepsilon_m\to0$. By Lemma~\ref{lem:block_entropy}, $\liminf_m\mathcal H_D(\alpha_m)\ge D-k>\mathcal H_D(\bar\alpha)$. Since $\mathcal H_D$ is increasing on $[0,\bar\alpha]$, it follows that $\alpha_m>\bar\alpha$ for all large $m$, and hence $s(\alpha_m)<\bar s$. By part~2 of Theorem~\ref{thm:guarantee} with $h(x)=1-|x|/m$, which does not require positive mass on every coset, the expected satisfied fraction that DQI reaches, averaged over the offset, is at most $s(\alpha_m)+2\sqrt{\varepsilon_m}<\bar s+o(1)$.

\emph{Part~2.} By Lemma~\ref{lem:cert_reconstruction} and Table~\ref{tab:certified}, $\beta_d\ge\underline\beta$, where $\underline\beta$ is $1.8365$ for $(3,6)$, $1.6350$ for $(3,7)$, $1.4878$ for $(3,8)$, and $1.7510$ for $(4,8)$. For each pair, $p_{\underline\beta}(0)=1/(1+e^{-\underline\beta})$ exceeds the value in the third column of the table by more than $3\times10^{-5}$, so by continuity there is $\beta<\underline\beta$ at which $p_\beta(0)$ also exceeds this value. The inverse temperatures without reconstruction form an interval that starts at zero (Corollary~\ref{cor:binary}), so the primal model has non-reconstruction at $\beta$. By Proposition~\ref{prop:sharp}, with probability $1-o(1)$ over the instance, the two-stage decoder has block success at least $1-2m^{-3}$ at $\beta$, and by part~2 of Theorem~\ref{thm:guarantee}, the expected satisfied fraction is at least $p_\beta(0)-2\sqrt2\,m^{-3/2}$, which exceeds the value in the third column for large $m$.
\end{proof}

\paragraph{The optimal estimate of one symbol.} Let $\mathcal B$ be a finite set of blocks, each with the channel of Lemma~\ref{lem:block_channel} for the same $\theta$, and let $\mathcal C$ be a subgroup of $\mathcal P^{\mathcal B}$, such as the restriction of $C$ to the symbols of these blocks. Write $\nu^{\mathcal B}$ for the product law on $\mathcal Q^{\mathcal B}$ whose factors are $\nu$, and let $\mathcal K=\{y\in\mathcal Q^{\mathcal B}:c\cdot y=0\text{ for all }c\in\mathcal C\}$. For a symbol $i$ of one of the blocks, let $[e_i]\in\mathcal Q^{\mathcal B}$ be the pair of its unit vector in that block, extended by zero to the other blocks, so that $c\cdot[e_i]=c_i$ for $c\in\mathcal C$.

\begin{lemma}[Optimal estimate of one symbol]\label{lem:block_helstrom}
Suppose that $c_i$ is not identically zero on $\mathcal C$, let $c$ be uniformly random on $\mathcal C$, and give a decoder the state $\bigotimes_{S\in\mathcal B}\ket{\psi_{c_S}}$. For each coset $X$ of $\mathcal K+\{0,[e_i]\}$, let $X_0$ and $X_1$ be the two cosets of $\mathcal K$ that it contains, and let $w_X=\nu^{\mathcal B}(X)$. The largest probability with which a measurement recovers $c_i$ is
\begin{equation}\label{eq:block_helstrom}
  P=\frac12+\sum_X\sqrt{\nu^{\mathcal B}(X_0)\,\nu^{\mathcal B}(X_1)} .
\end{equation}
With $\mathfrak m_X=\big(\nu^{\mathcal B}(X_0)-\nu^{\mathcal B}(X_1)\big)/w_X$ for $w_X>0$,
\begin{equation}\label{eq:block_bias}
  P=\sum_Xw_X\,\frac{1+\sqrt{1-\mathfrak m_X^2}}2,\qquad\frac14\sum_Xw_X\mathfrak m_X^2\le1-P\le\frac12\sum_Xw_X\mathfrak m_X^2 .
\end{equation}
\end{lemma}

\begin{proof}
Index the computational basis states of the blocks by $\mathcal Q^{\mathcal B}$, as in Lemma~\ref{lem:block_channel}, so that the state is $\sum_y\sqrt{\nu^{\mathcal B}(y)}\,(-1)^{c\cdot y}\ket y$. Since $c\mapsto c_i$ is a nonzero homomorphism on $\mathcal C$, the bit $c_i$ is uniform. Its kernel $\{c\in\mathcal C:c_i=0\}$ has annihilator $\mathcal K+\{0,[e_i]\}$ in $\mathcal Q^{\mathcal B}$, so by character orthogonality, the average of $(-1)^{c\cdot z}$ over the $c\in\mathcal C$ with $c_i=u$ is $1$ for $z\in\mathcal K$, $(-1)^u$ for $z\in[e_i]+\mathcal K$, and $0$ otherwise. For a coset $X$ and $u\in\{0,1\}$, let $\ket{v_{X,u}}=\sum_{y\in X_u}\sqrt{\nu^{\mathcal B}(y)}\,\ket y$. The state conditioned on $c_i=u$ is therefore
$$
  \rho_u=\sum_X\big(\ket{v_{X,0}}+(-1)^u\ket{v_{X,1}}\big)\big(\bra{v_{X,0}}+(-1)^u\bra{v_{X,1}}\big).
$$
The vectors $\ket{v_{X,u}}$ are orthogonal, with squared norms $\nu^{\mathcal B}(X_u)$. The summand of $\rho_0-\rho_1$ for $X$ is $2\big(\ket{v_{X,0}}\!\bra{v_{X,1}}+\ket{v_{X,1}}\!\bra{v_{X,0}}\big)$, whose nonzero eigenvalues are $\pm2\sqrt{\nu^{\mathcal B}(X_0)\,\nu^{\mathcal B}(X_1)}$, and the Helstrom formula $P=\tfrac12+\tfrac14\|\rho_0-\rho_1\|_1$ proves~\eqref{eq:block_helstrom}. Since $\sum_Xw_X=1$ and $\sqrt{\nu^{\mathcal B}(X_0)\,\nu^{\mathcal B}(X_1)}=\tfrac12w_X\sqrt{1-\mathfrak m_X^2}$, the identity in~\eqref{eq:block_bias} follows, and the inequalities follow from $r/2\le1-\sqrt{1-r}\le r$ for $r\in[0,1]$.
\end{proof}

The number $\mathfrak m_X$ is a posterior bias. Let $E$ have law $\nu^{\mathcal B}$, let $Z$ be uniform on $\mathcal K$, and let $\xi$ be a uniformly random bit, all independent, and set $Y=E+Z+\xi[e_i]$. Given $Y$, the posterior probabilities of $\xi=0$ and $\xi=1$ are proportional to $\nu^{\mathcal B}(Y+\mathcal K)$ and $\nu^{\mathcal B}(Y+[e_i]+\mathcal K)$, so the absolute posterior bias of $\xi$ is $|\mathfrak m_X|$ for the coset $X$ that contains $Y$, which has probability $w_X$. Thus~\eqref{eq:block_bias} relates the optimal error to the mean squared posterior bias in this classical inference problem, as Corollary~\ref{cor:binary} does for product amplitude functions. The noise $E$ is independent across blocks but not across the symbols of a block, so this problem differs from the primal reconstruction problem of Section~\ref{sec:primal_recon}, and Theorem~\ref{thm:duality} does not apply to it.

\paragraph{Trees of blocks.} The optimal local estimate is determined by trees in which some checks are marked. The marked checks play the role of the checks of the first layer, and every symbol lies in exactly one of them. For $h\ge0$, let $\mathcal T^{\mathrm{block}}_h$ be the following tree. Its root is a marked check at level $0$, and one of its $D$ symbols is the root symbol. Every symbol of a marked check at a level $r<h$ lies in $k-1$ unmarked checks below it, except the symbol through which the marked check hangs from its parent, which lies in $k-2$. Every unmarked check has $D-1$ child symbols, and each of them lies in a marked check at level $r+1$ together with $D-1$ further symbols. The symbols of the marked checks at level $h$ lie in no unmarked checks. Let $\mathcal C^{\mathrm{block}}_h$ be the code of all assignments to the symbols that satisfy every check, and let $P^{\mathrm{block}}_h(\theta)$ be the largest probability with which a measurement recovers the root symbol from the states $\ket{\psi_{c_S}}$ of the marked checks $S$, when $c$ is uniformly random on $\mathcal C^{\mathrm{block}}_h$. Lemma~\ref{lem:block_helstrom} gives a formula for $P^{\mathrm{block}}_h(\theta)$.

\begin{lemma}[Completion to whole blocks]\label{lem:block_completion}
For a symbol $i$ and $t\ge0$, let $R_t(i)$ be the union of the blocks that meet $N_t(i)$. Then $N_t(i)\subseteq R_t(i)\subseteq N_{t+1}(i)$. If $\mathrm{LT}_{t+1}(i)$ and $\mathrm{TP}_{t+1}(i)$ hold, then, for the block amplitude function with $\theta\in[0,1]$, every $t$-local estimate at $i$ equals $c_i$ with probability at most $P^{\mathrm{block}}_t(\theta)$.
\end{lemma}

\begin{proof}
All symbols of a block lie in its check, so they are within distance two of each other, and every symbol lies in a block. Hence $N_t(i)\subseteq R_t(i)\subseteq N_{t+1}(i)$, and the blocks in $R_t(i)$ are the supports of the checks of the first layer at distance at most $2t+1$ from $i$. Suppose that $\mathrm{LT}_{t+1}(i)$ and $\mathrm{TP}_{t+1}(i)$ hold, and mark the checks of the first layer in the tree $N_{t+1}(i)$. Every symbol lies in exactly one marked check, so the root lies in one marked and $k-1$ unmarked checks, a symbol whose parent check is marked has $k-1$ unmarked child checks, and every other symbol has one marked and $k-2$ unmarked child checks. Let $F'$ be the set of checks of $F_{t+1}(i)$ whose symbols all lie in $R_t(i)$. These are the marked checks at distance at most $2t+1$ from $i$ and the unmarked checks at distance at most $2t-1$. Indeed, a child of an unmarked check at distance $2t+1$ is at distance $2t+2$, and by $\mathrm{LT}_{t+1}(i)$ its only check at distance at most $2t+1$ from $i$ is this unmarked check, so its block does not meet $N_t(i)$. A marked check at level $r$ is reached from $i$ through $r$ unmarked checks, so its symbols are at distance at least $2r$ from $i$, and every block in $R_t(i)$ has level at most $t$. Hence the symbols of $R_t(i)$ and the checks of $F'$ form the tree obtained from $\mathcal T^{\mathrm{block}}_t$ by deleting some unmarked checks together with all checks and symbols below them, with $i$ as the root symbol.

Let $y$ be an assignment on $R_t(i)$ that satisfies the checks of $F'$. The other checks of $F_{t+1}(i)$ are unmarked checks at distance $2t+1$, and their children are distinct leaves of the tree $N_{t+1}(i)$ outside $R_t(i)$. Choosing the values of these children so that these checks are satisfied extends $y$ to an assignment on $N_{t+1}(i)$ that satisfies every check of $F_{t+1}(i)$, which is the restriction of a codeword by $\mathrm{TP}_{t+1}(i)$. Hence $C|_{R_t(i)}$ is the code of the assignments that satisfy the checks of $F'$, and the restriction of a uniformly random codeword to $R_t(i)$ is uniform on it, because restriction is a group homomorphism. In the same way, every such assignment extends to a codeword of $\mathcal C^{\mathrm{block}}_t$, by choosing values in the deleted parts of the tree from the top down, so the restriction of a uniformly random codeword of $\mathcal C^{\mathrm{block}}_t$ has the same law. Finally, a $t$-local estimate at $i$ is the outcome of a POVM on the qudits of $N_t(i)\subseteq R_t(i)$, and given $c$, the qudits of $R_t(i)$ are in the state $\bigotimes_S\ket{\psi_{c_S}}$ over the blocks $S$ in $R_t(i)$. The success probability of the estimate therefore equals that of the same POVM, applied to the corresponding blocks, in the problem that defines $P^{\mathrm{block}}_t(\theta)$, and it is at most $P^{\mathrm{block}}_t(\theta)$.
\end{proof}

The conditions $\mathrm{LT}_{t+1}(i)$ and $\mathrm{TP}_{t+1}(i)$ concern the ball of radius $2t+2$, and they do not follow from $\mathrm{LT}_t(i)$ and $\mathrm{TP}_t(i)$. For $t\le c_0\log m$ and large $m$, Theorem~\ref{thm:tp_typical}, applied at depth $t+1$, shows that they hold with probability $1-o(1)$ at a uniformly random symbol of a random instance from Gallager's ensemble, and by the symmetry argument at the end of the proof of Corollary~\ref{cor:typical_nogo}, also at every fixed symbol. They need not hold at a symbol that a decoder selects after inspecting the instance.

\paragraph{A lossless recursion on trees of blocks.} On $\mathcal T^{\mathrm{block}}_h$, the formula~\eqref{eq:block_helstrom} can be evaluated without enumerating the code, by decoding from the leaves to the root as BPQM does on a tree code. We describe a message about a bit by the absolute value $v\in[0,1]$ of the overlap of its two pure states (Section~\ref{sec:bpqm}). We represent each incoming message by its spectrum on $\mathcal Q$. Convolving these spectra with the block spectrum combines the messages with the block state; the resulting message about symbol $j$ has a random absolute overlap, which we denote by $\Xi_j$. For $j\in\{1,\dots,D\}$ and $v\in[0,1]$, let $\iota_j(v)=\tfrac{1+v}2\,\delta_0+\tfrac{1-v}2\,\delta_{[e_j]}$, a probability vector on $\mathcal Q$, where $e_j$ is the $j$-th unit vector of $\Z_2^D$. For $v_1,\dots,v_D\in[0,1]$, let $\Lambda=\nu*\iota_1(v_1)*\cdots*\iota_D(v_D)$, and let $\Xi_j(v_1,\dots,v_D)$ be the random variable that, for each pair $\{y,y+[e_j]\}$ of elements of $\mathcal Q$ with $\Lambda(y)+\Lambda(y+[e_j])>0$, equals $|\Lambda(y)-\Lambda(y+[e_j])|/(\Lambda(y)+\Lambda(y+[e_j]))$ with probability $\Lambda(y)+\Lambda(y+[e_j])$.

In the recursion below, $W_h$ describes the absolute overlap sent from a non-root block to its parent, $U_h$ the overlap sent from an unmarked check, and $V_h$ the final overlap at the root.

\begin{proposition}[A lossless recursion on trees of blocks]\label{prop:block_tree}
Let $\theta\in[0,1]$, and let $W_0$ and $V_0$ have the law of $\Xi_1(1,\dots,1)$. For $h\ge0$, let $U_h$ be the result of combining $D-1$ independent copies of $W_h$ with $\widetilde\Gamma$, as in the definition of $\Psi_s$ in Appendix~\ref{app:certified}, write $U_h^{[r]}$ for the product of $r$ independent copies of $U_h$, with $U_h^{[0]}=1$, and let
\begin{align*}
  W_{h+1}&=\Xi_1\big(U_h^{[k-2]},U_h^{[k-1]},\dots,U_h^{[k-1]}\big),\\
  V_{h+1}&=\Xi_1\big(U_h^{[k-1]},U_h^{[k-1]},\dots,U_h^{[k-1]}\big),
\end{align*}
with independent arguments. Then $P^{\mathrm{block}}_h(\theta)=\E\,\tfrac12\big(1+\sqrt{1-V_h^2}\big)$. Moreover, $P^{\mathrm{block}}_h(\theta)$ is nondecreasing in $h$ and nonincreasing in $\theta$.
\end{proposition}

\begin{proof}
For a probability vector $\mu$ on $\mathcal Q$ and $c\in\mathcal P$, let $\ket{\psi^\mu_c}=\sum_{y\in\mathcal Q}\sqrt{\mu(y)}\,(-1)^{c\cdot y}\ket y$. Then $\langle\psi^\mu_c|\psi^\mu_{c'}\rangle=\hat\mu(c+c')$, where $\hat\mu(c)=\sum_y\mu(y)(-1)^{c\cdot y}$, and the Fourier transform of a convolution is the product of the Fourier transforms. Hence the tensor product of families with spectra $\mu_1,\dots,\mu_r$ has the same inner products as the family with spectrum $\mu_1*\cdots*\mu_r$, and an isometry that does not depend on $c$ maps the one to the other, as in Lemma~\ref{lem:nodes}(1). The states of a block are $\ket{\psi^\nu_c}$ by Lemma~\ref{lem:block_channel}. A message about $c_j=c\cdot[e_j]$ with overlap $v$, which is the pair $\ket{\psi_{c_j}(v)}$ of Section~\ref{sec:bpqm}, is the family $\ket{\psi^{\iota_j(v)}_c}$ after $\ket0$ and $\ket1$ are relabeled as $\ket0$ and $\ket{[e_j]}$. Replacing $v$ by $-v$ multiplies this family by $(-1)^{c_j}$, up to a unitary that does not depend on $c$, and this translates the spectrum of the tensor product by $[e_j]$. Translation permutes the pairs $\{y,y+[e_{j'}]\}$ and does not change the laws below, so we may use absolute overlaps. Two messages at the same coordinate combine to one whose overlap is the product of theirs, because $\iota_j(v)*\iota_j(v')=\iota_j(vv')$.

We decode from the leaves to the root, as in the proof of Proposition~\ref{prop:tree_optimal}. Under the uniform law on $\mathcal C^{\mathrm{block}}_h$, the root symbol is uniform, and conditional on the value of a symbol, the parts of the tree below its unmarked child checks are independent of each other and of the rest of the tree; an unmarked check fixes only the sum of its children, and a marked check only the parity of its symbols. We claim, by induction from level $h$ upwards, that conditional on the value $u$ of the symbol through which a marked check at level $r\ge1$ hangs, the state of its block and of the part of the tree below it is, up to an isometry that does not depend on $u$, of the form~\eqref{eq:flagged} for $q=2$, with a record whose law does not depend on $u$ and an absolute overlap with the law of $W_{h-r}$. For $r=h$, the block receives no messages, and the claim follows from the block step below with $\Lambda=\nu$.

\emph{Block step.} Conditional on the records below a block and on the restriction $c\in\mathcal P$ of the codeword to the block, the block and the messages from its unmarked child checks are in the tensor product of $\ket{\psi^\nu_c}$ and of the families $\ket{\psi^{\iota_j(v)}_c}$ of the messages. An isometry that does not depend on $c$ maps this tensor product to $\ket{\psi^\Lambda_c}$, with $\Lambda=\nu*\iota_1(v_1)*\cdots*\iota_D(v_D)$ and $v_j$ the product of the overlaps of the messages at coordinate $j$. Given the value $u$ of the coordinate $j'$ through which the block hangs, or of the root symbol for the root block, $c$ is uniform on $\{c\in\mathcal P:c_{j'}=u\}$, a coset of the subgroup $\{c\in\mathcal P:c_{j'}=0\}$, whose annihilator in $\mathcal Q$ is $\{0,[e_{j'}]\}$. By character orthogonality, the averaged state is the direct sum, over the pairs $\{y,y+[e_{j'}]\}$, of the rank-one operators of the vectors $\sqrt{\Lambda(y)}\ket y+(-1)^u\sqrt{\Lambda(y+[e_{j'}])}\ket{y+[e_{j'}]}$. Measuring the pair does not change this state for either value of $u$, the law of its outcome does not depend on $u$, and given the outcome, the state is a pair of pure states about $u$ whose absolute overlap has the law of $\Xi_{j'}(v_1,\dots,v_D)$.

\emph{Check step.} At an unmarked check, the check-node operation of BPQM (Section~\ref{sec:bpqm}) combines the messages from its $D-1$ children two at a time without loss, with labels whose laws do not depend on the parity of the children, and the absolute overlap of the result has the law of $\widetilde\Gamma$ applied to their absolute overlaps (Appendix~\ref{app:certified}). By the induction hypothesis, the message from an unmarked check below a marked check at level $r-1$ therefore has the law of $U_{h-r}$, and the block step gives the law of $W_{h-r+1}$, because the symbol through which the marked check hangs lies in $k-2$ unmarked child checks and its other symbols lie in $k-1$.

At the root block, every symbol lies in $k-1$ unmarked child checks, and the block step for the root symbol leaves a record and a pair of pure states about $c_i$ whose absolute overlap has the law of $V_h$. Every step is an isometry that does not depend on the codeword or the measurement of a record that does not change the state for either value of the bit to be recovered, so no step changes the optimal success probability. As in the proof of Proposition~\ref{prop:tree_optimal}, the optimum for a state that is block diagonal in a record is the average of the optima given the record, and the Helstrom measurement for a pair with absolute overlap $v$ succeeds with probability $\tfrac12(1+\sqrt{1-v^2})$. This proves the formula for $P^{\mathrm{block}}_h(\theta)$.

The tree $\mathcal T^{\mathrm{block}}_h$ is obtained from $\mathcal T^{\mathrm{block}}_{h+1}$ by deleting the unmarked checks below the marked checks at level $h$, together with everything below them, and every codeword of $\mathcal C^{\mathrm{block}}_h$ extends to the same number of codewords of $\mathcal C^{\mathrm{block}}_{h+1}$. The problem for $h$ is therefore obtained from that for $h+1$ by discarding blocks, so $P^{\mathrm{block}}_h(\theta)\le P^{\mathrm{block}}_{h+1}(\theta)$. For $0\le\theta<\theta'\le1$, the matrix with ones on the diagonal and $\theta/\theta'$ elsewhere is positive semidefinite, so it is the Gram matrix of unit vectors $\ket{g_c}$, $c\in\mathcal P$. By Lemma~\ref{lem:block_channel}, the vectors $\ket{\psi_c}\otimes\ket{g_c}$ for the overlap $\theta'$ have the same inner products as the vectors $\ket{\psi_c}$ for the overlap $\theta$, so an isometry that does not depend on $c$ maps the states at $\theta$ to those at $\theta'$, together with an extra register on each block. Discarding these registers shows that $P^{\mathrm{block}}_h(\theta)\ge P^{\mathrm{block}}_h(\theta')$.
\end{proof}

\paragraph{Numerical evaluation.} We evaluated the recursion of Proposition~\ref{prop:block_tree} by population dynamics (Appendix~\ref{app:numerics}) for the fifteen benchmark pairs. The laws of $W_h$ and $U_h$ were represented by populations, the convolution $\Lambda$ was computed on all $2^{D-1}$ elements of $\mathcal Q$, and the pair in $\Xi_1$ and the labels in $\widetilde\Gamma$ were sampled. At each level, we estimated $1-P^{\mathrm{block}}_h(\theta)$ by the mean, over the population, of the root error given the sampled overlaps $U_{h-1}$, with the average over the pairs at the root computed without sampling. For each pair, we took $\theta=1-(1-s_d)D/\mu_D$, at which $s(\alpha)$ equals our numerical estimate of $s_d$ in Table~\ref{tab:benchmark}; for example, $\theta\approx0.601$ for $(3,6)$.

For every pair, we used four configurations, with population size and number of levels $(4000,60)$, $(8000,200)$, $(8000,400)$, and $(12000,200)$, using an independent random seed for each configuration. Across these evaluations, the estimated root errors at the last level of each run lay between $0.169$ and $0.316$. For a fixed degree pair, the largest difference between the estimates from these configurations was below $0.01$. These observations do not provide confidence intervals or a bound on sampling error. Since $P^{\mathrm{block}}_h(\theta)$ is nondecreasing in $h$, the error at a finite depth bounds its limit only from above, so the evaluations do not show that the error stays positive as $h\to\infty$. They provide evidence for Conjecture~\ref{conj:block}, not a proof of its infinite-depth bound.

\section{Approximation for costs that are integer-valued up to an affine rescaling}\label{app:common_cost}

Subtracting $\min f$ changes neither Gibbs weights nor dependence on total cost, so in the proof assume $\min f=0$ and $\max f=L_f\ge1$. Let $\mathcal A$ be a finite alphabet of size $q$, let $H_f(x)=\sum_j f(x_j)$, and let $\mathcal S_m$ consist of vectors whose coefficient on $x$ depends only on $H_f(x)$. Write $\Pi_m$ for its orthogonal projection. The counting and approximation arguments use only this finite alphabet, not its group law. Define the compact Gibbs family
$$
\mathcal P_f=\{p_\beta:\beta\in[-\infty,+\infty]\},\qquad p_\beta(z)=\frac{e^{-\beta f(z)}}{\sum_u e^{-\beta f(u)}}.
$$
Its endpoint distributions are uniform on the maximum and minimum cost levels. Negative parameters occur only as auxiliary product components. For an attainable total cost $E$, let $r_E\in\mathcal P_f$ have mean $E/m$. It is unique, because the Gibbs mean has derivative $-\operatorname{Var}_{p_\beta}(f)<0$ at finite $\beta$, and it maximizes entropy at that mean.

\subsection{Counting words at an exact cost}\label{app:common_count}

\begin{lemma}[Exact-cost rounding]\label{lem:common_rounding}
If $E$ is attainable by a word of length $m$ and $r$ has mean cost $E/m$, there is an integer vector $w\ge0$ with $\sum_z w_z=m$, $\sum_z f(z)w_z=E$, and $\|w-mr\|_1\le C_f$, where $C_f=2qL_f$.
\end{lemma}
\begin{proof}
Let $A_f$ have columns $(1,f(z))^\top$, and choose a feasible integer $w$ minimizing $\|w-mr\|_1$. Put $u=w-mr\in\ker A_f$. Decompose $u$ into positive multiples of primitive integer vectors on minimally dependent columns, with every summand having the sign of $u$ at nonzero coordinates and vanishing at its zero coordinates. To obtain this sign-compatible decomposition, choose a nonzero kernel vector of smallest support in the orthant of $u$, subtract its largest multiple that stays in the orthant, and repeat. Minimality makes its support minimally dependent: a dependence on a proper subset permits a perturbation until a coordinate vanishes. Each subtraction removes a coordinate, so at most $q$ summands are required.

The minimal dependencies have two or three columns. Equal-cost columns have primitive coefficients $1,-1$. For distinct costs $e_i<e_j<e_s$, the coefficients are a common-divisor reduction of $(e_s-e_j,-(e_s-e_i),e_j-e_i)$ and have norm at most $2L_f$. Thus $u=\sum_{j=1}^J\lambda_j b_j$ with $J\le q$, $\lambda_j>0$, $\|b_j\|_1\le2L_f$, and each summand sign-compatible with $u$.

If $\lambda_j\ge1$, then $w-b_j$ is an integer feasible vector closer to $mr$. It is nonnegative: where $(b_j)_z>0$, sign compatibility gives $w_z-(b_j)_z\ge mr_z\ge0$; other coordinates are immediate. This contradicts minimality. Hence every $\lambda_j<1$, and $\|u\|_1\le2qL_f$. The case $u=0$ is immediate.
\end{proof}

\begin{lemma}[Number of exact-cost words]\label{lem:common_shell}
For every attainable $E$, the count $N_m(E)=|\{x:H_f(x)=E\}|$ satisfies
$$
N_m(E)\ge(e q)^{-C_f}(m+1)^{-(C_f+q)}e^{mH(r_E)}.
$$
\end{lemma}
\begin{proof}
Round $mr_E$ by Lemma~\ref{lem:common_rounding} and put $\tau=w/m$. The standard type bound gives at least $(m+1)^{-q}e^{mH(\tau)}$ words of type $\tau$~\cite{cover2006elements}. For completeness, under the multinomial law with probabilities $\tau$, its own integer count vector $w$ is a mode: the ratios at increased coordinates are products of $w_z/(w_z+j)\le1$, and those at decreased coordinates are products of $(w_z-j)/w_z\le1$. A positive count at a zero-probability coordinate has probability zero. Since there are at most $(m+1)^q$ count vectors, this proves the type bound.

Entropy continuity gives $m|H(\tau)-H(r_E)|\le C_f\log(e q m)$. One may couple the two laws with disagreement probability $u=\|\tau-r_E\|_1/2$ and use $|H(\tau)-H(r_E)|\le h_2(u)+u\log(q-1)\le s\log(e q/s)$ for $s=2u\le1$. Apply this with $s\le C_f/m$ when $m\ge C_f$. When $m<C_f$, the bound $m\log q\le C_f\log(e q m)$ suffices. All words of the rounded type have cost $E$, giving the claimed estimate.
\end{proof}

Write $d_H(p,r)^2=\sum_z(\sqrt{p(z)}-\sqrt{r(z)})^2$ and $p_\xi(z)=|\xi_z|^2$. The next estimate explains why cost symmetry constrains the product components.
\begin{lemma}[Product overlap with the cost subspace]\label{lem:common_overlap}
For a unit vector $\xi$, put $\delta(\xi)=\operatorname{dist}_{d_H}(p_\xi,\mathcal P_f)$. Then
$$
\|\Pi_m\xi^{\otimes m}\|^2\le(e q)^{C_f}(m+1)^{C_f+3q}(mL_f+1)e^{-m\delta(\xi)^2}.
$$
\end{lemma}
\begin{proof}
The normalized vectors $N_m(E)^{-1/2}\sum_{H_f(x)=E}\ket x$ form an orthonormal basis of $\mathcal S_m$. Expand their overlaps with $\xi^{\otimes m}$ by types. Taking absolute values and applying Cauchy--Schwarz over the at most $T=(m+1)^q$ types bounds the squared overlap for cost $E$ by
$$
\frac{T}{N_m(E)}\sum_{\tau:\langle f\rangle_\tau=E/m}\exp\left(m\left[2H(\tau)+\sum_z\tau_z\log p_\xi(z)\right]\right).
$$
Since $\log r_E$ is affine in $f$, the exponent after subtracting $mH(r_E)$ is $-m[D_{\mathrm{KL}}(\tau\Vert p_\xi)+D_{\mathrm{KL}}(\tau\Vert r_E)]$. At an endpoint, every feasible type is supported on that endpoint level, where $r_E$ is uniform, so the same identity holds. Zero entries either exclude a type or make its contribution zero.

For any probability vectors $p,r$, normalization of $(\sqrt{p_zr_z})_z$ gives
$$
\inf_\tau[D_{\mathrm{KL}}(\tau\Vert p)+D_{\mathrm{KL}}(\tau\Vert r)]=-2\log\sum_z\sqrt{p_zr_z}\ge d_H(p,r)^2.
$$
For disjoint supports the infimum is infinite; otherwise this is nonnegativity of relative entropy, followed by $-\log u\ge1-u$. Apply Lemma~\ref{lem:common_shell} and the bound $T$ on the remaining type sum. Each cost contributes at most $(e q)^{C_f}(m+1)^{C_f+3q}e^{-m\delta(\xi)^2}$. There are at most $mL_f+1$ costs.
\end{proof}

\subsection{A quantitative mixture of Gibbs product states}\label{app:common_mixture}

Let $d\xi$ be Haar probability on the unit sphere of $\mathbb C^q$, put $d_m=\binom{m+q-1}{q-1}$, and define
$$
d\mu_m(\xi)=d_m|\langle\xi^{\otimes m},\Phi\rangle|^2d\xi,\qquad \sigma_v=\int(\ket\xi\!\bra\xi)^{\otimes v}d\mu_m(\xi).
$$
The symmetric-projector identity $\int(\ket\xi\!\bra\xi)^{\otimes m}d\xi=P_{\mathrm{sym},m}/d_m$ makes $\mu_m$ a probability measure.

\begin{lemma}[Coherent-state marginal approximation]\label{lem:common_husimi}
For the reduced state $\rho_v$ of $\ket\Phi$ on $v\le m$ factors,
$$
\frac12\|\rho_v-\sigma_v\|_1\le1-a_{m,v}\le\frac{v(q+v-1)}m,\qquad a_{m,v}=\prod_{j=0}^{v-1}\frac{m-j}{m+q+j}.
$$
\end{lemma}
\begin{proof}
Put $\rho_0=1$ and use falling and rising factorials. The Haar moment formula gives the positive expansion
\begin{equation}\label{eq:common_positive_moment}
\sigma_v=\frac{1}{(m+q)^{\overline v}}\sum_{j=0}^v\binom vj^2j!\,(m)_{\underline{v-j}}P_{\mathrm{sym},v}(\rho_{v-j}\otimes I^{\otimes j})P_{\mathrm{sym},v}.
\end{equation}
To count the contractions, pair the $m+v$ ket coordinates with the conjugate coordinates in the Haar moment. If $j$ retained coordinates pair directly with retained coordinates, choosing and pairing them gives $\binom vj^2j!$. The other $s=v-j$ coordinates pair with distinct input coordinates in $(m)_{\underline s}^2$ ways, and the remaining inputs pair in $(m-s)!$ ways. Their product is $m!(m)_{\underline s}$. Symmetry makes the contracted input state its marginal $\rho_s$; direct retained pairings contribute identities. Averaging their placements gives the two symmetric projections. Finally, $d_m=q(q+1)\cdots(q+m-1)/m!$ cancels the input factor and the first $m$ terms of the Haar denominator, proving~\eqref{eq:common_positive_moment}.

The $j=0$ term is $a_{m,v}\rho_v$, and all other terms are positive. Since $\operatorname{tr}\sigma_v=1$, their normalized sum is a state, so the trace distance is at most $1-a_{m,v}$. Apply $1-\prod_j(1-b_j)\le\sum_jb_j$ with $b_j=(q+2j)/(m+q+j)$ to obtain the second bound.
\end{proof}

Define the polynomial factor
$$
M_m=(L_f+1)(e q)^{C_f}(m+1)^{C_f+4q}.
$$
Because $\Phi\in\mathcal S_m$, Lemma~\ref{lem:common_overlap} bounds its overlap with a product by the product's projection norm. Using $d_m\le(m+1)^{q-1}$ therefore gives
$$
\mu_m\{\xi:\operatorname{dist}_{d_H}(p_\xi,\mathcal P_f)\ge\delta\}\le M_me^{-m\delta^2}.
$$
For every $\xi$, choose a nearest Gibbs vector $r(\xi)\in\mathcal P_f$ and set $\eta_z=(\xi_z/|\xi_z|)\sqrt{r(\xi)_z}$, with phase one at zero coordinates. The choice is measurable by parametrizing the compact Gibbs curve by $\arctan\beta$ and taking its smallest distance minimizer. If $b=\langle\xi,\eta\rangle=\sum_z\sqrt{p_\xi(z)r(\xi)_z}$, then
$$
\frac12\|(\ket\xi\!\bra\xi)^{\otimes v}-(\ket\eta\!\bra\eta)^{\otimes v}\|_1=\sqrt{1-b^{2v}}\le\sqrt v\,d_H(p_\xi,r(\xi)).
$$
Use $\delta_m^2=(\log M_m+2\log(m+1))/m$. The exceptional set has mass at most $(m+1)^{-2}$, and the complementary set has product-state distance at most $\sqrt v\delta_m$. Convexity and Lemma~\ref{lem:common_husimi} prove that one product measure, after this replacement, approximates all marginals on at most $v$ factors with error
\begin{equation}\label{eq:common_marginal_error}
\gamma_m(v)=\frac{v(q+v-1)}m+\sqrt{\frac{v(\log M_m+2\log(m+1))}m}+\frac1{(m+1)^2}.
\end{equation}
Every product factor has Gibbs magnitudes. Since $q,L_f$ are fixed, $\gamma_m(v_n)\to0$ when $v_n^2/m_n\to0$.

\subsection{A uniform error under reconstruction}\label{app:common_gap}

\begin{lemma}[Uniform tree error]\label{lem:common_gap}
Suppose $k\ge3$ and $D\ge2$. Set $c_{\mathrm{tree}}=(q-1)(D-1)2^{D-2}$ and
\begin{equation}\label{eq:common_tree_gap}
\kappa_{q,k,D}=\frac{c_{\mathrm{tree}}^{-2k/(k-2)}}{4(q-1)^{2+2/(k-1)}}.
\end{equation}
For every spectrum $a$ with $P_\infty(a)<1$, $P_t(a)\le1-\kappa_{q,k,D}$ for every $t$. If $\beta_d<\infty$, this applies to every finite $\beta>\beta_d$ and to $p_{+\infty}$.
\end{lemma}
\begin{proof}
The check-node estimate in Lemma~\ref{lem:checkbound} gives $\E d_F(C^{(t)})\le c_{\mathrm{tree}}v_t$, where $v_t=\E d_F(V^{(t)})$. Fourier multiplication gives $v_{t+1}\le(c_{\mathrm{tree}}v_t)^{k-1}$. If any $v_t<c_{\mathrm{tree}}^{-(k-1)/(k-2)}$, rescaling this inequality makes its iterates tend to zero. The root distance is at most the corresponding cavity distance, so $P_t\to1$. Therefore, when $P_\infty<1$, every $v_t$ is at least this constant.

For nontrivial characters put $a_s=|\hat a(s)|$ and $b_s=\E|\hat C^{(t-1)}(s)|$. With $h=k-1$,
$$
v_t=\sum_{s\ne0}a_sb_s^h,\qquad R_t=\E d_F(N^{(t)})=\sum_{s\ne0}a_sb_s^k\ge\frac{v_t^{k/h}}{(q-1)^{1/h}}.
$$
The last inequality is the weighted power-mean inequality, using $\sum_sa_s\le q-1$; a vanishing weight sum would give $v_t=0$. Also $d_F(b)\le(q-1)d_1(b)\le2(q-1)\sqrt{1-S(b)}$. Taking expectations and applying Cauchy--Schwarz gives $1-P_t\ge R_t^2/[4(q-1)^2]\ge\kappa_{q,k,D}$ for $t\ge1$. Monotonicity covers $t=0$.

Every finite $\beta>\beta_d$ has reconstruction. At fixed depth the ensemble of tree states and its optimal success depend continuously on the spectrum, including boundary spectra, since the state and measurement spaces are finite-dimensional and the measurement set is compact. Letting $\beta\to+\infty$ extends the same bound to $p_{+\infty}$.
\end{proof}

\subsection{The expected-cost bound}\label{app:common_cost_bound}

\begin{proof}[Proof of Theorem~\ref{thm:common_cost_local}]
Use the Gibbs product measure obtained above at the largest admissible neighborhood size $v$, and write $\gamma=\gamma_m(v)$. Permutation invariance and partial trace give the same approximation, with the same measure, on every neighborhood selected by an admissible seed. Diagonal phases of each factor commute with the codeword phase encoding and do not depend on the codeword. Thus its best local success is the tree success $P_{t_r}(p_\beta)$.

Let $w$ be the measure of factors with Gibbs parameter in $(\beta_d,+\infty]$. Lemma~\ref{lem:common_gap} bounds their success by $1-\kappa_{q,k,D}$. Splitting block recovery according to the seed and the correctness of the initial estimate gives
$$
1-\varepsilon\le1-g_{\mathrm{init}}+g_{\mathrm{init}}(1-\kappa_{q,k,D}w+\gamma)+a_{\mathrm{rec}}.
$$
Consequently $w\le[(\varepsilon+a_{\mathrm{rec}})/g_{\mathrm{init}}+\gamma]/\kappa_{q,k,D}$.

The Gibbs mean decreases on the full extended real line. Every remaining component therefore has mean at least $e_d$, including negative parameters, and every component has cost at least zero under our normalization. Their average cost is at least $e_d-L_fw$. The one-factor trace-distance error loses at most $L_f\gamma$. Theorem~\ref{thm:guarantee}, applied to $H_f/(mL_f)$, loses at most a further $2L_f\sqrt\varepsilon$. Restoring $\min f$ gives~\eqref{eq:common_cost_local}. The limit follows from~\eqref{eq:common_marginal_error} and the stated seed-error hypothesis.
\end{proof}

\subsection{Finite-field decoding, projection and attainment}\label{app:common_field}

The character, geometry and attainment arguments in this subsection allow any fixed real cost. Integer-valued costs are required only for the final application of Theorem~\ref{thm:common_cost_local}. We keep the field group law distinct from addition modulo $q$.

\begin{proof}[Proof of Corollaries~\ref{cor:field_sharp} and~\ref{cor:common_cost_field}]
\emph{Characters and tree decoding.} For $q=p^r$, use the characters
$$
  \chi_y(z)=\exp\!\left(\frac{2\pi i}{p}\operatorname{Tr}_{\mathbb F_q/\mathbb F_p}(yz)\right).
$$
Nondegeneracy of the trace pairing makes these all additive characters of the field, with $\sum_y\chi_y(z)=q\mathbf1[z=0]$. Characters multiply under addition, Fourier transforms turn convolution into products, and Fourier magnitudes are translation-invariant.

For $K=\operatorname{im}B$ and $C=\ker B^\top$, the same pairing gives $C=K^\perp$. Encode $c$ by $Z^c\ket\Phi=\sum_x\Phi(x)\chi_{c\cdot x}(1)\ket x$. The node proofs of Lemma~\ref{lem:nodes} use only the preceding character identities: symbol nodes convolve spectra, and a check state's blocks are indexed by the difference of the child values. Within a block the spectrum is the same normalized product. The final Fourier measurement has success $q^{-1}(\sum_z\sqrt{a(z)})^2$, where $a$ is the outgoing spectrum. These are the additive-group BPQM operations~\cite{mandal2026abelian}. The tree induction and posterior-translation proof of Theorem~\ref{thm:duality} therefore apply over the field, as do the local optimum and the reconstruction equivalence.

\emph{Two-valued costs and degradation.} Write $f(z)=c+\Delta\mathbf1[z\notin A]$, with $\Delta>0$ and $A$ a nonempty proper subset of the field. For $a=|A|$, the Gibbs law is $p_\beta=\theta_\beta U_A+(1-\theta_\beta)\unif$, where
$$
  \theta_\beta=\frac{a(1-e^{-\beta\Delta})}{a+(q-a)e^{-\beta\Delta}}.
$$
This coefficient increases from zero to one. Uniform noise is absorbing under convolution on the additive field group, so for $0\le\beta'<\beta$,
$$
  p_{\beta'}=p_\beta*\left[\frac{\theta_{\beta'}}{\theta_\beta}\delta_0+
  \left(1-\frac{\theta_{\beta'}}{\theta_\beta}\right)\unif\right].
$$
The bracket is a probability law. The node-isometry proof of Proposition~\ref{prop:degrade} therefore gives $P_t(\beta')\ge P_t(\beta)$ at every tree depth. Together with the field tree correspondence, this proves the interval assertion without requiring $A$ to be a subgroup. The Potts case is $A=\{0\}$ and $\Delta=1$.

\emph{Projection and distance.} Choose an $\mathbb F_p$ basis of $\mathbb F_q$. Because the matrix entries are zero or one, its linear constraints act separately on each basis coordinate. This identifies the field code with $r$ copies of the prime-field code, and does the same for its restriction and for each local tree code. Thus TP over $\mathbb F_q$ is equivalent to TP over $\mathbb F_p$; LT depends only on the graph. Theorem~\ref{thm:locality} and the typicality conclusions of Corollary~\ref{cor:typical_nogo} transfer to the field model.

A nonzero field codeword has a nonzero prime-field component, whose support is contained in its support. Its weight is therefore at least the prime-field minimum distance. Lemma~\ref{lem:distance} gives a lower bound $\gamma_0m$ with high probability in both ensembles. This basis argument factors the constraints, not the distribution of a field-valued symbol: an arbitrary cost $f$ may couple all its basis coordinates.

\emph{Tree-error decay and sequential decoding.} Define $d_F$ by summing Fourier magnitudes over the nontrivial trace characters. The inequalities $d_1\le d_F\le(q-1)d_1$ and the bijection $(\delta,z)\mapsto(z+\delta,z)$ used in Lemma~\ref{lem:checkbound} hold on the additive field group. Consequently the expected Fourier distance obeys the same recursion~\eqref{eq:decay_recursion}. Non-reconstruction implies its convergence to zero, and $k-1\ge2$ then gives $1-P_{t_m}\le m^{-8}$ as in Lemma~\ref{lem:decay}.

The sequential projection argument in Appendix~\ref{app:sequential} applies to arbitrary received states and is unchanged. It bounds the probability of any wrong retained estimate by $2m\sqrt{m^{-8}}=2m^{-3}$, for every input-independent processing order. The unmeasured set has size $o(m)$ and the field code has linear distance, so the retained values have a unique codeword completion. Gaussian elimination over the fixed field finds it in polynomial time. Disjoint-first schedules and random schedules have the same locality proofs, because those arguments use only graph neighborhoods, independent local work registers and retained records.

\emph{Finite precision.} At fixed finite $\beta$, every entry of $p_\beta$ is positive. All node probabilities and denominators therefore have the same positive lower bounds used in Appendix~\ref{app:sequential}. The field Fourier transform has fixed dimension, and field arithmetic has constant cost for fixed $q$. Using $b(m)=\lceil\log_2^2(m+1)\rceil$ bits makes the accumulated error $o(1)$ by the same Lipschitz and circuit-size estimates. Computability of $p_\beta$ to polynomial-time precision gives a polynomial-time implementation without changing any input neighborhood or retained record.

\emph{Output and approximation.} Character orthogonality gives
$$
  \frac1{|C|}\sum_{c\in C}\chi_{c\cdot(x-x_0)}(1)=\mathbf1[x\in x_0+K].
$$
This identity and orthogonality over a uniformly random field offset are the only group facts in Theorem~\ref{thm:guarantee}. Its vector comparison, quadratic-form bound and fixed fallback therefore prove both output conclusions in the field model, consistent with~\cite{shutty2026locallyquantumdecoders}. For product amplitudes every coset has positive mass. The reliable field decoder thus gives expected total-variation error $o(1)$ and expected cost $\langle f\rangle_{p_\beta}+o(1)$, completing Corollary~\ref{cor:field_sharp}.

For integer-valued costs, the counting and coherent-state arguments are independent of the alphabet's group law. The tree-error gap of Lemma~\ref{lem:common_gap} uses only the same check and Fourier identities, so the initial-estimate averaging proof of Theorem~\ref{thm:common_cost_local} applies. Finally, typical field LT/TP neighborhoods have size $v=O(m^{1/2-\delta})$ for every prescribed logarithmic-radius sequence, so $v^2/m\to0$. This proves Corollary~\ref{cor:common_cost_field}.
\end{proof}

\section*{Notation}\label{app:notation}

The following table lists the main notation in order of first use.
\begin{center}
\small
\begin{tabular}{@{}p{0.24\textwidth}p{0.7\textwidth}@{}}
\toprule
Symbol & Meaning \\
\midrule
$m$, $n$ & number of constraints (code symbols) and of variables (checks)\\
$B$, $(k,D)$ & $m\times n$ matrix of the instance, with $k$ ones in each row and $D$ ones in each column\\
$q$, $\omega$ & alphabet size, and $\omega=e^{2\pi i/q}$\\
$z$, $x_0$, $x$ & assignment in $\Z_q^n$, offset, and violation pattern $x=Bz+x_0$\\
$f$, $H$, $\beta$ & cost of one constraint, objective, and inverse temperature\\
$p=p_\beta$ & single-symbol weights $p(z)\propto e^{-\beta f(z)}$\\
$\mu_\beta$, $\pi_\beta$ & primal Gibbs measure on assignments, and its image on violation patterns\\
$K$, $C$ & $K=\{Bz:z\in\Z_q^n\}$, and the code $C=K^\perp$ that DQI decodes\\
$\Phi$, $\ket\Phi$ & amplitude function, and the state $\sum_x\Phi(x)\ket x$\\
$\Phi_\beta$ & product amplitude function $\prod_j\sqrt{p_\beta(x_j)}$\\
$\ket{\psi^\Phi_c}$ & state that the decoder receives when the codeword is $c$\\
$\sigma_\beta$ & overlap $\tanh(\beta/2)$ of the two single-symbol states for $q=2$\\
$\alpha$, $\alpha^\perp$ & violation probability $p(1)$, and error rate of the decoding problem, for $q=2$\\
$P_t$ & largest probability of recovering the root of the depth-$t$ tree code\\
$\beta_d$, $\rho_d$, $s_d$ & reconstruction threshold, and the error rate and satisfied fraction at $\beta_d$\\
$\beta_{\mathrm H}$, $\rho_H$, $s_H$ & largest inverse temperature allowed by the Holevo bound, and the corresponding error rate and satisfied fraction\\
$\beta_c$, $\rho_c$, $s_c$ & condensation threshold (Section~\ref{sec:condensation}), and the corresponding error rate and satisfied fraction\\
$W_{B,\alpha}$, $F_B(\alpha)$ & law of the coset $E+K$ when $E$ has independent Bernoulli$(\alpha)$ entries, and its relative entropy with respect to the uniform law $U_B$ on cosets\\
$\bar F$, $\mathcal B_\sigma$ & limit of $\frac1n\E_BF_B$ on the configuration model, and the functional~\eqref{eq:functional} whose supremum is $\bar F$\\
$P_{\mathrm{PGM}}$ & block success of the pretty-good measurement (Lemma~\ref{lem:pgm_success})\\
$N_t(i)$ & symbols within distance $2t$ of $i$ in the factor graph\\
$\mathrm{LT}_t(i)$, $\mathrm{TP}_t(i)$ & local tree condition and tree projection condition\\
$c_0$ & constant in the radius bound $t\le c_0\log m$, with $0<c_0<1/(2\log((k-1)(D-1)))$\\
$\bar {\mathfrak m}_t$ & mean posterior bias of the root, for $q=2$\\
$\lambda_\beta$, $T_Q$ & probability of an optimum under $\mu_\beta$, and the cost $\lambda_\beta^{-1/2}$ of quantum minimum finding\\
$\lambda_\Phi$, $T_\Phi$ & probability of an optimum in the target state of $\Phi$, and preparation count $\lambda_\Phi^{-1/2}$\\
$\beta_\rho$ & inverse temperature whose product amplitude function has decoding error rate $\rho$ (Section~\ref{sec:poly_exact})\\
$\beta_{\mathrm P}$, $\beta_U$ & $2\operatorname{atanh}(k/D)$, at which the expected number of retained clauses of Prange's algorithm equals the number of variables, and the certified inverse temperature of Section~\ref{sec:prange}\\
$\sigma^\ast(k,D)$ & zero of $\sigma-\frac k D\big(1-(1-\sigma)^D\big)$ in $(0,1)$; retained systems with $\sigma_\beta>\sigma^\ast$ are rank deficient with high probability (Lemma~\ref{lem:rank})\\
$h^\star$ & fraction of constraints violated by an optimal assignment\\
\bottomrule
\end{tabular}
\end{center}

\bibliographystyle{alpha}
\bibliography{refs}

@article{jordan2024optimization,
  title={Optimization by decoded quantum interferometry},
  author={Jordan, Stephen P. and Shutty, Noah and Wootters, Mary and Zalcman, Adam and Schmidhuber, Alexander and King, Robbie and Isakov, Sergei V. and Khattar, Tanuj and Babbush, Ryan},
  journal={Nature},
  volume={646},
  number={8086},
  pages={831--836},
  year={2025},
  doi={10.1038/s41586-025-09527-5}
}

@misc{JNZ26,
    author = {Stephen P.~Jordan and Quynh T.~Nguyen and Leo Zhou},
    title = {{Regev's Reduction Reaches Reconstruction}},
    note = {To appear},
    year = 2026}

@article{zhou2026gibbs,
  title={Gibbs Sampling in the Shattered Phase by Decoded Quantum Interferometry},
  author={Zhou, Leo and Shutty, Noah and Sellke, Mark and Jordan, Stephen P.},
  journal={arXiv preprint arXiv:2609.40345},
  year={2026}
}

@article{piveteau2026gibbs,
  title={Gibbs State Preparation through Quantum Decoding},
  author={Piveteau, Christophe and Quek, Yihui and Franco Garrido, M. Isabel},
  journal={arXiv preprint arXiv:2610.03566},
  year={2026}
}

@article{piveteau2022messagepassing,
  title={Quantum Message-Passing Algorithm for Optimal and Efficient Decoding},
  author={Piveteau, Christophe and Renes, Joseph M.},
  journal={Quantum},
  volume={6},
  pages={784},
  year={2022},
  doi={10.22331/q-2022-08-23-784}
}

@inproceedings{brandsen2022bpqm,
  title={Belief Propagation with Quantum Messages for Symmetric Classical--Quantum Channels},
  author={Brandsen, Sarah and Mandal, Avijit and Pfister, Henry D.},
  booktitle={2022 IEEE Information Theory Workshop (ITW)},
  pages={494--499},
  year={2022},
  doi={10.1109/ITW54588.2022.9965841}
}

@article{mandal2026blockerror,
  title={Quantum Message Passing Convergence and Vanishing Block-Error Probability for Random {LDPC} Codes},
  author={Mandal, Avijit and Piveteau, Christophe and Renes, Joseph M. and Pfister, Henry D.},
  journal={arXiv preprint arXiv:2609.05857},
  year={2026}
}

@article{renes2018duality,
  title={Duality of Channels and Codes},
  author={Renes, Joseph M.},
  journal={IEEE Transactions on Information Theory},
  volume={64},
  number={1},
  pages={577--592},
  year={2018},
  doi={10.1109/TIT.2017.2754921}
}

@article{krzakala2009hiding,
  title={Hiding Quiet Solutions in Random Constraint Satisfaction Problems},
  author={Krzakala, Florent and Zdeborov{\'a}, Lenka},
  journal={Physical Review Letters},
  volume={102},
  number={23},
  pages={238701},
  year={2009},
  doi={10.1103/PhysRevLett.102.238701}
}

@article{krzakala2011melting,
  title={On Melting Dynamics and the Glass Transition. {II}. {G}lassy Dynamics as a Melting Process},
  author={Krzakala, Florent and Zdeborov{\'a}, Lenka},
  journal={The Journal of Chemical Physics},
  volume={134},
  number={3},
  pages={034513},
  year={2011},
  doi={10.1063/1.3506843}
}

@article{mezard2006reconstruction,
  title={Reconstruction on Trees and Spin Glass Transition},
  author={M{\'e}zard, Marc and Montanari, Andrea},
  journal={Journal of Statistical Physics},
  volume={124},
  number={6},
  pages={1317--1350},
  year={2006}
}

@article{eldar2001square,
  title={On Quantum Detection and the Square-Root Measurement},
  author={Eldar, Yonina C. and Forney, Jr., G. David},
  journal={IEEE Transactions on Information Theory},
  volume={47},
  number={3},
  pages={858--872},
  year={2001},
  doi={10.1109/18.915636}
}

@book{richardson2008modern,
  title={Modern Coding Theory},
  author={Richardson, Tom and Urbanke, R{\"u}diger},
  publisher={Cambridge University Press},
  year={2008}
}

@article{rizzo2022stochastic,
  title={Stochastic Equations and Dynamics Beyond Mean-Field Theory},
  author={Rizzo, Tommaso},
  journal={arXiv preprint arXiv:2209.09689},
  year={2022}
}

@inproceedings{chailloux2024quantumdecoding,
  title={The Quantum Decoding Problem},
  author={Chailloux, Andr{\'e} and Tillich, Jean-Pierre},
  booktitle={19th Conference on the Theory of Quantum Computation, Communication and Cryptography (TQC 2024)},
  series={LIPIcs},
  volume={310},
  pages={6:1--6:14},
  year={2024},
  doi={10.4230/LIPIcs.TQC.2024.6},
  note={Full version: arXiv:2310.20651}
}

@article{shutty2026locallyquantumdecoders,
  title={Optimization Using Locally-Quantum Decoders},
  author={Shutty, Noah and Mandal, Avijit and Ragavan, Seyoon and Buzet, Quentin and Chailloux, Andr{\'e} and Rubin, Nicholas C. and Khan, Abid and Boulebnane, Sami and Shaydulin, Ruslan and Azariah, John and Jordan, Stephen P.},
  journal={arXiv preprint arXiv:2604.24633},
  year={2026}
}

@article{zalka1999grover,
  title={Grover's quantum searching algorithm is optimal},
  author={Zalka, Christof},
  journal={Physical Review A},
  volume={60},
  number={4},
  pages={2746},
  year={1999}
}

@article{eldan2020spectral,
  title={A Spectral Condition for Spectral Gap: Fast Mixing in High-Temperature {I}sing Models},
  author={Eldan, Ronen and Koehler, Frederic and Zeitouni, Ofer},
  journal={Probability Theory and Related Fields},
  volume={182},
  number={3-4},
  pages={1035--1051},
  year={2022},
  doi={10.1007/s00440-021-01085-x}
}

@inproceedings{anari2024trickle,
  title={Trickle-Down in Localization Schemes and Applications},
  author={Anari, Nima and Koehler, Frederic and Vuong, Thuy-Duong},
  booktitle={Proceedings of the 56th Annual ACM Symposium on Theory of Computing (STOC)},
  pages={1094--1105},
  year={2024},
  doi={10.1145/3618260.3649622}
}

@inproceedings{deshpande2019threshold,
  title={The threshold for {SDP}-refutation of random regular {NAE}-3{SAT}},
  author={Deshpande, Yash and Montanari, Andrea and O'Donnell, Ryan and Schramm, Tselil and Sen, Subhabrata},
  booktitle={Proceedings of the Thirtieth Annual ACM-SIAM Symposium on Discrete Algorithms (SODA)},
  pages={2305--2321},
  year={2019}
}

@inproceedings{liu2024sparse,
  title={Fast Mixing in Sparse Random {I}sing Models},
  author={Liu, Kuikui and Mohanty, Sidhanth and Rajaraman, Amit and Wu, David X.},
  booktitle={2024 IEEE 65th Annual Symposium on Foundations of Computer Science (FOCS)},
  pages={120--128},
  year={2024},
  doi={10.1109/FOCS61266.2024.00018}
}

@article{efthymiou2026unbounded,
  title={On sampling diluted Spin-Glasses with unbounded interactions},
  author={Efthymiou, Charilaos and Zampetakis, Kostas},
  journal={arXiv preprint arXiv:2603.22432},
  year={2026}
}

@article{renes2017bpqm,
  title={Belief propagation decoding of quantum channels by passing quantum messages},
  author={Renes, Joseph M.},
  journal={New Journal of Physics},
  volume={19},
  number={7},
  pages={072001},
  year={2017},
  doi={10.1088/1367-2630/aa7c78}
}

@article{rengaswamy2021bpqm,
  title={Belief propagation with quantum messages for quantum-enhanced classical communications},
  author={Rengaswamy, Narayanan and Seshadreesan, Kaushik P. and Guha, Saikat and Pfister, Henry D.},
  journal={npj Quantum Information},
  volume={7},
  pages={97},
  year={2021},
  doi={10.1038/s41534-021-00422-1}
}

@article{garrido2026regev,
  title={Regev's reduction as a candidate quantum algorithm for the discrete logarithm problem in finite abelian groups},
  author={Franco Garrido, M. Isabel and Chailloux, Andr{\'e}},
  journal={arXiv preprint arXiv:2605.03972v2},
  year={2026}
}

@article{schmidhuber2025quartic,
  title={Quartic quantum speedups for planted inference},
  author={Schmidhuber, Alexander and O'Donnell, Ryan and Kothari, Robin and Babbush, Ryan},
  journal={Physical Review X},
  volume={15},
  number={2},
  pages={021077},
  year={2025},
  doi={10.1103/PhysRevX.15.021077}
}

@inproceedings{raghavendra2017strongly,
  title={Strongly refuting random {CSPs} below the spectral threshold},
  author={Raghavendra, Prasad and Rao, Satish and Schramm, Tselil},
  booktitle={Proceedings of the 49th Annual ACM SIGACT Symposium on Theory of Computing (STOC 2017)},
  pages={121--131},
  publisher={ACM},
  year={2017},
  doi={10.1145/3055399.3055417}
}

@inproceedings{kothari2017sum,
  title={Sum of squares lower bounds for refuting any {CSP}},
  author={Kothari, Pravesh K. and Mori, Ryuhei and O'Donnell, Ryan and Witmer, David},
  booktitle={Proceedings of the 49th Annual ACM SIGACT Symposium on Theory of Computing (STOC 2017)},
  pages={132--145},
  publisher={ACM},
  year={2017},
  doi={10.1145/3055399.3055485}
}

@article{feldman2018complexity,
  title={On the complexity of random satisfiability problems with planted solutions},
  author={Feldman, Vitaly and Perkins, Will and Vempala, Santosh},
  journal={SIAM Journal on Computing},
  volume={47},
  number={4},
  pages={1294--1338},
  year={2018},
  doi={10.1137/16M1078471}
}

@article{alekhnovich2011more,
  title={More on average case vs approximation complexity},
  author={Alekhnovich, Michael},
  journal={Computational Complexity},
  volume={20},
  number={4},
  pages={755--786},
  year={2011},
  doi={10.1007/s00037-011-0029-x}
}

@inproceedings{applebaum2010public,
  title={Public-key cryptography from different assumptions},
  author={Applebaum, Benny and Barak, Boaz and Wigderson, Avi},
  booktitle={Proceedings of the 42nd ACM Symposium on Theory of Computing (STOC 2010)},
  pages={171--180},
  publisher={ACM},
  year={2010},
  doi={10.1145/1806689.1806715}
}

@article{boyer1998tight,
  title={Tight bounds on quantum searching},
  author={Boyer, Michel and Brassard, Gilles and H{\o}yer, Peter and Tapp, Alain},
  journal={Fortschritte der Physik},
  volume={46},
  number={4--5},
  pages={493--505},
  year={1998}
}

@article{gallager1962ldpc,
  title={Low-Density Parity-Check Codes},
  author={Gallager, Robert G.},
  journal={IRE Transactions on Information Theory},
  volume={8},
  number={1},
  pages={21--28},
  year={1962},
  doi={10.1109/TIT.1962.1057683}
}

@book{cover2006elements,
  title={Elements of Information Theory},
  author={Cover, Thomas M. and Thomas, Joy A.},
  edition={2},
  publisher={Wiley},
  year={2006}
}

@article{anschuetz2025decoded,
  title={Spin Glass Transitions Obstruct Decoded Quantum Interferometry},
  author={Anschuetz, Eric R. and Gamarnik, David and Lu, Jonathan Z.},
  journal={arXiv preprint arXiv:2509.14509v2},
  year={2026}
}

@article{buzet2025fine,
  title={Fine-Grained Unambiguous Measurements},
  author={Buzet, Quentin and Chailloux, Andr{\'e}},
  journal={arXiv preprint arXiv:2510.07298},
  year={2025}
}

@article{durr1996minimum,
  title={A quantum algorithm for finding the minimum},
  author={D{\"u}rr, Christoph and H{\o}yer, Peter},
  journal={arXiv preprint quant-ph/9607014},
  year={1996}
}

@article{christandl2007oneandahalf,
  title={One-and-a-half quantum de {F}inetti theorems},
  author={Christandl, Matthias and K{\"o}nig, Robert and Mitchison, Graeme and Renner, Renato},
  journal={Communications in Mathematical Physics},
  volume={273},
  number={2},
  pages={473--498},
  year={2007},
  doi={10.1007/s00220-007-0189-3}
}

@inproceedings{mandal2026qary,
  title={Belief Propagation with Quantum Messages for Symmetric $q$-ary Pure-State Channels},
  author={Mandal, Avijit and Pfister, Henry D.},
  booktitle={2026 IEEE International Symposium on Information Theory (ISIT)},
  pages={1--6},
  year={2026},
  doi={10.1109/ISIT62367.2026.11653951}
}

@article{mandal2026abelian,
  title={Quantum Message Passing for Factor Graphs over Finite Abelian Groups},
  author={Mandal, Avijit and Pfister, Henry D.},
  journal={arXiv preprint arXiv:2604.12186},
  year={2026}
}

@article{regev2009lattices,
  title={On lattices, learning with errors, random linear codes, and cryptography},
  author={Regev, Oded},
  journal={Journal of the ACM},
  volume={56},
  number={6},
  pages={34:1--34:40},
  year={2009},
  doi={10.1145/1568318.1568324}
}

@article{eldar2003inconclusive,
  title={Mixed-quantum-state detection with inconclusive results},
  author={Eldar, Yonina C.},
  journal={Physical Review A},
  volume={67},
  number={4},
  pages={042309},
  year={2003},
  doi={10.1103/PhysRevA.67.042309}
}

@article{kramer2026provable,
  title={A provable quantum advantage for approximate optimization via decoded quantum interferometry},
  author={Kramer, Maximilian J. and Gil-Fuster, Elies and Jones, Benjamin D. M. and Eisert, Jens and Schreiber, Franz J.},
  journal={arXiv preprint arXiv:2610.02145},
  year={2026}
}

@article{gilfuster2026approximate,
  title={Approximate sampling from decoded quantum interferometry via {M}arkov chain {M}onte {C}arlo methods},
  author={Gil-Fuster, Elies and Ninio, Matan and Bittel, Lennart and Shimoni, Yishai and Eisert, Jens and Woerner, Stefan and Carrera V{\'a}zquez, Almudena},
  journal={arXiv preprint arXiv:2607.28120},
  year={2026}
}

@article{montanari2006rigorous,
  title={Rigorous Inequalities Between Length and Time Scales in Glassy Systems},
  author={Montanari, Andrea and Semerjian, Guilhem},
  journal={Journal of Statistical Physics},
  volume={125},
  number={1},
  pages={23--54},
  year={2006},
  doi={10.1007/s10955-006-9175-y}
}

@article{mohanty2022explicit,
  title={Explicit Near-{R}amanujan Graphs of Every Degree},
  author={Mohanty, Sidhanth and O'Donnell, Ryan and Paredes, Pedro},
  journal={SIAM Journal on Computing},
  volume={51},
  number={3},
  pages={STOC20-1--STOC20-23},
  year={2022},
  doi={10.1137/20M1342112}
}

@article{cojaoghlan2018information,
  title={Information-Theoretic Thresholds from the Cavity Method},
  author={Coja-Oghlan, Amin and Krzakala, Florent and Perkins, Will and Zdeborov{\'a}, Lenka},
  journal={Advances in Mathematics},
  volume={333},
  pages={694--795},
  year={2018},
  doi={10.1016/j.aim.2018.05.029}
}

@article{krzakala2007gibbs,
  title={Gibbs States and the Set of Solutions of Random Constraint Satisfaction Problems},
  author={Krzakala, Florent and Montanari, Andrea and Ricci-Tersenghi, Federico and Semerjian, Guilhem and Zdeborov{\'a}, Lenka},
  journal={Proceedings of the National Academy of Sciences},
  volume={104},
  number={25},
  pages={10318--10323},
  year={2007},
  doi={10.1073/pnas.0703685104}
}

@incollection{mcdiarmid1998concentration,
  title={Concentration},
  author={McDiarmid, Colin},
  booktitle={Probabilistic Methods for Algorithmic Discrete Mathematics},
  publisher={Springer},
  address={Berlin, Heidelberg},
  pages={195--248},
  year={1998},
  doi={10.1007/978-3-662-12788-9_6}
}

@incollection{wormald1999models,
  title={Models of Random Regular Graphs},
  author={Wormald, Nicholas C.},
  booktitle={Surveys in Combinatorics, 1999},
  publisher={Cambridge University Press},
  pages={239--298},
  year={1999},
  doi={10.1017/CBO9780511721335.010}
}

@article{vandenbrand2017mutual,
  title={The Mutual Information of {LDGM} Codes},
  author={van den Brand, Jan and Jaafari, Nor},
  journal={arXiv preprint arXiv:1707.04413},
  year={2017}
}

@inproceedings{rengaswamy2021duality,
  title={On the Duality Between the {BSC} and Quantum {PSC}},
  author={Rengaswamy, Narayanan and Pfister, Henry D.},
  booktitle={2021 IEEE International Symposium on Information Theory (ISIT)},
  pages={2232--2237},
  year={2021},
  doi={10.1109/ISIT45174.2021.9518034},
  note={Full version: arXiv:2103.09225}
}

@article{piveteau2025efficient,
  title={Efficient and Optimal Quantum State Discrimination via Quantum Belief Propagation},
  author={Piveteau, Christophe and Renes, Joseph M.},
  journal={arXiv preprint arXiv:2509.19441},
  year={2025}
}

@inproceedings{cojaoghlan2021inference,
  title={Inference and Mutual Information on Random Factor Graphs},
  author={Coja-Oghlan, Amin and Hahn-Klimroth, Max and Loick, Philipp and M{\"u}ller, Noela and Panagiotou, Konstantinos and Pasch, Matija},
  booktitle={38th International Symposium on Theoretical Aspects of Computer Science (STACS 2021)},
  series={LIPIcs},
  volume={187},
  pages={24:1--24:15},
  year={2021},
  doi={10.4230/LIPIcs.STACS.2021.24},
  note={Full version: arXiv:2007.07494}
}

@article{franz2002dynamic,
  title={Dynamic Phase Transition for Decoding Algorithms},
  author={Franz, Silvio and Leone, Michele and Montanari, Andrea and Ricci-Tersenghi, Federico},
  journal={Physical Review E},
  volume={66},
  number={4},
  pages={046120},
  year={2002},
  doi={10.1103/PhysRevE.66.046120}
}

@article{blanvillain2026tight,
  author = {Blanvillain, Agathe and Chailloux, Andr{\'e} and Tillich, Jean-Pierre},
  title = {The Quantum Decoding Problem: Tight Achievability Bounds and Application to {Regev's} Reduction},
  journal = {IEEE Transactions on Information Theory},
  volume = {72},
  number = {7},
  pages = {4980--4999},
  year = {2026},
  doi = {10.1109/TIT.2026.3689737},
  eprint = {2509.24796},
  archivePrefix = {arXiv}
}

@inproceedings{chailloux2026equivalence,
  author = {Chailloux, Andr{\'e} and Hermouet, Paul},
  title = {On the Quantum Equivalence between {$S|\mathrm{LWE}\rangle$} and {$\mathrm{ISIS}$}},
  booktitle = {Advances in Cryptology -- CRYPTO 2026},
  pages = {391--428},
  year = {2026},
  doi = {10.1007/978-3-032-35377-1_13},
  eprint = {2510.06097},
  archivePrefix = {arXiv}
}

@article{como2008spectra,
  author = {Como, Giacomo and Fagnani, Fabio},
  title = {Average Spectra and Minimum Distances of Low-Density Parity-Check Codes over {Abelian} Groups},
  journal = {SIAM Journal on Discrete Mathematics},
  volume = {23},
  number = {1},
  pages = {19--53},
  year = {2008},
  doi = {10.1137/070686615}
}

@article{yang2011weights,
  author = {Yang, Shengtian and Honold, Thomas and Chen, Yan and Zhang, Zhaoyang and Qiu, Peiliang},
  title = {Weight Distributions of Regular Low-Density Parity-Check Codes over Finite Fields},
  journal = {IEEE Transactions on Information Theory},
  volume = {57},
  number = {11},
  pages = {7507--7521},
  year = {2011},
  doi = {10.1109/TIT.2011.2162642},
  eprint = {1010.2030},
  archivePrefix = {arXiv}
}

@article{montanari2004cooling,
  title={Cooling-Schedule Dependence of the Dynamics of Mean-Field Glasses},
  author={Montanari, Andrea and Ricci-Tersenghi, Federico},
  journal={Physical Review B},
  volume={70},
  number={13},
  pages={134406},
  year={2004},
  doi={10.1103/PhysRevB.70.134406}
}

@article{krzakala2010following,
  title={Following {G}ibbs States Adiabatically: The Energy Landscape of Mean-Field Glassy Systems},
  author={Krzakala, Florent and Zdeborov{\'a}, Lenka},
  journal={EPL (Europhysics Letters)},
  volume={90},
  number={6},
  pages={66002},
  year={2010},
  doi={10.1209/0295-5075/90/66002}
}
\end{document}